\documentclass{elsarticle}

\usepackage{arydshln}
\usepackage{xspace}
\newcommand{\etal}{\textit{et al.\xspace}}
\usepackage{graphicx}
\usepackage{fullpage}
\usepackage[algo2e,vlined]{algorithm2e} 
\usepackage{url}
\usepackage{booktabs}
\usepackage{microtype}
\usepackage{relsize}
\usepackage{dsfont}
\usepackage{float}
\restylefloat{table}

\usepackage{color}

\newcommand{\comment}[1]{}
\title{Computational Analysis of Perfect-Information Position Auctions}
\author{David R.M. Thompson \qquad Kevin Leyton-Brown\\\smaller Computer Science Department, University of British Columbia\\\smaller\texttt{\{daveth; kevinlb\}@cs.ubc.ca}}
\date{}

\begin{document}
\begin{abstract}
After experimentation with other designs, the major search engines converged on the weighted, generalized second-price auction (wGSP) for selling keyword advertisements. Notably, this convergence occurred before position auctions were well understood (or, indeed, widely studied) theoretically. While much progress has been made since, theoretical analysis is still not able to settle the question of why search engines found wGSP preferable to other position auctions. We approach this question in a new way, adopting a new analytical paradigm we dub ``computational mechanism analysis." By sampling position auction games from a given distribution, encoding them in a computationally efficient representation language, computing their Nash equilibria, and then calculating economic quantities of interest, we can quantitatively answer questions that theoretical methods have not. We considered seven widely studied valuation models from the literature and three position auction variants (generalized first price, unweighted generalized second price, and wGSP). We found that wGSP consistently showed the best ads of any position auction, measured both by social welfare and by relevance (expected number of
clicks). Even in models where wGSP was already known to have bad worse-case efficiency, we found that it almost always performed well on average. In contrast, we found that revenue was extremely variable across auction mechanisms, and was highly sensitive to equilibrium selection, the preference model, and the valuation distribution.
\end{abstract}
\maketitle

\newcommand{\paperonly}[1]{#1}
\newcommand{\chapteronly}[1]{}

\newcommand{\whiskerBP}{The ``whiskers'' in this plot indicate equilibrium selection effects: the bar is the median equilibrium, while the top and bottom whiskers correspond to best- and worst-case equilibria.} 

\section{Introduction}

\chapteronly{This chapter covers the first application domain for my computational mechanism analysis techniques, position auctions.}
Position auctions are a relatively new family of mechanisms in which bidders place a single bid for a set of goods of varying quality, and the $i$th-highest bidder wins the $i$th-most-desirable good. Each year, these auctions yield billions of dollars selling advertising space on search-engine results pages. Various position auction designs have been considered over the years; e.g.,
advertisers who attract more clicks may be given advantages over weaker bidders, and bidders may have to pay their own bid amounts or a smaller amount computed from the bids of others.
After some initial experimentation with these design dimensions, the major search engines have converged on a single design: the weighted, generalized second-price auction (which we denote wGSP; we define it formally in what follows). The main question that this \paperonly{paper}\chapteronly{chapter} seeks to address is whether wGSP represents a good choice, as compared both to the auctions it has replaced and to theoretical benchmarks. Specifically, we ask whether wGSP is more economically efficient, whether it generates more revenue, whether it yields results that users find more relevant, and whether it produces low-envy allocations.

There is an enormous literature on auction analysis that seeks to answer such questions. Overwhelmingly, this literature proceeds by modeling a setting as a (Bayesian; perfect-information) game and then using theoretical analysis to describe what occurs in (Bayes--Nash; dominant-strategy; locally envy-free) equilibrium. There is much to like about this approach: it is often capable of determining that a given mechanism optimizes an objective function of interest or proving that no mechanism can satisfy a given set of properties. Indeed, most of what we know about mechanism design was established through such analysis. However, the approach also has limitations: in order to obtain clean theoretical results it can be necessary to make strong assumptions about bidder preferences and to simplify tie-breaking and bid discretization rules. Even when considering such a simplified version of a given problem, it is often extremely difficult to make quantitative comparisons between non-optimal mechanisms (e.g., which non-revenue-optimizing mechanism yields higher revenue on expectation)? The current state of affairs in the literature is thus that we know a great deal about auctions, but that many open questions appear to be resistant to analysis by known techniques.

In the case of position auctions, most research has used perfect-information Nash equilibrium as the solution concept of choice.  This choice is justified by the fact that advertisers interact repeatedly: nearly identical goods---user views of a particular search result page---are sold up to millions of times per day, and advertisers can continuously adjust their bids and observe the effects.
A variety of other technical assumptions are commonly made, characterizing advertiser preferences (e.g., a click has the same value regardless of position, and regardless of which other ads are shown), user behavior (e.g., a user's response to an ad is independent of which other ads she has seen), and advertiser behavior (e.g., an advertiser will act to reduce his envy even when doing so does not increase his utility).  With all these assumptions in place, various strong results have been obtained (e.g., wGSP is efficient and generates weakly more revenue than VCG), but other important questions remain open (e.g., does wGSP generate more revenue than other position auctions?).  Relaxing any one of these assumptions can lead to many further open questions.

\paperonly{This paper introduces \emph{computational} techniques for the analysis of mechanisms and shows that these techniques}
\chapteronly{This chapter shows that my computational mechanism analysis techniques }%
are able to address a wide variety of open questions that have not proven amenable to theoretical analysis. We maintain the approach of modeling a mechanism as a game and of reasoning about (exact) solution concepts of interest; we depart from traditional analysis by allowing only a discrete set of bids and by answering questions by providing statistical evidence rather than theorems. Specifically, we sample advertiser preferences from a given distribution, 
compute an exact Nash equilibrium of the resulting game between advertisers, and reason about this equilibrium to compute properties of the outcome, such as expected revenue or social welfare.  By repeatedly sampling, we can make quantitative statistical claims about our position auction setting (e.g., that one auction design generates significantly more expected revenue than another).
\paperonly{While our approach is not specific to position auctions, we focus on that domain here to demonstrate that our} %
\chapteronly{The results of this chapter demonstrate that my}%
methods are able to yield qualitatively new findings about a widely studied setting.

Computational mechanism analysis differs in many ways from theoretical methods.  As already mentioned, one clear disadvantage is that our approach only produces statistical results (e.g., given distribution $D$, $A$ performs significantly better than $B$ on expectation) rather than theorems (e.g., $A$ always performs better than $B$).  Conversely, our methods have the advantage that they are able to produce results in settings for which such simple patterns do not exist. For example, we have observed distributions over advertiser preferences under which the wGSP auction sometimes generates far less revenue than its predecessor, uGSP, and sometimes generates far more.  Thus, we know that any comparison of these auctions must necessarily be statistical and distribution dependent, rather than guaranteeing that one of these auctions always yields more revenue.  Our computational approach also allows us to consider arbitrary preference distributions, possibly derived from real-world data, as opposed to being restricted to distributions with convenient theoretical properties like monotone hazard rates. A further property of computational mechanism analysis is both a benefit and a weakness: bidders must be restricted to a finite set of discrete bids, unlike the vast majority of literature on auction theory, which assumes that bids are continuous.  While we depart from this tradition, we do not see discreteness as necessarily disadvantageous; real-world position auctions tend to be rather coarsely discrete.  For example, position auctions often clear for tens of cents per click, while bids are required to be placed in integer numbers of cents.  Finally, some auction features, like rules for tie-breaking and rounding, are difficult to analyze in the continuous case, but pose no obstacle to our approach.
\enlargethispage{.4em}

\paperonly{Given all of these advantages, the reader might wonder why the computational analysis of mechanisms is not already commonplace. The answer is that to date, the scale of the problem has made computational equilibrium analysis infeasible for virtually all games of interesting size. Normal-form games grow exponentially in the number of players, making all but the simplest problems too big even to store on the most powerful of modern computers.  Even when representation is not a problem, finding a sample Nash equilibrium of a general-sum game is computationally hard, with every known algorithm requiring time that grows exponentially in the size of the normal form.  These problems compound, meaning that overall, equilibrium computation requires time doubly exponential in the number of players!  Luckily, recent advances in algorithmic game theory (surveyed in Section~\ref{sec:background}) offer a way around this roadblock: games can be encoded in exponentially less space and equilibria can be identified exponentially more quickly when players' payoffs are suitably structured. In Section~\ref{sec:represent} we describe algorithms for encoding many position auction variants into such efficient representations, guaranteeing exponential improvements in both representation size and runtime.  Previously, it was infeasible to compute exact Nash equilibria of position auction games even using a supercomputer.  Using our approach, exact Nash equilibria can often be found on a desktop PC in less than a second.}

The bulk of this \paperonly{paper}\chapteronly{chapter} shows the effectiveness of computational mechanism analysis by demonstrating what it can tell us about position auctions. In Section~\ref{sec:results} we consider both simple models (such as those of Varian and Edelman \etal) and richer models (such as cascade) in which advertisers have position-dependent valuations and externalities, in each case using our techniques to shed light on open problems.\footnote{We previously presented results for the no-externalities models in a conference paper \cite{TLb09position}. The representation for models with externalities is new to the current paper, as are all experimental results.}  Some high-level findings emerged from our analysis. Most strikingly, we found that wGSP consistently showed the best ads of any position auction, measured both by social welfare and by relevance (expected number of clicks).  Further, even in models where wGSP was already known to have bad worse-case efficiency (either in terms of price of stability or price of anarchy), we found that it almost always had very good average-case performance.  In contrast, we found that revenue was extremely variable across auction mechanisms, and furthermore was highly sensitive to equilibrium selection, the preference model, and the valuation distribution. In Section~\ref{sec:sensitivity} we consider the extent to which our findings are sensitive to the bid discretization used, the number of bidders, and the number of slots sold.

\section{Background}\label{sec:background}

Although a variety of position auction variants have been proposed, only three have seen large-scale use in practice. We describe them here and also provide short form names that we will use throughout the  \paperonly{paper}\chapteronly{chapter}. All auctions are pay-per-click; that is, bidders pay every time an end-user clicks on an advertisement, not every time an advertisement is displayed.
\begin{description}
\item[GFP] The generalized first-price auction, used by Overture and by Yahoo!\ from 1997--2002. Each bidder submits a single bid; the highest bidder's advertisement is displayed in the highest position, the second bidder gets the second-highest position, and so on. Each bidder pays the amount of his bid.
\item[uGSP] The unweighted, generalized second-price auction, used by Yahoo!\ from 2002--2007. As before, bidders are ranked by their bids; the bidder winning the $i$th position pays the $i+1$st-highest bid.
\item[wGSP] The weighted, generalized second-price auction, used by Google AdWords and Microsoft adCenter, and by Yahoo!\ since 2007. The search engine assigns each bidder a weight or ``quality score''; we model this as the probability that an end-user would click on each bidder's advertisement if it were shown in the highest position. Bidders are scored by the products of their weights and their bids, and are ranked in descending order of their scores. Each bidder pays the smallest bid amount that would have been sufficient to cause him to maintain his position.
\end{description}
\noindent These auctions have all received theoretical analysis under a variety of models, typically with the assumption that bidders will converge, in repeated play, to an equilibrium of the full-information, one-shot game.

\subsection{Metrics for evaluating auction outcomes}

It is tempting to believe that search engines have settled on wGSP because it is a better auction design. To claim this, we need to decide what we mean by ``better.'' In this  \paperonly{paper}\chapteronly{chapter}, we will consider four such metrics.
\begin{enumerate}
	\item Most straightforwardly, perhaps search engines gravitated to an auction design that maximizes their (short-term) interests: \textit{revenue}.
	\item Perhaps search engines are better off maximizing the \textit{welfare} of advertisers, to ensure advertisers' ongoing participation, and thus the search-engine's longer-term revenue. If so, we should assess a mechanism according to the sum of advertisers' valuations for the allocation achieved in equilibrium.
	\item Both of these measures neglect a third group of agents: the search engine's end-users, without whom there would be no profits for advertisers and thus no revenues for the search engine. Some researchers believe that our second metric, social welfare among advertisers, is a good proxy for end-user payoffs, because advertisers only get revenue from clicks that satisfy users' needs \cite{AE2011consumer}.  Others have argued that click-through rates are a more direct measure of whether advertisements are interesting to users, measuring the \textit{relevance} of a page of search ads by the expected number of clicks it receives \cite{LP07squash}.
	\item Finally, \textit{envy} is another important measure of the quality of a multi-good allocation. One agent $i$ envies another agent $j$ if $i$'s expected utility could be increased by exchanging $i$'s and $j$'s allocations and payments \cite{HY11envy}.  Allocations that do not give rise to such envy have been considered desirable in themselves. In the position auction literature, however, it is more common for envy to be used as a tool for equilibrium selection.  Many researchers have restricted their attention to envy-free Nash equilibria (i.e., Nash equilibria in which the total envy across all bidders is zero) \cite{EOS07internet,V07position}.\footnote{Although these researchers focused on locally envy-free equilibria---under which no bidder envies the bidders in adjacent positions---this distinction is not important for our purposes.  Under their models, local envy-freeness implies global envy-freeness, using the more general definition of envy from \cite{HY11envy}.  We need this richer definition of envy---which allows for randomization---to study cases involving randomized tie-breaking or mixed strategies.  Even so, this richer definition does not cover externalities---where an advertiser's utility depends not just on his own allocation, but also on which agents receive which other goods.  Thus, we do not consider envy when working with settings involving externalities.}  In this  \paperonly{paper}\chapteronly{chapter}, we also use envy as a tool for equilibrium selection, contrasting envy-free and general Nash equilibria.
\end{enumerate}


\subsection{Models of Bidder Valuations}

A wide range of different models have been proposed for bidder valuations in position auctions.  There are broadly two classes of models: those that make a no-externalities assumption---holding an ad's position constant, it will generate the same expected number of clicks and same expected value regardless of which ads are shown in other positions---and those that do not make such an assumption.

\subsubsection{Models without externalities}

We consider four no-externalities models.  The first two models (which we call EOS and V, after the researchers who introduced them) have in common the assumption that each advertiser values clicks independently of their advertisement's position.  The next two models (which we call BHN and BSS) allow for ``position preferences'': an advertiser might have different values for clicks when the advertisement appears in different positions. In each case, we describe important results from the literature, as well as open questions that we will address.

\newlength{\modelspace}
\setlength{\modelspace}{.5em}
\newcommand{\model}[1]{\vspace{\modelspace}\noindent\textbf{#1 Model.}~}
\newcommand{\question}[1]{\vspace{\modelspace}\noindent\textbf{Question #1:}~}
\model{EOS} Edelman, Ostrovsky and Schwarz \cite{EOS07internet} analyzed the GSP under a preference model in which each bidder's expected value per click is independent of position. The click-through rate is the same for all ads in a given position (making uGSP equivalent to wGSP), and decreasing in position (ads that appear lower on the screen get fewer clicks).  EOS defined locally envy-free equilibria as Nash equilibria in which no bidder envies the allocation received by a bidder in a neighboring position, and showed that in such equilibria, uGSP is efficient and revenue dominates the truthful equilibrium of VCG.  Caragiannis et al \cite{CKKKLPLT13efficiency} showed that other, lower-efficiency Nash equilibria exist, but that none is worse in terms of social welfare by a factor greater than 1.259.  (In other words, 1.259 is an upper bound on the \emph{price of anarchy}.) Given that locally envy-free equilibria are only guaranteed to exist in the continuous case, while real wGSP uses discrete increments, some natural open questions about this model follow.

\newcommand{\qi}{\question{1} Under EOS preferences, how often does wGSP give rise to envy-free (efficient, VCG-revenue-dominating) Nash equilibria?  What happens in other equilibria, and how often do they occur?}
\qi

\model{V} Varian \cite{V07position} analyzed wGSP under a more general model, in which each bidder's value per click is still independent of position, but click-through rates are decreasing and ``separable.'' Separability means that for any position/bidder pair, the click-through rate can be factored into a position-specific component that is independent of bidder identity and a bidder-specific component that is independent of position (corresponding to wGSP's weights).
Varian showed that in any ``symmetric'' (globally envy-free) equilibrium, wGSP is efficient and revenue dominates VCG.  The price of anarchy result of Caragiannis et al.\ also applies to the V model.  Lahaie and Pennock studied the problem of what happens in this model in uGSP and wGSP (and points in between) \cite{LP07squash}.  Under additional assumptions about the valuations, they found that uGSP was less efficient than wGSP but generated more revenue.  These findings suggest some natural open questions about the V model.

\newcommand{\qii}{\question{2} Under V preferences, how often does wGSP have envy-free (efficient, VCG-revenue-dominating) Nash equilibria?  What happens in other equilibria, and how often do they occur?  Is uGSP generally better than wGSP for revenue, or does their relative performance depend on equilibrium selection, the valuation distribution, and/or other properties of the game? Is wGSP generally better for efficiency?}
\qii

\model{BHN} Blumrosen, Hartline and Nong \cite{BHN08nonuniform} proposed a model that (like the one that follows) allows for the possibility that not all clicks are equally valuable to an advertiser. In this model, click-through rates are still decreasing and separable.  However, a bidder's expected value per click increases with the bidder's rank in a separable fashion, subject to the constraint that a bidder's expected value \emph{per impression} is weakly decreasing.  The authors support their generalization by describing empirical evidence that conversions (e.g., sales) occur for a higher proportion of clicks on lower-ranked ads.  They show that preference profiles exist under their valuation model in which wGSP has no efficient, pure-strategy Nash equilibrium.  However, while we know that such preference profiles exist, we do not know how much of a problem they pose on average.

\newcommand{\qiii}{\question{3} Under BHN preferences, how often does wGSP have no efficient Nash equilibrium?  How much social welfare is lost in such equilibria?}
\qiii

\model{BSS} Benisch, Sadeh and Sandholm \cite{BSS08inexpressive} proposed another position-preference model that generalizes EOS. In this model, click-through rates are decreasing in position but independent of bidder identities.  However, bidders' values are single peaked in position and strictly decreasing from that peak. For example, ``brand'' bidders might prefer the prestige of top positions, while ``value'' bidders prefer positions further down.  Benisch, Sadeh and Sandholm analyzed this model in an imperfect-information setting, and showed both that uGSP and wGSP ranking rules can be arbitrarily inefficient for such models and that more expressive bidding languages can improve efficiency.  For different valuation distributions consistent with their model, they bounded the loss of efficiency in the best-case Bayes-Nash equilibrium.  We observe that this bound derives entirely from GSP's inexpressiveness---the fact that agents lack the ability to communicate their true preferences---rather than from agents' incentives.  Under the more common assumption of perfect-information Nash equilibria, expressiveness cannot cause an inefficient outcome (because for any order of the bidders, strategy profiles exist that will rank the bidders in that order), but incentives can (some bidder might want to deviate from any efficient strategy profile).

\newcommand{\qiv}{\question{4} Under BSS preferences, how often does wGSP have no efficient perfect-information Nash equilibrium?  How much social welfare is lost in such equilibria?}
\qiv

\subsubsection{Models with Externalities}

So far we have assumed that advertisers do not care about \emph{which} ads appear above and below their own. We now consider three models that relax this assumption: cascade and two further models that generalize cascade (hybrid, GIM).

\model{Cascade} The most widely studied model of position auction preferences with externalities is the ``cascade model '' \cite{GM08cascade,KM08cascade,AFMP08markovian}.
It captures the idea that users scan and click ads in the order they appear; a good ad can make lower ones less desirable, while a bad ad can cause a user to give up entirely. More specifically, users scan the ads starting from the top, and each time the user looks at (and possibly clicks on) an ad, she may subsequently decide to stop scanning.  In this model, it is possible for wGSP to have low-efficiency equilibria, with price of anarchy 4 \cite{RT12externalities}.  If weights are modified to take into account the probability that a user will continue reading ads, wGSP has a revenue-optimal equilibrium, but this equilibrium may require bidders to bid above their own valuations, which is a weakly dominated strategy.\footnote{This reweighted wGSP mechanism is only optimal among mechanisms that do not use reserve prices.  Also, because some agents' strategies may be dominated---bidding strictly more than their valuations---the equilibria tend to not be rationalizable.}

\newcommand{\qv}{\question{5}Under cascade preferences, how often are there low-efficiency equilibria?  In cases where low-efficiency equilibria exist, are these the only equilibria?  When agents play only undominated strategies, how much revenue can wGSP generate?  How do the modified weights affect wGSP's efficiency?}
\qv

\model{Hybrid} A richer model, which we call ``hybrid'', combines features of the separable (V) and cascade models.  Users decide whether or not to continue scanning ads based on the number of ads they have already scanned (as in V), and the content of each of those ads (as in cascade) \cite{KM08cascade}.  Under this model, auctions that allocate according to greedy heuristics---including GSP---are not economically efficient; indeed,  no economically efficient, polynomial-time allocation algorithm is known \cite{KM08cascade}.  Further, strategic behavior in GSP can lead to greater efficiency loss.  When $k$ slots are being sold, the worst Nash equilibrium can be as little as $1/k$-efficient, while the best can be as little as $2/k$-efficient \cite{GK08cascade}.

\newcommand{\qvi}{\question{6} With hybrid preferences, how much social welfare is lost?  On typical instances (as opposed to worst-case ones), how much difference is there between best- and worst-case equilibria?}
\qvi

\model{GIM} Beyond their work on cascade, Gomes, Immorlica and Markakis \cite{GIM09externalities} also introduced an even richer model (which we call GIM) which generalizes the hybrid model in two important ways.  First, it allows for arbitrary pairwise externalities.  For example, after looking at---and possibly clicking on---the first ad, a user might update her beliefs about which ads are promising, affecting some ads positively and others negatively.  Second, this model allows for the possibility that a user's behavior might depend on which sets of ads she has seen and clicked on before.  Its authors argue that this richer model should have similar properties to the cascade model.\footnote{This is clearly true for the issue of whether or not clicking on an ad affects the click-through rates of subsequent ads: before the user first clicks, the auction must already have allocated all the ad space.  Nevertheless, GIM's increased range of possible externalities could lead to dramatic differences from cascade in terms of equilibrium outcomes.}

\newcommand{\qvii}{\question{7} Are the efficiency and revenue achieved by wGSP substantially different under the cascade and GIM models?}
\qvii


\paperonly{
\subsection{Action-Graph Games}

Since position auctions are typically analyzed as perfect-information games, if we discretize bid amounts (as search engines indeed do) then regardless of the preference model, every position auction corresponds to a normal-form game. This suggests the possibility of computational analysis: we could explicitly construct games and then identify Nash equilibria using standard computational tools such as Gambit \cite{Gambit}. The catch, of course, is that the normal-form representation of a realistic ad auction problem is unmanageably large. For example, the normal-form representation of a relatively small game with 10 agents and 10 bid amounts per agent consists of 100 billion values, too many to store in RAM on most modern computers---and even on some of these computers' hard drives. Furthermore, as described earlier, existing equilibrium-finding algorithms all require time exponential in the size of their inputs, and theoretical evidence suggests that no polynomial-time algorithm exists. Thus, to have any hope of tackling position auctions computationally, it is necessary to work with a representation language that allows the corresponding games to be compactly described.

The action-graph game (AGG) representation \cite{JLbB11agg} is uniquely well-suited to compactly representing position auction games.  Action-graph games are similar to the (more widely known) graphical game representation \cite{KLS01graphical} in that they exploit utility independencies, but are strictly more powerful in the sense that they are compact not only for games with ``strict utility independencies'' (the property that one agent's payoff never depends on some second agent's action) but also ``context-specific independencies'' (the property that one agent's payoff is independent of a second agent's action, given some action of the first agent and some subset of actions of the second). Further, AGGs allow compact representation of ``anonymity'': roughly, the property that agents do not care \emph{which} agents take which actions, but only \emph{how many}.
Both of these distinctions help us to compactly model position auctions. Because any bidder can affect any other bidder's payoff (e.g., by outbidding him), the graphical game representation of a perfect-information position auction is a clique, and thus no more compact than the normal form. However, position auctions have considerable structure. To give one simple example, in a no-externalities GFP auction, bidder $i$'s utility is independent of bidder $j$'s \textit{bid}, conditional on $j$ bidding less than $i$ (context-specific independence), and  bidder $i$'s utility is independent of bidder $j$'s \textit{identity}, given both of their bids (anonymity). 


The main idea behind action-graph games is the action graph, so called because nodes in this directed graph represent actions. Each agent chooses his action from an arbitrary subset of the nodes; agents' subsets may overlap or coincide. Play of the game can be visualized as each agent simultaneously placing a single, identical token on one of the nodes in the graph. An agent's utility can then be computed as a function only of the \emph{number of tokens} in the \emph{neighborhood} of his chosen node. (The neighborhood of a node $v$ is the set of all nodes having outgoing edges that point to $v$; self-edges are allowed, and so a node can belong to its own neighborhood.)
The set of possible ``configurations'' (counts of tokens) in the neighborhood of any node is weakly smaller than the set of pure strategy profiles. Thus, AGGs can represent utility functions in weakly less space than the normal form, and (roughly speaking) the sparser the graph, the greater the reduction in representation size.


Compact size is not the only interesting thing about AGGs. Crucially, AGG structure can be leveraged computationally, and hence game-theoretic computations can be performed dramatically more quickly for AGGs than for games represented in normal form. For example, given action graphs with bounded in-degree, a polynomial-time dynamic programming algorithm can be used to compute an agent's expected utility under an arbitrary mixed strategy profile \cite{JLbB11agg}. This is interesting because the standard method of computing expected utility---summing over pure strategy profiles---requires time polynomial in the size of the normal form, but potentially exponential in the size of more compact representations like AGGs. This computational problem is important because it constitutes the inner loop of many game-theoretic algorithms, including state-of-the-art algorithms for computing Nash equilibria like Simplicial Subdivision \cite{ss} and Govindan and Wilson's continuation method \cite{GW03newton}.  An exponential speedup to the solution of the expected utility problem therefore translates directly to an exponential speedup of such algorithms, without altering the solution obtained.\footnote{It is easy to conceive of even more succinct game representations that represent payoffs using algebraic expressions. We can now see why these representations are not useful computationally: quantities like expected utility cannot generally be computed in time polynomial in the description length of such games.}  Similarly, AGG structure can be leveraged to achieve exponential speedups when identifying dominated strategies and encoding best-response constraints as systems of polynomial inequations \cite{TLLb11sem}.  These optimizations make it possible to use the support enumeration method (SEM) of Porter, et al.\ \cite{PNS08sem} on AGGs.  This algorithm has the advantage that it can enumerate Nash equilibria.

Although it is beyond the scope of this paper to describe AGGs in detail, there \emph{is} one further element of the representation that we must describe here. Specifically, it is possible to add so-called \emph{function nodes} to the action graph, which are nodes that belong to no agent's action set. Instead, the ``action count'' at a function node is calculated as an (arbitrary) deterministic function of the counts at the function node's parents. For example, when an agent's payoff for playing $a$ depends on how many agents play any of $b, c$ or $d$, we can add a summation node to reduce the in-degree of $a$. Function nodes can dramatically reduce representation size when (for example) many actions affect a given action in the same way; note that the space required to represent an arbitrary utility function is exponential in the maximum number of possible configurations over any node's neighborhood. As long as their functions are well-behaved (``contribution-independent''; roughly, commutative and associative), function nodes can be used with the algorithms just mentioned \cite{JLbB11agg,TLLb11sem}. Since the variable most important to the asymptotic running time of both algorithms is the game's description length---which can be drastically reduced by the introduction of high-in-degree function nodes---function nodes can thus also lead to substantial computational savings.
}

\section{Representing Position Auctions}
\label{sec:represent}

We now present algorithms for succinctly encoding various position auction settings as AGGs.
Implementations of all our algorithms are freely available at {\smaller\url{http://www.cs.ubc.ca/research/position_auctions}}.

\subsection{Representing No-Externality GFPs as AGGs}

We can naively represent a no-externality GFP as an AGG as follows. We begin by creating a distinct action set for each agent $i$ containing nodes for each of his possible bids $b_i$. Other agents will push agent $i$ below the top position if they bid strictly more than $i$'s bid $b_i$ or if they bid exactly $b_i$ and are favored by the tie-breaking rule. Thus, we create directed edges in the action graph pointing to each $b_i$ from every $b_j$ that could potentially be awarded a higher position: every $b_j$ for which $b_j \geq b_i$ and $j \neq i$. (If the tie-breaking rule is randomized, we consider all of the different positions the agent could achieve and return his expected utility.)  The price an agent pays is simply the amount of his bid, and so can be determined without adding any more edges.

How much space does this representation require? Let the number of agents be $n$ and the number of bid amounts for each agent be $k$. Consider the action node $b_i^0$ corresponding to $i$'s lowest bid amount. This node has edges incoming from each action node belonging to another agent, meaning that the utility table corresponding to $b_i^0$ must store a value for every configuration over these nodes. There are $k$ possible configurations over each agent $j$'s action nodes, and $n-1$ other agents who bid independently, so a total of $k^{n-1}$ such configurations. This expression is exponential, and so we consider the AGG to be intractably large.

\begin{figure}[t]
\centering
\includegraphics[width=0.45\hsize]{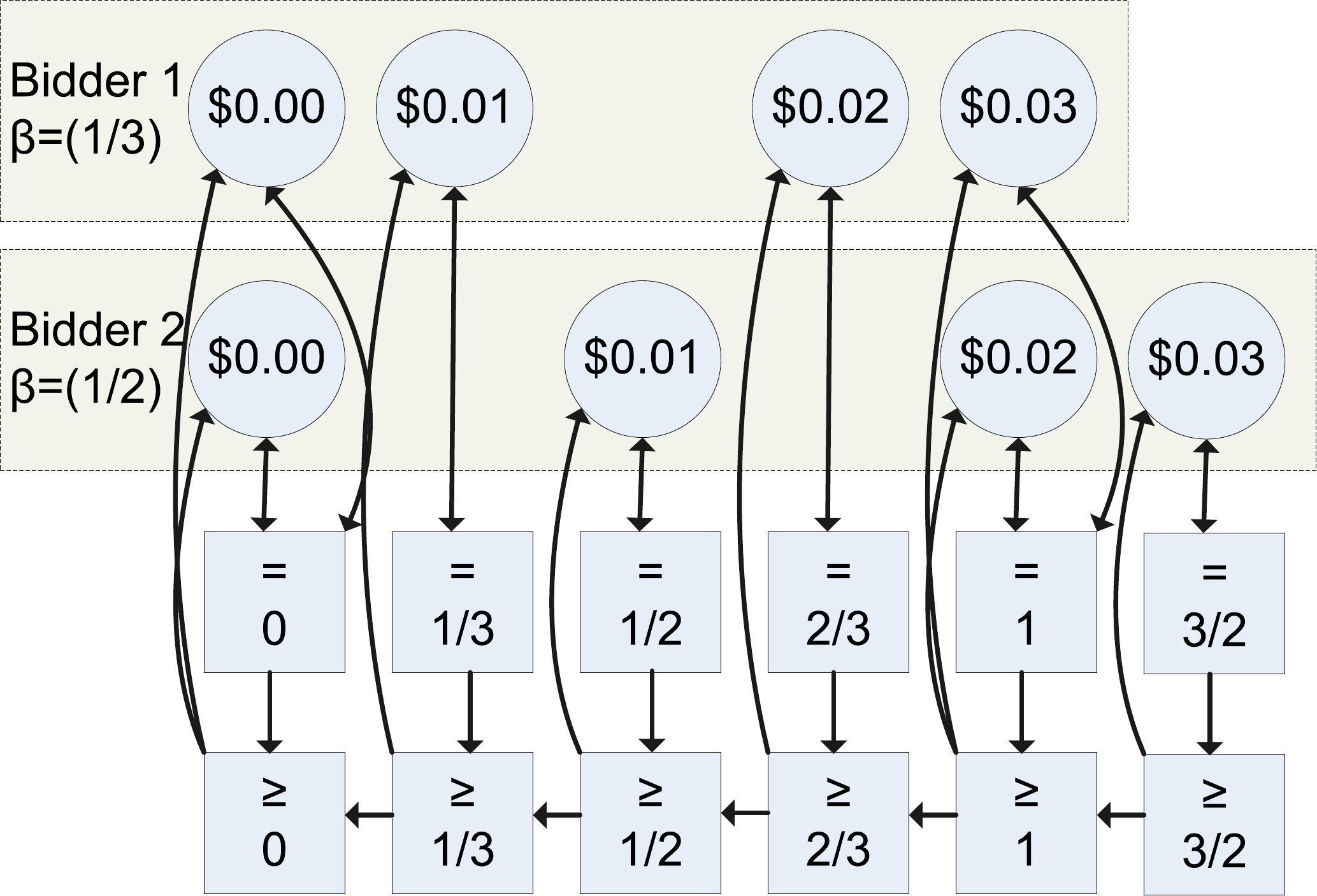}
\caption[A weighted GFP represented as an AGG]{A weighted GFP represented as an AGG. (Square nodes represent summation function nodes.)}
\label{fig:gfp-to-agg}
\end{figure}

However, notice that our naive representation failed to capture key regularities of the no-externalities GFP setting. Bidder $i$ does not care about the amount(s) by which he is outbid, nor does he care about the identities of the bidders who outbid him. We can capture these regularities by introducing function nodes. For every action node $b_i$, we can create one summation function node (formally, a node whose count is defined as the sum of the counts of its parents in the action graph) counting the number of other bids greater than or equal to $b_i$, and another counting the number equal to $b_i$. These two quantities are sufficient for computing the position of an agent who bids $b_i$. Each node in the action graph is thus connected to only two other nodes. Consider again the table encoding the utility for agent $i$'s bid $b_i^0$. There are $n-1$ possible configurations over the $\geq$ function node, and for each of these configurations, up to $n-1$ possible configurations over the $=$ function node, yielding a total of $O((n-1)^2)$ configurations---a polynomial number. There are $nk$ nodes in the action graph (and none has a larger table), so the graph's total representation size is $O(n^3k)$.
An example of such an AGG is given in Figure~\ref{fig:gfp-to-agg}. (As it does not complicate the representation, and is useful both to our exposition and in Section~\ref{sec:weighted-GFP}, we build \emph{weighted} GFPs, which we define in the natural way. Of course, setting all bidders' weights to 1 returns us to the unweighted case.)

More formally, we define a no-externalities position auction setting as a 4-tuple $\langle N,v,c,q \rangle$  where $N$ is a set of agents numbered $1,\dots,n$;  $v_{i,j}$ is agent $i$'s value per click in position $j$; $c_{i,j}$ is agent $i$'s probability of receiving a click in position $j$; and $q_i$ is agent $i$'s quality (typically, the probability that $i$ will receive a click in the top position). To specify a position auction game we additionally specify the range of allowed bid amounts $K$ and the tie breaking rule $T$. (For now we will consider only the case where $T$ is uniform-random; in Section~\ref{sec:tie-break} we revisit this choice.) We can now more precisely define our construction as Algorithm~\ref{algm:gfp-to-agg}.
%

\begin{algorithm2e}[t]
\ForEach{\textnormal{agent} $i \in N$}{
	\ForEach{\textnormal{bid} $k \in K$}{
		create an action node representing $i$ bidding $k$\;
	}
}
\ForEach{\textnormal{effective bid} $e \in \{k \cdot q_i\;|\; \forall i \in N, \forall k \in K\}$}{
	create a summation function node ($=,e$) counting the bidders bidding exactly $e$\;
	create a summation function node ($\geq,e$) counting the bidders bidding above $e$\;
	add an arc from ($=,e$) to ($\geq,e$) \;
	\If{$e>0$}{
		add an arc from ($\geq,e$) to ($\geq,e'$) (where $e'$ is the next largest effective bid)\;
	}
}
\ForEach{\textnormal{action node} $a_{i,k}$ \textnormal{representing $i$ bidding} $k$}{
	add an arc from $a_{i,k}$ to ($=,k\cdot q_i$)\;
	add an arc from ($=,k\cdot q_i$) to $a_{i,k}$, and denote the value of ($=,k\cdot q_i$) as $\ell$\;
	add an arc from ($\geq,k\cdot q_i$) to $a_{i,k}$, and denote the value of ($\geq,k\cdot q_i$) as $g$\;
	instantiate the utility table for $a_{i,k}$ as \[u_{a_{i,k}}(\ell,g) = \frac{1}{\ell}\sum_{j=g+1}^{g+\ell} c_{i,j}(v_{i,j}-k).\;\]	
}
\caption[An algorithm for converting a no-externality auction setting into an action graph representing a (weighted) GFP]{An algorithm for converting a no-externality auction setting into an action graph representing a (weighted) GFP.  Inputs are an AGG setting $(N,v,c,q)$, a set of allowable bid values $K$ and the tie-breaking rule $T$; this algorithm assumes that $T$ corresponds to uniform randomization.}
\label{algm:gfp-to-agg}%
\end{algorithm2e}

\subsection{Representing No-Externality uGSPs and wGSPs as AGGs}

GSPs are similar to GFPs in that each agent's payoff depends on a small number of values.  To determine an agent's position (or possible range of positions under randomized tie breaking), we start with a graph structure similar to the one we built for GFPs.  The first difference is that we must allow for bidder weights. We do this by creating function nodes for each ``effective bid''---each distinct value that can be obtained by multiplying a bid amount by a bidder weight---rather than one for each bid amount.\footnote{We can also derive AGG representations of Lahaie and Pennock's ranking rules \cite{LP07squash} by adjusting the values of $q$ appropriately.} Of course, we can recover the unweighted case (and thus represent uGSP) by setting all weights to 1. Because effective bids are real numbers while the set of possible bids and payments $K$ is discrete, our definition of a position auction game must now also include a ``rounding rule'' $R$ that is used to determine payments. For most of the  \paperonly{paper}\chapteronly{chapter} we will only consider the case where $R$ corresponds to rounding up (meaning that agents pay the minimum amount that they could have bid to maintain their positions); we will revisit this choice in Section \ref{sec:rounding}.

\begin{figure}[!t]
\centering
\includegraphics[width=0.45\hsize]{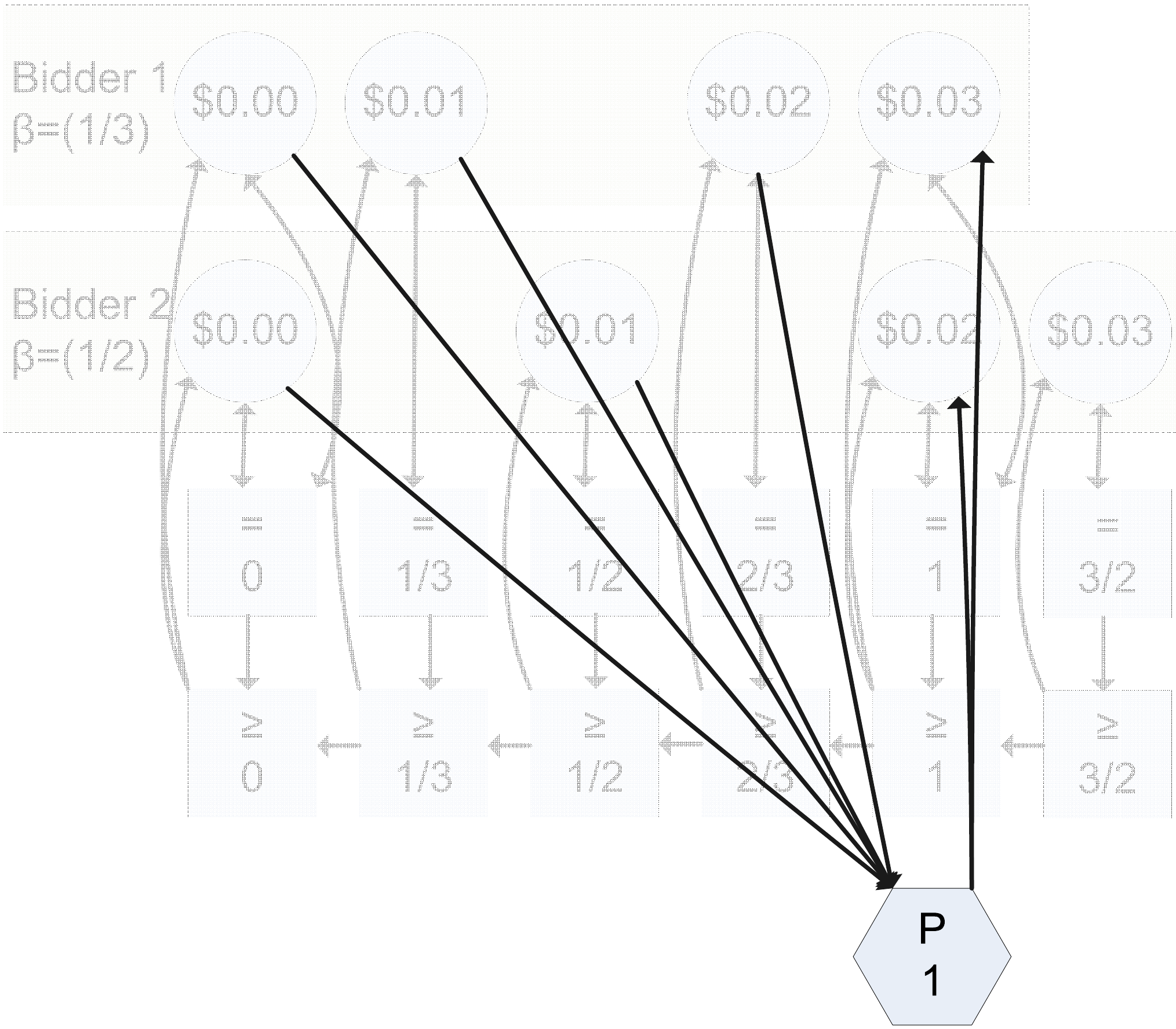}
\caption[A weighted GSP represented as an AGG]{To represent a GSP as an AGG, we add price nodes (argmax nodes denoted by hexagons) to a GFP representation.  For clarity only one price node is pictured; a full GSP representation requires one price node for each effective bid.}
\label{fig:gsp-to-agg}
\end{figure}

We also need to augment the action graph to capture the GSP pricing rule.  We do this by adding ``price nodes'': function nodes that identify the next-highest bid below the bid amount encoded by each given action node.  We use the term \emph{argmax node} to refer to a function node whose value is equal to the largest in-arc carrying a non-zero value, based on a given, fixed ordering of the in-arcs.  By ordering action nodes according to the values of their effective bids (i.e., bids multiplied by bidder weights), an argmax node identifies the highest effective bid among the subset of action nodes connected to it.  Our encoding is defined more precisely in Algorithm~\ref{algm:gsp-to-agg}.  An example of the resulting action graph is illustrated in Figure~\ref{fig:gsp-to-agg}.\footnote{Note that although the in-degree of the argmax nodes can get large---$O(nm)$---the computational complexity of solving an AGG only depends on the in-degrees of the \textit{action} nodes.}
This representation results in a graph containing $nm$ action nodes, each of which stores a payoff table with at most $O(n^2|E|)$  entries, where $E$ is the set of effective bids and $|E| \leq nm$.  Thus, this representation requires $O(n^4m^2)$ space.

\newcommand{\price}{\rho}

\begin{algorithm2e}[t]
\ForEach{\textnormal{agent} $i \in N$}{
	\ForEach{\textnormal{bid} $k \in K$}{
		create an action node representing $i$ bidding $k$\;
	}
}
\ForEach{\textnormal{effective bid} $e \in \{k \cdot q_i\;|\; \forall i \in N, \forall k \in K\}$}{
	create a summation function node ($=,e$) counting the bidders bidding exactly $e$\;
	create a summation function node ($\geq,e$) counting the bidders bidding above $e$\;
	add an arc from ($=,e$) to ($\geq,e$) \;
	\If{$e>0$}{
		add an arc from ($\geq,e$) to ($\geq,e'$) (where $e'$ is the next largest effective bid)\;
	}
  create a weighted argmax function node ($\arg\max,e$) identifying the next-highest effective bid below $e$\;
  \ForEach{\textnormal{action node} $a$ \textnormal{corresponding to effective bid} $e'$}{
  	\If{$e' < e$}
  	{
  		add an arc from $a$ to ($\arg\max,e$) with arc weight of $e'$ \;
  	}
  	\If{$e' = e$}
  	{
  		add an arc from ($\arg\max,e$) to $a$, and denote the value of ($\arg\max,e$) as $\price$ \;
  	}
  }
}
\ForEach{\textnormal{action node} $a_{i,k}$ \textnormal{representing $i$ bidding $k$}}{
	add an arc from $a_{i,k}$ to ($=,k \cdot q_i$)\;
	add an arc from ($=,k \cdot q_i$) to $a_{i,k}$, and denote the value of ($=,k \cdot q_i$) as $\ell$\;
	add an arc from ($\geq,k \cdot q_i$) to $a_{i,k}$, and denote the value of ($\geq,k \cdot q_i$) as $g$\;
	instantiate the utility table for $a_{i,k}$ as \[u_{a_{i,k}}(\ell,g,\price) = \frac{1}{\ell}\left( c_{i,g+\ell}(v_{i,g+\ell}-\lceil \price/q_i \rceil)\right)+\sum_{j=g+1}^{g+\ell-1} c_{i,j}(v_{i,j}-k).\;\]		
}
\caption[An algorithm for converting an auction setting into an action graph representing a wGSP]{An algorithm for converting an auction setting into an action graph representing a wGSP.  (Specifically, this algorithm encodes wGSP with uniform random tie-breaking and prices that are rounded up to whole increments.  These choices are revisited in Sections~\ref{sec:tie-break} and~\ref{sec:rounding}, respectively.)}
\label{algm:gsp-to-agg}%
\end{algorithm2e}


\subsection{Representing Auctions with Externalities}

We now describe an algorithm for representing GIM auctions as AGGs; this also suffices for cascade and hybrid, which GIM generalize. We define a GIM position auction setting by the 4-tuple $\langle N,v,q,f \rangle$ where
$N$ is a set of agents numbered $1,\dots,n$;
$v_i$ is agent $i$'s value per click;
$q_i$ is agent $i$'s quality (the probability that $i$ will receive a click in the top position); and
$f_i: 2^N \rightarrow \mathds{R}$ encodes the externalities that affect $i$.  When agent $i$'s ad is shown below the ads of the agents in set $S$, his probability of receiving a click is $q_if_i(S)$.  We assume that $f$ is monotone decreasing (i.e., $S' \subseteq S$ implies $f(S') \geq f(S)$), and normalized so that $f(\emptyset)=1$.

A GSP auction in a GIM setting can be converted to an AGG using Algorithm~\ref{algm:gsp-to-agg-gim}.  
 This representation results in a graph containing $nk$ action nodes, where each node has $2n-1$ in-arcs.  The total size of a payoff table for any node is $O(2^{2n}nk)$, so the full representation requires $O(2^{2n}n^2k^2)$ space.  However, storing a single agent's preferences requires $O(2^n)$ values (to encode $f(\cdot)$), so the AGG representation is only quadratically larger than the input.  Nevertheless, $GIM$ settings with large $n$ are impractical.

\begin{algorithm2e}
\ForEach{\textnormal{agent} $i \in N$}{
  \ForEach{\textnormal{bid} $k \in K$}{
    create an action node representing $i$ bidding $k$\;
  }
}
\ForEach{\textnormal{pair of agents} $i,j \in N$}{
    \ForEach{$a_i \in A_i$}{
      create an OR function note ($=,a_i,j$) representing that $j$ having the same effective bid as $i$\;
      create an OR function node ($>,a_i,j$) representing that $j$ having a strictly greater effective bid than $i$\;
      create a weighted argmax function node ($\price,a_i$) representing the price $i$ pays given $a_i$\;
      create arcs from all three function nodes to $a_i$\;
     \ForEach{$a_j \in A_j$}{
         \If{$a_j > a_i$}{create an arc from $a_j$ to ($>,a_i,j$)\;}
         \If{$a_j = a_i$}{create an arc from $a_j$ to ($=,a_i,j$)\;}
         \If{$a_j < a_i$}{create an arc from $a_j$ to ($\price,a_i$)\;}
      }
   }
}
\ForEach{\textnormal{agent} $i \in N$}{
    \ForEach{\textnormal{action node} $a_{i,k}$ \textnormal{representing $i$ bidding} $k$}{
        Let $L$ denote the set of agents with the same effective bid as $a_{i,k}$\;
        Let $G$ denote the set of agents with greater effective bids than $a_{i,k}$\;
        Let $\price$ denote the next highest effective bid after $a_{i,k}$\;
        instantiate the utility table for $a_{i,k}$ as \[u_{a_{i,k}}(L,G,\price) = \frac{f_i(G\cup L)}{2^{|L|}} (v_{i,g+\ell}-\lceil \price/q_i \rceil\})+\sum_{S\in 2^L\setminus L} \frac{f_i(G\cup S)}{2^{|L|}}(v_{i,j}-k).\;\]
    }
}
\caption[Creating the action graph for a GIM position auction]{Creating the action graph for a GIM position auction}
\label{algm:gsp-to-agg-gim}%
\end{algorithm2e}


%
%

\section{Experimental Setup}
\label{sec:prelim}
Broadly speaking, the method we employ in this  \paperonly{paper}\chapteronly{chapter} is to generate many preference-profile instances from distributions over each of the preference models, to build AGGs encoding the corresponding perfect-information auction problems for each auction design, to solve these AGG computationally, and then to compare the outcomes against each other and against VCG.

\subsection{Problem instances}

\begin{table}
\centering
\begin{tabular}{cc}
	\toprule
	Without externalities & With externalities \\ \midrule
	       EOS-UNI        &    Cascade-UNI     \\
	       EOS-LN         &     Cascade-LN     \\
	        V-UNI         &     Hybrid-UNI     \\
	        V-LN          &     Hybrid-LN      \\
	       BHN-UNI        &      GIM-UNI       \\
	       BHN-LN         &       GIM-LN       \\
	         BSS          &  \\ \bottomrule
\end{tabular}
\caption[The distributions we considered for our position-auctionn experiments]{The distributions we considered for our experiments.  UNI denotes distributions that are uniform across all values.  LN denotes distributions with parameters log-normal distributed following \cite{LP07squash}, with additional parameters not discussed in that work again uniformly distributed.}
\label{tbl:distros}
\end{table}

We generated our preference profiles by (1) imposing a probability distribution over a preference model and then (2) drawing instances from those distributions.  Except for BSS, whose definition already includes a distribution, we take two approaches to imposing distributions: one is to assume that all variables are uniformly distributed over their acceptable ranges, and the other is to draw variables from the log-normal distributions of Lahaie and Pennock \cite{LP07squash} that have been fitted to real-world data.  Thus, we used a total of 13 distributions (see Table~\ref{tbl:distros}). For each we generated 200 preference-profile instances for each of the three position-auction types, GFP, uGSP and wGSP, yielding  $13\times 200\times 3=7800$ perfect-information games in total.
Each of these games had 5 bidders, 5 positions and 30 bid increments. In Section~\ref{sec:sensitivity} we describe further experiments in which we investigated the our findings' sensitivity to these and other parameters.

We normalized values in each game so that the highest was equal to the highest possible bid, to ensure that the full number of bid increments was potentially useful.  We also removed weakly dominated strategies from each game, both for computational reasons and to select against implausible equilibria. Specifically, we eliminated strategies in which agents bid strictly more than (the ceilings of) their valuations.\footnote{In past work, for computational reasons, we also eliminated \textit{very weakly} dominated strategies.  In particular, this included bids that led to identical outcomes regardless of the actions of other agents (e.g., from an agent with an extremely low quality score).  We did not do this here because we now enumerate equilibria: although the elimination of very weakly dominated strategies does not change the set of equilibrium outcomes, it can change the relative frequency of these outcomes.}.

\subsection{Equilibrium Computation}

\newcommand{\simpdiv}{{\small{\texttt{simpdiv}}}}
\newcommand{\gnm}{{\small{\texttt{gnm}}}}
\newcommand{\sem}{{\small{\texttt{sem}}}}

We performed our experiments on WestGrid's Orcinus cluster---384 machines with dual Intel Xeon E5450 3.0GHz CPUs, 12MB cache and 16GB RAM, running 64-bit Red Hat Enterprise Linux	Server 5.3.
To compute Nash equilibria, we used three algorithms:\footnote{Implementations of all three algorithms are available at \url{http://agg.cs.ubc.ca}.}
(1)~\simpdiv, the simplicial subdivision algorithm of  van der Laan \textit{et al} \cite{ss}, adapted to AGGs by Jiang, Bhat and Leyton-Brown \cite{JLbB11agg};
(2)~\gnm, the global Newton method of Govindan and Wilson \cite{GW03newton}, adapted to AGGs by Jiang, Bhat and Leyton-Brown \cite{JLbB11agg}; and
(3)~\sem, the support-enumeration method of Porter, Nudelman and Shoham \cite{PNS08sem}, adapted to AGGs by Thompson, Leung and Leyton-Brown \cite{TLLb11sem}.
We ran \sem{} to enumerate all pure-strategy Nash equilibria.  For finding sample mixed-Nash equilibria, we ran \simpdiv{} from ten different pure-strategy-profile starting points chosen uniformly at random, and also ran \gnm{} ten times with random seeds one through ten.  We limited runs of \simpdiv{} and \gnm{} to five CPU minutes.  In total, we spent about 15 CPU months on equilibrium computation.

\subsection{Benchmarks: VCG and Discretized VCG}

As well as comparing GSP and GFP to each other, we also compared these position auctions to VCG. There are two ways of doing this. First, we considered VCG's truthful equilibrium given agents' actual (i.e., non-discretized) preferences. However, this prevents us from determining whether differences arise because of the auction mechanism or discretization. To answer this question, we also compared to an alternate version of VCG in which we discretized bids to the same number of increments as in the position auctions.
In this case, we assume that bidders report the discrete value nearest to their true value. (Observe that this is always an $\epsilon$-Nash equilibrium with $\epsilon$ equal to half a bid increment.)


\subsection{Statistical Methods}

In order to justify claims that one auction achieved better performance than another according to a given metric (e.g., revenue), we ensure that this difference was judged significant by a statistical test. Specifically, we performed blocking, means-of-means, bootstrapping tests \cite{bootstrap} as follows:
\begin{enumerate}
\item For each setting instance, find the difference in the metric across that pair of auctions on that instance.  Each value is normalized by the achievable social welfare in that instance.  Call this set of values $S$.
\item Draw $|S|$ samples from $S$ (with replacement), and compute the mean. Perform this procedure 20,000 times. Let $M$ denote the set of means thus computed.
\item Our estimated performance difference is the mean of $M$ (the mean-of-means of $S$).
\item This difference has significance level $\alpha$ if the $\alpha^{th}$ quantile of $M$ is weakly greater than zero.
\end{enumerate}
When reporting results, we use the symbol $^*$ to denote that a result has a significance level of at least $\alpha=0.05$ and $^{**}$ to denote that a result has a significance level of at least $\alpha=0.01$.  For each group of data points (i.e., for a specific size and preference model) we perform many simultaneous tests, comparing revenue, welfare, relevance and envy between all pairs of auctions. To avoid spurious claims of significance, we thus perform Bonferroni multiple-testing correction (effectively, dividing the desired significance level by the number of tests performed) \cite{M81simultaneous}.

To avoid undermining the statistical reliability of our data, we did not drop individual games that we could not solve within our chosen time budget: we worried that the features that made some instances hard to solve could also make their equilibrium outcomes qualitatively different from those of easier-to-solve games.  Instead, when we were not able to identify any Nash equilibria of a particular game (typically involving GFP auctions), we replaced the metric value of interest with an suitable upper or lower bound (e.g., a position auction's revenue is trivially guaranteed to be between 0.0 and the maximum possible social welfare).  When we have incomplete data about one or both of auctions $A$ and $B$, we do not claim that auction $A$ achieves significantly better performance according to some metric than auction $B$ unless the lower bound on $A$'s performance is significantly better than the upper bound on $B$'s performance.

\section{Results}
\label{sec:results}

We now turn to our experimental results. Our goal is  to demonstrate the effectiveness of computational mechanism analysis in general and to shed light on open questions about position auctions in particular. In the latter vein our main aim is to justify the search industry's convergence on the wGSP auction; we also consider the seven model-specific questions we asked in the introduction. This section thus begins with a broad comparison of the different position auction mechanisms, followed by a more detailed examination of each individual model. We also describe some follow-up experiments prompted by model-specific claims in the research literature. \paperonly{The appendix at the end of the paper}\chapteronly{The last section of this chapter} gives all of our results in tabular form (i.e., providing numerical values rather than just graphs) and also reports the results of statistical significance tests for all comparisons.

\subsection{Main Comparison}

We begin by looking at the relative performance of GFP, uGSP, and wGSP position auctions, averaged across all of our different models and distributions, and compared to the VCG and discrete VCG baselines as appropriate (see Figure~\ref{fig:ga-comparison}). In a nutshell, the industry's choice of wGSP appears to be justified in terms of efficiency, relevance and (to a lesser extent) envy, but not in terms of revenue. More specifically, in terms of both efficiency and relevance, we found that wGSP was clearly the best position auction design; it also exhibited relatively little variation in these metrics (less than 10\%) across equilibria. wGSP ranked lower on efficiency and relevance than VCG (with or without discretization), but was surprisingly close, given that in many of our models wGSP is known to be inefficient.  wGSP also clearly outperformed GFP and uGSP in terms of envy, but also exhibited very substantial variation across equilibria. Revenue comparisons were much more ambiguous: all auctions achieved fairly similar median revenues, and variation from one equilibrium to another could be very large for both GSP variants (with best-case equilibrium revenues of about twice worst-case equilibrium revenues). We also saw substantial revenue variation across models, distributions, and equilibrium selection criteria, which we will discuss in detail in what follows.

\newcommand{\kcaption}[2]{\begin{minipage}{#2}\centering{}#1\vspace{.3em plus .2em}\end{minipage}}
\begin{figure}
\centering
\includegraphics[width=0.45\hsize]{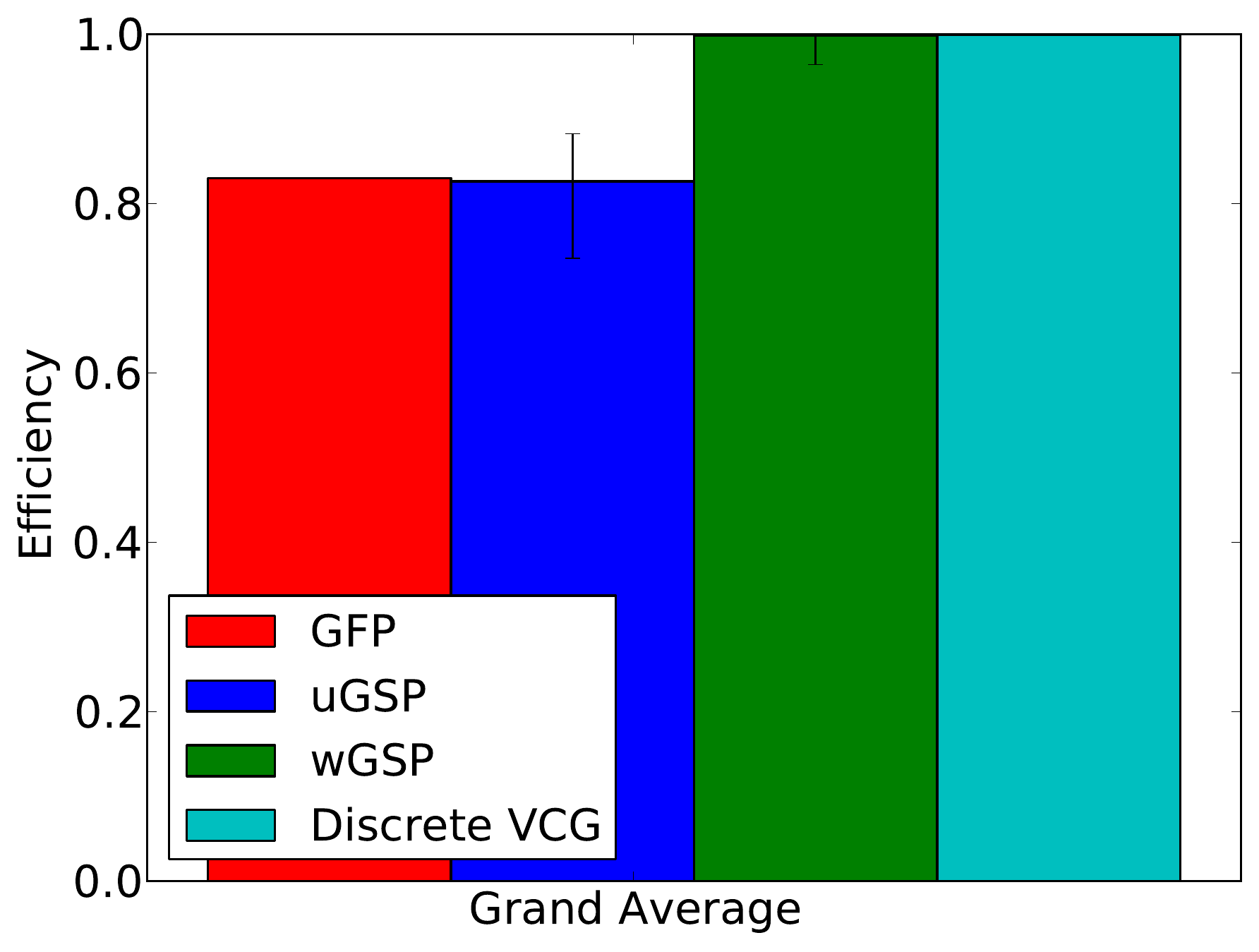}
\includegraphics[width=0.45\hsize]{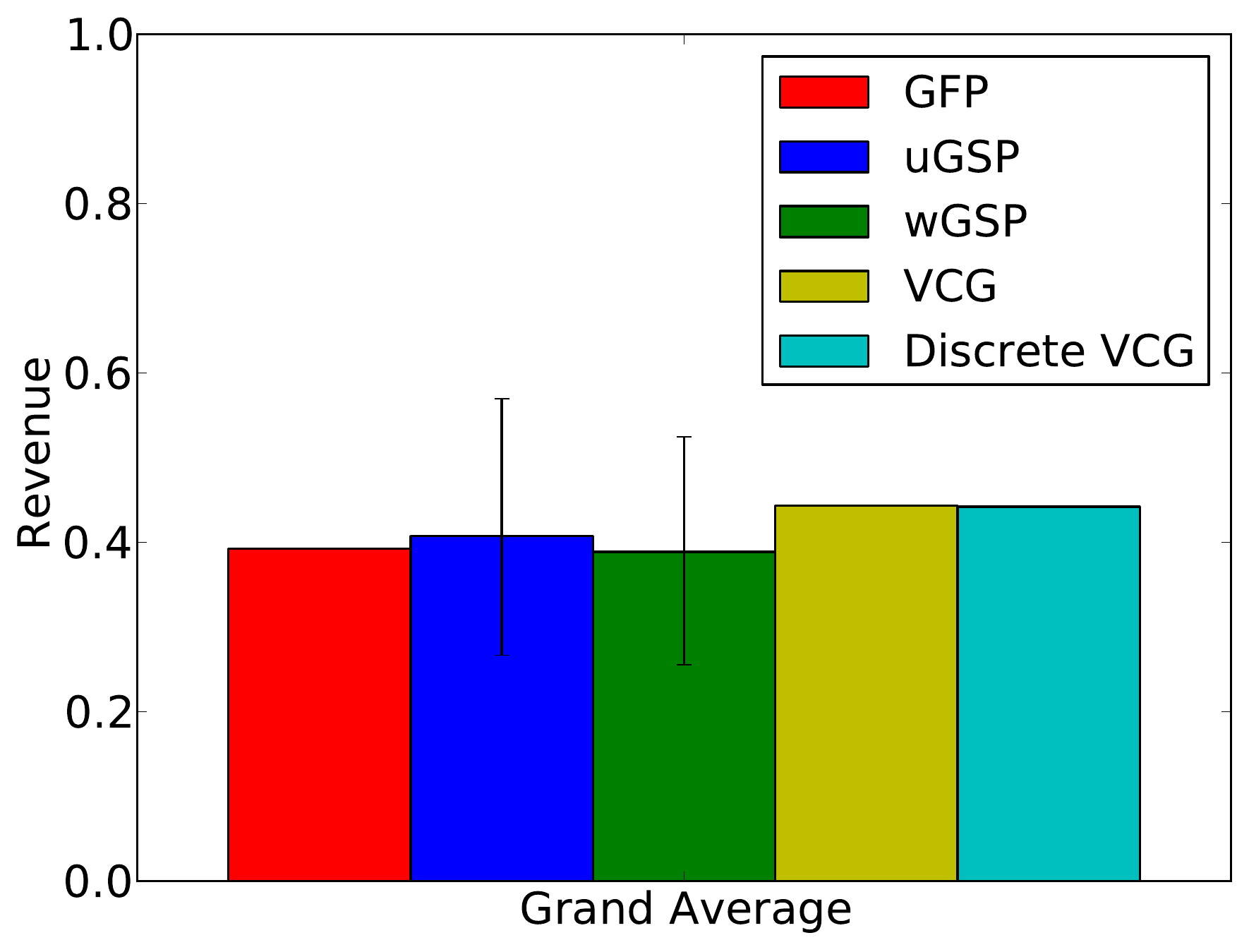}
\kcaption{(a) Efficiency}{.45\hsize}
\kcaption{(b) Revenue}{.45\hsize}
\includegraphics[width=0.45\hsize]{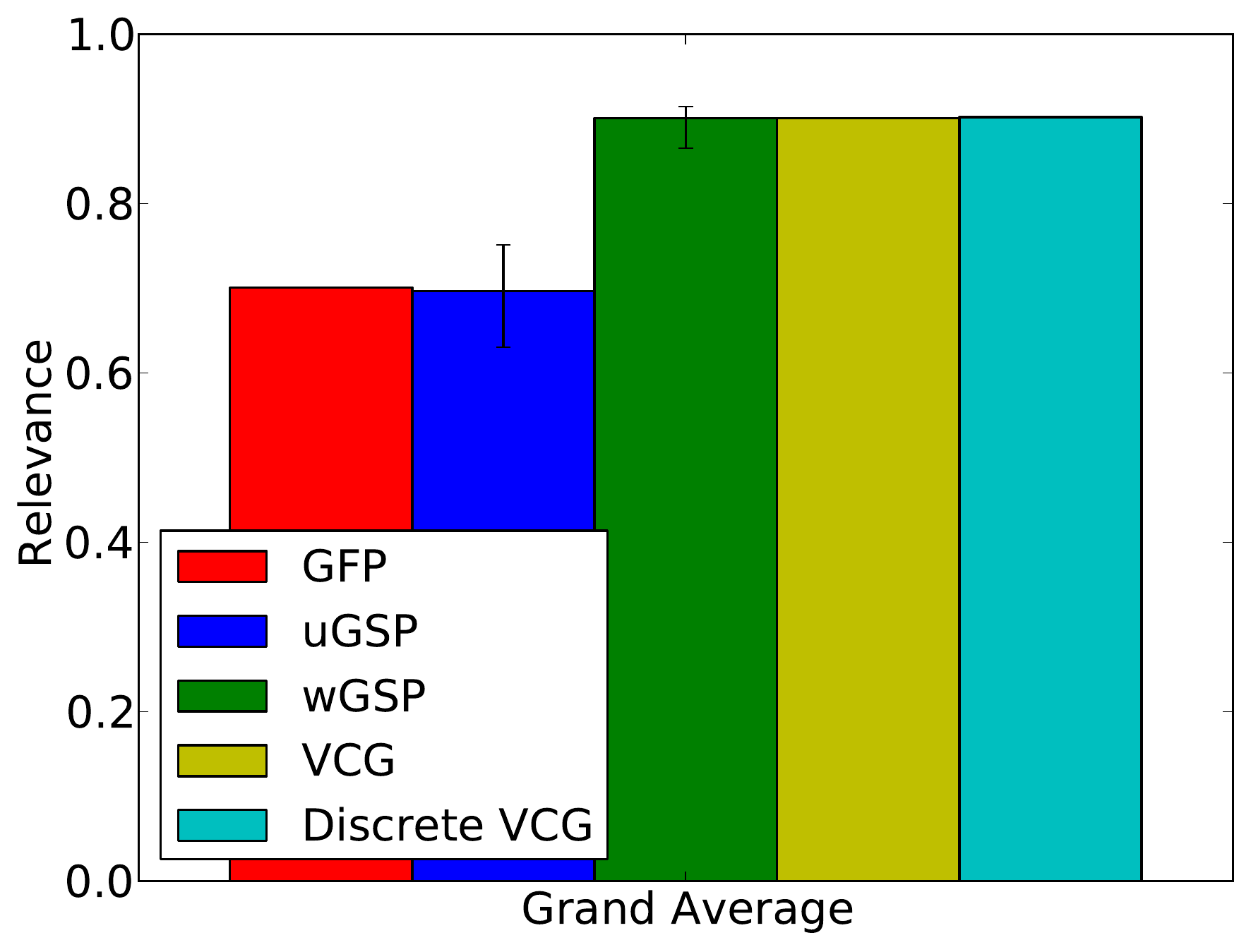}
\includegraphics[width=0.45\hsize]{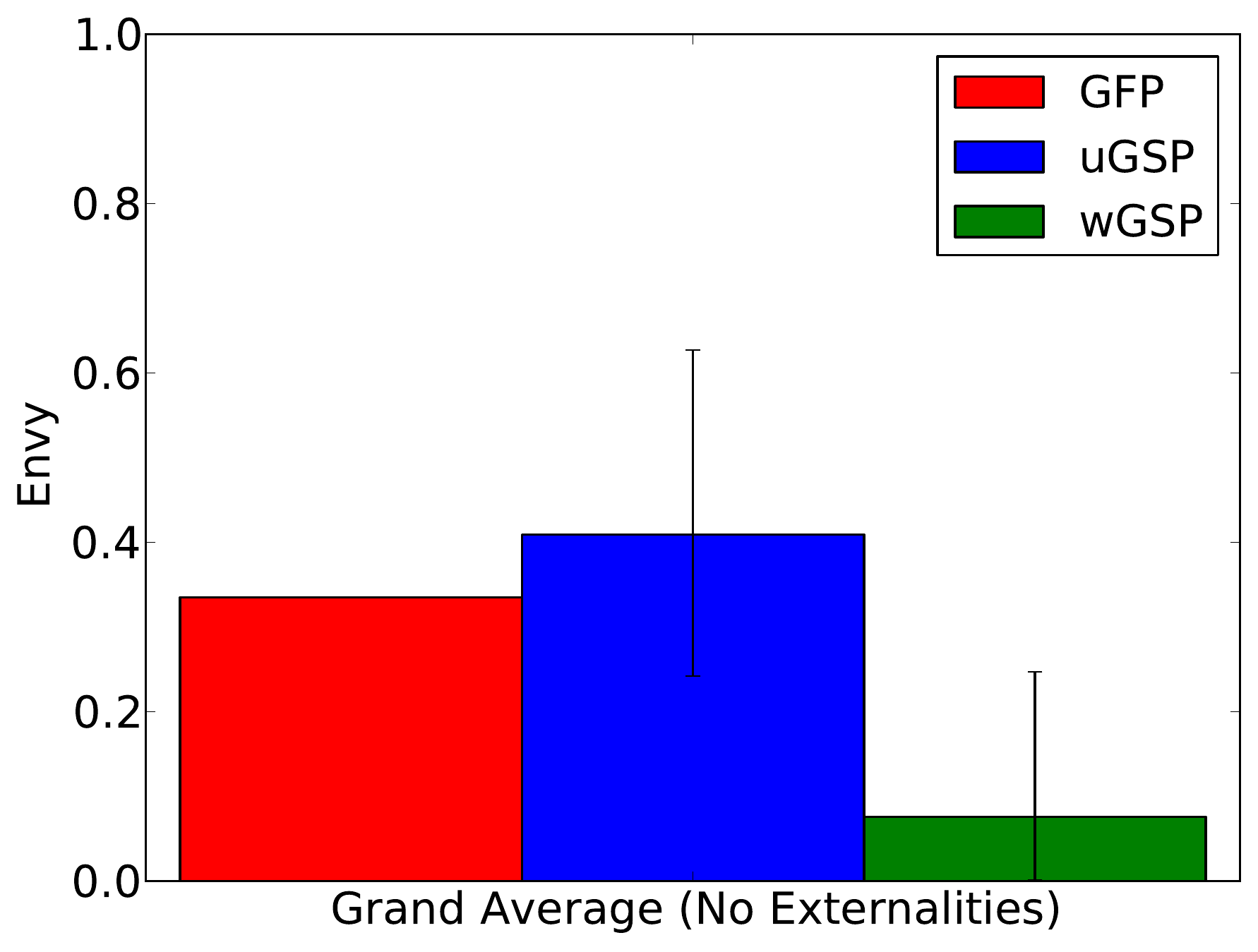}
\kcaption{(c) Relevance}{.45\hsize}
\kcaption{(d) Envy}{.45\hsize}
\caption[Performance of different position auction types, averaged across all 13 distributions ]{Performance of different position auction types, averaged across all 13 distributions (efficiency, revenue, and relevance) and the 7 no-externality distributions (envy, which does not apply to settings with externalities).  \whiskerBP}
\label{fig:ga-comparison}
\end{figure}

\subsection{Basic models: EOS and V}

We consider the EOS and V models together because they are very similar, both in terms of what was known from previous work, and in terms of our findings.  The main difference between EOS and V is that V introduces quality scores, which can be used to weight advertisers' bids.  Thus, in EOS, wGSP and uGSP are identical, while in V they are not. Earlier we asked the following two questions.

\qi

\qii\vspace{\modelspace}

\begin{figure}[pt]
	\centering
	\includegraphics[width=0.6\hsize]{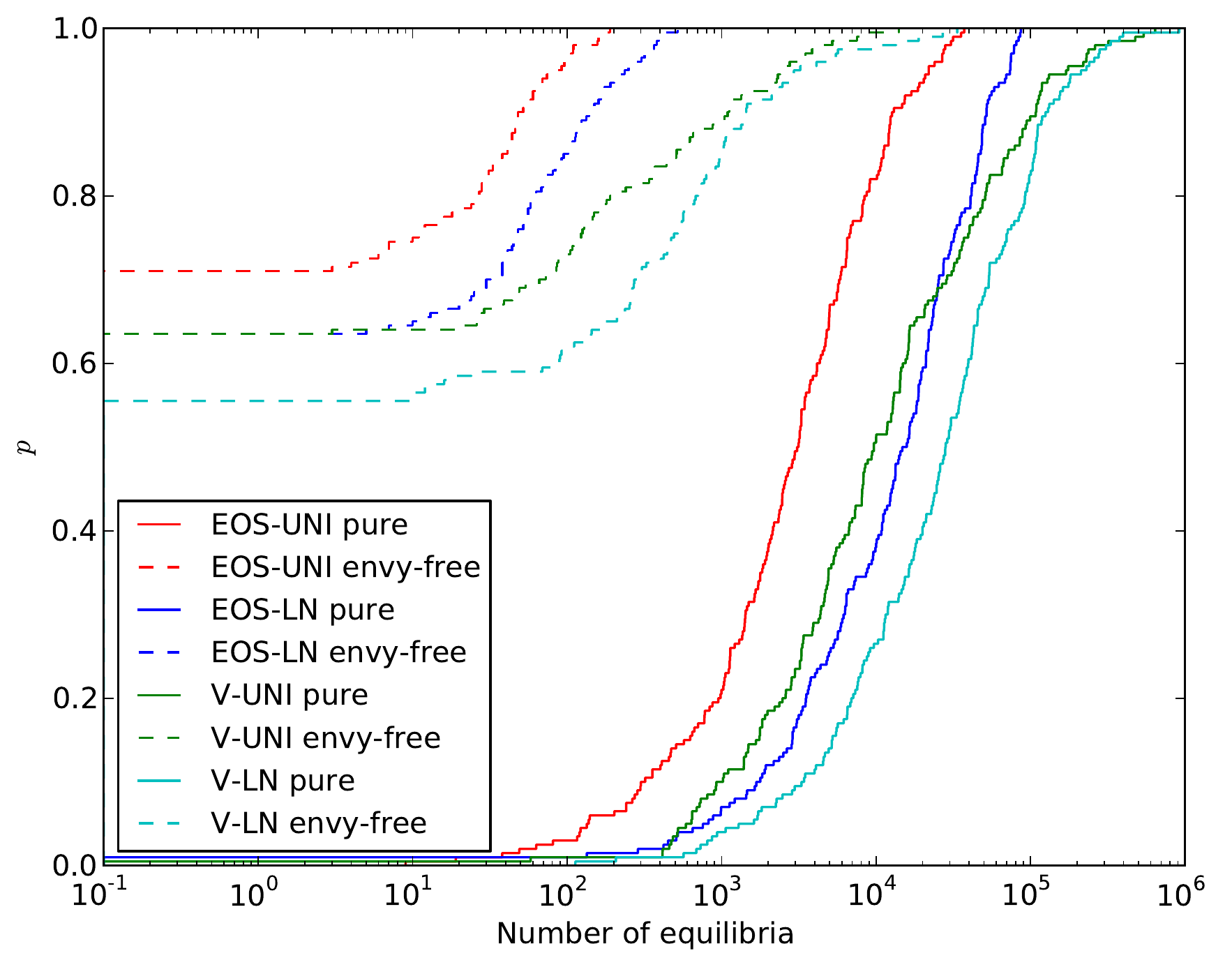}
	\caption[Empirical cumulative probability distributions over the number of equilibria in EOS and V models]{Empirical cumulative probability distributions over the number of equilibria.  Many lines do not begin at $p=0$, due to our use of a log scale; a line beginning at $p=0.6$ indicates that $60\%$ of games had zero envy-free Nash equilibria.}%
	\label{fig:veos-efnes}%
\end{figure}

\begin{figure}[pt]
	\centering
	\includegraphics[width=0.95\hsize]{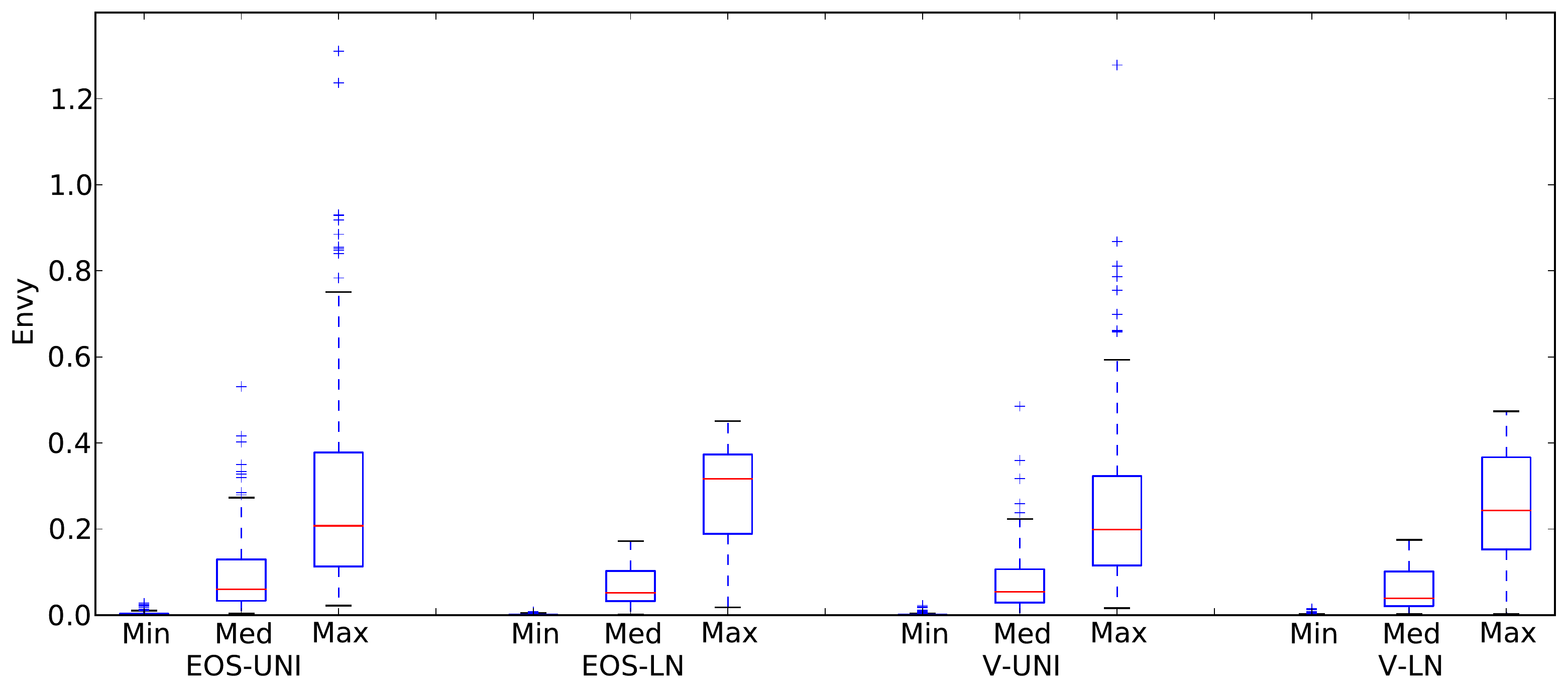}
	\caption[Empirical box plot of wGSP's envy under different equilibrium selection criteria ]{Empirical box plot of wGSP's envy under different equilibrium selection criteria (minimum, median and maximum).  The envy-minimizing Nash equilibrium always had very little envy, but other equilibria can have substantial envy.  Note that we report envy normalized by total possible social welfare.  Even normalized, envy can occasionally exceed 1, as it is possible for an agent to be envied by many other agents.}
	\label{fig:veos-wgsp-envybox}
\end{figure}

We begin by investigating how often envy-free equilibria exist in wGSP (with discrete bids) and how many such equilibria there are, as compared to pure Nash equilibria.  Surprisingly, we found that envy-free equilibria did not exist in the majority of games, despite most games having hundreds or thousands of pure-strategy Nash equilibria (see Figure~\ref{fig:veos-efnes}), and that even when they did exist there were orders of magnitude more Nash equilibria than envy-free equilibria.  This led us to investigate the \emph{amount} of envy present in wGSP equilibria. We found that this quantity varied substantially across equilibria, but that envy-minimizing Nash equilibria tended to get very close to zero envy  (see Figure~\ref{fig:veos-wgsp-envybox}).

\begin{figure}[tp]
	\centering
	\includegraphics[width=0.95\hsize]{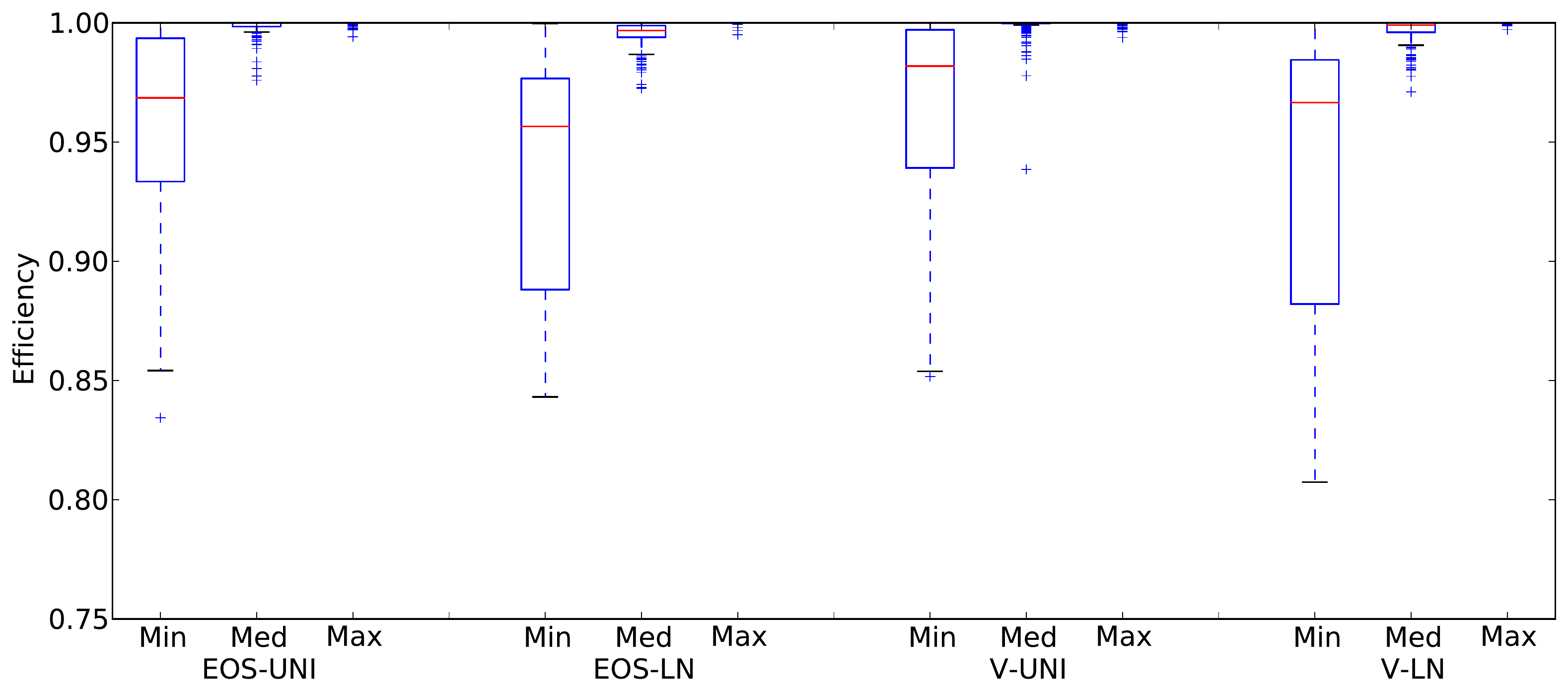}
	\caption[Empirical box plot of wGSP's social welfare under different equilibrium selection criteria ]{Empirical box plot of wGSP's social welfare under different equilibrium selection criteria (minimum, median and maximum).  Note that the majority of equilibria (indicated by the median) tend to be very nearly efficient.}
	\label{fig:veos-wgsp-effbox}
\end{figure}

We know that in the EOS and V models, envy-free equilibria are economically efficient and yield high revenue; however, we have just observed that such envy-free equilibria did not reliably exist in our games. We next investigated the efficiency and revenue properties of general Nash equilibria.
We found that wGSP was very efficient, not just in the best-case equilibrium, but in the majority of equilibria; however, the worst-case equilibria could be substantially less efficient  (see Figure~\ref{fig:veos-wgsp-effbox}).\footnote{We note that in our previous work we found almost no instances with such low efficiency \cite{TLb09position}; we attribute this discrepancy to the fact that we did not previously have a practical algorithm for enumerating Nash equilibria, and so we previously reported minimum and maximum efficiency based only on the most extreme equilibria our sample-Nash-finding algorithms could find.  In contrast, our current work uses the SEM algorithm, and hence guarantees that we have truly found the best/worse case PSNEs, and furthermore allows us to identify the median equilibrium (which we could not do before).}

\begin{figure}[tp]
	\centering
	\includegraphics[width=0.95\hsize]{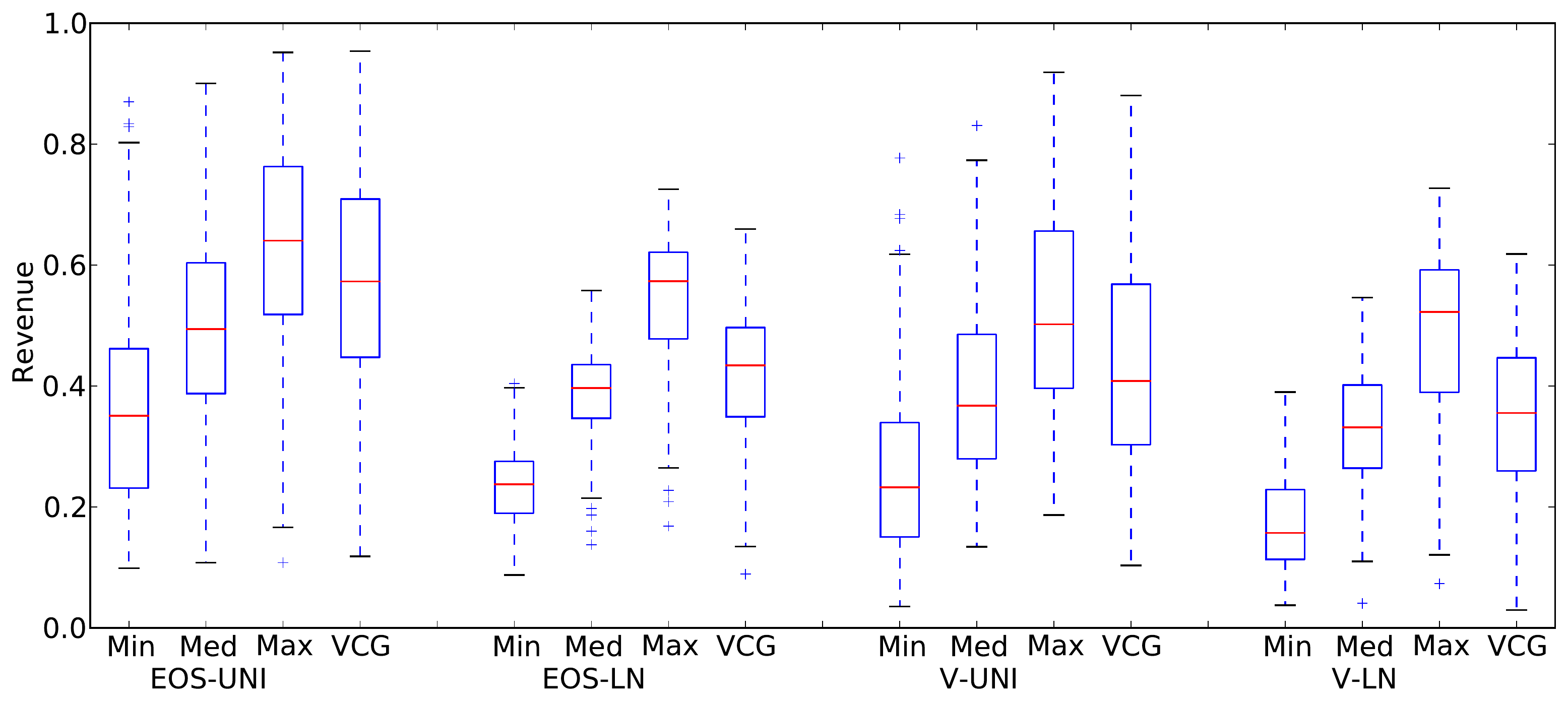}
	\caption[Empirical box plot of wGSP's revenue under different equilibrium selection criteria ]{Empirical box plot of wGSP's revenue under different equilibrium selection criteria (minimum, median and maximum).  The median equilibrium of wGSP performs similarly to VCG, while the worst-case and best-case equilibria are substantially different.}
	\label{fig:veos-wgsp-revbox}
\end{figure}

Concerning revenue, we found that equilibrium selection has a very large effect.  Further, this effect is not merely confined to a few very bad equilibria.  In every distribution, we found that wGSP's best-case revenue was better than VCG, worst-case was much worse than VCG and median was slightly worse in EOS settings and approximately equal in V settings (see Figure~\ref{fig:veos-wgsp-revbox}).

\begin{figure}[tp]
	\centering
	\includegraphics[width=0.95\hsize]{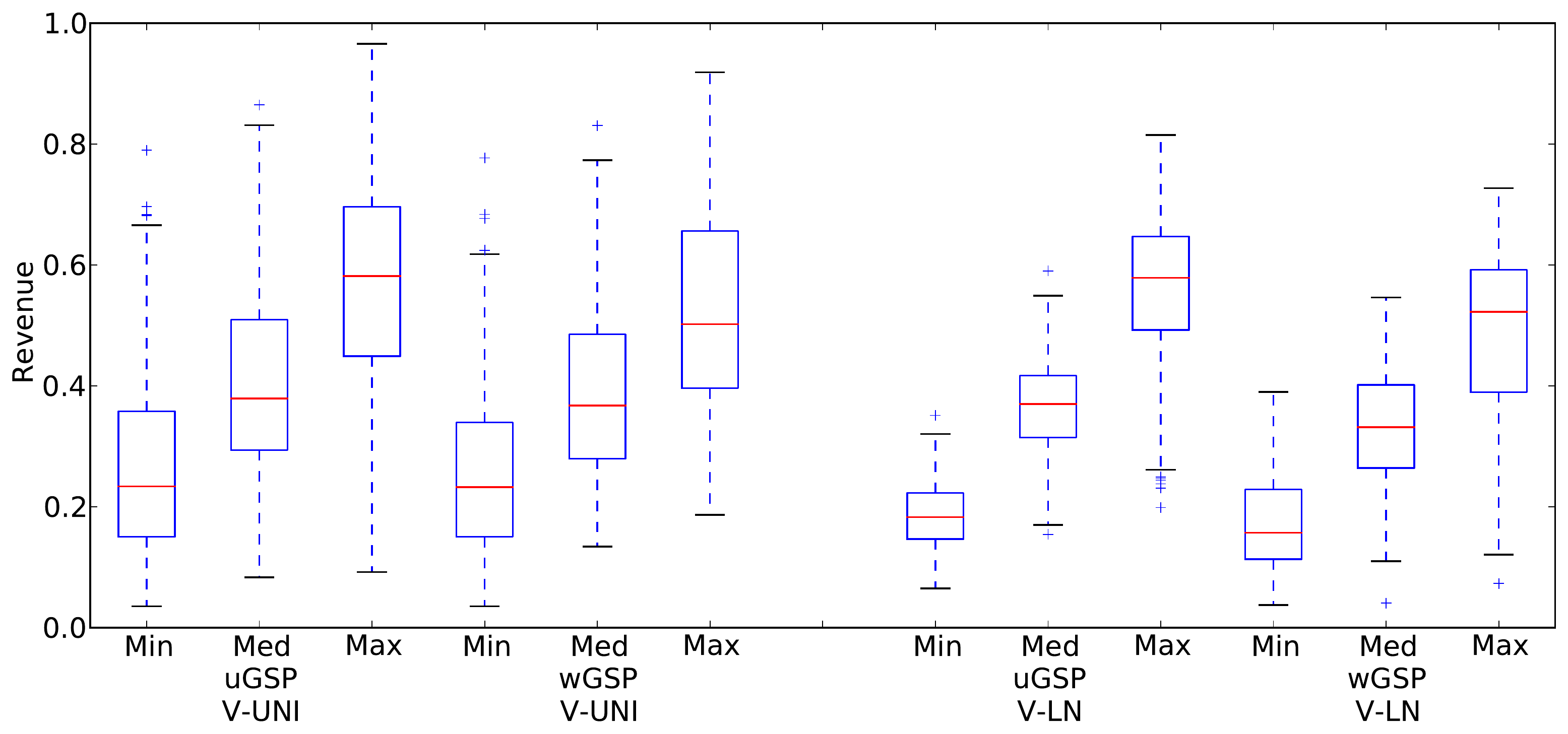}
	\caption[Empirical box plot of uGSP and wGSP's revenue under different equilibrium selection criteria ]{Empirical box plot of uGSP and wGSP's revenue under different equilibrium selection criteria (minimum, median and maximum).}
	\label{fig:veos-uwgsp-revbox}
\end{figure}

\begin{figure}[tp]
	\centering
	\includegraphics[width=0.95\hsize]{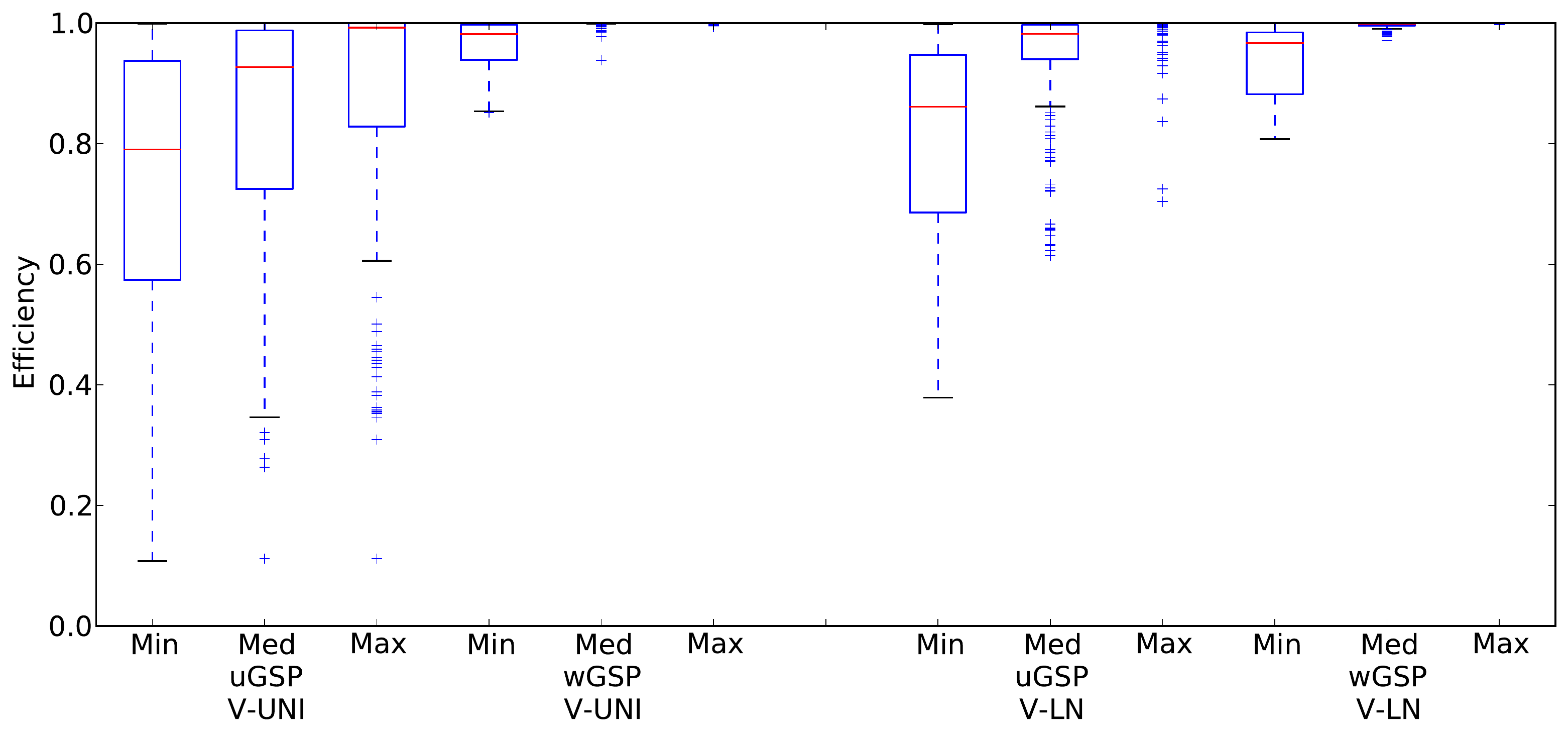}
	\caption[Empirical box plot of uGSP and wGSP's social welfare under different equilibrium selection criteria]{Empirical box plot of uGSP and wGSP's social welfare under different equilibrium selection criteria (minimum, median and maximum).}
	\label{fig:veos-uwgsp-effbox}
\end{figure}

It remains to compare uGSP with wGSP under the V model in terms of both revenue and efficiency.  First, recall that Lahaie and Pennock \cite{LP07squash} found that uGSP generated much more revenue but moderately less social welfare, but only for the specific case of ``symmetric equilibria'' and log-normal distributions.  For both $V$ distributions,
we found that uGSP was better than wGSP for revenue (provided that we compared the mechanisms under the same equilibrium selection criterion, e.g., median), but that both mechanisms were very sensitive to equilibrium selection: the difference between mechanisms ($\sim 1.2\times$) was very small compared to the difference between good and bad equilibria (at least $\sim2\times$)   (see Figure~\ref{fig:veos-uwgsp-revbox}).  The gains from uGSP were much larger in the log-normal distribution than in the uniform distribution.  Concerning efficiency, wGSP was indeed better overall. We also found that uGSP's social welfare was far more sensitive to equilibrium selection (at least $\sim 1.2\times$) than wGSP's ($\sim 1.04\times$), and even uGSP's best equilibria were sometimes very inefficient  (see Figure~\ref{fig:veos-uwgsp-effbox}).

\begin{figure}[tp]
\centering
\includegraphics[width=0.45\hsize]{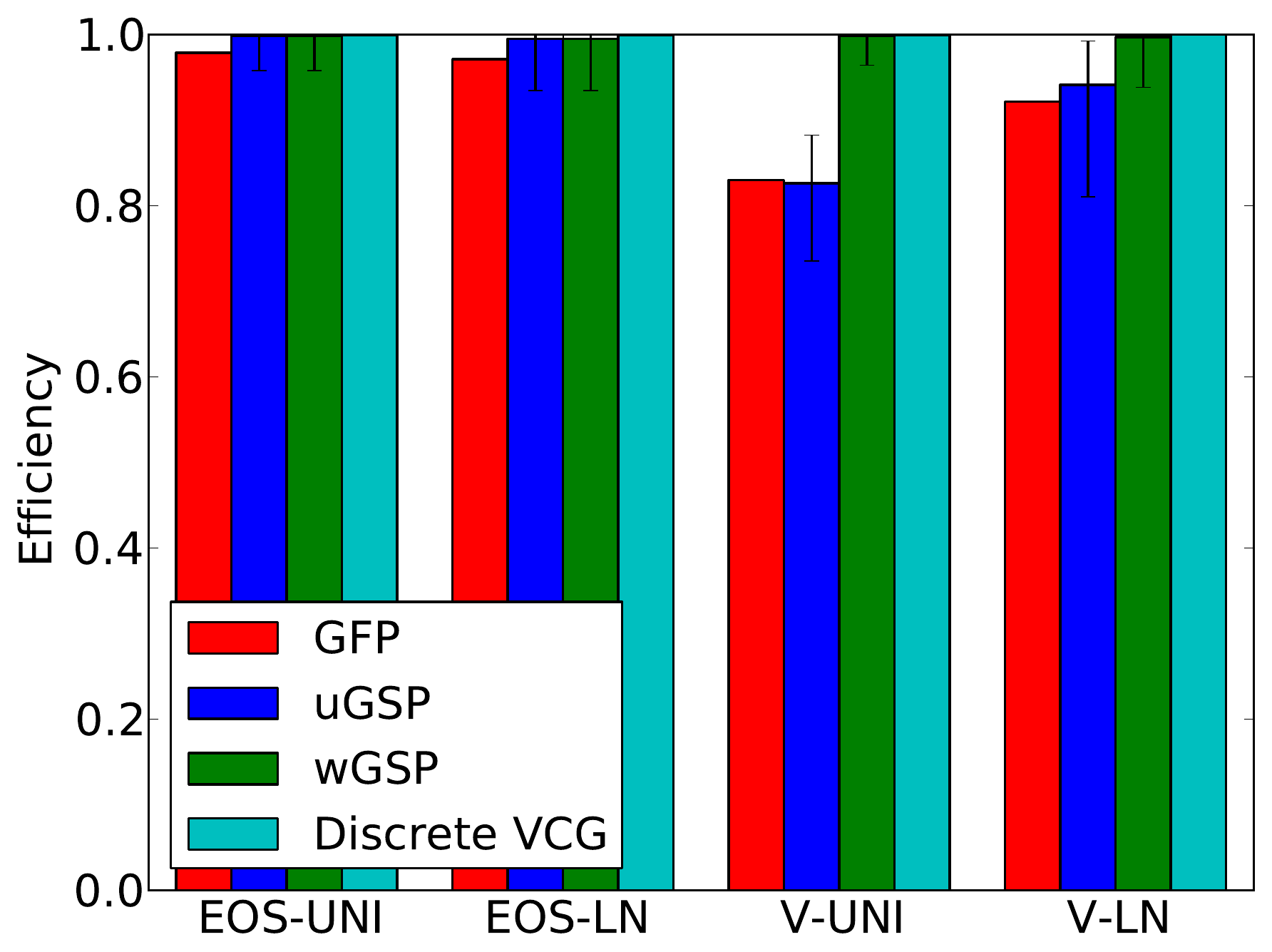}
\includegraphics[width=0.45\hsize]{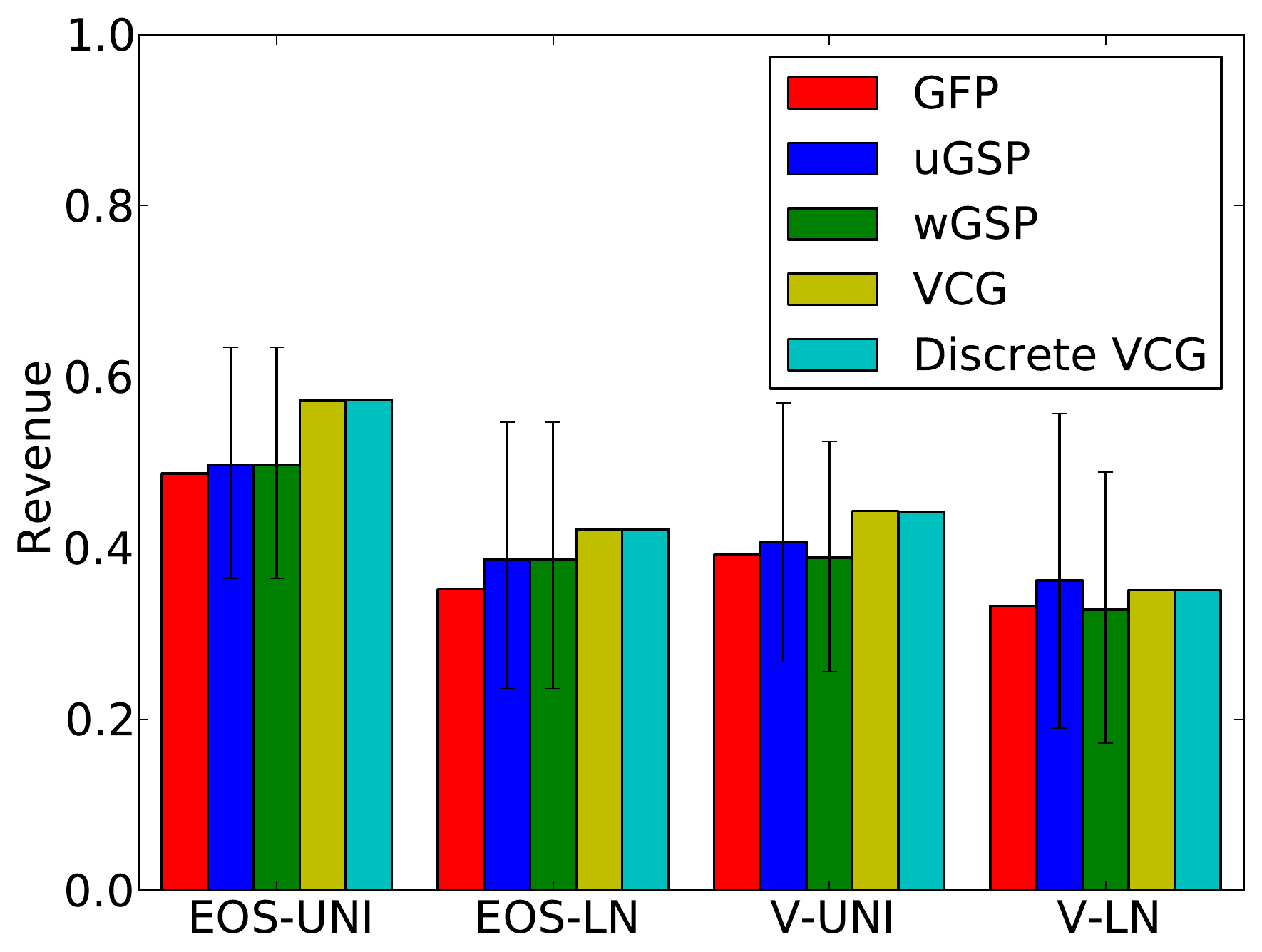}
\kcaption{(a) Efficiency}{.45\hsize}
\kcaption{(b) Revenue}{.45\hsize}
\includegraphics[width=0.45\hsize]{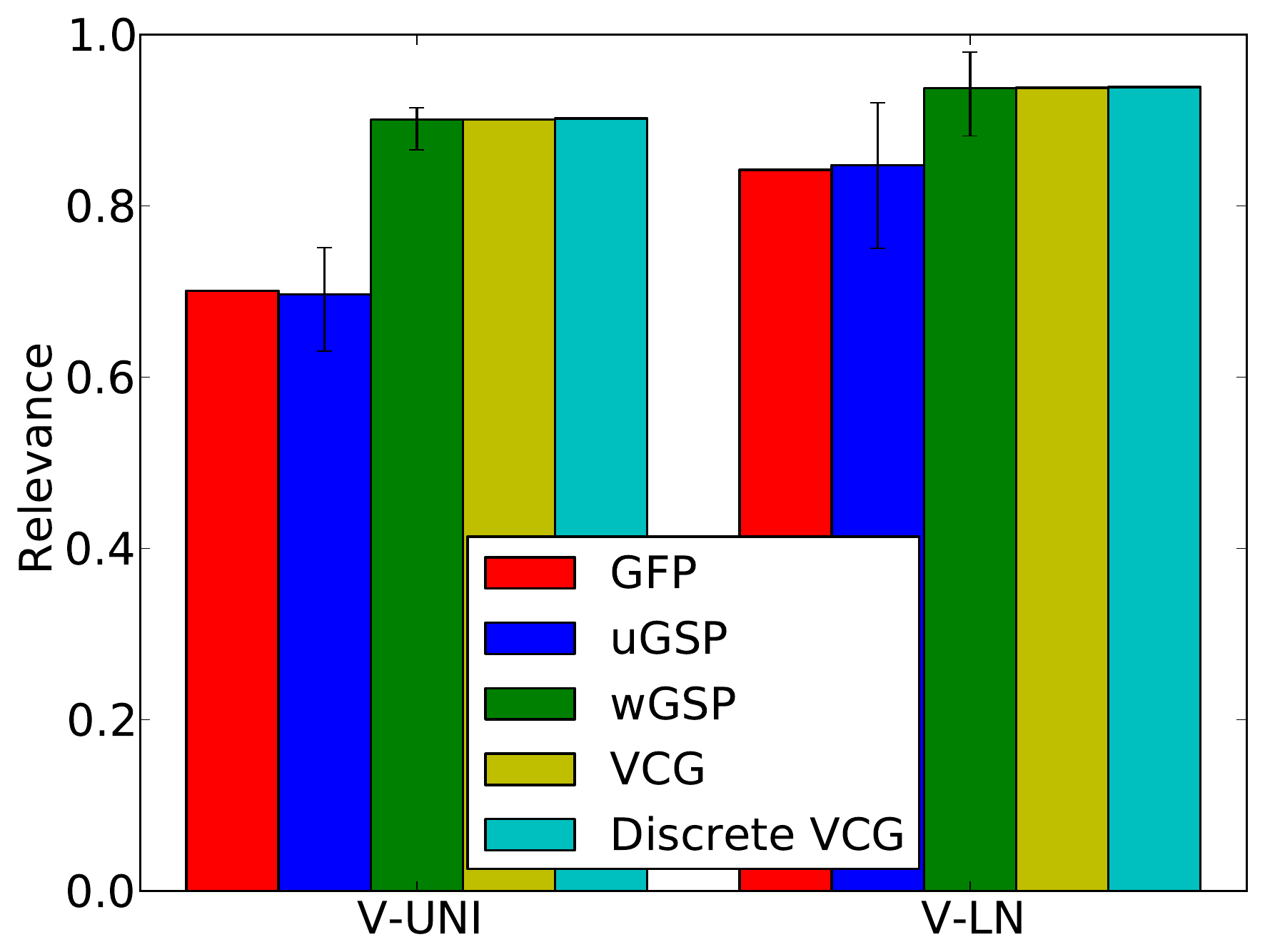}
\includegraphics[width=0.45\hsize]{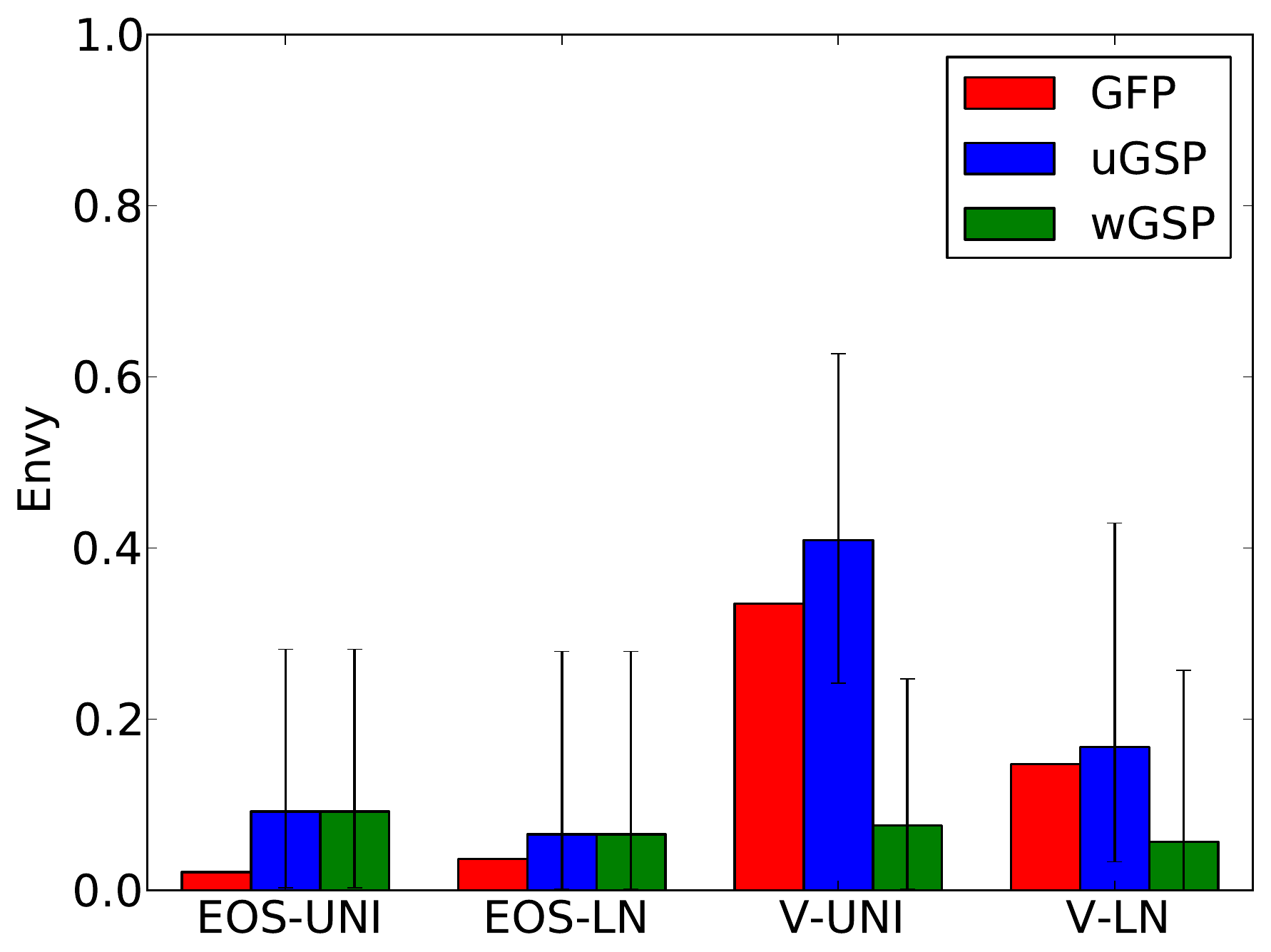}
\kcaption{(c) Relevance}{.45\hsize}
\kcaption{(d) Envy}{.45\hsize}
\vspace{-.2em}\caption[Comparing the average performance of different position auction types in EOS and V settings]{Comparing the average performance of different position auction types in EOS and V settings.  \whiskerBP}
\label{fig:veos-compare}
\end{figure}

\enlargethispage{.95em}
Finally, we come to the overarching question of this  \paperonly{paper}\chapteronly{chapter}, ``which position auction design is best?''
We conducted a general 4-way comparison between GFP, uGSP, wGSP and VCG for each metric.  In terms of social welfare, we found that wGSP was clearly the most efficient position design in the V model, regardless of equilibrium selection.  In the EOS model, u/wGSP was more efficient than GFP in best- and median-case equilibria, but not in the worst case (see Figure~\ref{fig:veos-compare}(a)). Comparing revenue, we found that equilibrium-selection effects were very large for GSP auctions---much larger than differences between auction designs (see Figure~\ref{fig:veos-compare}(b)). In terms of relevance (i.e., expected number of clicks),\footnote{We do not consider the EOS model for relevance: because it is unweighted, it predicts that every complete allocation---where all positions are sold---will produce the same expected number of clicks.} we see that wGSP was consistently better than the other designs and roughly comparable to VCG (see Figure~\ref{fig:veos-compare}(c)). Finally, treating envy as an objective to minimize, in EOS settings we see that in GFP performed very well in all equilibria, while GSP designs were more sensitive to equilibrium selection. In the V model, unweighted designs performed very poorly compared to wGSP (see Figure~\ref{fig:veos-compare}(d)).

\subsubsection{Weighted GFP}\label{sec:weighted-GFP}

Under the EOS model, we found that GFP had high efficiency in its worst-case equilibria, compared to the other two position auctions.  This finding nicely compliments another by Chawla and Hartline \cite{CH13unique}: GFP's worst-case Bayes-Nash equilibrium is efficient in games with independent, identically distributed private values.  Because the EOS model does not include bidder-specific click probabilities, GFP's unweighted allocation is not a problem.  We thus wondered whether adding weights to GFP would produce a mechanism with good worst-case Nash equilibria under the V model, which does include bidder-specific click probabilities.  Recall that it is straightforward to include weights in our AGG reduction for GFP position auctions. We performed two experiments, one at the same scale as our others (where enumerating all mixed-strategy Nash equilibria (MSNEs) of weighted GFP is computationally infeasible) and one at a scale small enough to permit such enumeration.

\begin{figure}
\centering
\includegraphics[width=0.45\hsize]{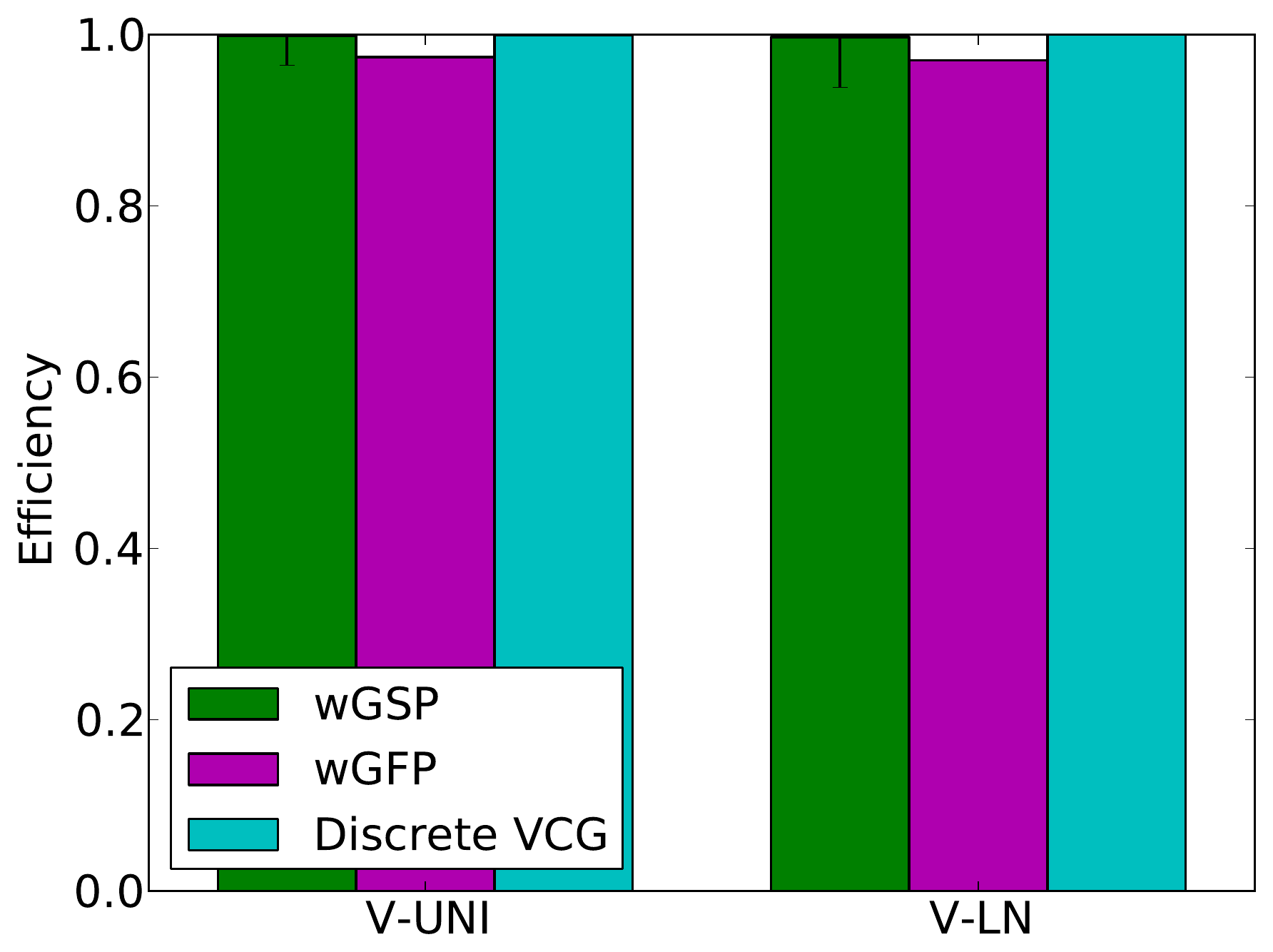}
\includegraphics[width=0.45\hsize]{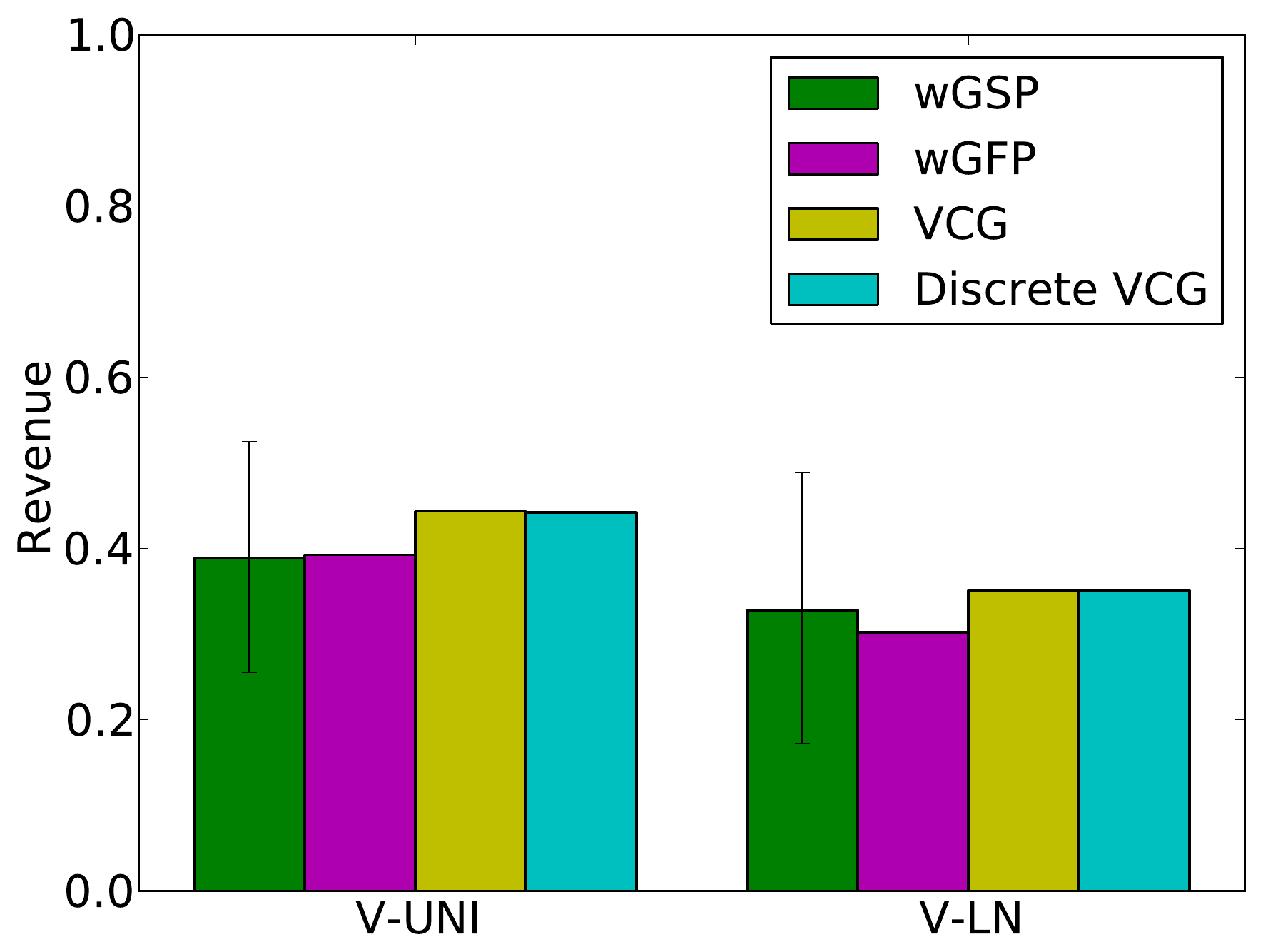}
\kcaption{(a) Efficiency}{.45\hsize}
\kcaption{(b) Revenue}{.45\hsize}
\caption[Comparing wGFP to wGSP]{Comparing wGFP to wGSP.  We found that wGFP's efficiency varied extremely little across equilibria, but that its worst case was not better than wGSP's.  wGFP's revenue again varied little, with  worst-case revenue much larger than wGSP's worst case and roughly comparable to wGSP's median case.}
\label{fig:wgfp}
\end{figure}

In the smaller experiment, we found that 98.5\% of games had a single mixed Nash equilibrium.  When multiple equilibria existed, they tended all to have nearly identical revenue and social welfare (within $4\%$ and $1\%$ respectively).
In our full-sized experiments, we also observed little variability across equilibria.  In 97\% of games, revenue and social welfare varied less than 0.1\% from the best case to the worst case. In the remaining 3\% of games, social welfare never differed by more than 0.3\% across  equilibria while revenue never differed by more than 3.5\%.   Comparing wGFP to wGSP, we found that wGFP \emph{did not} yield significantly better worst-case efficiency.  However, wGFP did achieve much higher worst-case revenue, roughly equal to wGSP's median case  (see Figure~\ref{fig:wgfp}).

\subsection{Position-Preference Models}

We now consider the richer BHN and BSS models, which allow advertisers' values to depend on ad positions.

\subsubsection{The BHN Model}

Blumrosen, Hartline and Nong's \cite{BHN08nonuniform} key motivation was that advertisers may prefer
clicks in the lowest position, because the few users who actually click on low-position ads are likely to buy.  They found that wGSP sometimes has no efficient Nash equilibrium under their model.

\qiii\vspace{\modelspace}

\begin{figure}
	\centering
	\includegraphics[width=0.45\hsize]{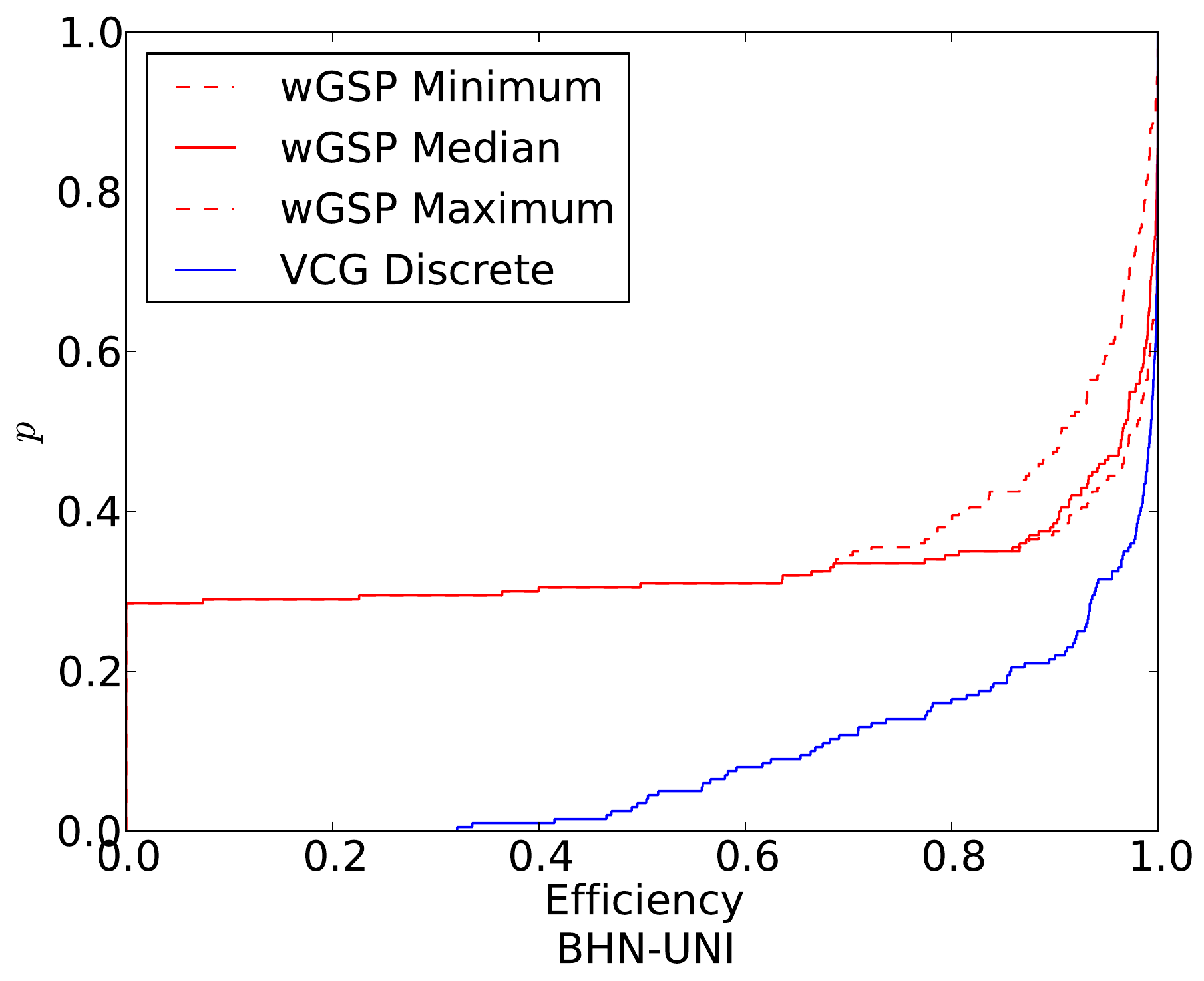}
	\includegraphics[width=0.45\hsize]{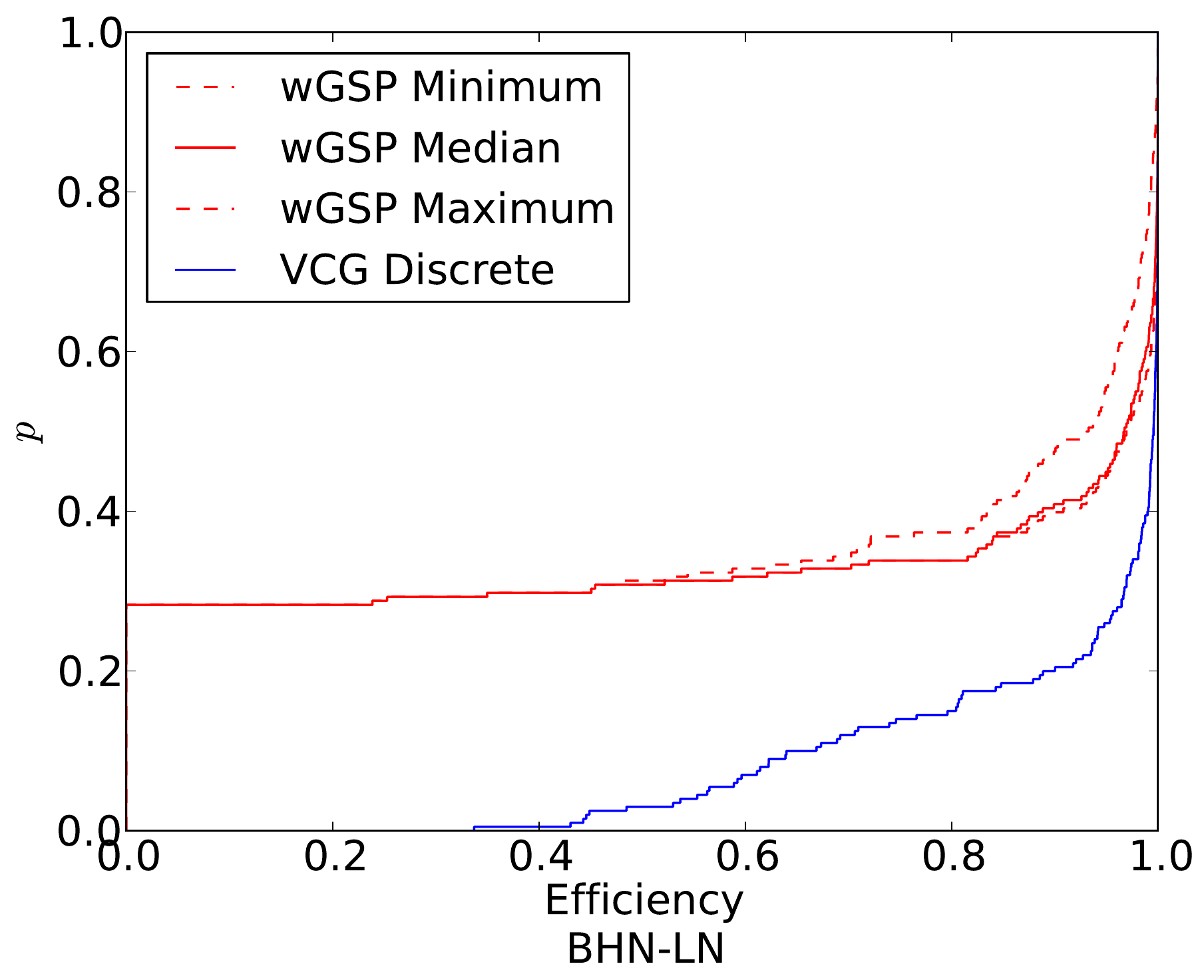}
	\kcaption{(a) Uniform Distribution}{.45\hsize}
	\kcaption{(b) Log-Normal Distribution}{.45\hsize}
	\caption[Empirical CDF of economic efficiency in BHN models]{Empirical CDF of economic efficiency in BHN models.}
	\label{fig:bhn-welfare-cdfs}
\end{figure}

We observed two kinds of efficiency failures  (see Figure~\ref{fig:bhn-welfare-cdfs}). The first kind involved a complete failure to allocate, with every agent bidding zero.  This equilibrium arises when the top-most position has so many spurious (non-converting) clicks that paying even one increment per click is too much.  (This scenario also leads to an interesting outcome under discrete VCG: the advertiser who gets the top position can actually be paid, since the externality he imposes on other bidders is to allow them to reach the higher-value lower positions.)  A similar phenomenon can arise in lower positions as well: e.g., if exactly one advertiser values the second position (and by monotonicity, the first as well) at weakly more than one bid increment, then in any pure equilibrium all the other bidders bid zero.
The second kind of efficiency failure involves mis-ordering the advertisers (e.g., placing the highest-valuation advertiser in some slot other than the highest), as we already saw in the V and EOS settings.
However, the magnitude of inefficiency can be more than we saw in those models, because different positions have different conversion rates.  In equilibria of EOS and V models, the high-valuation bidder might prefer to avoid the top position because the top positions' greater numbers of clicks are offset by disproportionately greater prices per click.  In the BHN model, top-position clicks can also be too expensive, and this is compounded by the fact that top-position clicks are least likely to convert.
Discretization can exacerbate this problem. Because per-click valuations for the top positions are small compared to lower positions, discretization has a greater effect on top positions, leading to inefficiencies due to rounding. Because high positions get the majority of clicks, this causes a large loss of welfare.

\begin{figure}
\centering
\includegraphics[width=0.45\hsize]{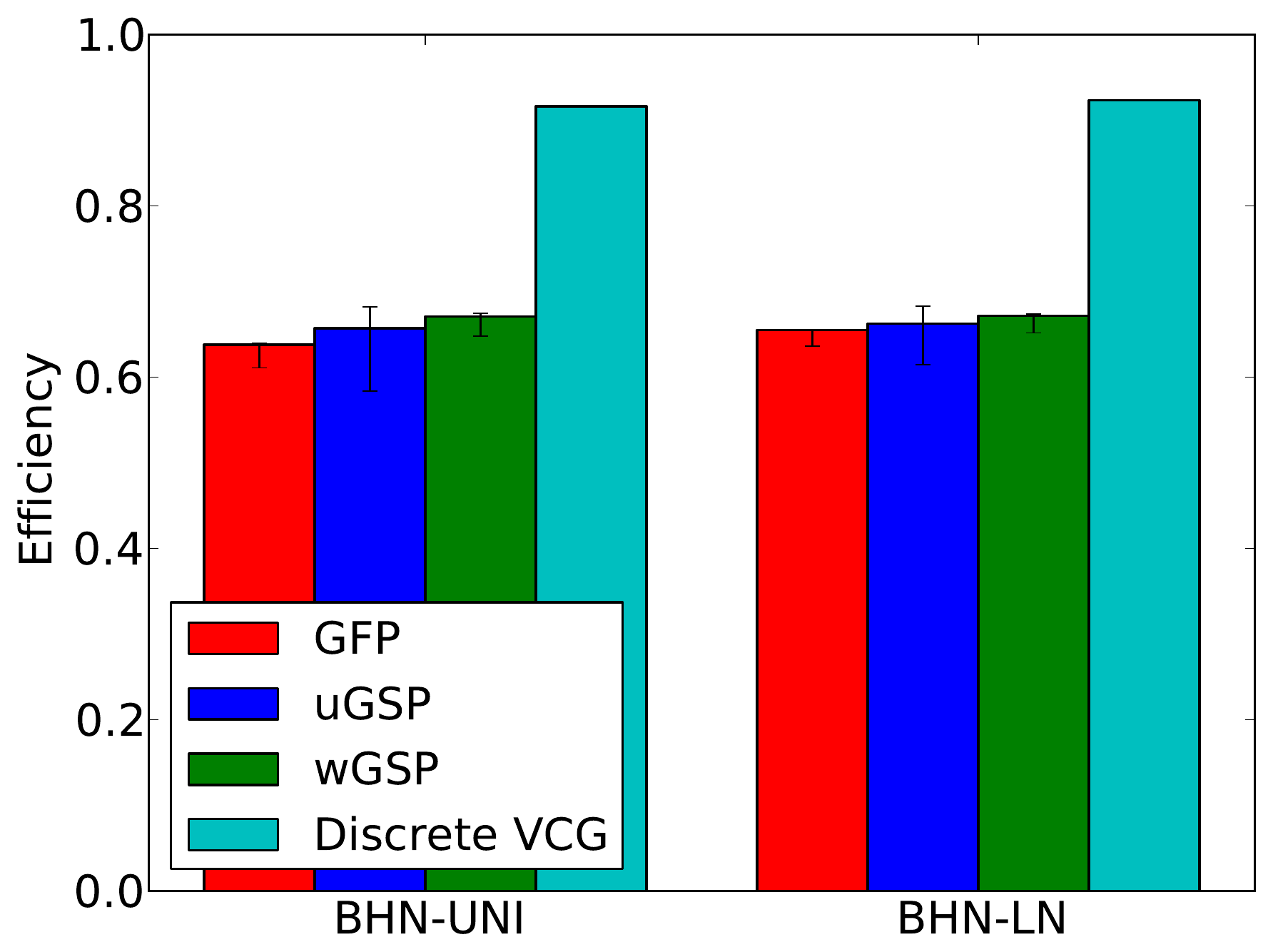}
\includegraphics[width=0.45\hsize]{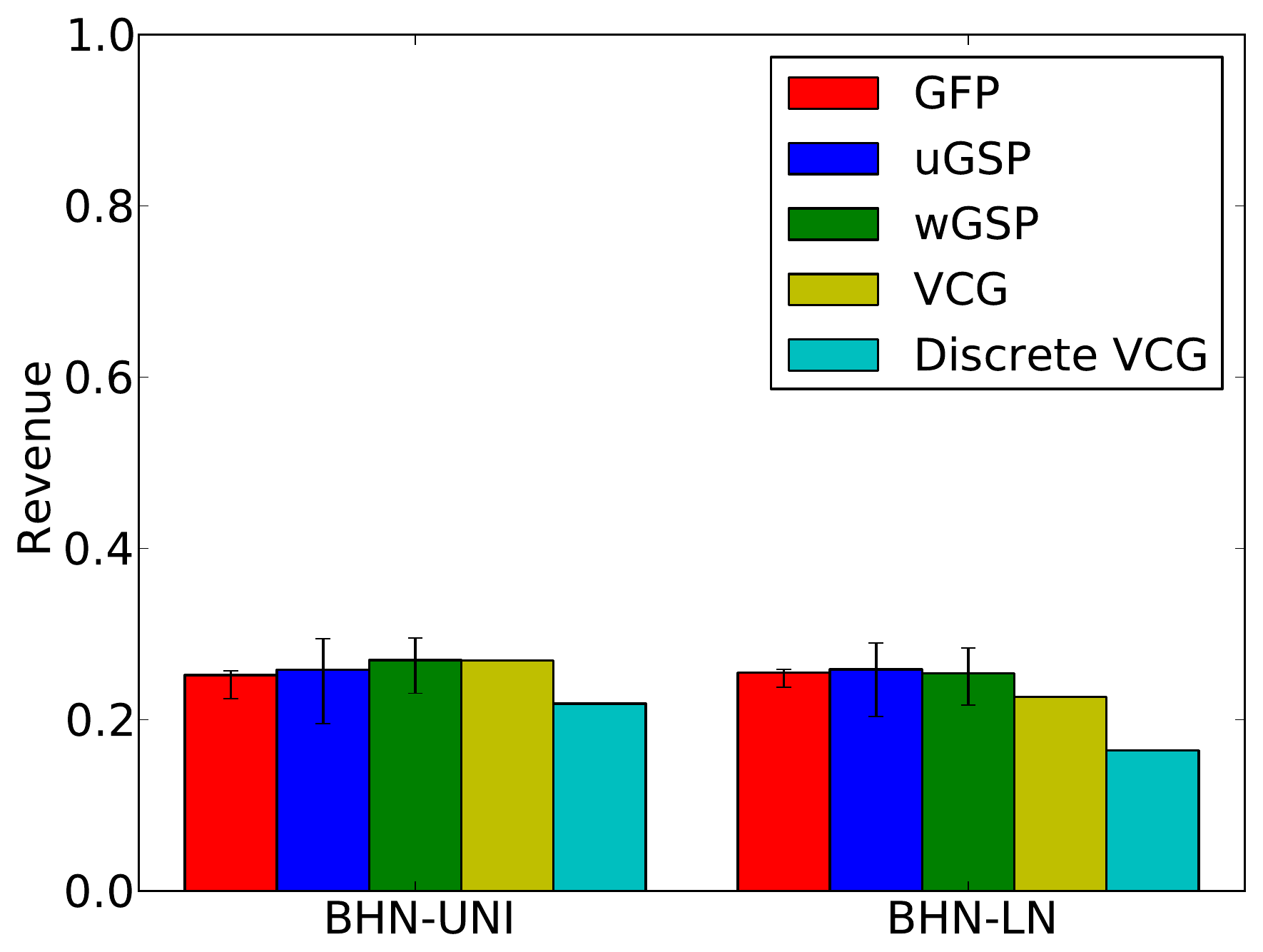}
\kcaption{(a) Efficiency}{.45\hsize}
\kcaption{(b) Revenue}{.45\hsize}
\includegraphics[width=0.45\hsize]{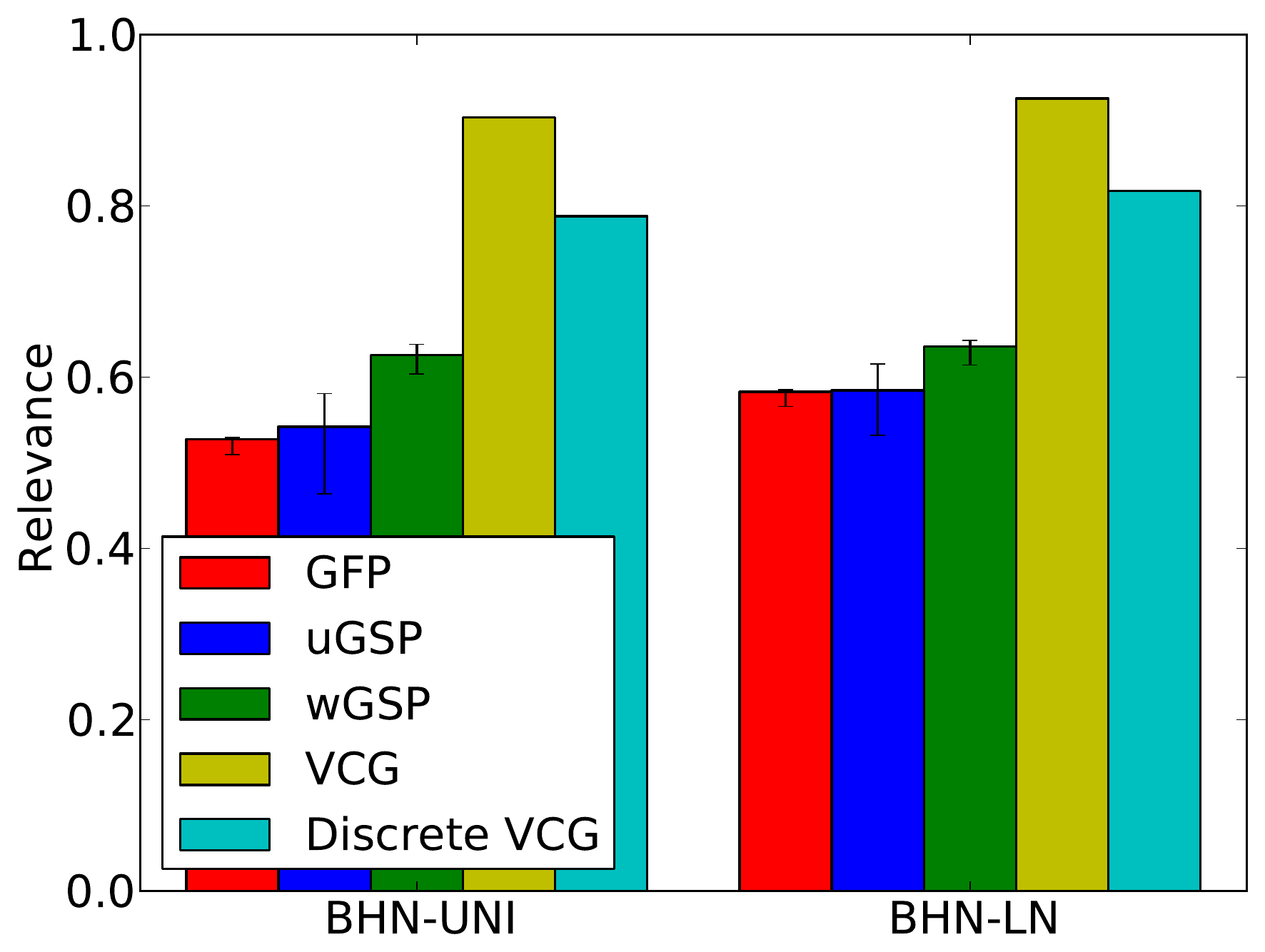}
\includegraphics[width=0.45\hsize]{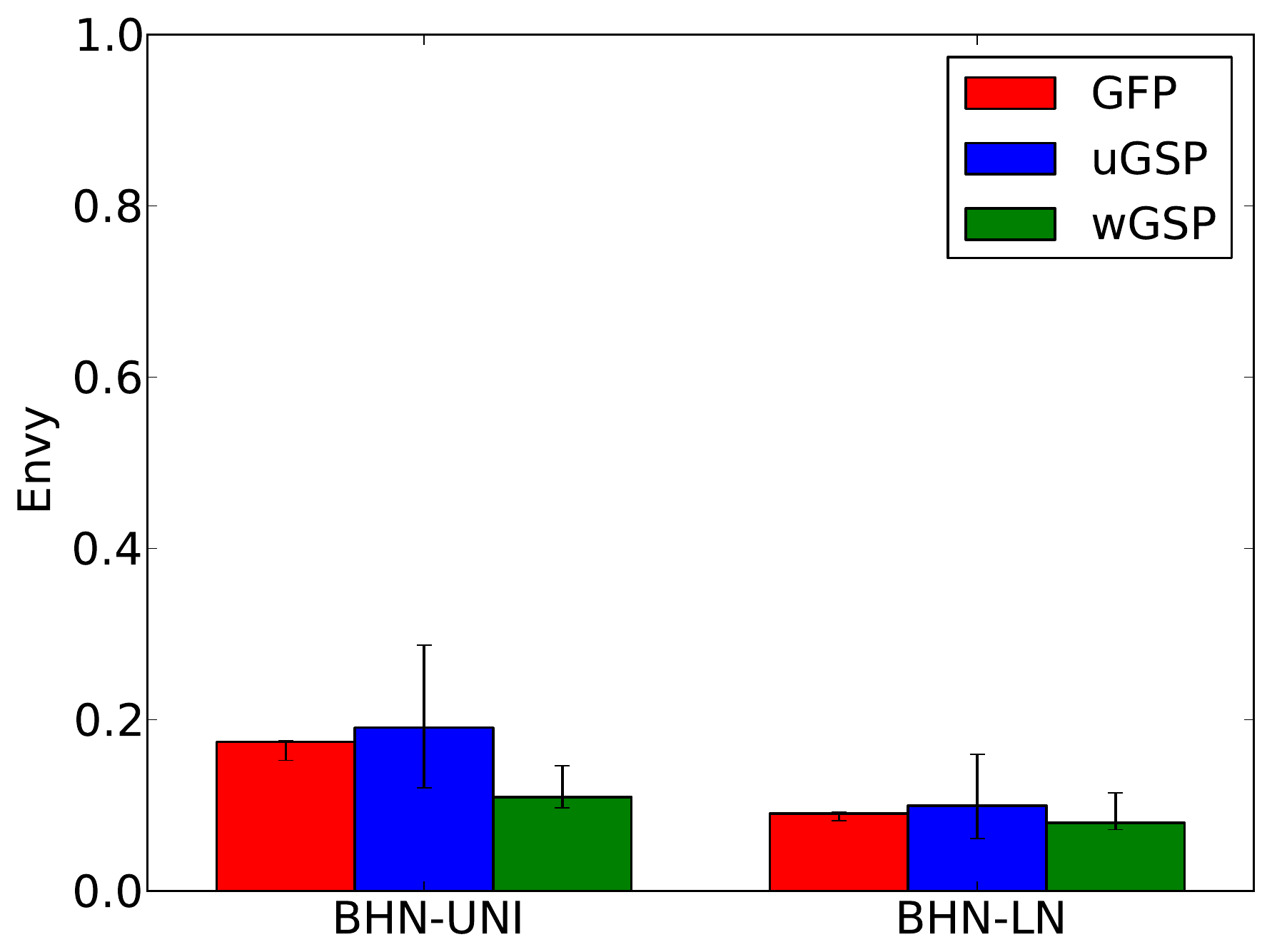}
\kcaption{(c) Relevance}{.45\hsize}
\kcaption{(d) Envy}{.45\hsize}
\caption[Comparing the average performance of different position auction types in BHN settings]{Comparing the average performance of different position auction types in BHN settings.  \whiskerBP}
\label{fig:bhn-comparison}
\end{figure}

We now consider how different auction designs performed on BHN (see Figure~\ref{fig:bhn-comparison}). First, we observed that all position auctions designs were very inefficient ($\leq 69\%$), and substantially less efficient  than discrete VCG (which achieved $91.6\%$ and $90.5\%$ efficiency in BHN-UNI and BHN-LN respectively.)  wGSP was dramatically better than GFP; wGSP's worst-case equilibria were more efficient that GFP's best case.  The comparison between uGSP and wGSP was less clear, because uGSP's efficiency varied much more across equilibria. In the case of BHN-UNI, this made uGSP's best-case equilibria  slightly (but not significantly) better than wGSP's.  In other distributions and for other criteria, uGSP was clearly worse.

Turning to revenue, we found that no auction design extracted more than $30\%$ of the surplus, but that every position auction design was dramatically better at generating revenue than VCG.
This finding has a straightforward explanation: in BHN models, conversion rates are higher in low positions.
In VCG, an agent pays his externality, i.e., the amount he decreases the welfare of the lower-ranked agents by displacing them one position.  In GSP, an agent pays the amount of the next highest bid, as if he had entirely removed the next highest bidder entirely.  In EOS, V and BHN models, the VCG payment (for a given bid profile) is always weakly less than GSP, because the VCG payment takes into account that next highest agent gets a position of value.  In BHN, the displaced agents are moved into positions that are likely to be closer to the value of pre-displacement positions (compared with EOS and V settings).
We also observe that, as we saw before, revenue variation across equilibria was dramatically larger than revenue variation across position auction designs.  Unlike the V model, under BHN uGSP did not provide dramatically more revenue than wGSP when values were lognormally distributed.

All position auction designs achieved dramatically worse relevance than VCG (both with and without discretization). wGSP was clearly the best position auction in terms of relevance..

Envy varied substantially among both GSP designs, but nevertheless wGSP outperformed the other position auction designs in minimizing this metric.

\subsubsection{The BSS Model}

Next we investigated to the model of Benish, Sadeh and Sandholm \cite{BSS08inexpressive}.  This model is similar to the BHN model in that different advertisers can have different per-click valuations.  However, in this case,  valuations are ``single-peaked'' in the sense that each advertiser has a most preferred position and their per-click valuation decreases as the ad is moved further from that position.  Recall that BSS is an unweighted model (i.e., all the advertisers have identical quality scores and click probabilities).  Thus, wGSP and uGSP are equivalent in this model.

\qiv\vspace{\modelspace}

\begin{figure}[tp]
	\centering
	\includegraphics[width=0.45\hsize]{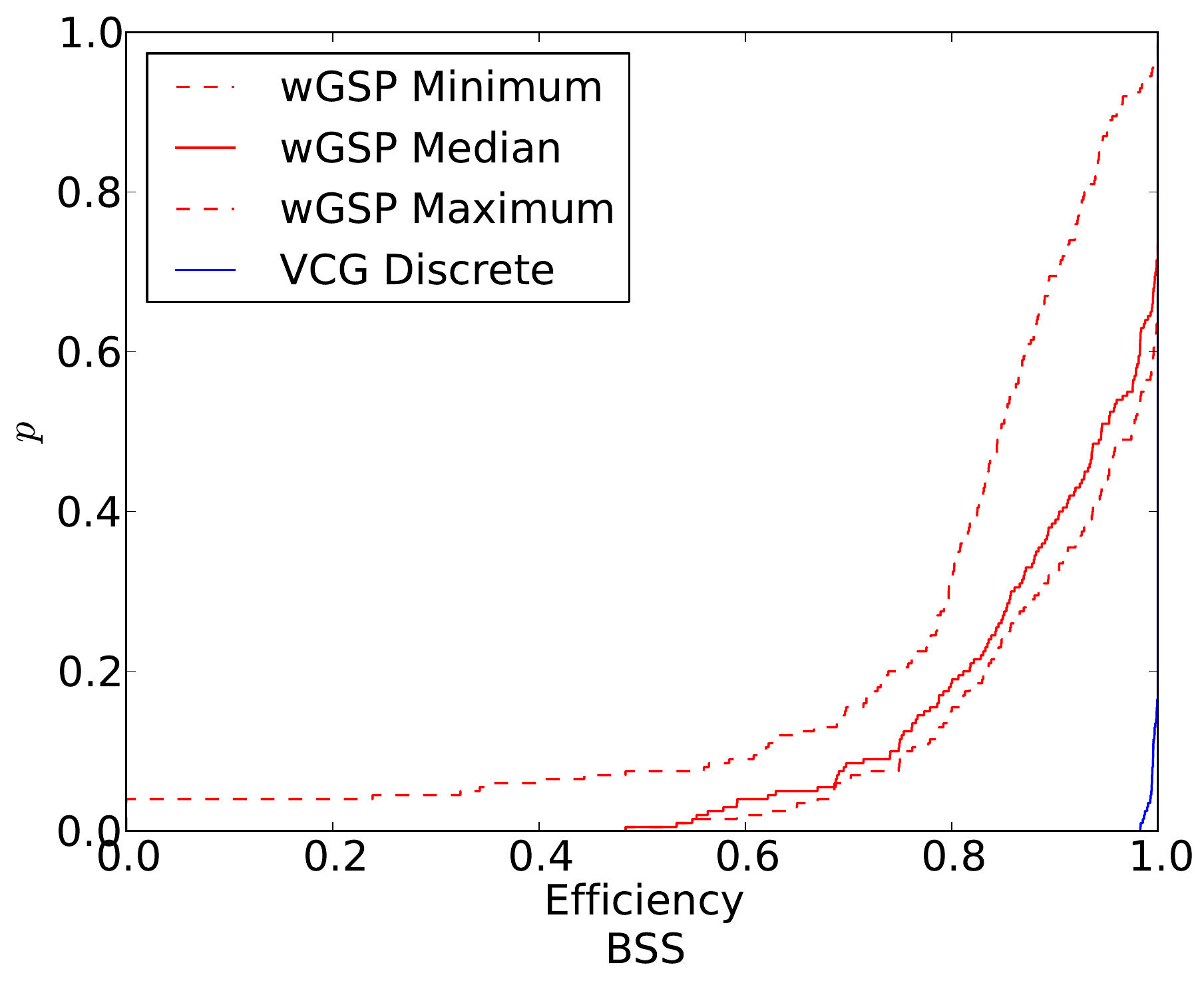}
	\caption{Empirical CDF of economic efficiency in BSS models}
	\label{fig:bss-welfare-cdf}
\end{figure}

We found that efficiency failures were very common under BSS, and could be very large, though zero-efficiency outcomes were rare and never happened in best-case equilibria (see Figure~\ref{fig:bss-welfare-cdf}). This was due in part to randomized tie-breaking.  When none of the agents values the top slot, all the agents can bid zero in equilibrium.  However, they can also all bid a small but nonzero amount and tie for top bid; this leads them all to get more valuable intermediate  positions often enough to justify the cost.

\begin{figure}
\centering
\includegraphics[width=0.45\hsize]{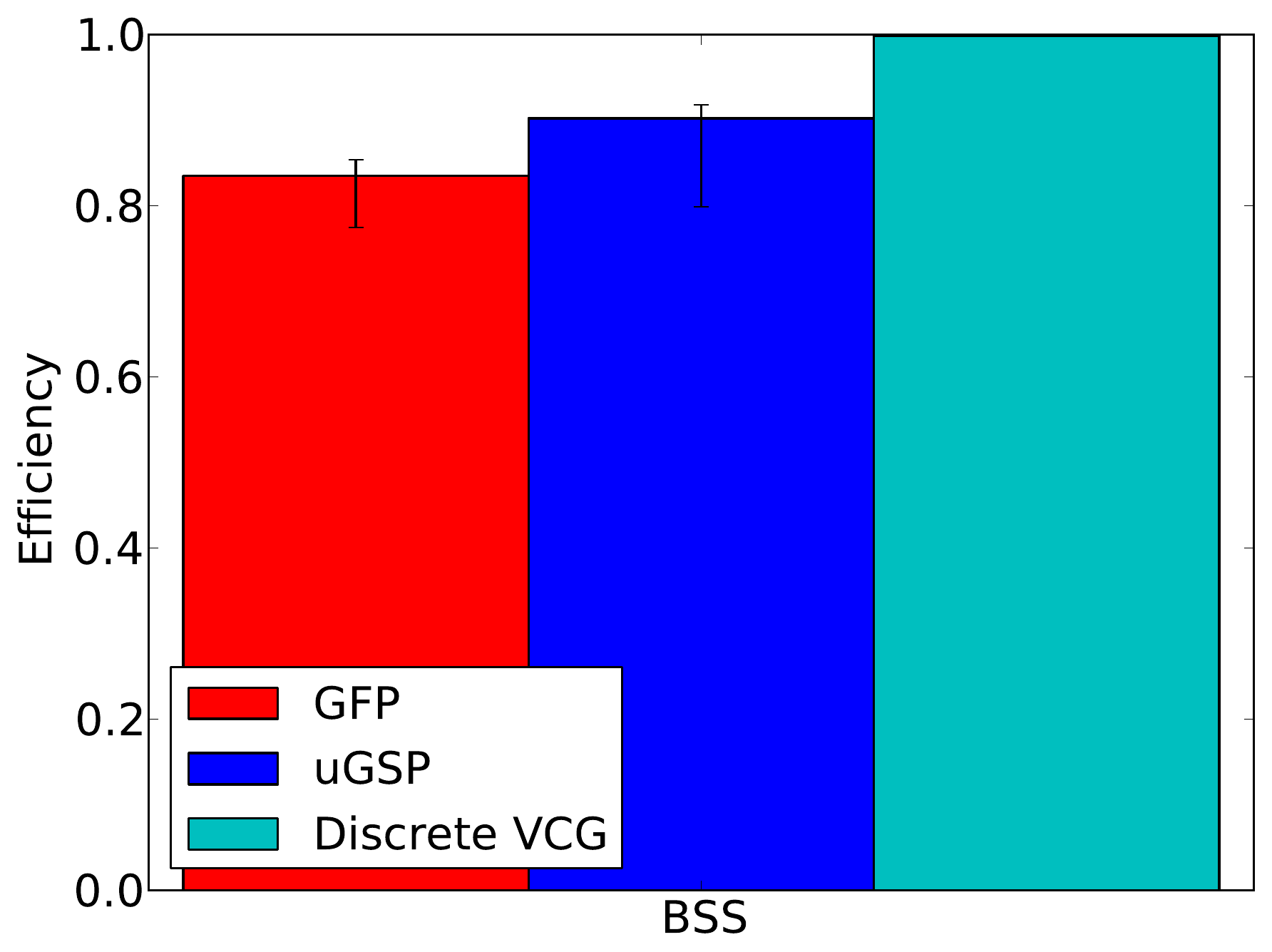}
\includegraphics[width=0.45\hsize]{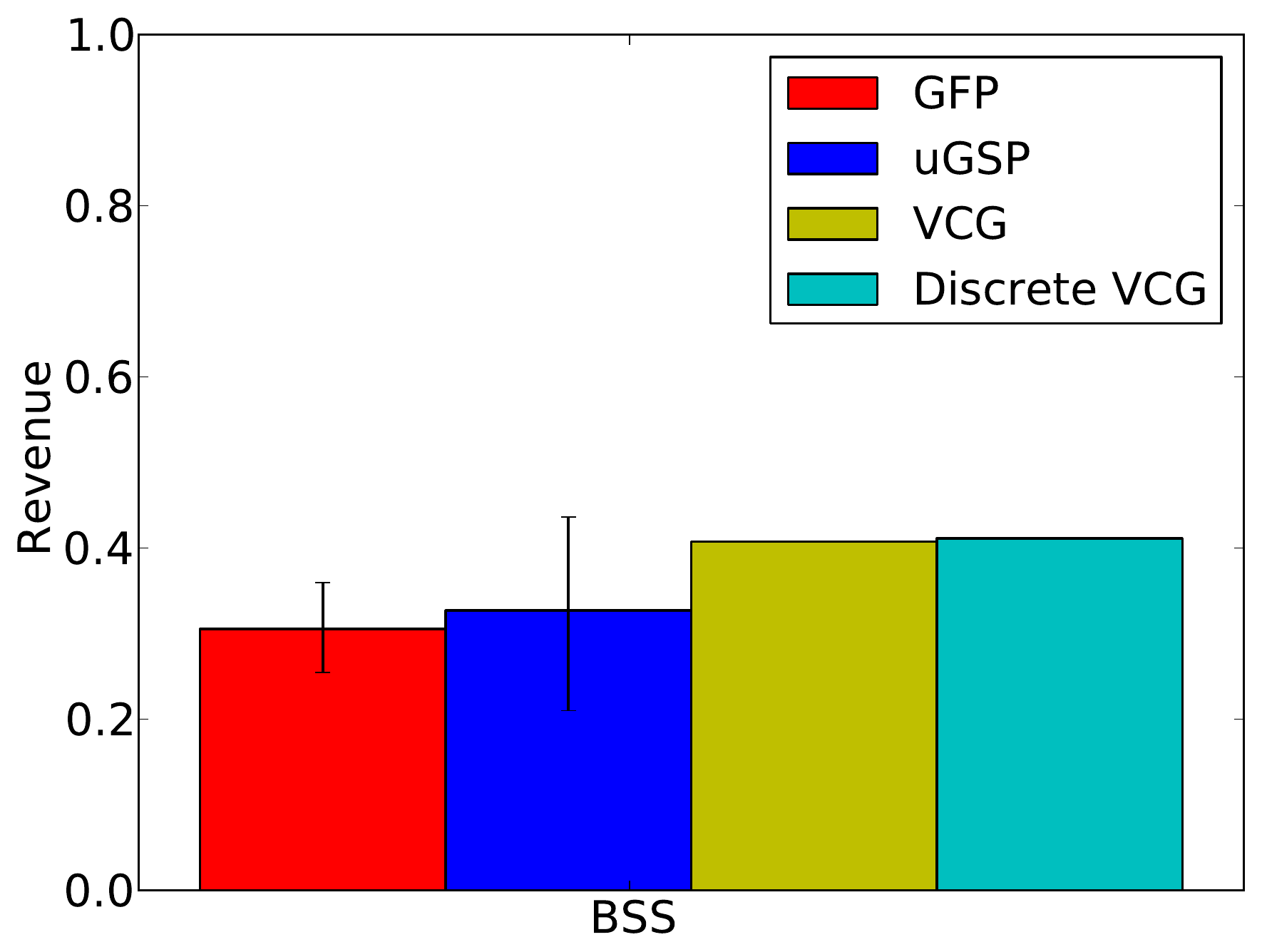}
\kcaption{(a) Efficiency}{.45\hsize}
\kcaption{(b) Revenue}{.45\hsize}
\includegraphics[width=0.45\hsize]{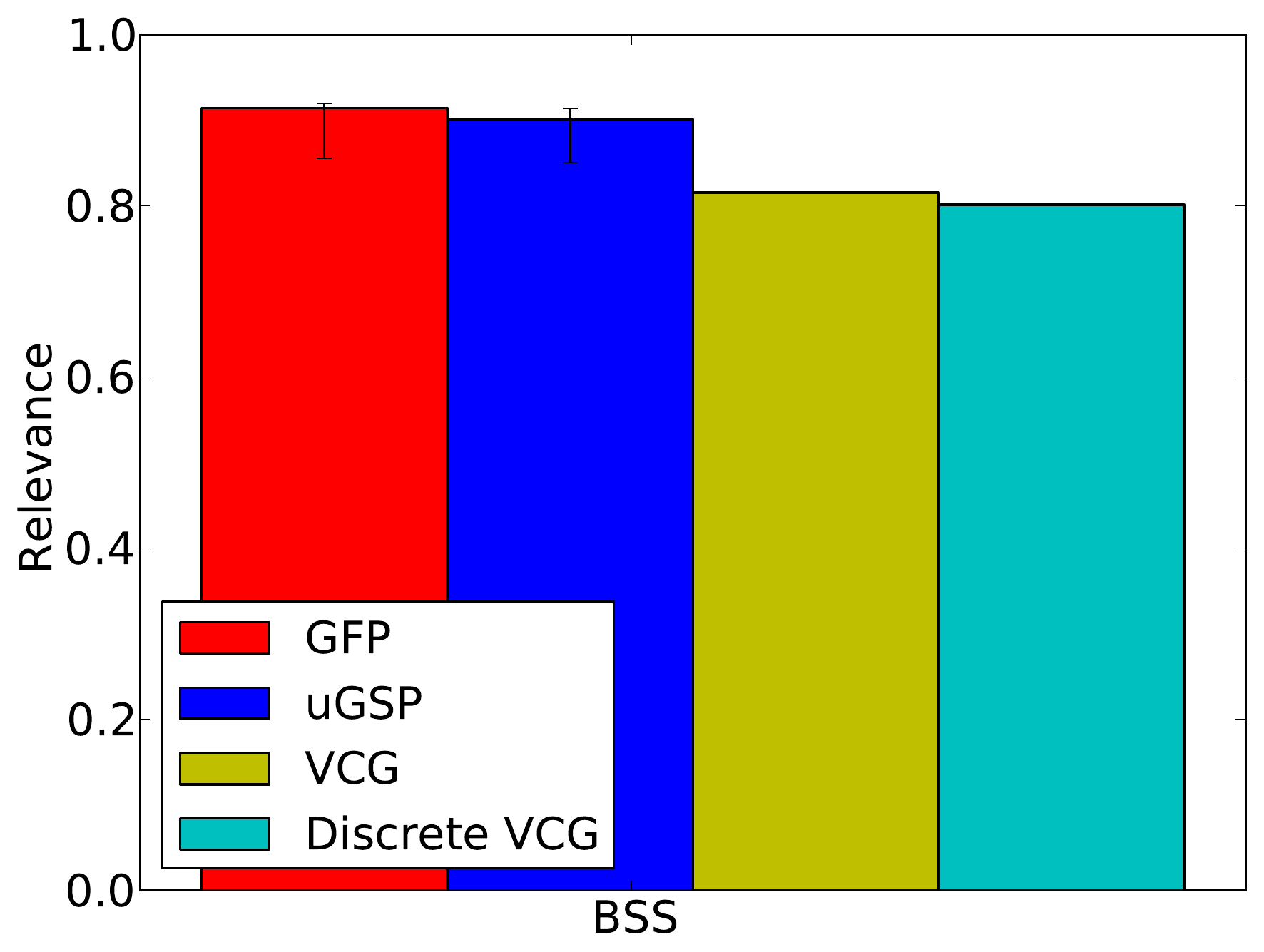}
\includegraphics[width=0.45\hsize]{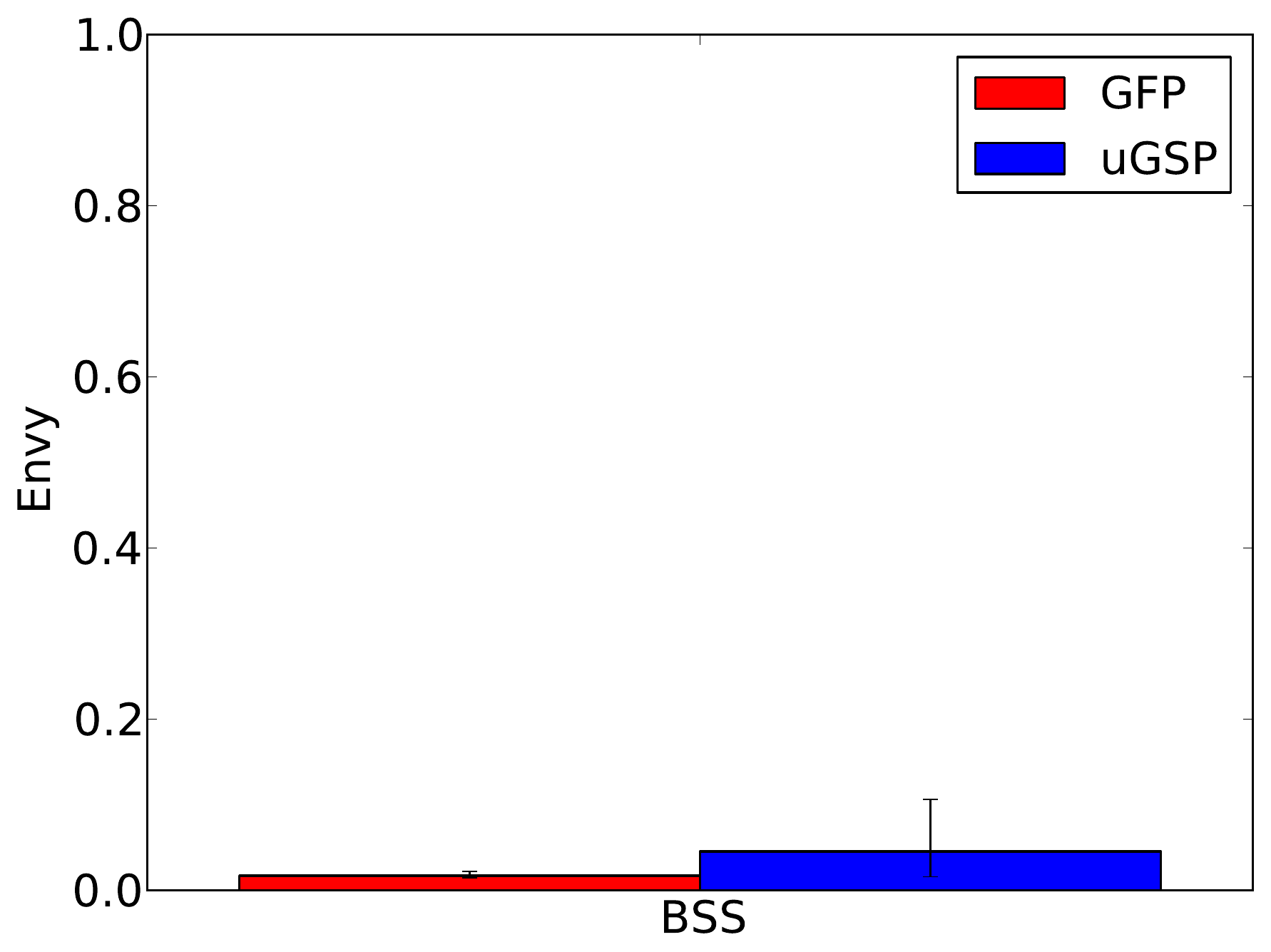}
\kcaption{(c) Relevance}{.45\hsize}
\kcaption{(d) Envy}{.45\hsize}
\caption[Comparing the average performance of different position auction types in BSS settings.]{Comparing the average performance of different position auction types in BSS settings. \whiskerBP}
\label{fig:bss-comparison}
\end{figure}

Comparing auction designs (see Figure~\ref{fig:bss-comparison}),
 we observed that uGSP was significantly more efficient than GFP, but fell well behind discrete VCG.  In revenue, uGSP and GFP both varied substantially across equilibria, and both were significantly worse than VCG in their median equilibria.  uGSP slightly outperformed VCH in best-case equilibrium, but not significantly so.   Both uGSP and GFP achieved more relevant outcomes than VCG, largely due to the tieing effect described earlier, where agents were often sold clicks that they did not particularly value.  Both auctions had little envy in absolute terms, though GSP was worse than GFP.

\chapteronly{\clearpage}

\subsection{Externality Models}

We now consider Cascade, Hybrid and GIM, three models in which bidders can care about the positions awarded to other agents.

\subsubsection{The Cascade Model}

Cascade is the simplest of our models involving externalities.  Cascade settings are similar to V settings in that each bidder has a quality (proportional to click probability) and a per-click valuation.  The difference is that each advertiser also has a continuation probability: the probability that a user looks at subsequent ads.  Thus, in both models, lower positions are less valuable because they receive fewer clicks.  However, in cascade models, this reduction depends on what is shown in higher positions.

\qv\vspace{\modelspace}

\begin{figure}
	\centering
	\includegraphics[width=0.45\hsize]{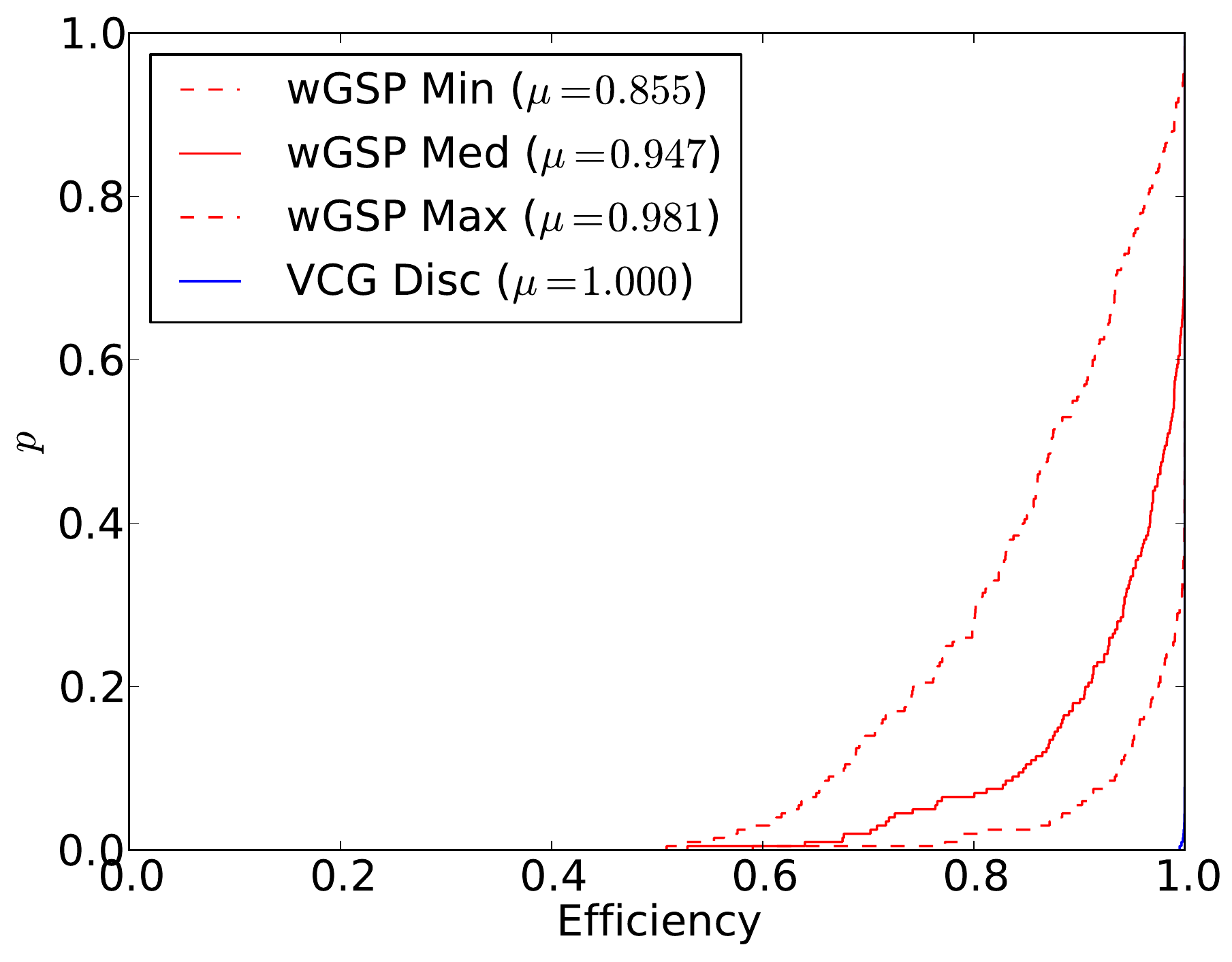}
	\includegraphics[width=0.45\hsize]{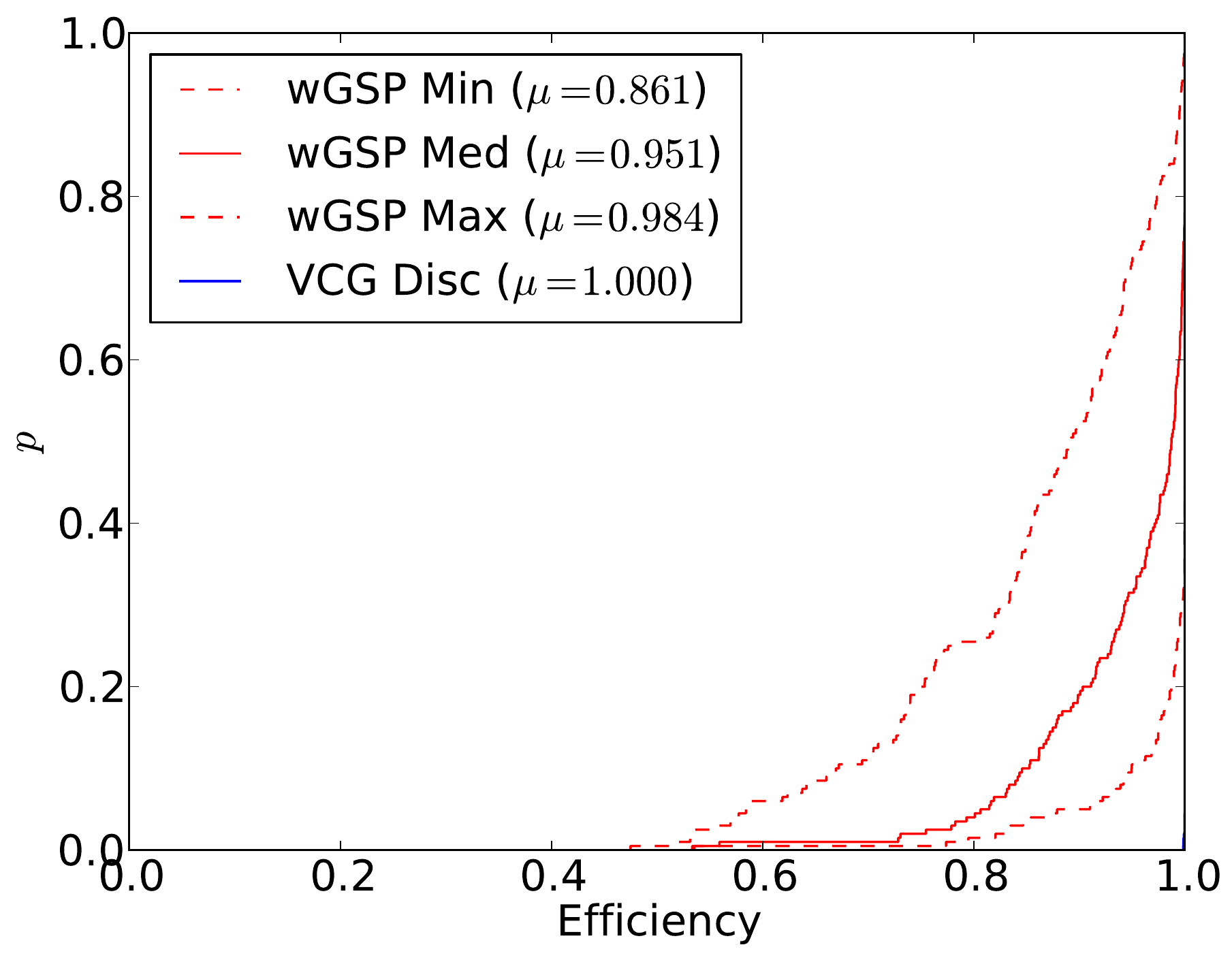}
	\kcaption{(a) Uniform Distribution}{.45\hsize}
	\kcaption{(b) Log-Normal Distribution}{.45\hsize}
	\caption{Empirical CDF of economic efficiency in the cascade models.}
	\label{fig:cas-welfare-cdf}
\end{figure}

\begin{figure}
\centering
\includegraphics[width=0.45\hsize]{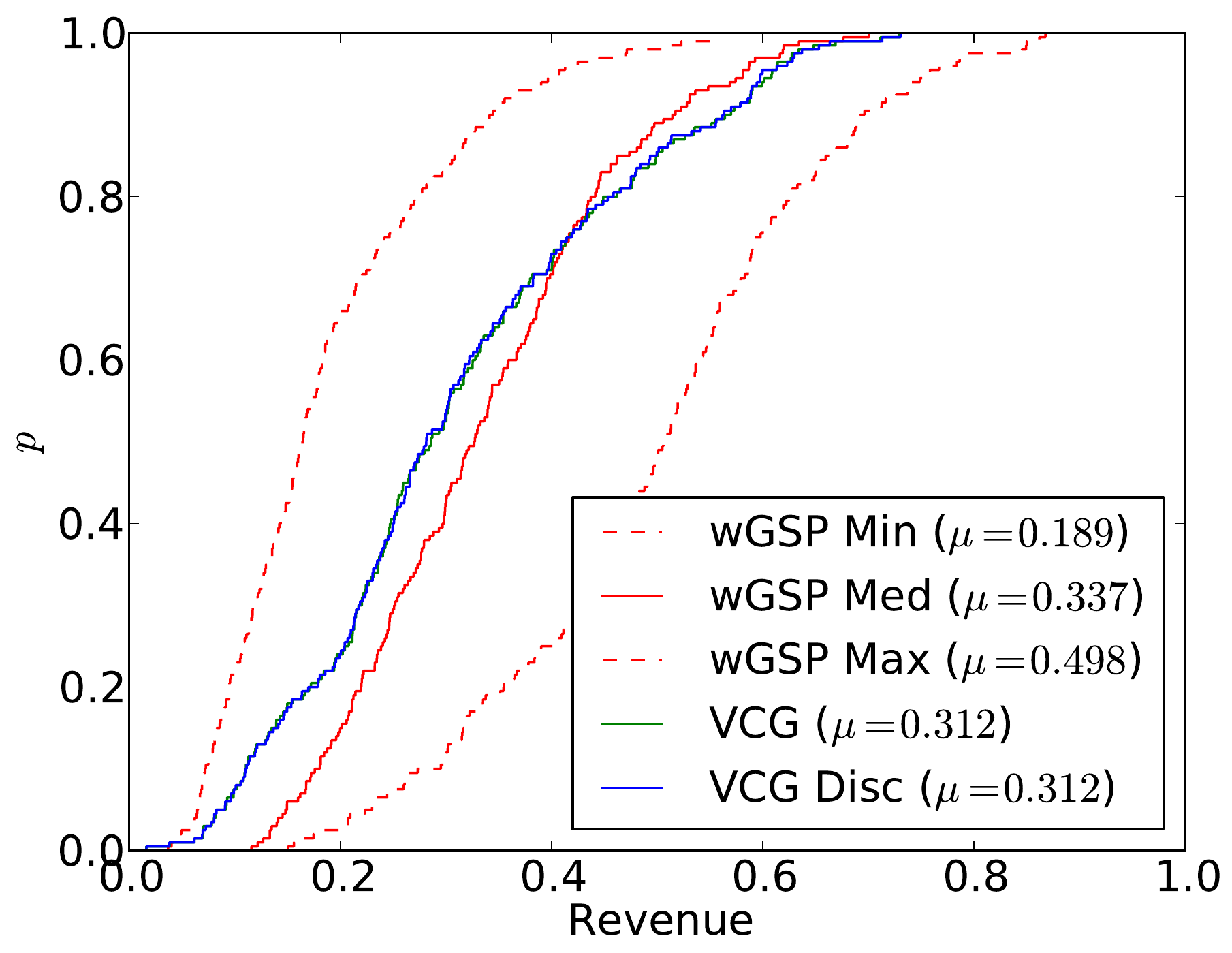}
\includegraphics[width=0.45\hsize]{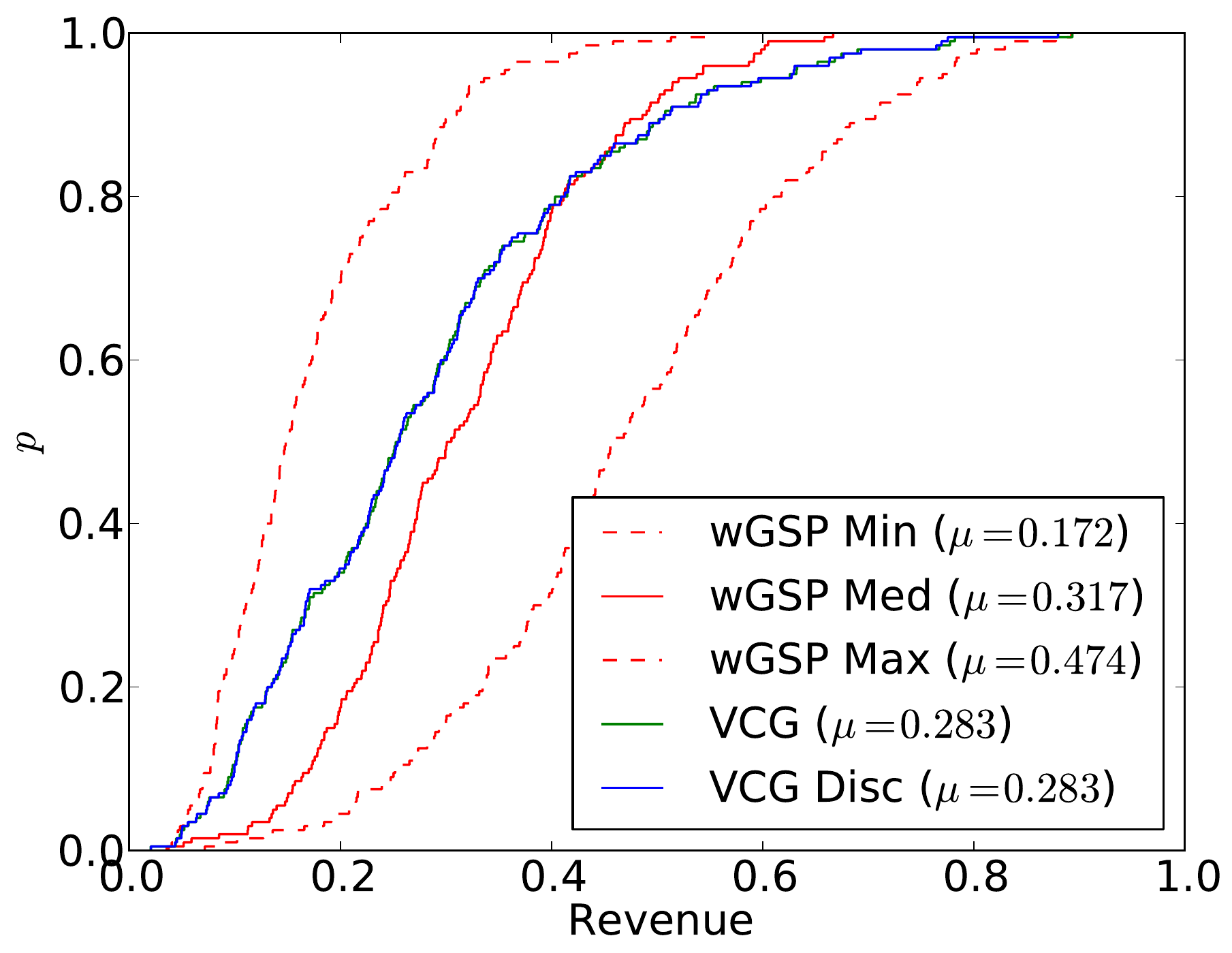}
\kcaption{(a) Uniform Distribution}{.45\hsize}
\kcaption{(b) Log-Normal Distribution}{.45\hsize}
\caption{Empirical CDF of revenue in the cascade models.}
\label{fig:cas-revenue-cdf}
\end{figure}

We found that inefficiency was common, though the magnitudes of efficiency losses were smaller than we expected (see Figure~\ref{fig:cas-revenue-cdf}).  We found that wGSP's efficiency was always greater than $50\%$ and was almost always greater than $80\%$.  In contrast, Giotis and Karlin \cite{GK08cascade} showed that the price of anarchy is $k$ and thus, for $k=5$ there exist instances (which we evidently did not sample) with as little as $20\%$ efficiency.\footnote{Although they studied the more general model that we refer to as hybrid, the construction of their worst-case example uses a part of the hybrid model space that is also consistent with the more restrictive cascade model.}
The gap between best- and worst-case equilibria was typically quite large, though a few instances had poor ($\sim 60\%$) efficiency even in the best case.

wGSP's revenue varied substantially across equilibria (see Figure~\ref{fig:cas-revenue-cdf}), with the best-case equilibrium generating at least $2.5$ times as much revenue as the worst case.

\begin{figure}
\centering
\includegraphics[width=0.32\hsize]{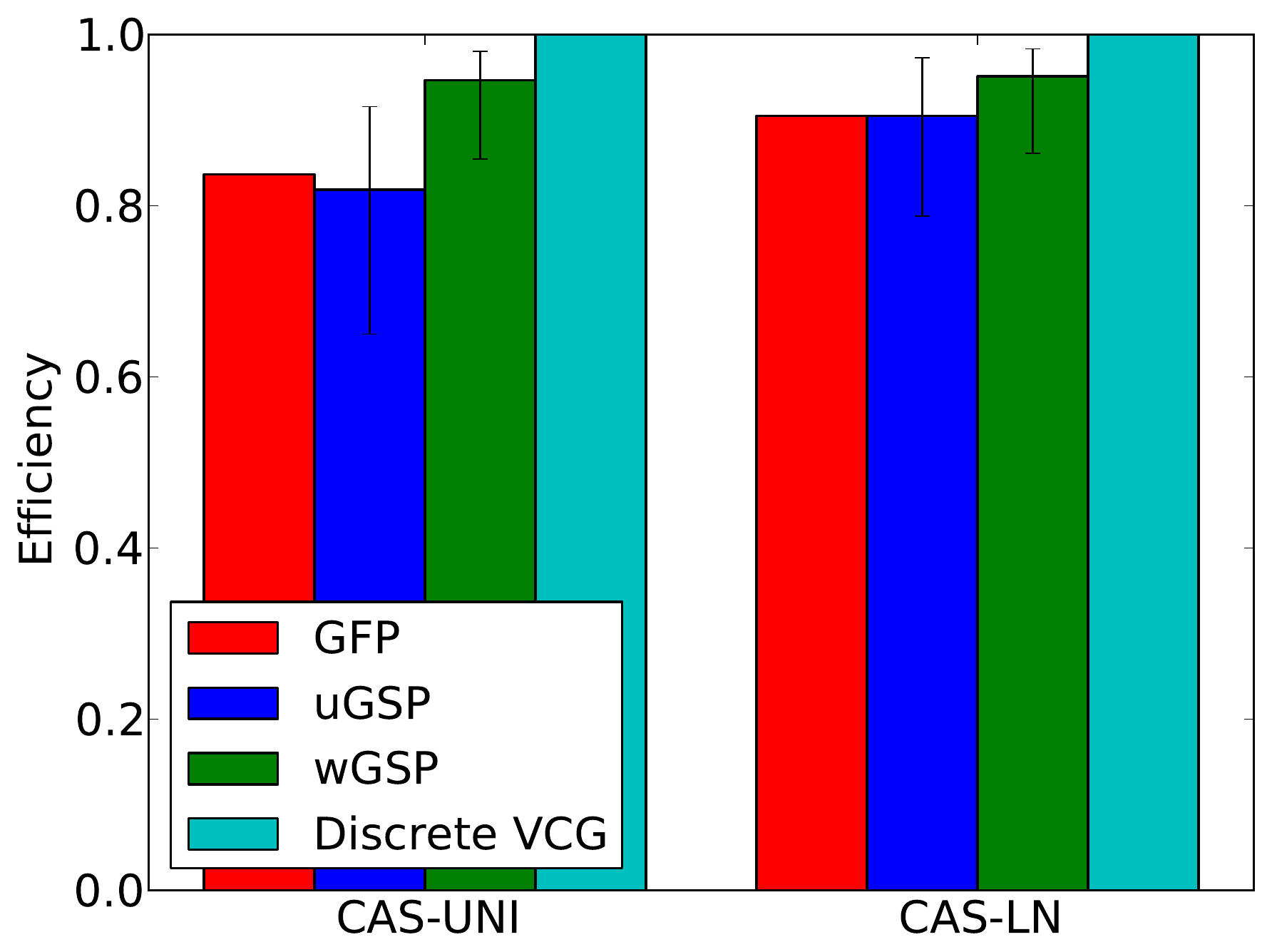}
\includegraphics[width=0.32\hsize]{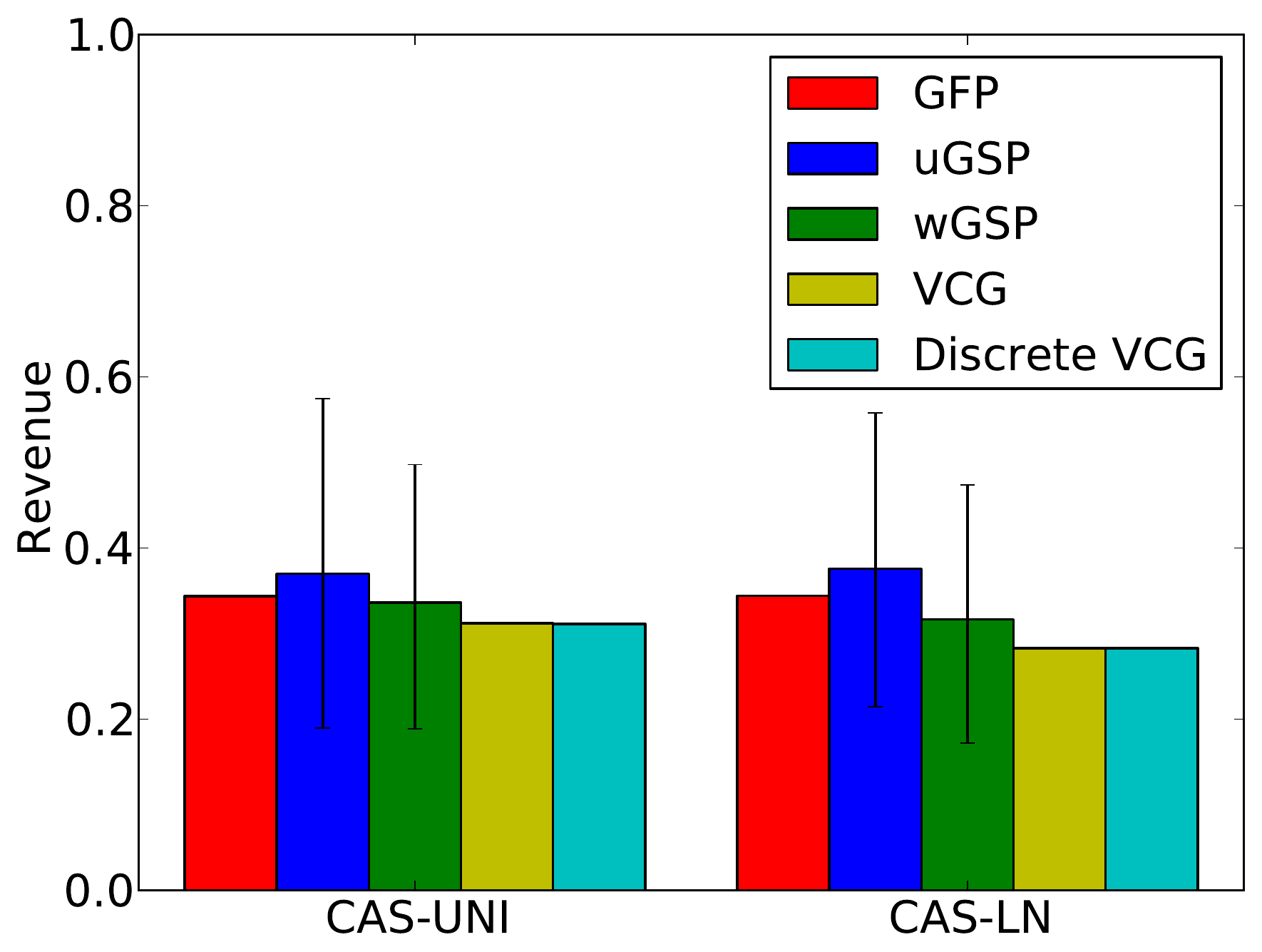}
\includegraphics[width=0.32\hsize]{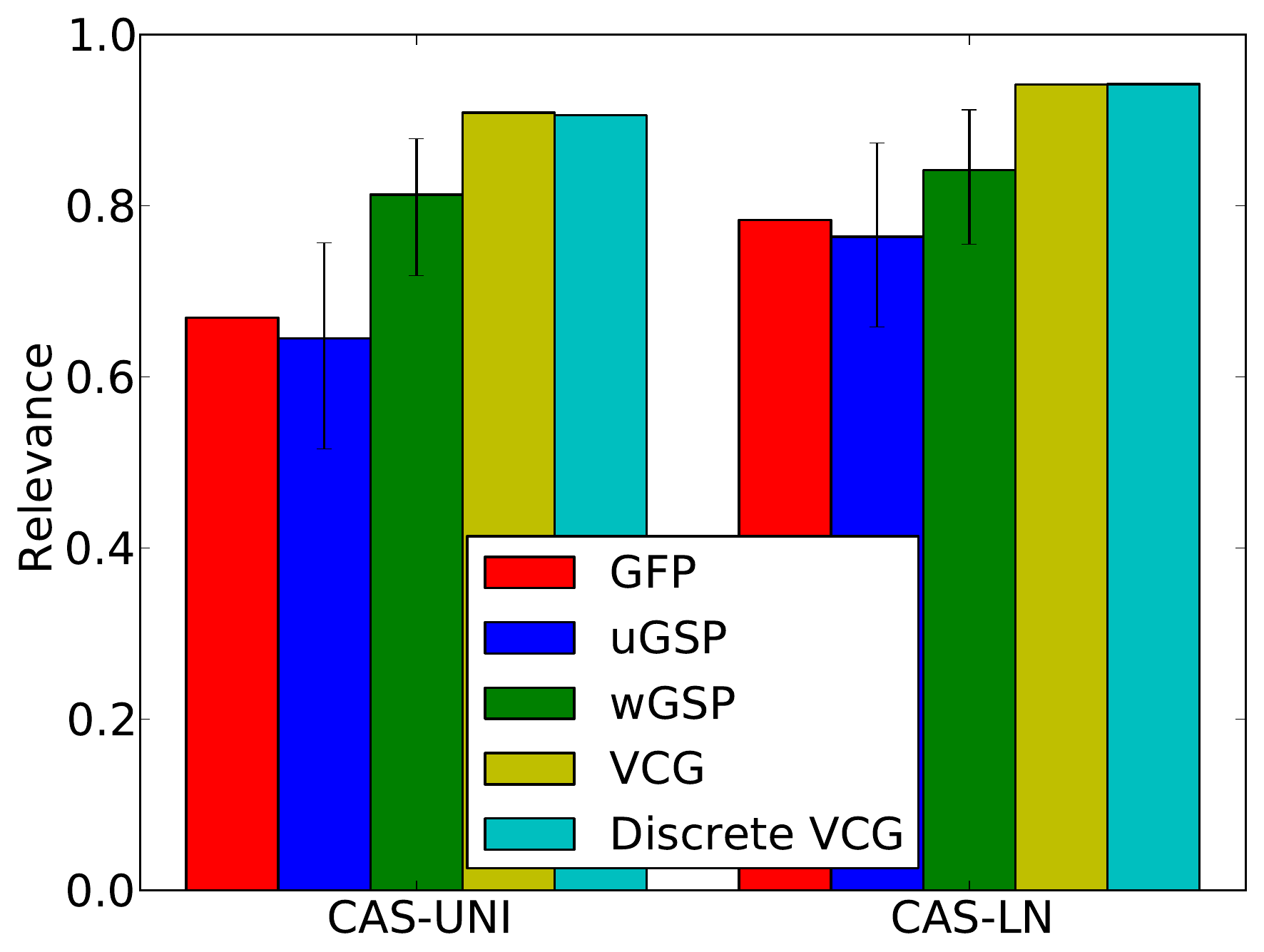}\\
\kcaption{(a) Efficiency}{.32\hsize}
\kcaption{(b) Revenue}{.32\hsize}
\kcaption{(c) Relevance}{.32\hsize}
\caption[Comparing the average performance of different position auction types in cascade settings]{Comparing the average performance of different position auction types in cascade settings.  \whiskerBP}
\label{fig:cas-comparison}
\end{figure}

Comparing position auctions, we observed relative performance similar to that under the no-externality models  (see Figure~\ref{fig:cas-comparison}).  First, we found that wGSP was the most efficient position auction design.  Surprisingly, wGSP's best-case equilibria tended to be close to fully efficient (at least $\sim 95\%$).  However, variation across equilibria was greater than in no-externality settings like EOS and V.  Both uGSP and wGSP varied substantially in their revenue across equilibria, with wGSP's median equilibrium achieving slightly more revenue than VCG.  Like in the V model, wGSP was roughly comparable to uGSP in terms of revenue when values were uniformly distributed, and noticeably worse when values were log-normally distributed.  Concerning relevance, we found that wGSP was clearly superior to other position auction designs, but not quite as good as VCG.  Recall that envy is not well defined in settings with externalities; thus, we do not discuss it here or in what follows.

\subsubsection{wGSP with Cascade-Specific Weights}

To address of some of the shortcomings of wGSP in cascade, Gomes \etal\ proposed an alternative way of calculating advertiser qualities for wGSP.  In their alternative weighting, an advertiser's weight is equal to her top-position click probability divided by the probability that a user will stop looking at ads after seeing hers (i.e., one minus her continuation probability).  This simple reweighting scheme has the nice property that an advertiser with a continuation probability of one will always be ranked first, a desirable property given that the efficient ranking will always rank such bidders first.  Gomes \etal\ also showed that this weighting gives rise to a revenue-optimal equilibrium.  (There are two important caveats about this equilibrium. (1) It is only revenue-optimal among mechanisms that always allocate every position; mechanisms that use reserve prices can get strictly more revenue.  (2) It might not be rationalizable, requiring some agents to play the weakly dominated strategy of bidding strictly above their own valuations.)

\begin{figure}
\centering
\includegraphics[width=0.45\hsize]{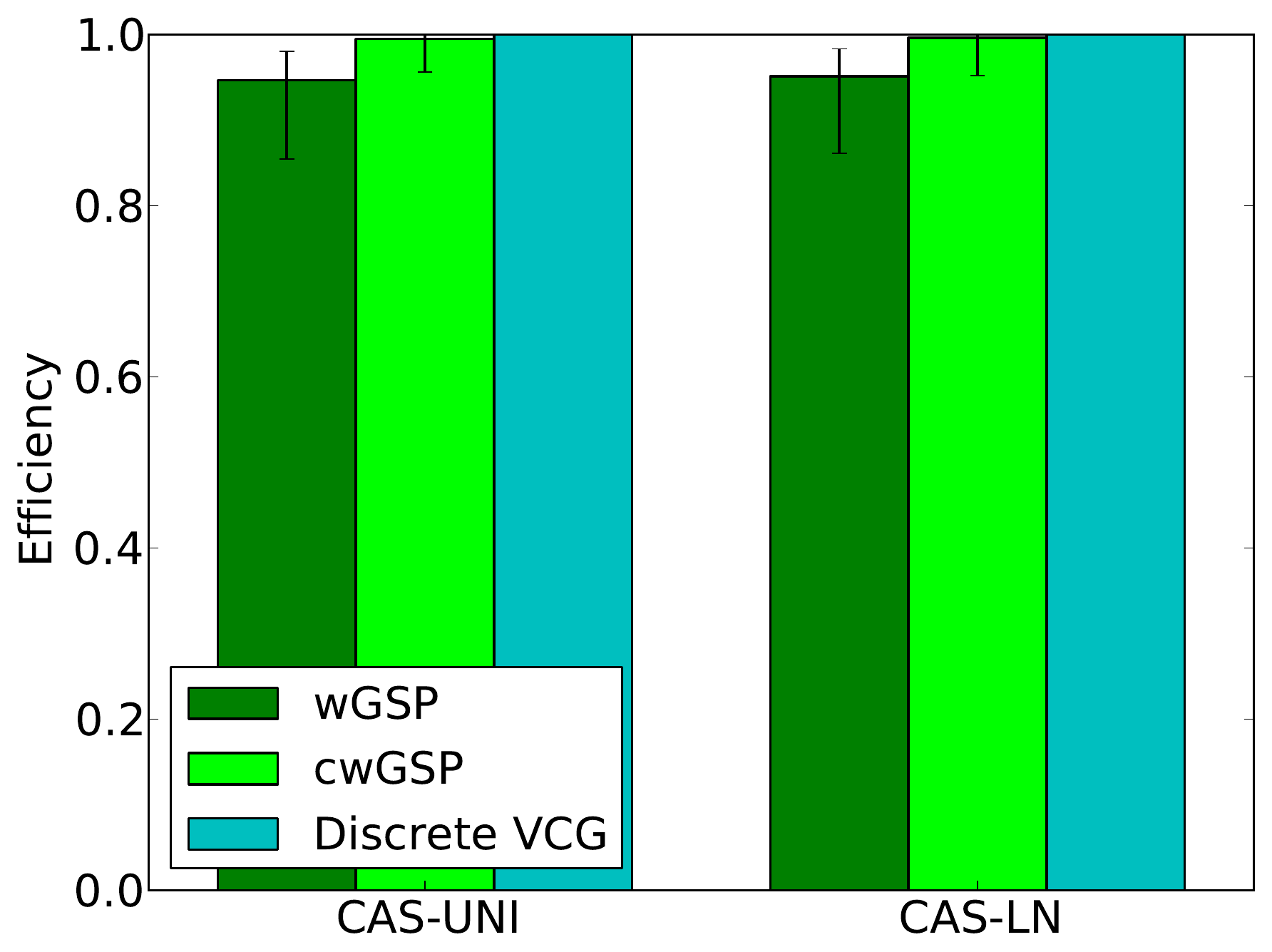}
\includegraphics[width=0.45\hsize]{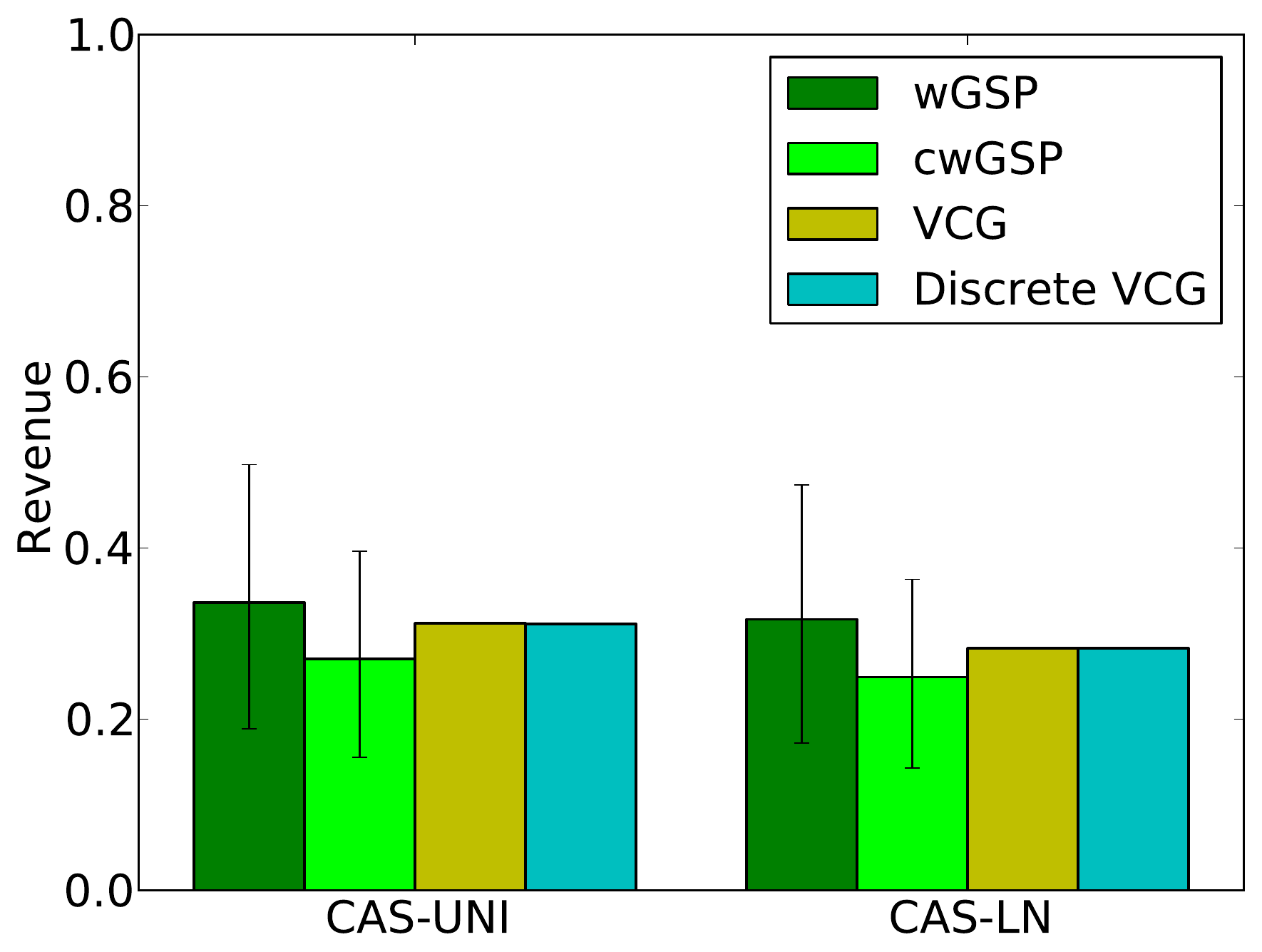}
\kcaption{(a) Efficiency}{.45\hsize}
\kcaption{(b) Revenue}{.45\hsize}
\caption[Comparing the average performance of wGSP and cwGSP in cascade settings]{Comparing the average performance of wGSP and cwGSP in cascade settings. \whiskerBP}
\label{fig:cwgsp-comparison}
\end{figure}

We experimented with this alternative mechanism, which we call cwGSP, and found that it was dramatically more efficient that wGSP (see Figure~\ref{fig:cwgsp-comparison}).  However, we also found that it was  noticeably worse than wGSP in terms of revenue.  Thus, we conclude that Gomes \etal's revenue claims about cwGSP stem mainly from their unusual equilibrium-selection criteria.


\subsubsection{The Hybrid Model}

The hybrid model generalizes both the V and cascade models; lower positions can get fewer clicks either due to the number of higher-placed ads (as in V) or to the content of those ads (as in cascade).  Hybrid settings can have a very large price of stability and an even larger price of anarchy \cite{KM08cascade}.

\qvi\vspace{\modelspace}

\begin{figure}
	\centering
	\includegraphics[width=0.45\hsize]{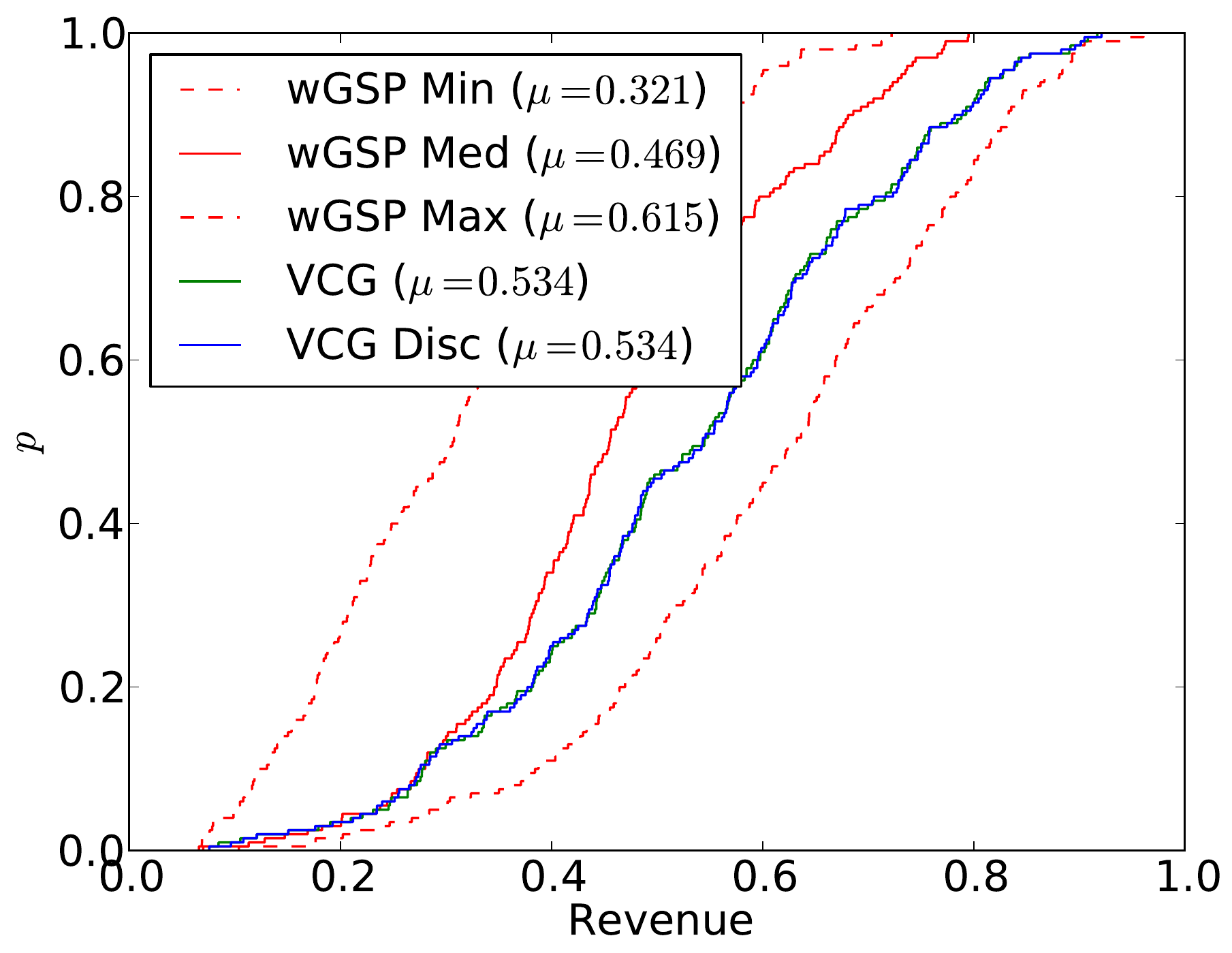}
	\includegraphics[width=0.45\hsize]{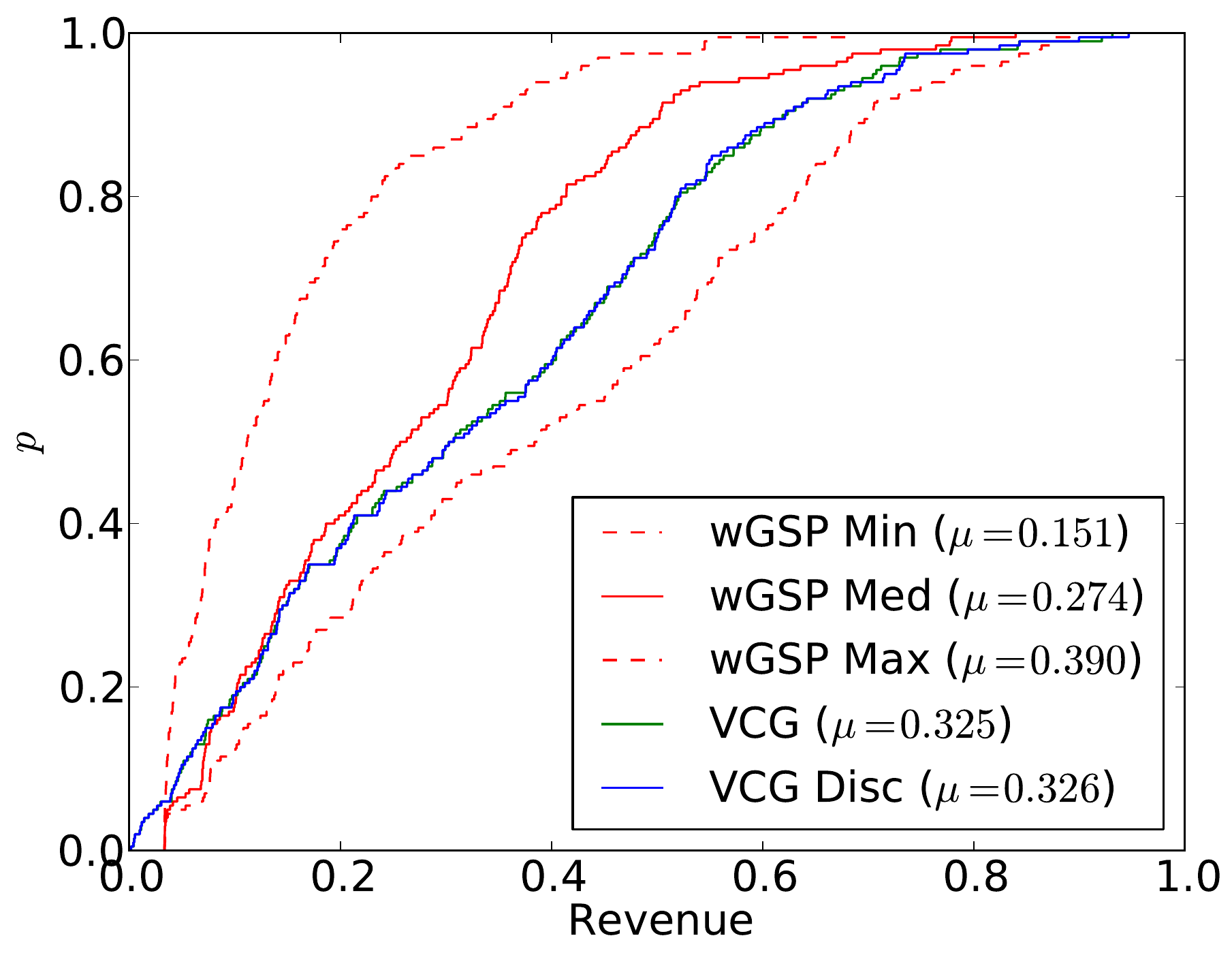}
	\kcaption{(a) Uniform Distribution}{.45\hsize}
	\kcaption{(b) Log-Normal Distribution}{.45\hsize}
	\caption{Empirical CDF of revenue in the hybrid model.}
	\label{fig:hyb-revenue-cdf}
\end{figure}

\begin{figure}
	\centering
	\includegraphics[width=0.45\hsize]{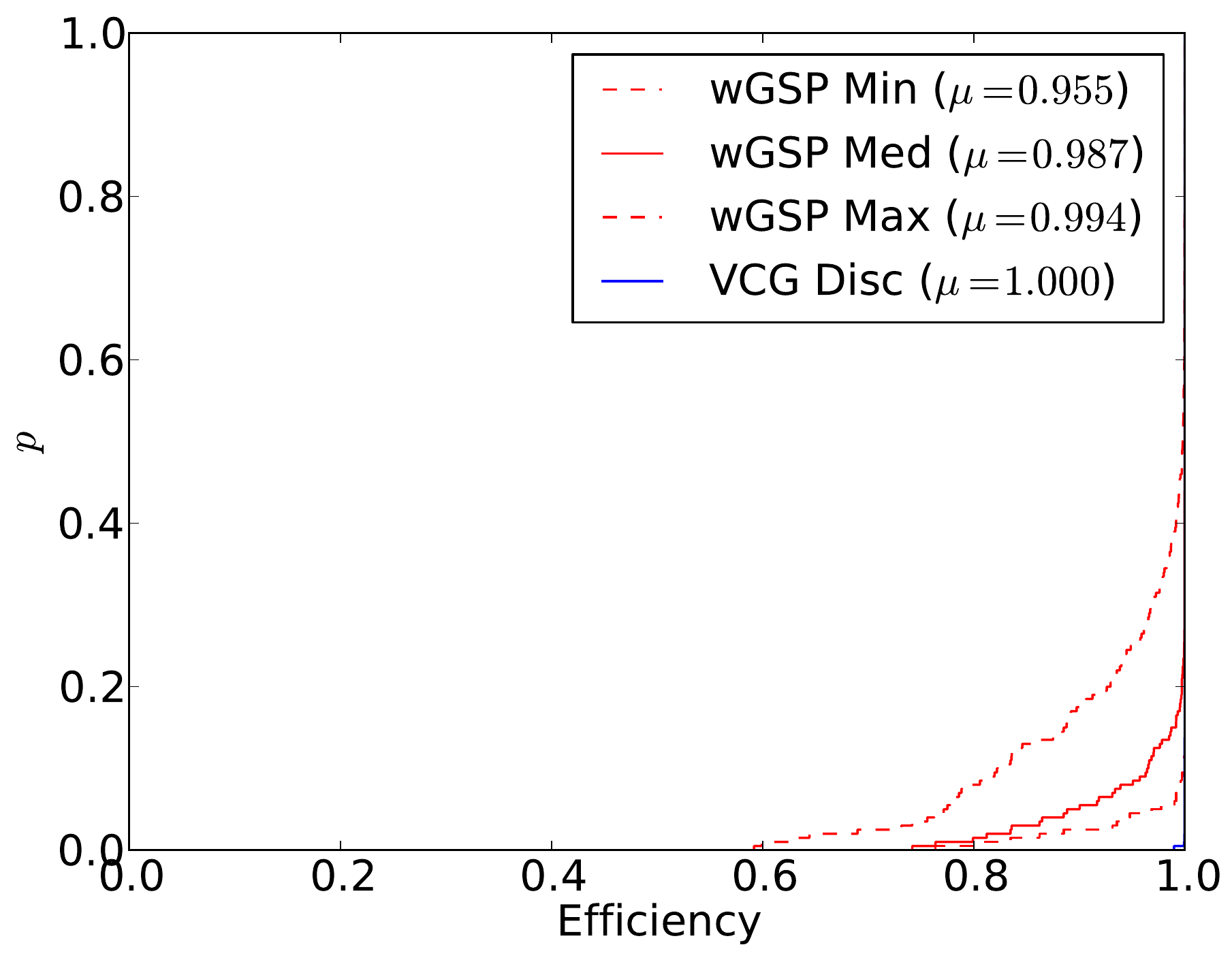}
	\includegraphics[width=0.45\hsize]{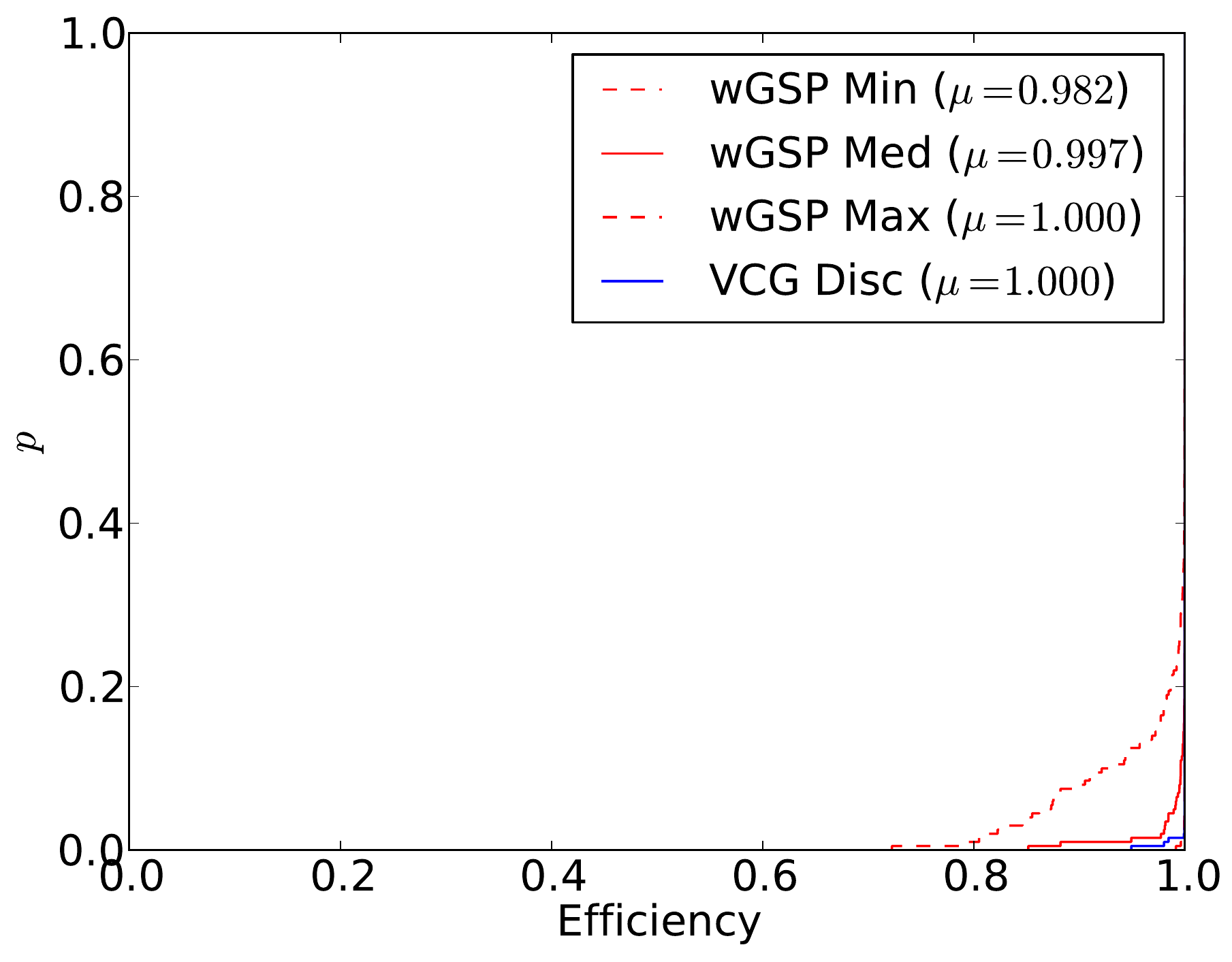}
	\kcaption{(a) Uniform Distribution}{.45\hsize}
	\kcaption{(b) Log-Normal Distribution}{.45\hsize}
	\caption{Empirical CDF of economic efficiency in the hybrid model.}
	\label{fig:hyb-welfare-cdf}
\end{figure}

We again found that wGSP's revenue varied substantially across equilibria (see Figure~\ref{fig:hyb-revenue-cdf}). However, we were surprised to find that while wGSP's efficiency was quite low ($\sim60\%$) in a few games, wGSP was often efficient, even in its worst-case equilibria   (see Figure~\ref{fig:hyb-welfare-cdf}).   This result stands in contrast with our previous findings on cascade. The very small difference between best- and worst-case equilibria did not arise because games had few Nash equilibira: wGSP still had many Nash equilibria in just about every game.  In the next subsection, we compare all three externality models and investigate this anomaly more deeply.

\begin{figure}
\centering
\includegraphics[width=0.32\hsize]{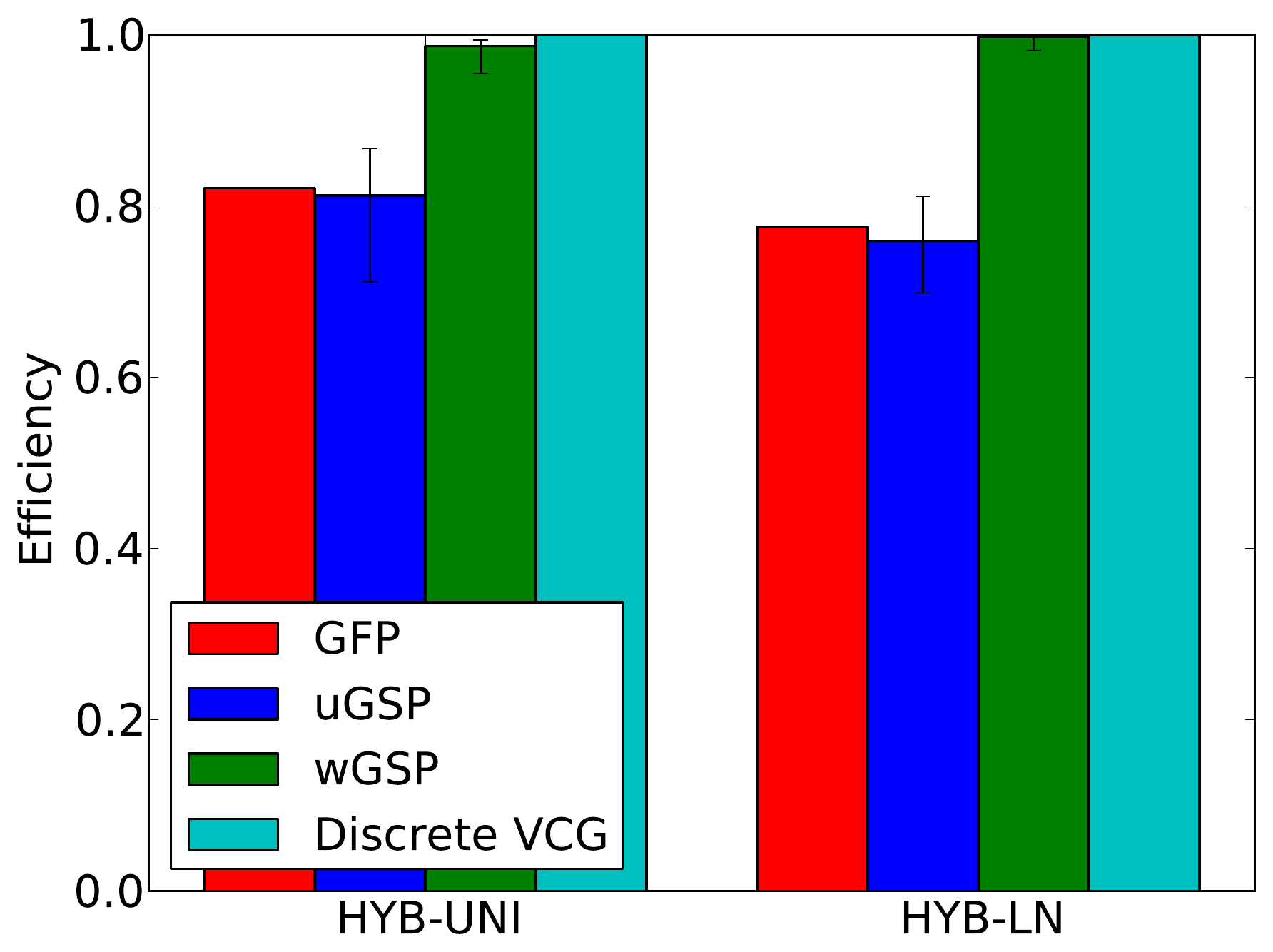}
\includegraphics[width=0.32\hsize]{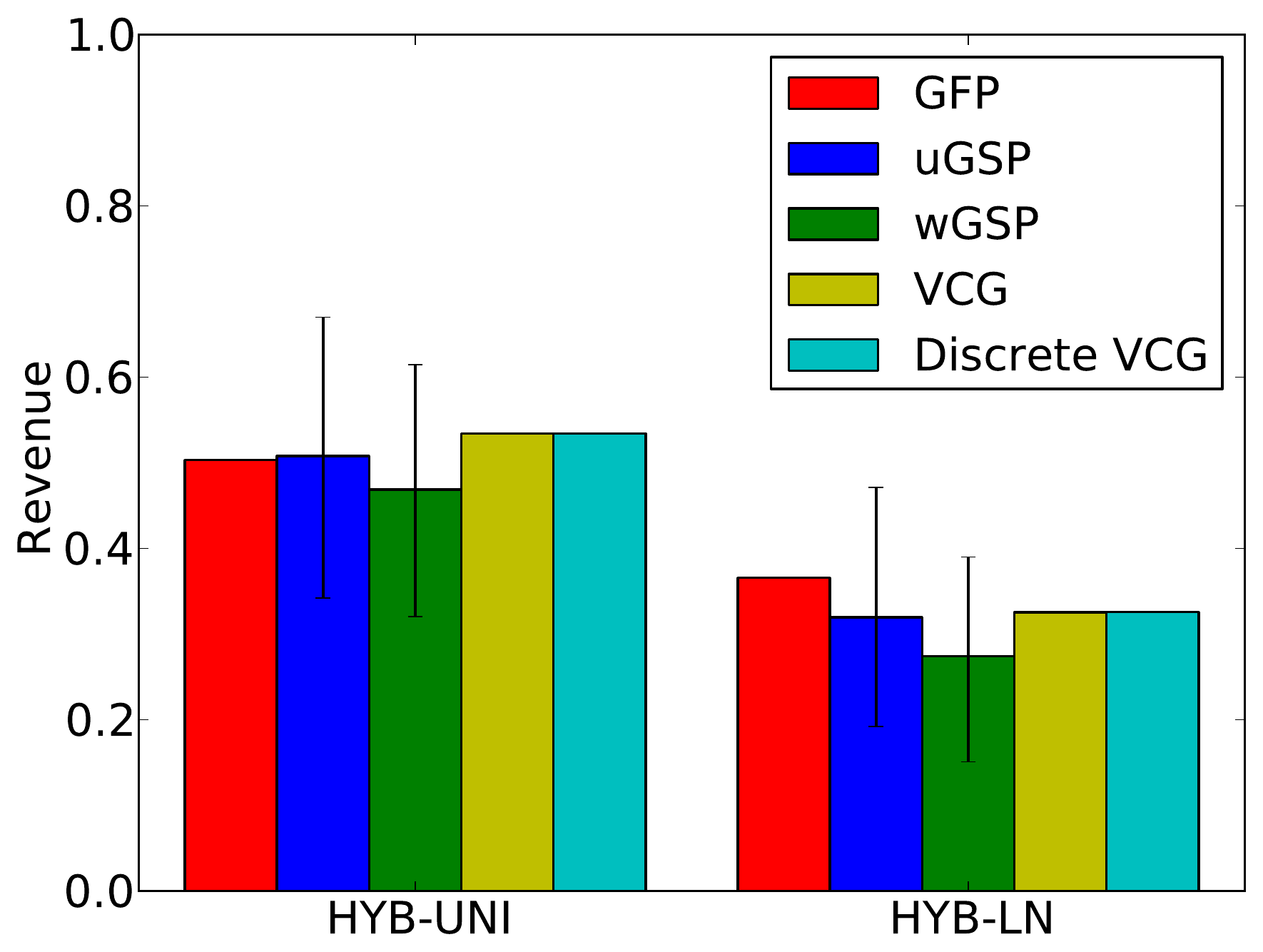}
\includegraphics[width=0.32\hsize]{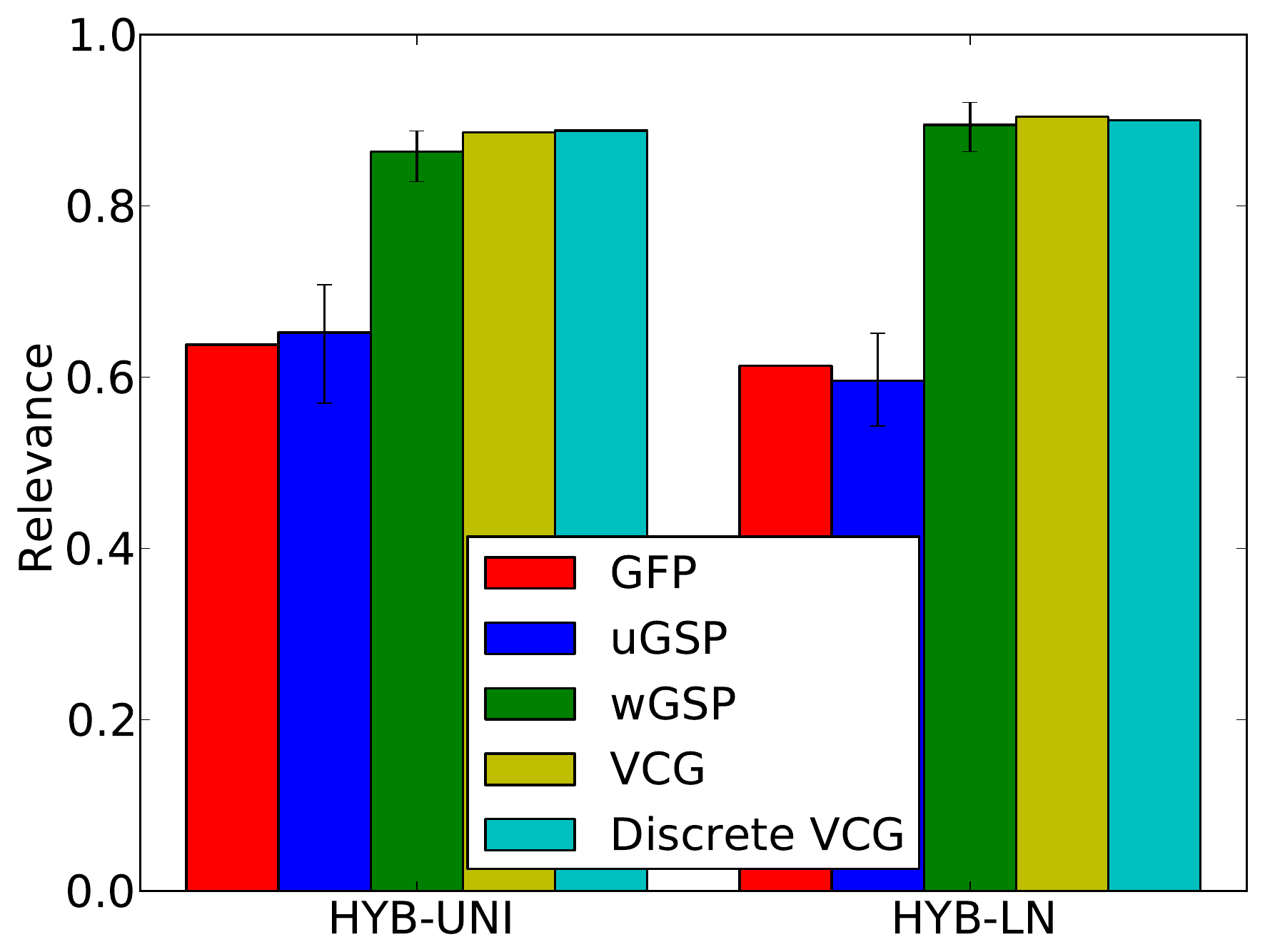}\\
\kcaption{(a) Efficiency}{.32\hsize}
\kcaption{(b) Revenue}{.32\hsize}
\kcaption{(c) Relevance}{.32\hsize}
\caption[Comparing the average performance of different position auction types in hybrid settings]{Comparing the average performance of different position auction types in hybrid settings.  \whiskerBP}
\label{fig:hyb-comparison}
\end{figure}

Comparing the different position auction designs (see Figure~\ref{fig:hyb-comparison}), we again found that wGSP was the most efficient and produced the most relevant results.  Concerning revenue, we found that both uGSP and wGSP varied substantially across their equilibria, but that uGSP was consistently better than wGSP.  wGSP substantially outperformed GFP and uGSP in terms of relevance.

\subsubsection{The GIM Model}

Finally, we turn to the model of Gomes, Immorlica and Markakis, which is the most general of our externality models.  In the GIM model, every bidder has a quality score, which corresponds to her probability of getting a click in the top position, but the click probability in lower positions can depend arbitrarily on which advertisers are shown in higher positions as long as this probability weakly decreases with position.  Gomes \etal\ conjectured that this model was similar to cascade, which motivated our questions.

\qvii\vspace{\modelspace}

\begin{figure}
	\centering
	\includegraphics[width=0.45\hsize]{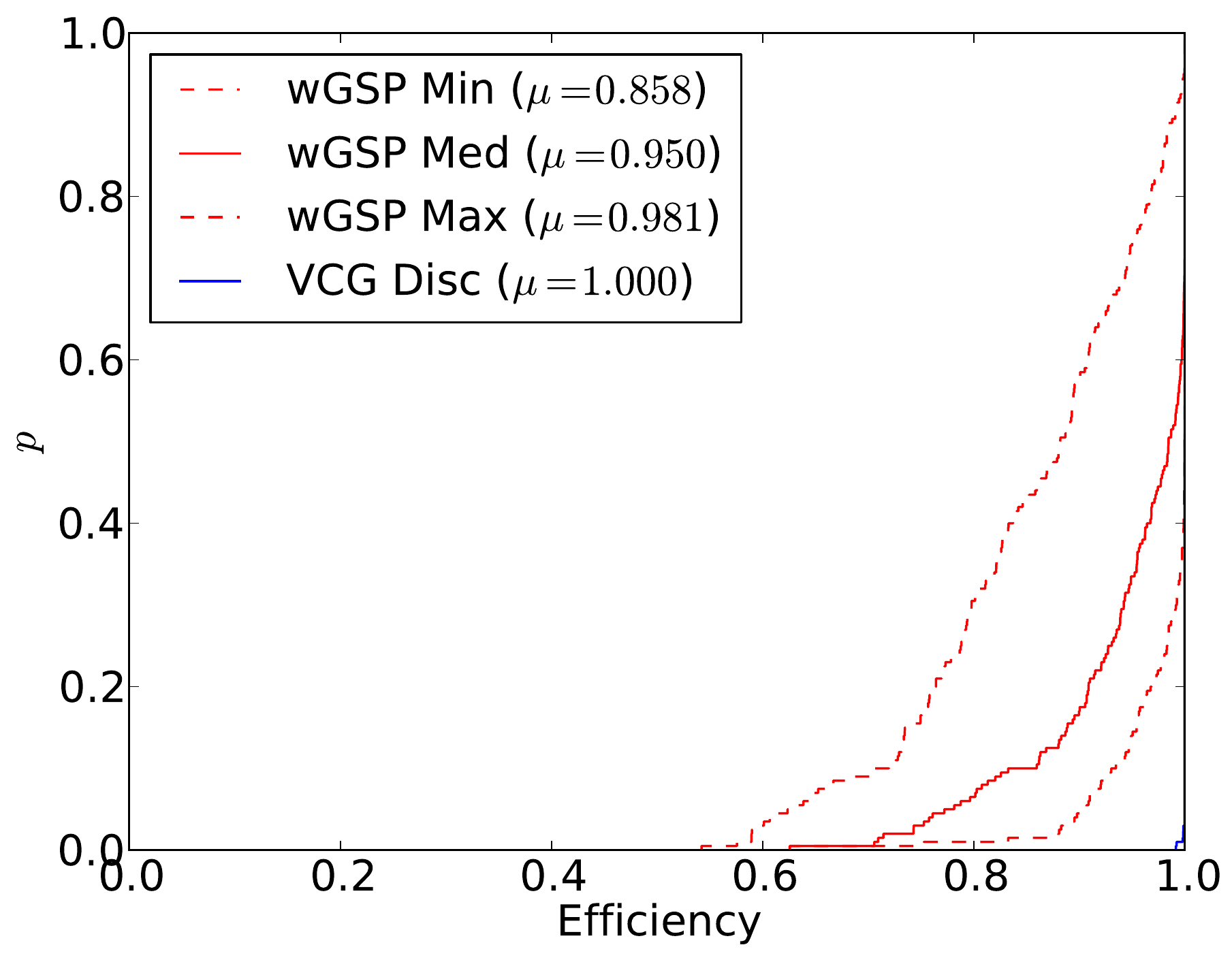}
	\includegraphics[width=0.45\hsize]{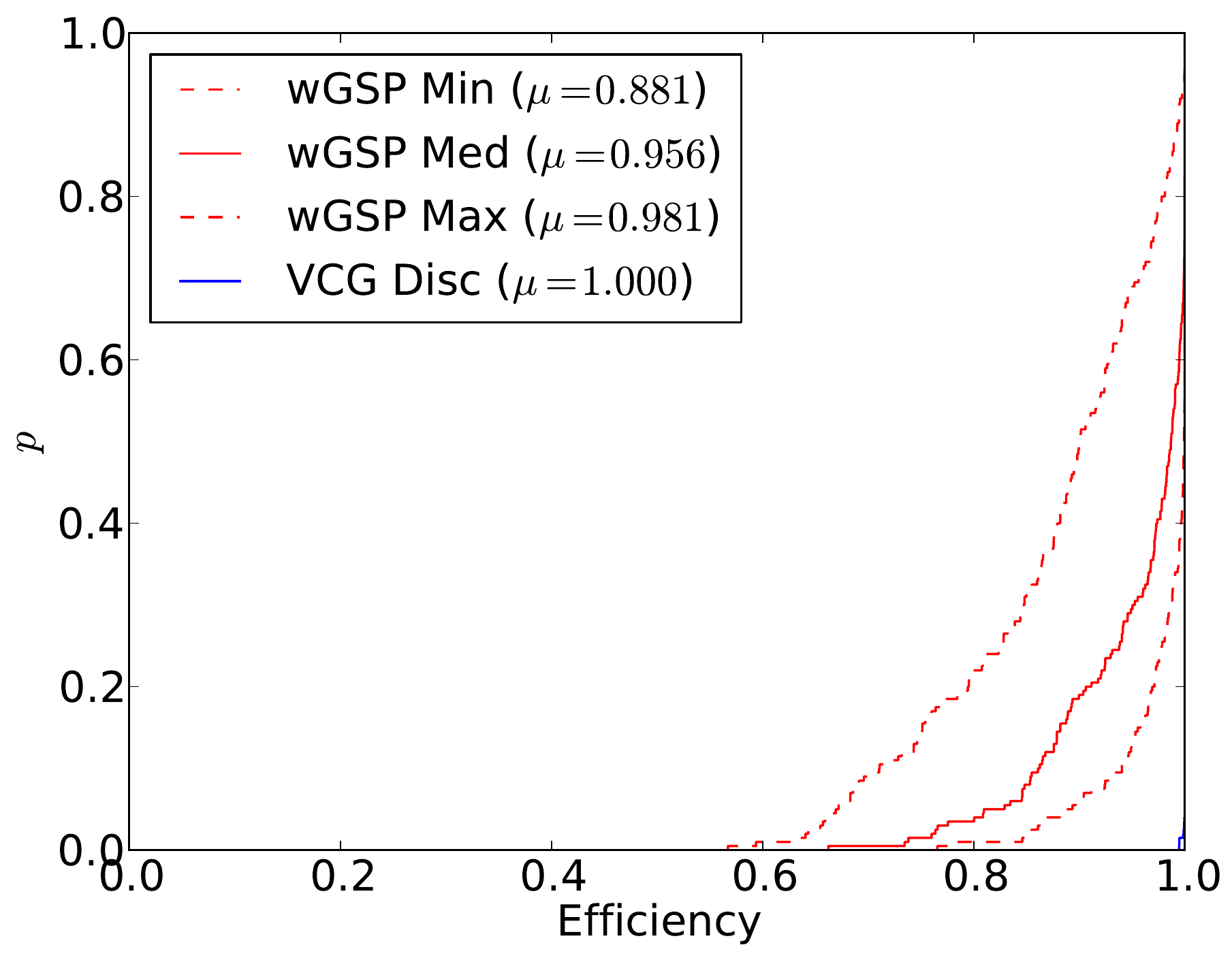}
	\kcaption{(a) Uniform Distribution}{.45\hsize}
	\kcaption{(b) Log-Normal Distribution}{.45\hsize}
	\caption{Empirical CDF of economic efficiency in the GIM model.}
	\label{fig:gim-welfare-cdf}
\end{figure}

\begin{figure}
	\centering
	\includegraphics[width=0.45\hsize]{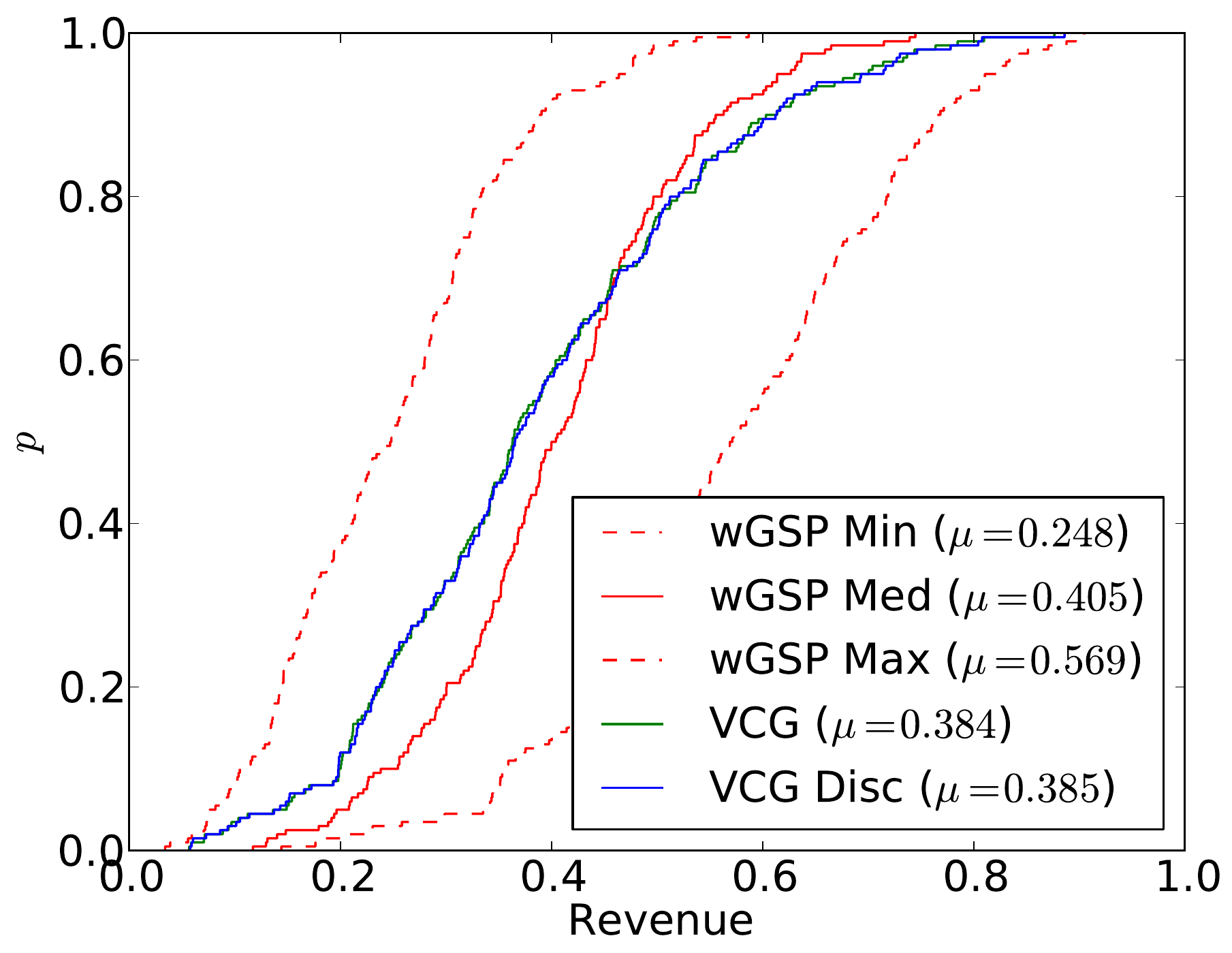}
	\includegraphics[width=0.45\hsize]{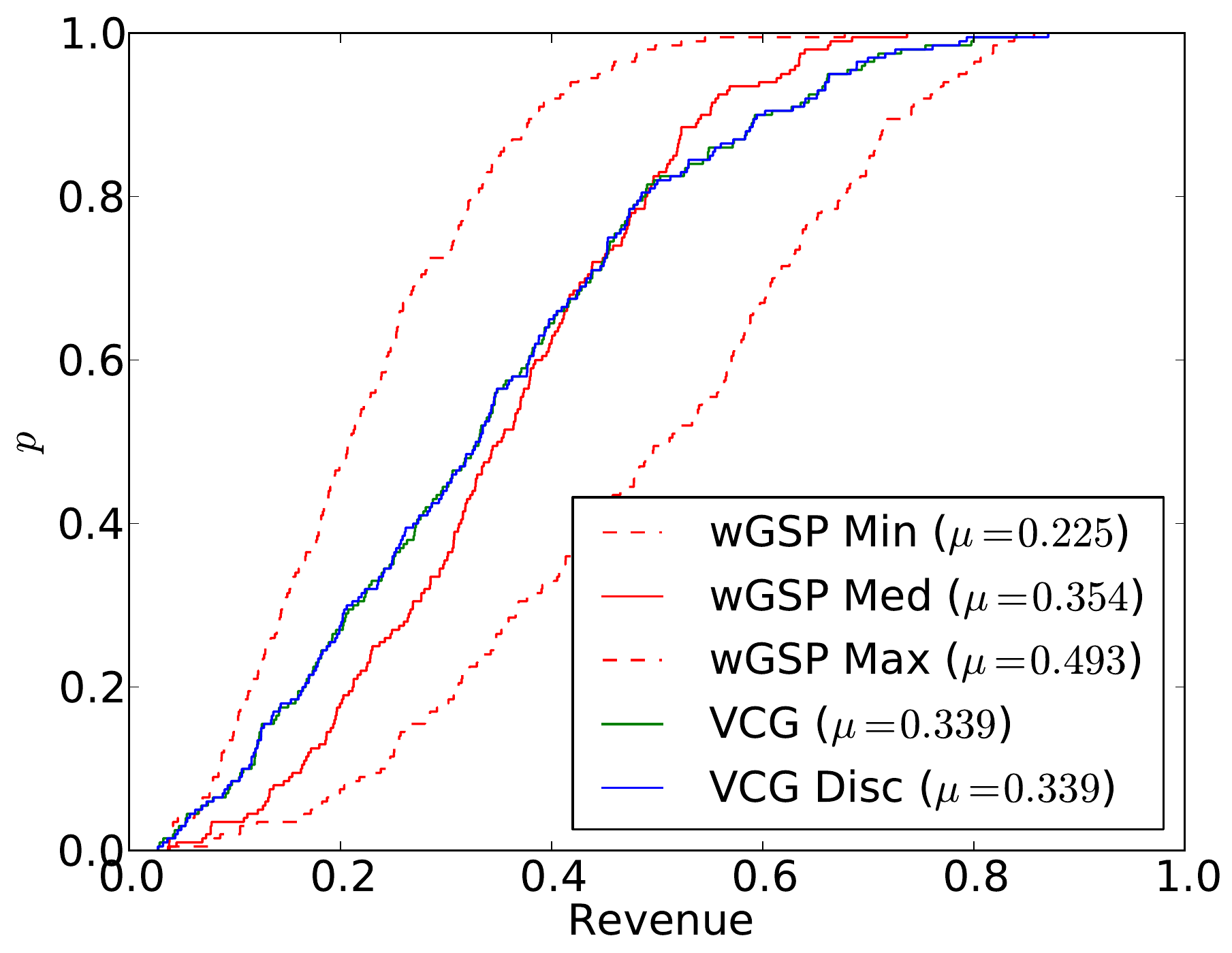}
	\kcaption{(a) Uniform Distribution}{.45\hsize}
	\kcaption{(b) Log-Normal Distribution}{.45\hsize}
	\caption{Empirical CDF of revenue in the GIM model.}
	\label{fig:gim-revenue-cdf}
\end{figure}

\begin{figure}
	\centering
	\includegraphics[width=0.30\hsize]{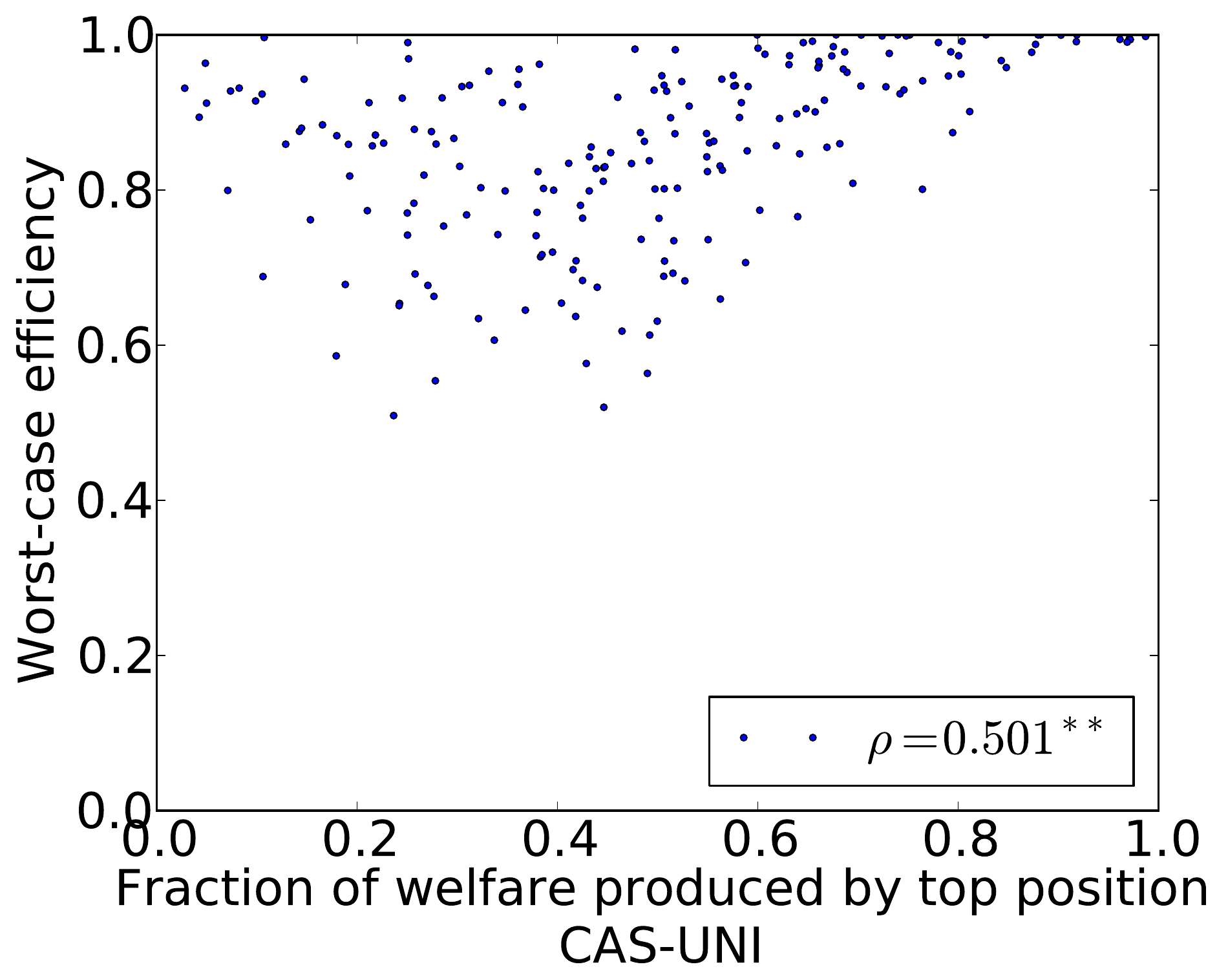}
	\includegraphics[width=0.30\hsize]{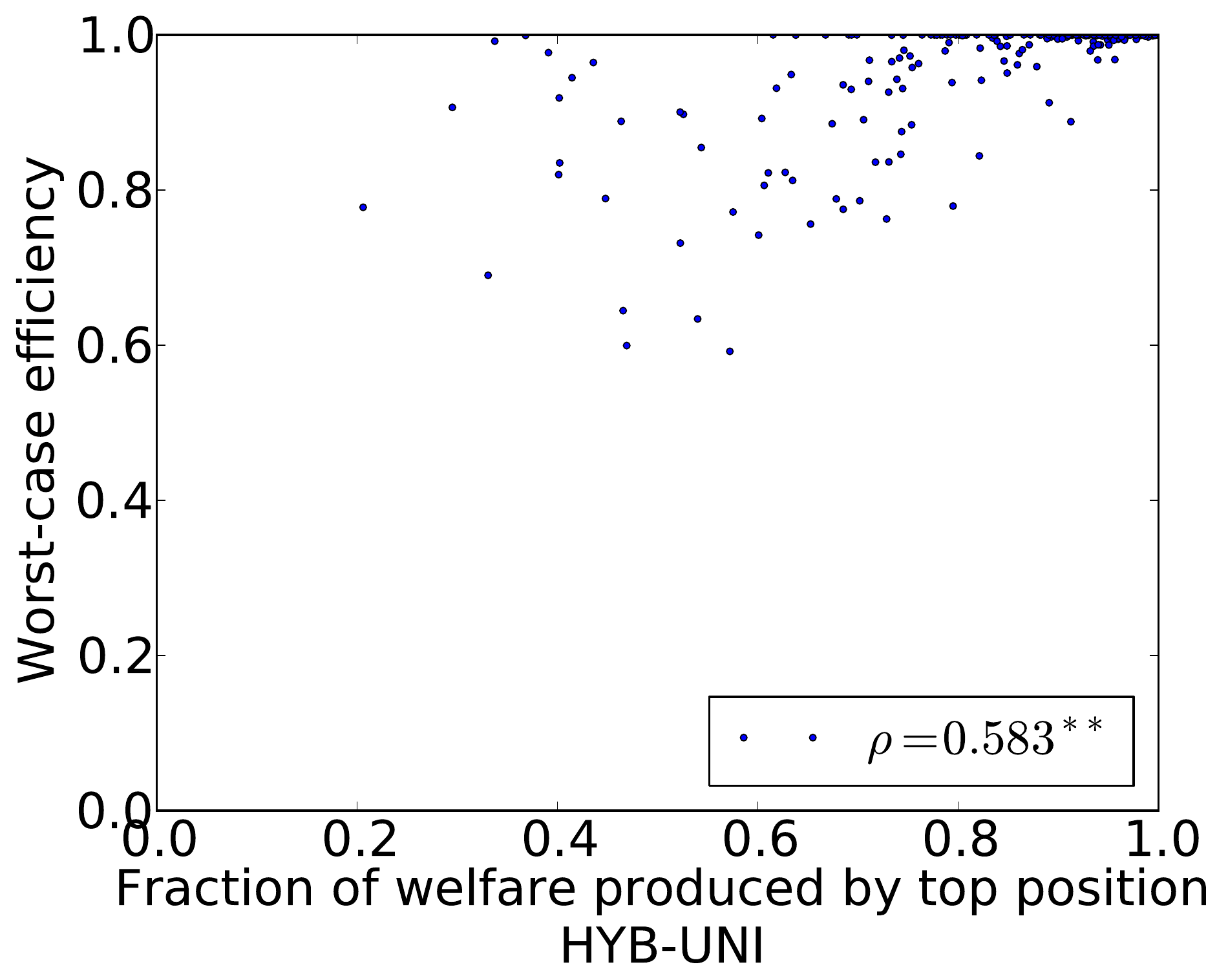}
	\includegraphics[width=0.30\hsize]{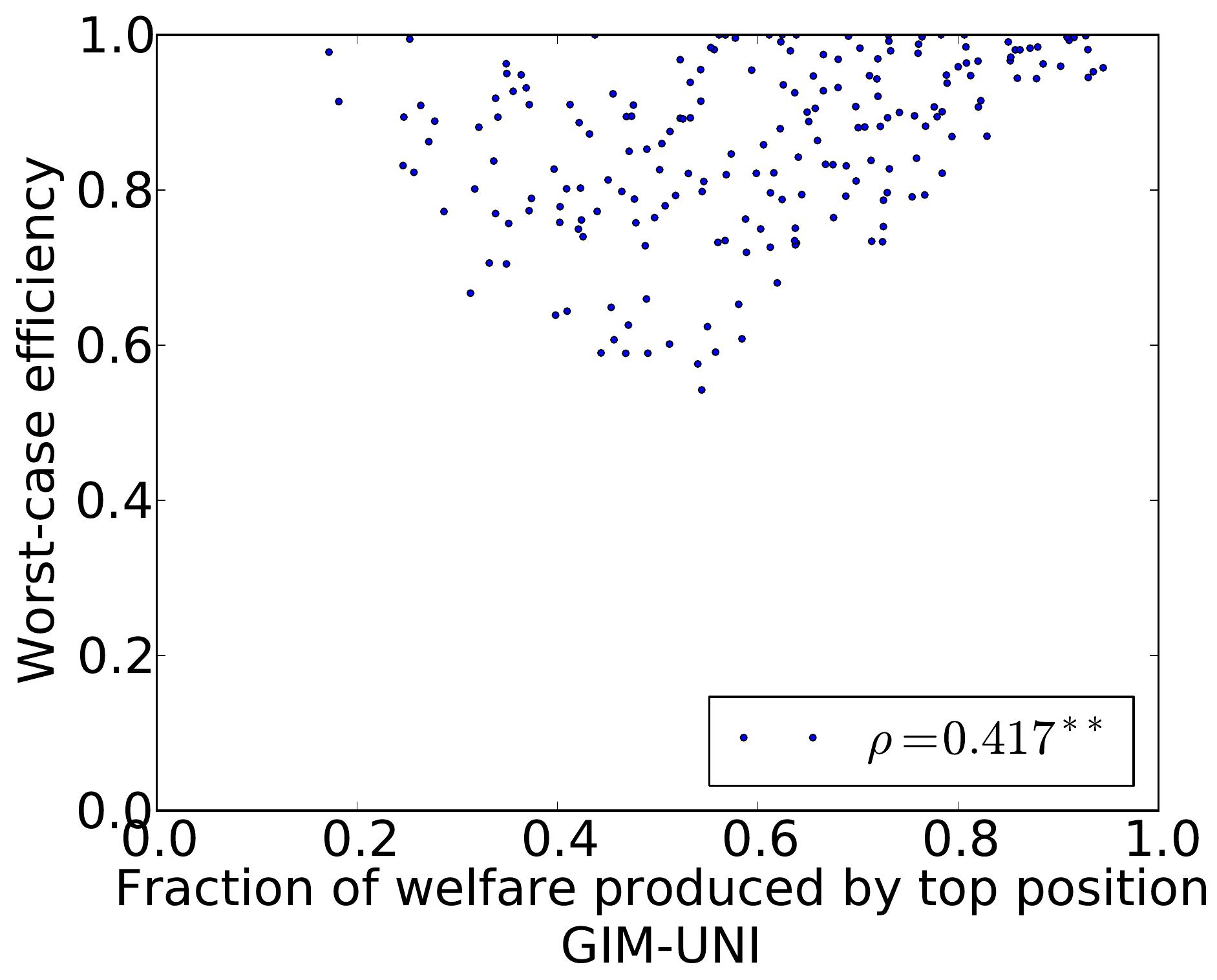}
	\kcaption{(a) Cascade: Uniform}{.3\hsize}
	\kcaption{(b) Hybrid: Uniform}{.3\hsize}
	\kcaption{(c) GIM: Uniform}{.3\hsize}
	
	\includegraphics[width=0.30\hsize]{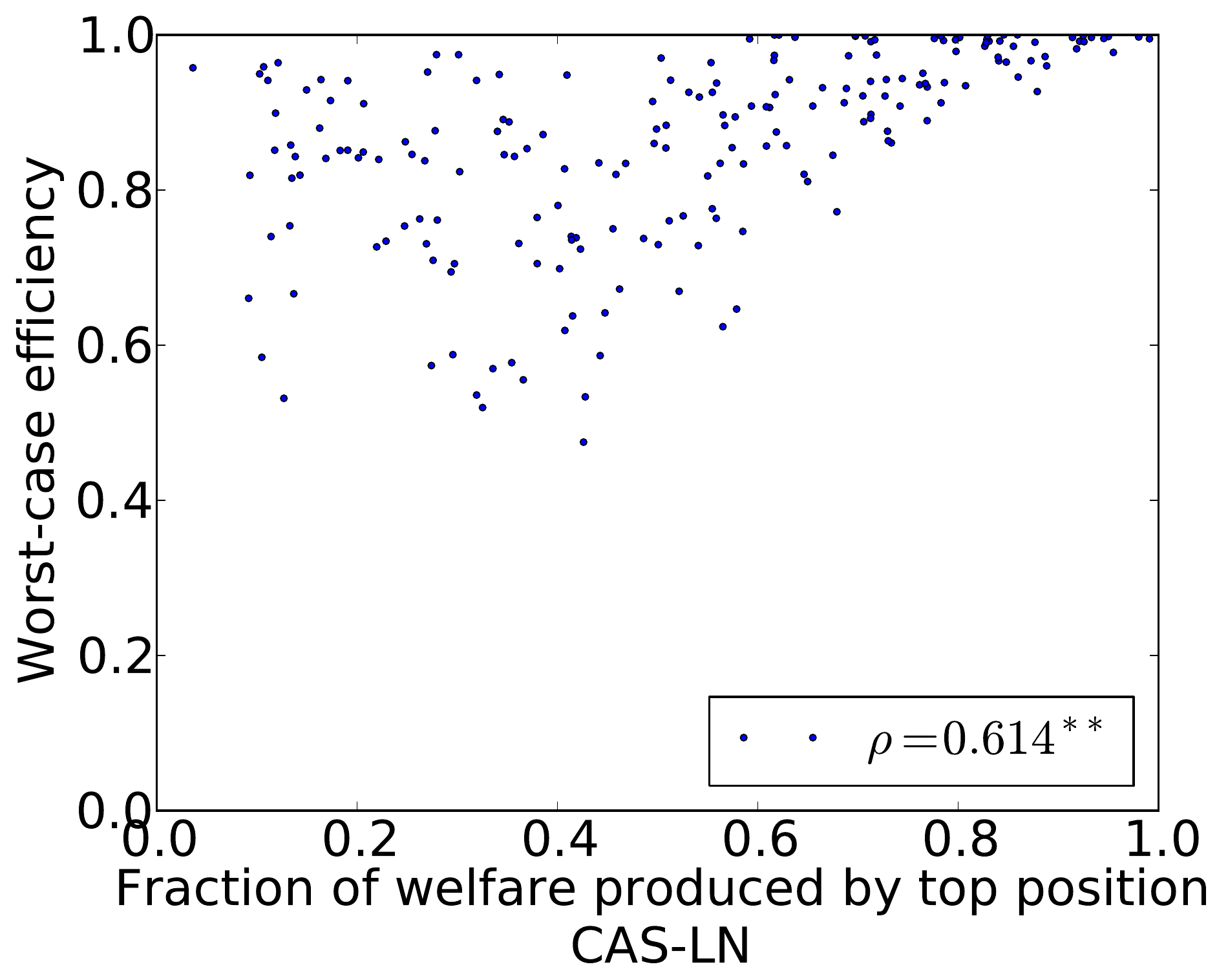}
	\includegraphics[width=0.30\hsize]{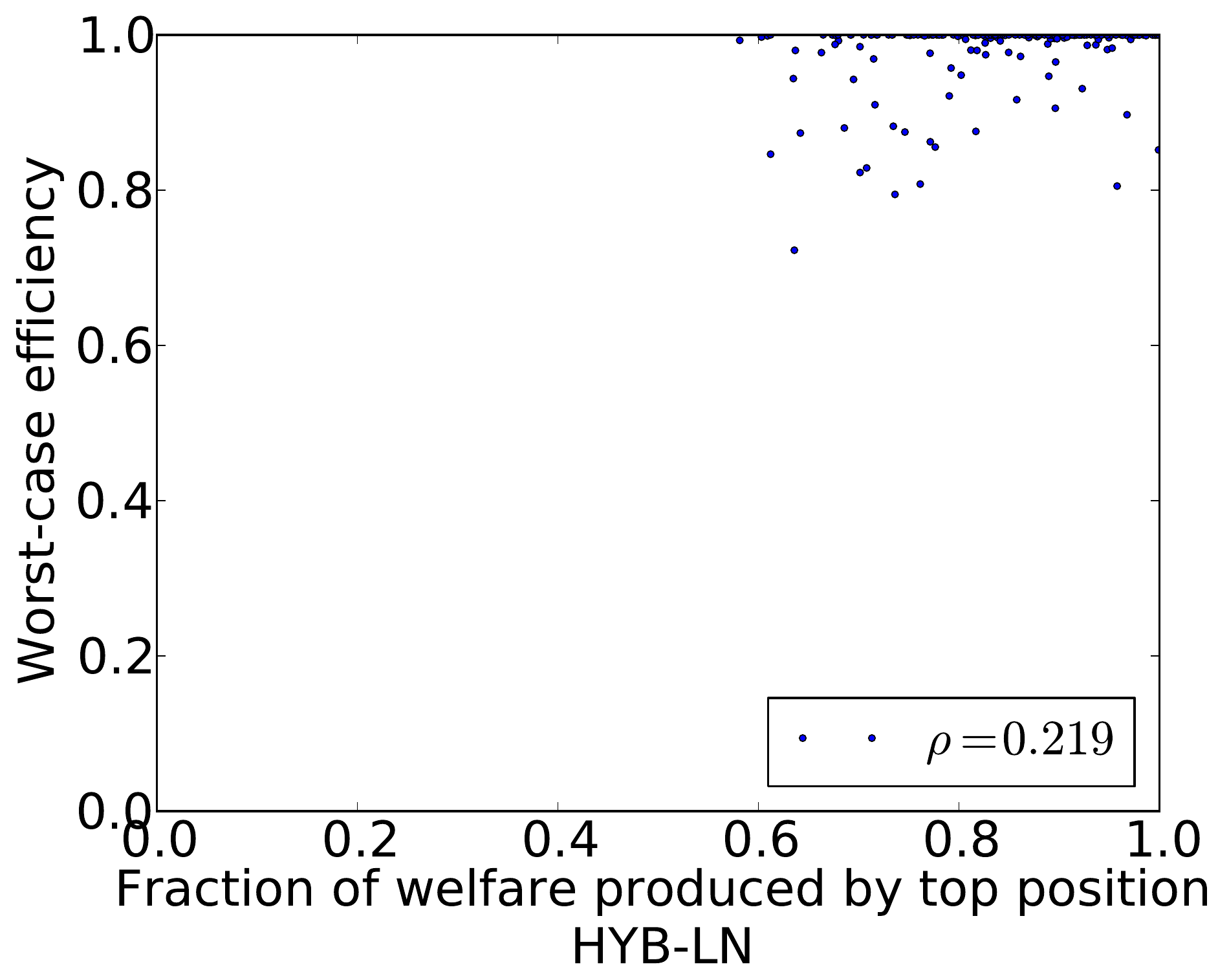}
	\includegraphics[width=0.30\hsize]{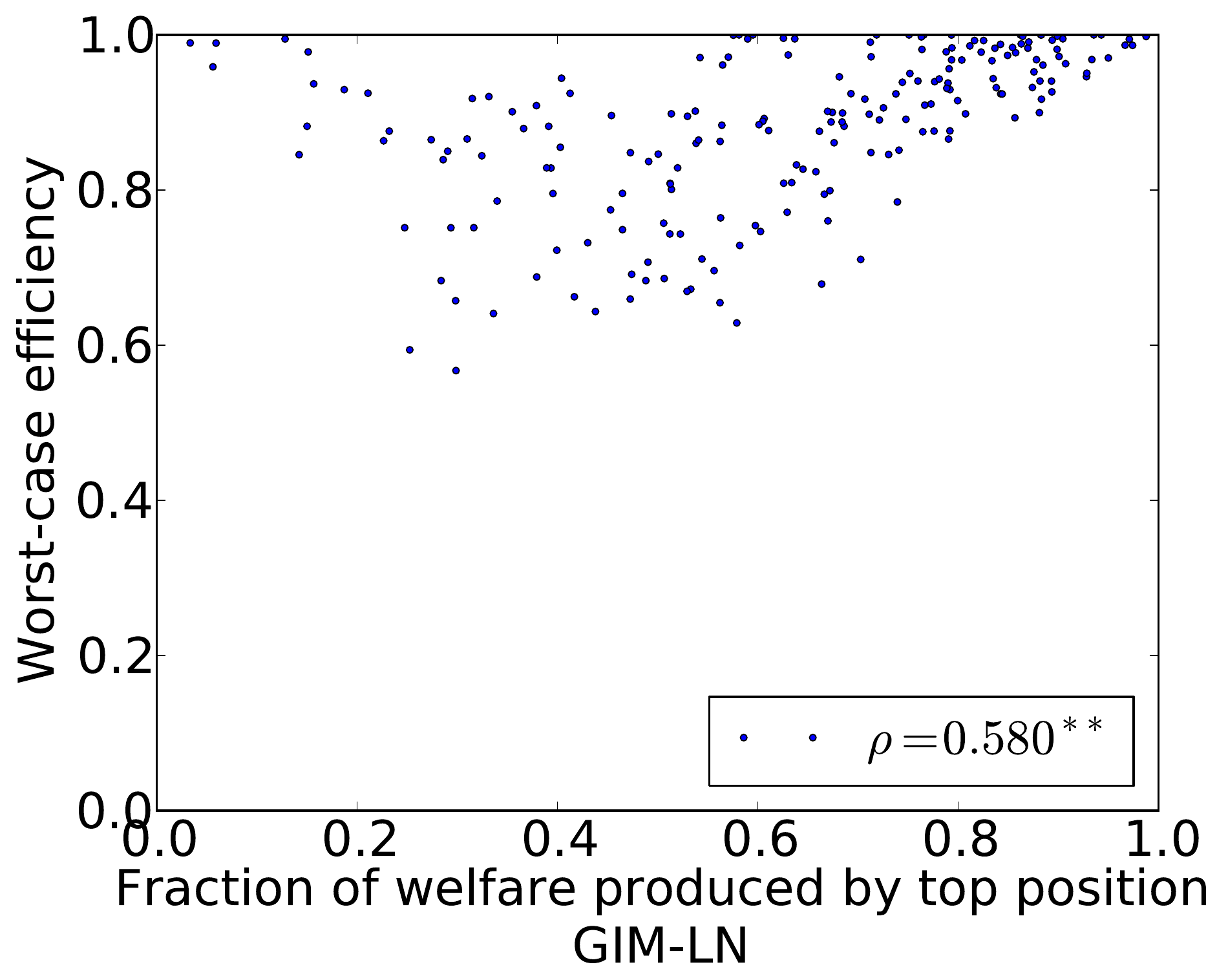}
	\kcaption{(d) Cascade: Log-Normal}{.3\hsize}
	\kcaption{(e) Hybrid: Log-Normal}{.3\hsize}
	\kcaption{(f) GIM: Log-Normal}{.3\hsize}
	\caption[wGSP tended to have good worst-case efficiency when the top position produced the majority of the surplus]{wGSP tended to have good worst-case efficiency when the top position produced the majority of the surplus (in an efficient allocation).  Thus, in hybrid distributions, where this occurred extremely frequently, wGSP tended to have better worst-case efficiency than in cascade or GIM.}
	\label{fig:poa-ext}
\end{figure}

Recall that in the cascade model, revenue and efficiency both varied substantially across equilibria,  and that the majority of games exhibited moderately large inefficiency ($\sim90\%$) in worst-case equilibrium.  In GIM, we observed similar effects of similar magnitude (see Figures~\ref{fig:gim-welfare-cdf} and \ref{fig:gim-revenue-cdf}). We found it surprising that GIM, which generalizes the hybrid model (which in turn generalizes cascade), should be have outcomes that are more similar to cascade than to hybrid.  To explain the anomaly, we looked for features that were common to cascade and GIM, but not to hybrid.  There are many such features (e.g., in wGSP's equilibrium rankings, the effect of position on CTR is much greater under hybrid). In the end we concluded that the feature that best explains our findings is the relative value of the top position---which is higher in hybrid settings---because in all three models, instances where this fraction was large tended to have very good worst-case efficiency in wGSP (see Figure~\ref{fig:poa-ext}). This correlation was moderately strong ($\rho>0.4$ using Spearman's rank correlation) and highly significant ($p<0.01$ with Bonferroni correction) for every distribution except BHN-LN, in which the majority of equilibria achieved almost perfect efficiency.

\begin{figure}
	\centering
	\includegraphics[width=0.32\hsize]{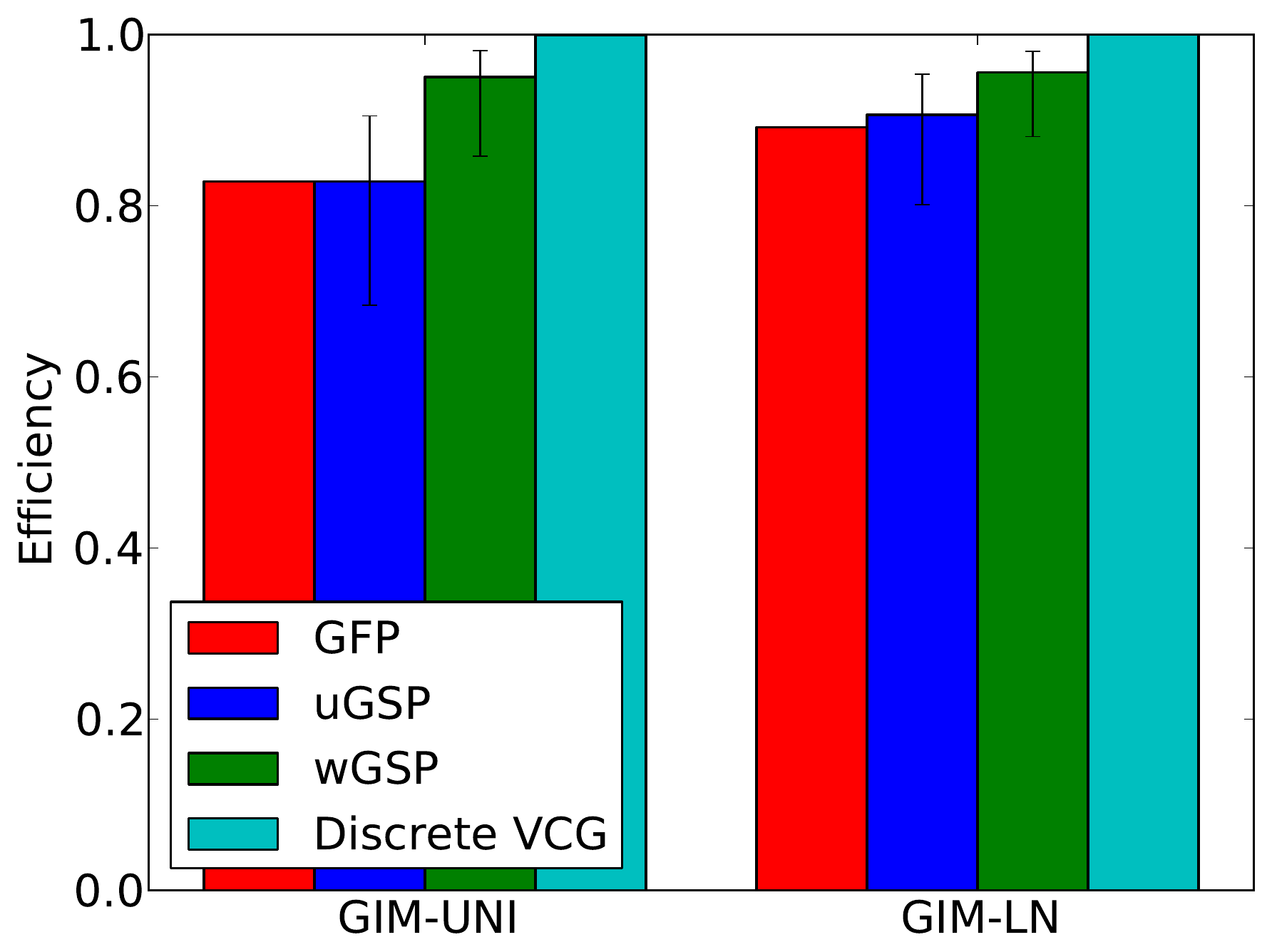}
	\includegraphics[width=0.32\hsize]{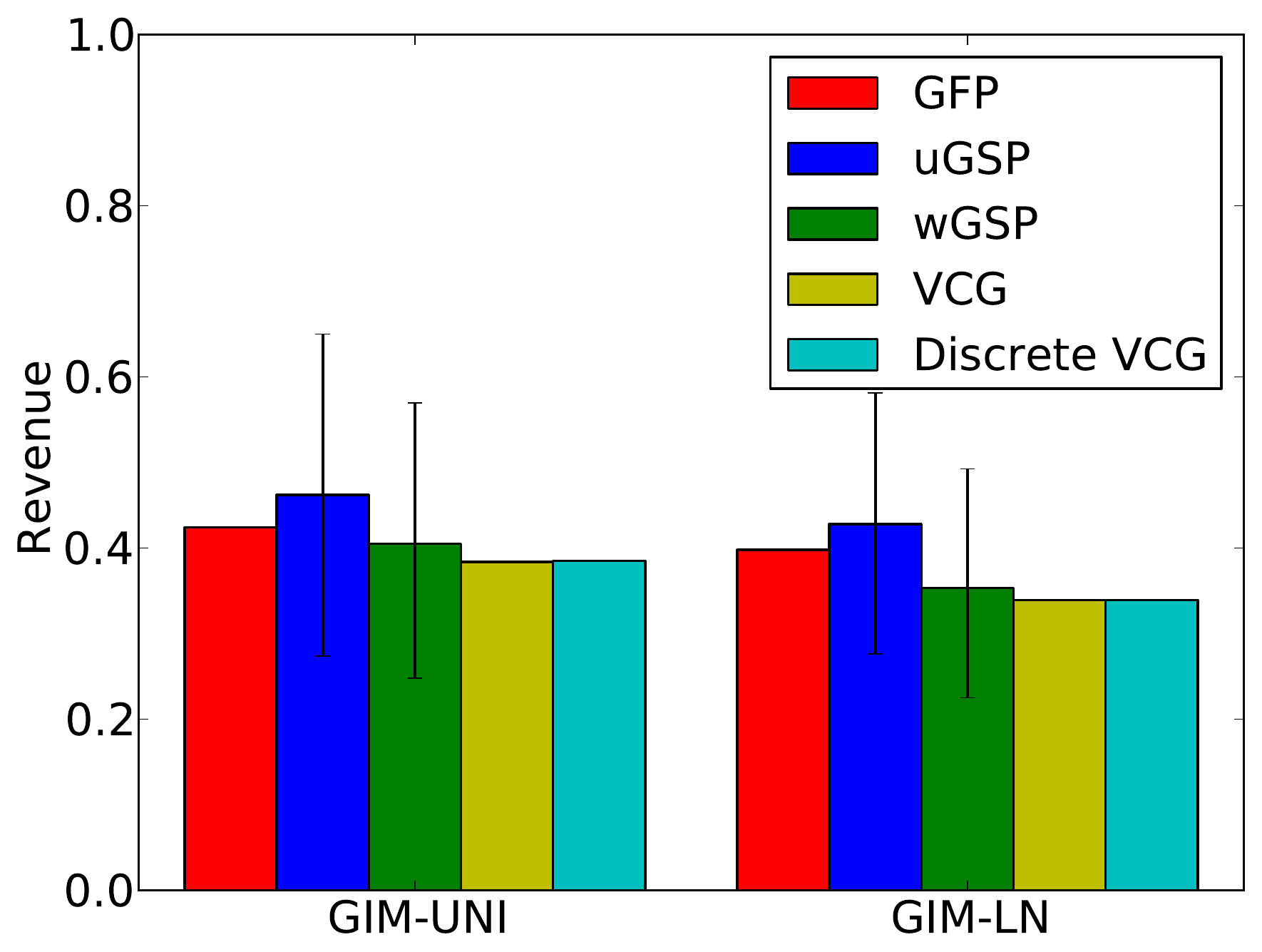}
	\includegraphics[width=0.32\hsize]{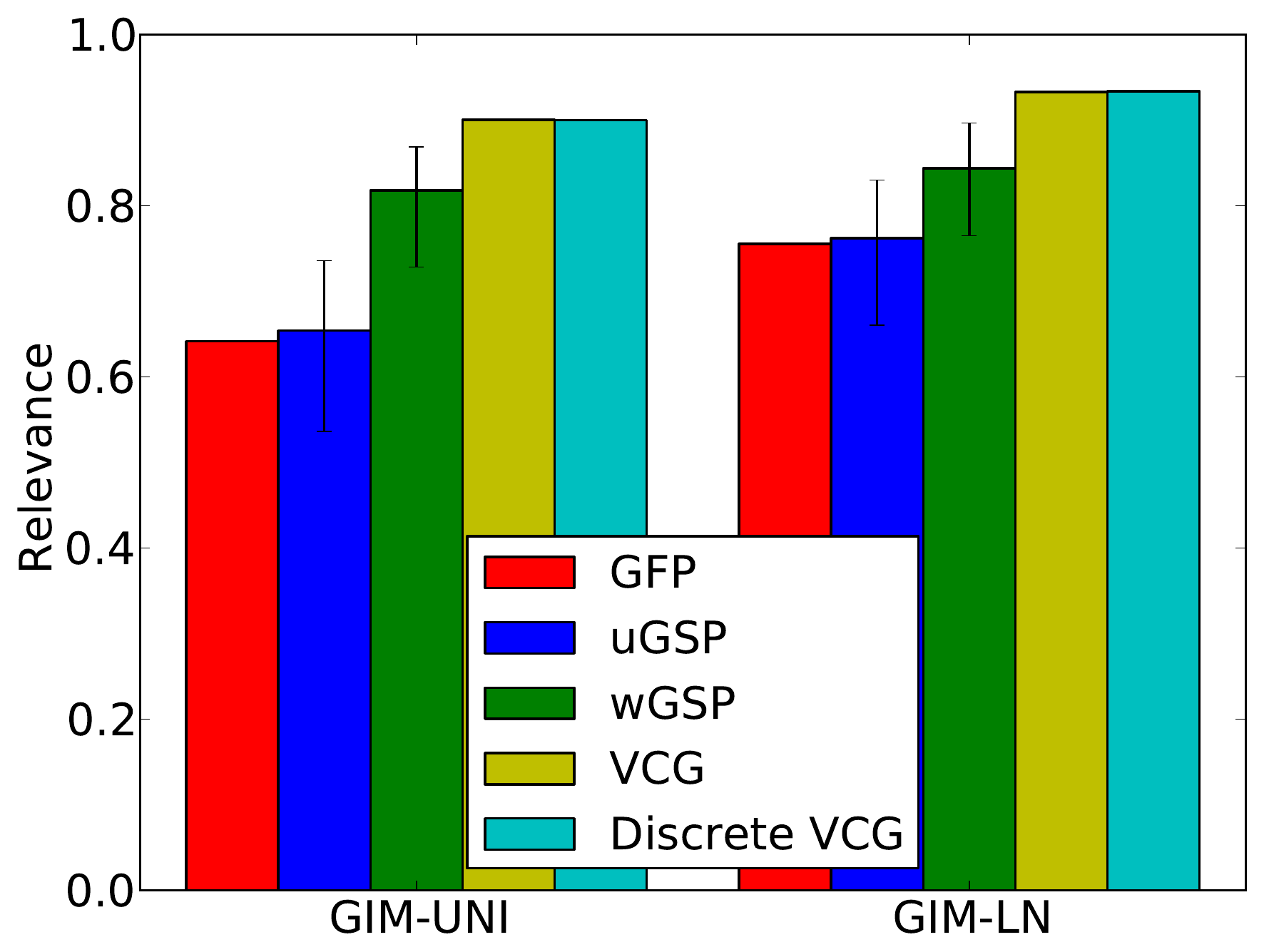}
	\kcaption{(a) Efficiency}{.32\hsize}
	\kcaption{(b) Revenue}{.32\hsize}
	\kcaption{(c) Relevance}{.32\hsize}
	\caption[Comparing the average performance of different position auction types in GIM settings]{Comparing the average performance of different position auction types in GIM settings.  \whiskerBP}
	\label{fig:gim-comparison}
\end{figure}

Lastly, we compare different position auctions in the GIM setting (see Figure~\ref{fig:gim-comparison}).  As under other externality models, we found that wGSP produced more efficient and relevant rankings than GFP and uGSP, but performed significantly worse than VCG.  Again, revenue was ambiguous, with GSP revenue varying substantially across equilibria and all median revenues fairly close to VCG's revenue.

\section{Sensitivity Analysis}\label{sec:sensitivity}

Computational analysis techniques oblige us to make concrete choices about discretization and the sizes of problems that we study. This section examines the extent to which our findings varied under five such details about the position auction setting that we have held constant until now. (1)~Although real-world auctions are discrete, the coarseness of the discretization can vary greatly from keyword to keyword or advertiser to advertiser, depending on the size of each advertiser's valuation relative to the bid increment. Thus, it is important to understand how sensitive our findings are to the scale of the discretization. When bids and payments are discrete, issues also arise that are not significant in continuous games, such as (2)~how ties will be broken and (3)~how payments will be rounded or aggregated.  Furthermore, computational analysis requires setting many parameters that can often be left unspecified in a theoretical analysis, such as (4)~the number of bidders and (5)~the number of slots for sale. We chose two widely studied distributions to consider in our scaling studies: one simple (V) and one with externalities (cascade). In both cases we used the more realistic valuation distribution, log-normal.

\subsection{Sensitivity to Bid Increment Size}

\paperonly{
\begin{figure}
	\includegraphics[width=0.32\hsize]{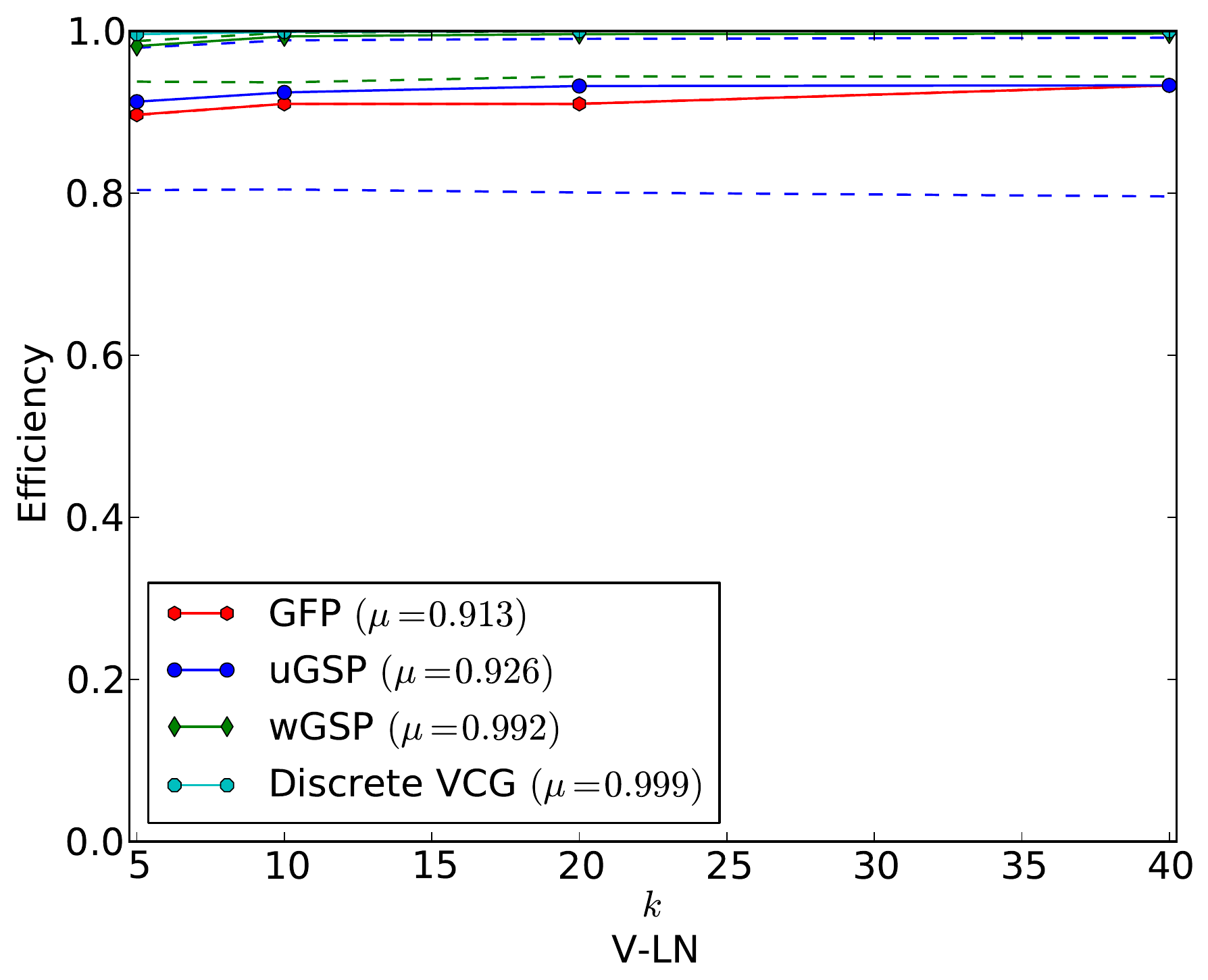}
	\includegraphics[width=0.32\hsize]{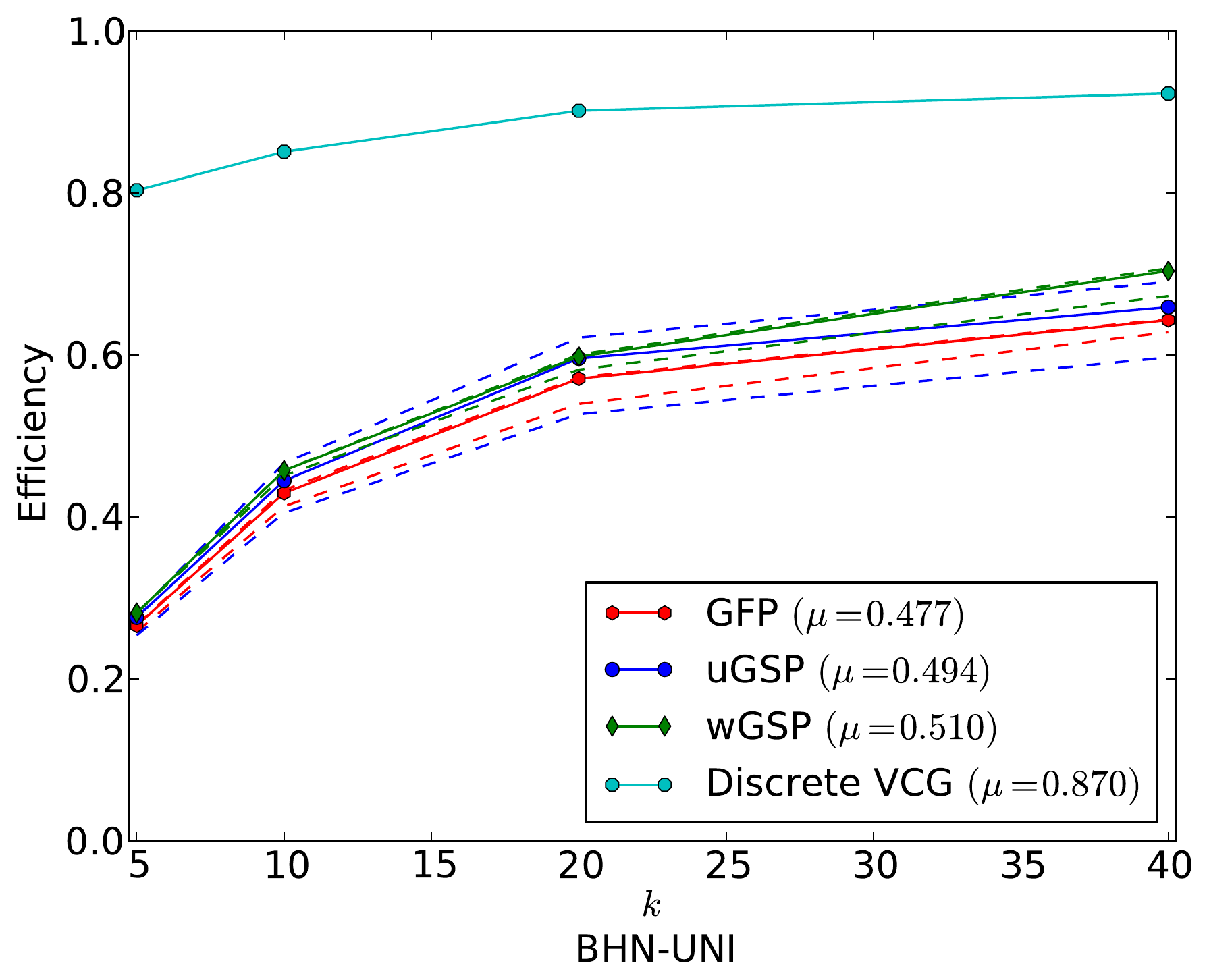}
	\includegraphics[width=0.32\hsize]{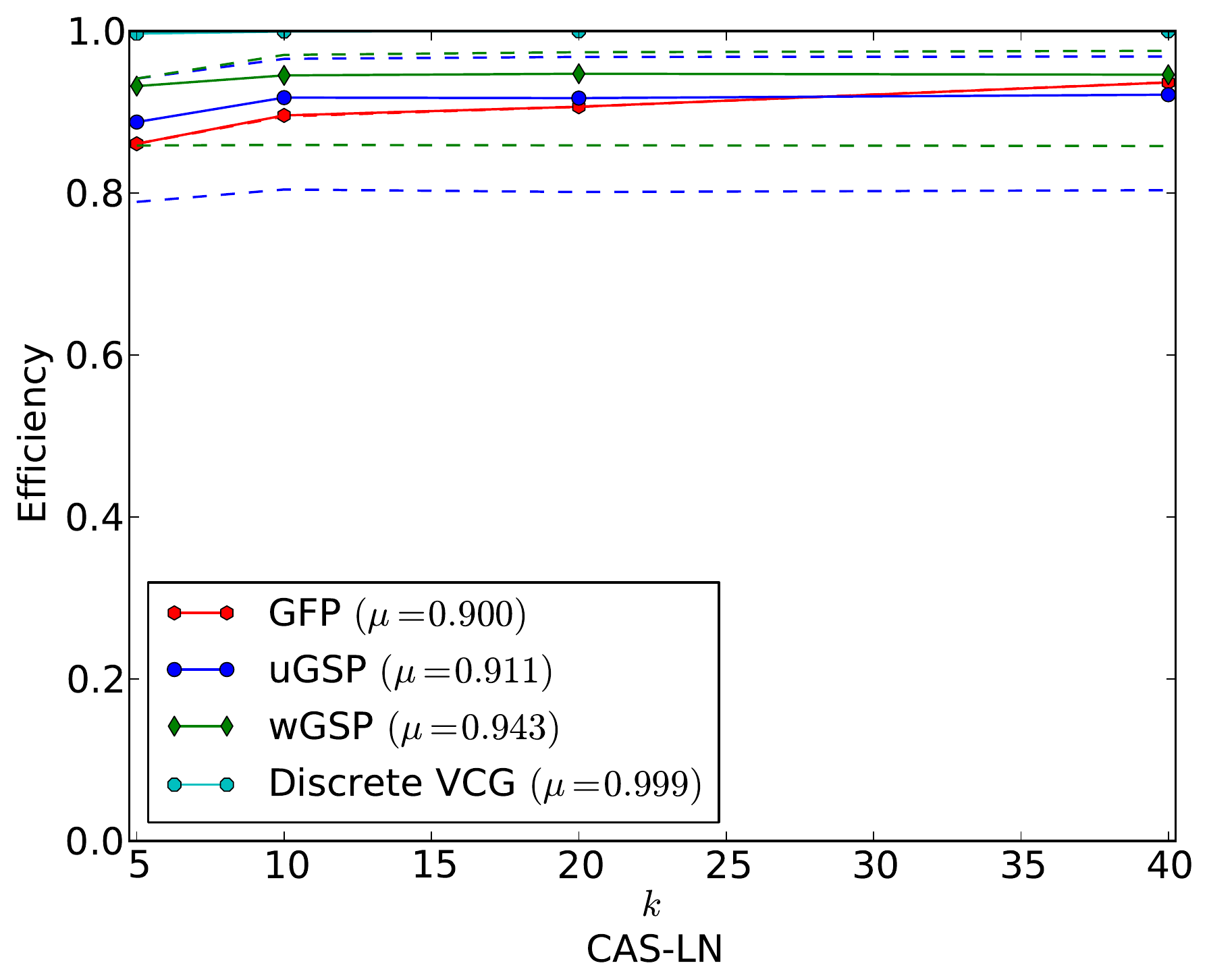}
	\kcaption{(a) V: Log-Normal, Efficiency}{.32\hsize}
	\kcaption{(b) BHN: Uniform, Efficiency}{.32\hsize}
	\kcaption{(c) Cascade: Log-Normal, Efficiency}{.32\hsize}
	
	\includegraphics[width=0.32\hsize]{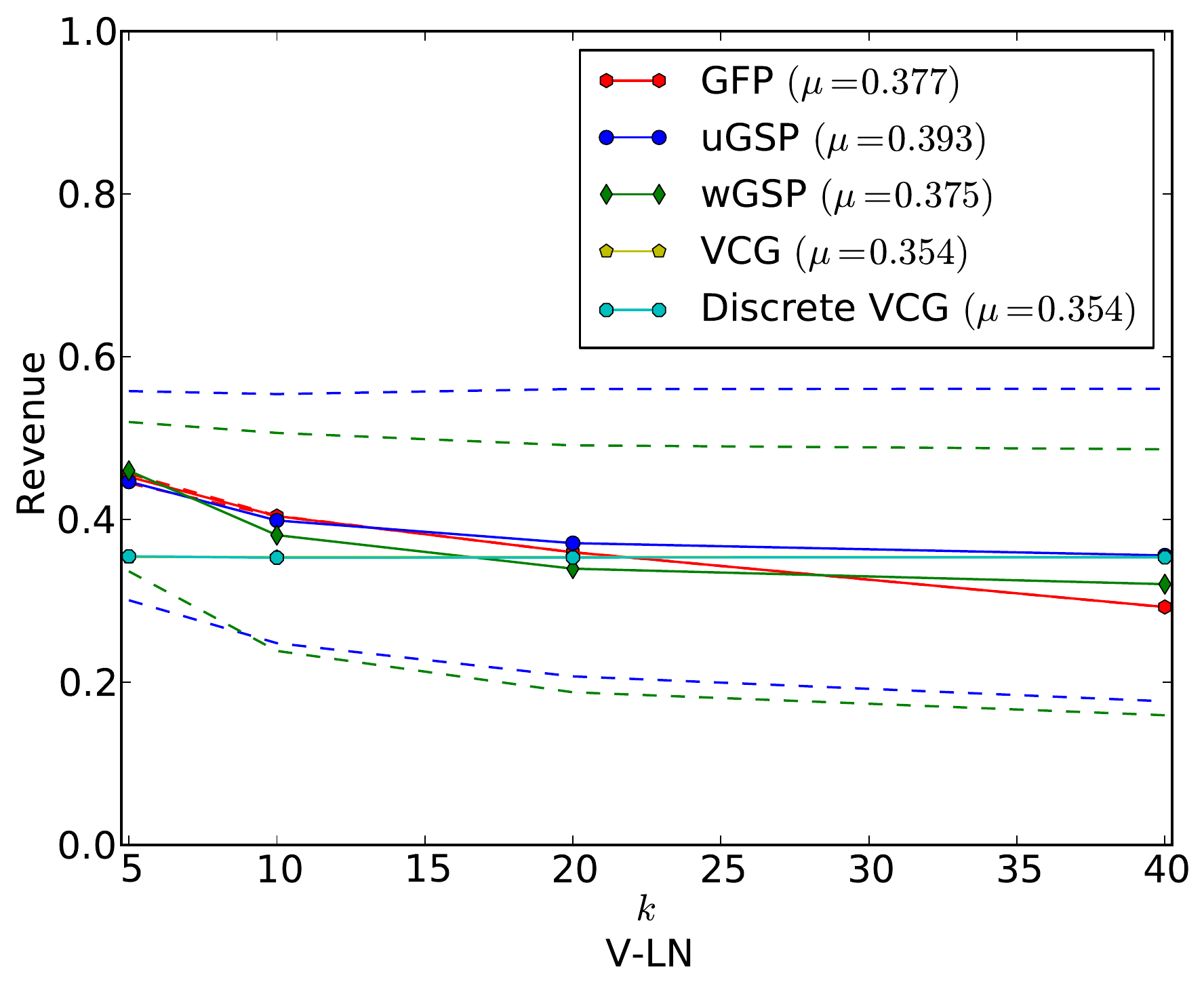}
	\includegraphics[width=0.32\hsize]{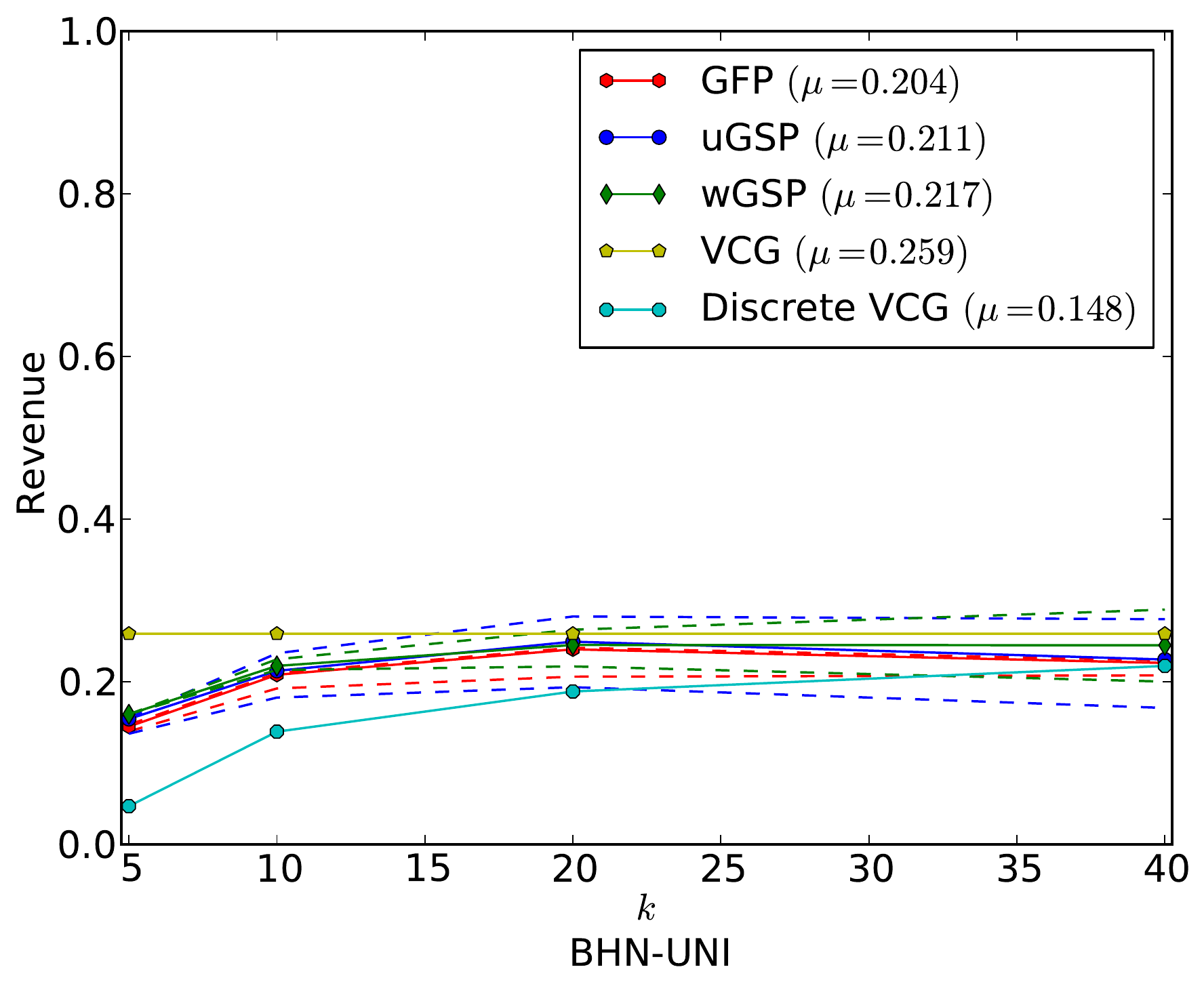}
	\includegraphics[width=0.32\hsize]{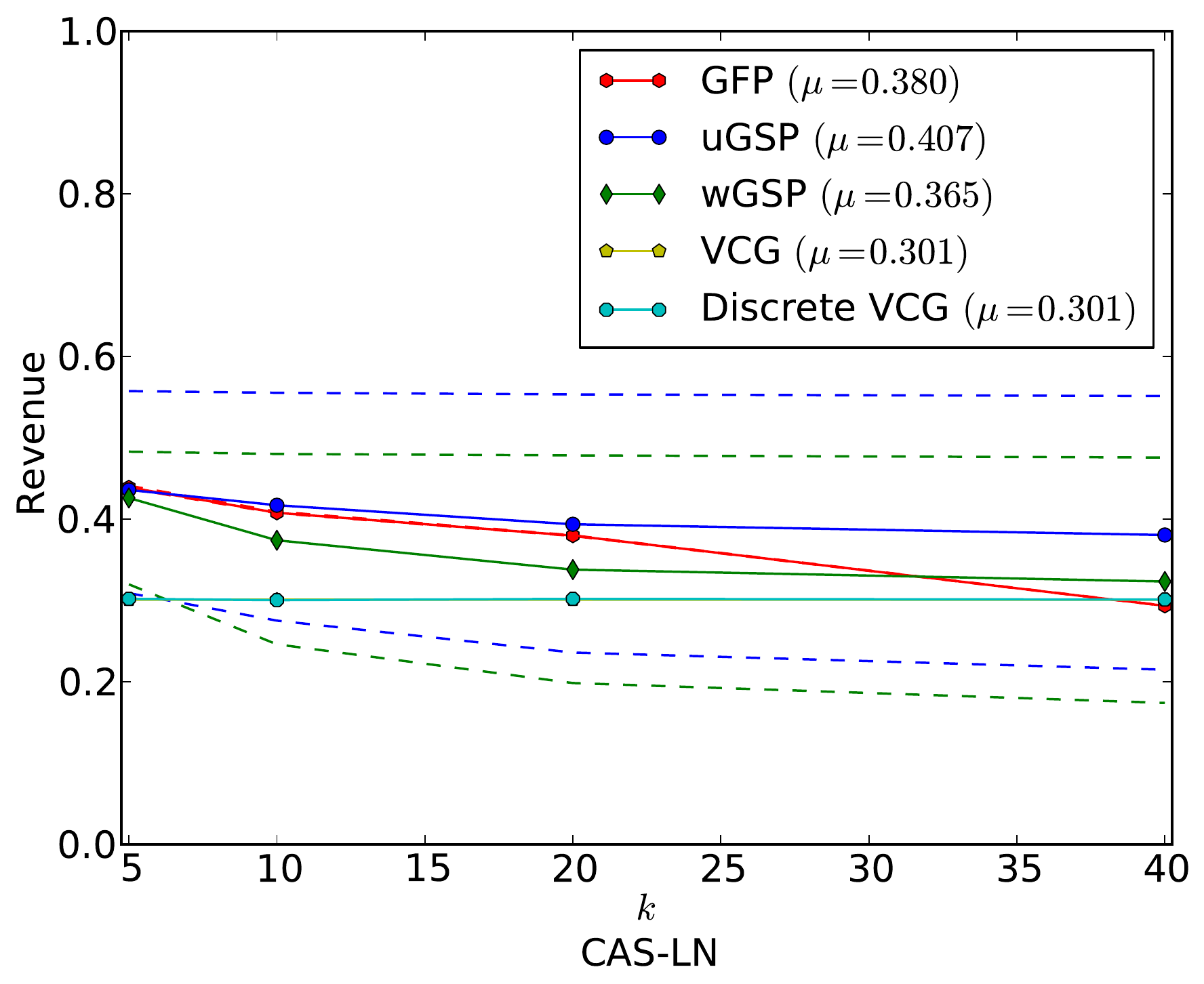}
	\kcaption{(d) V: Log-Normal, Revenue}{.32\hsize}
	\kcaption{(e) BHN: Uniform, Revenue}{.32\hsize}
	\kcaption{(f) Cascade: Log-Normal, Revenue}{.32\hsize}
	
	\includegraphics[width=0.32\hsize]{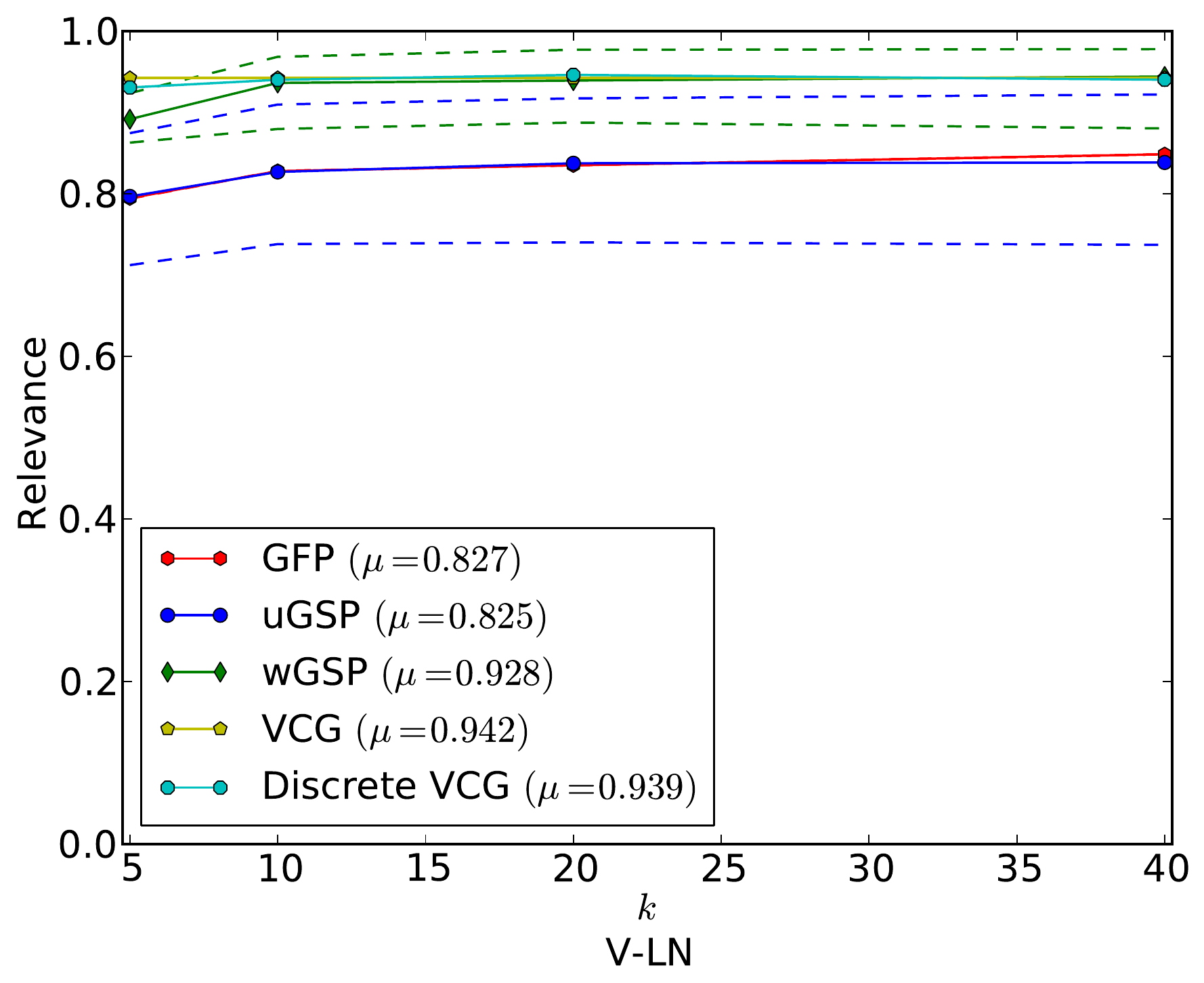}
	\includegraphics[width=0.32\hsize]{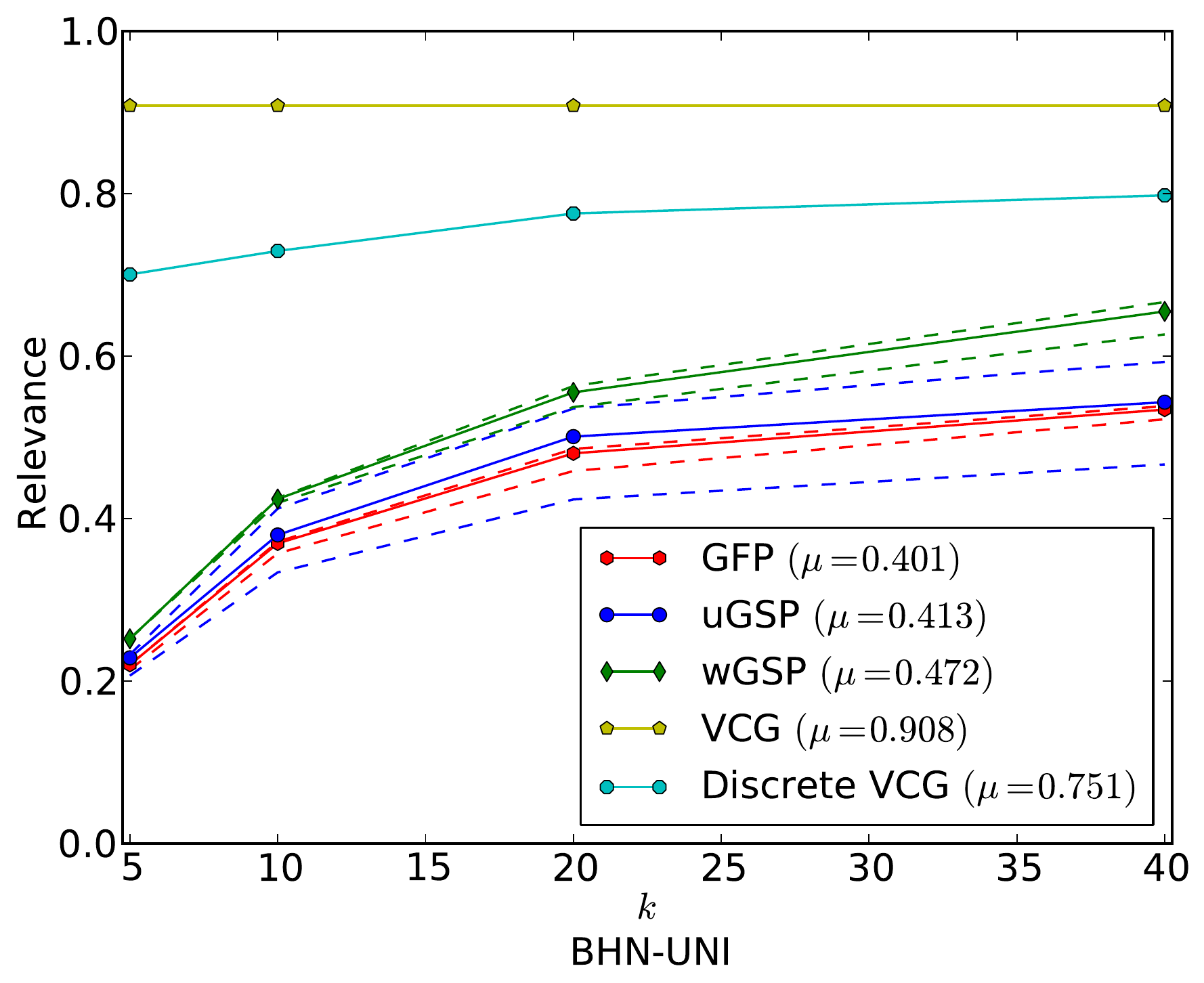}
	\includegraphics[width=0.32\hsize]{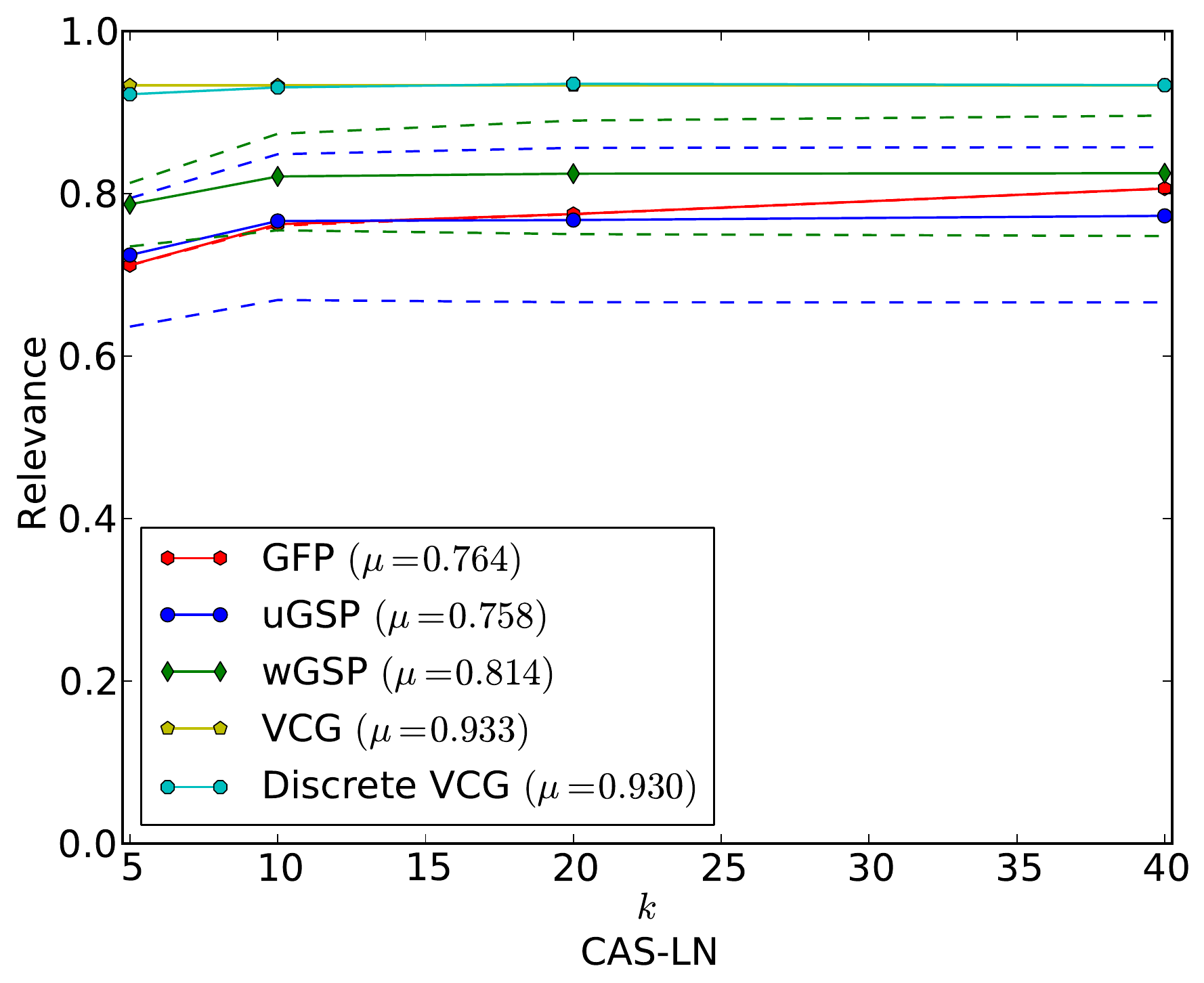}
	\kcaption{(g) V: Log-Normal, Relevance}{.32\hsize}
	\kcaption{(h) BHN: Uniform, Relevance}{.32\hsize}
	\kcaption{(i) Cascade: Log-Normal, Relevance}{.32\hsize}
	
	\hspace{0.165\hsize}
	\includegraphics[width=0.32\hsize]{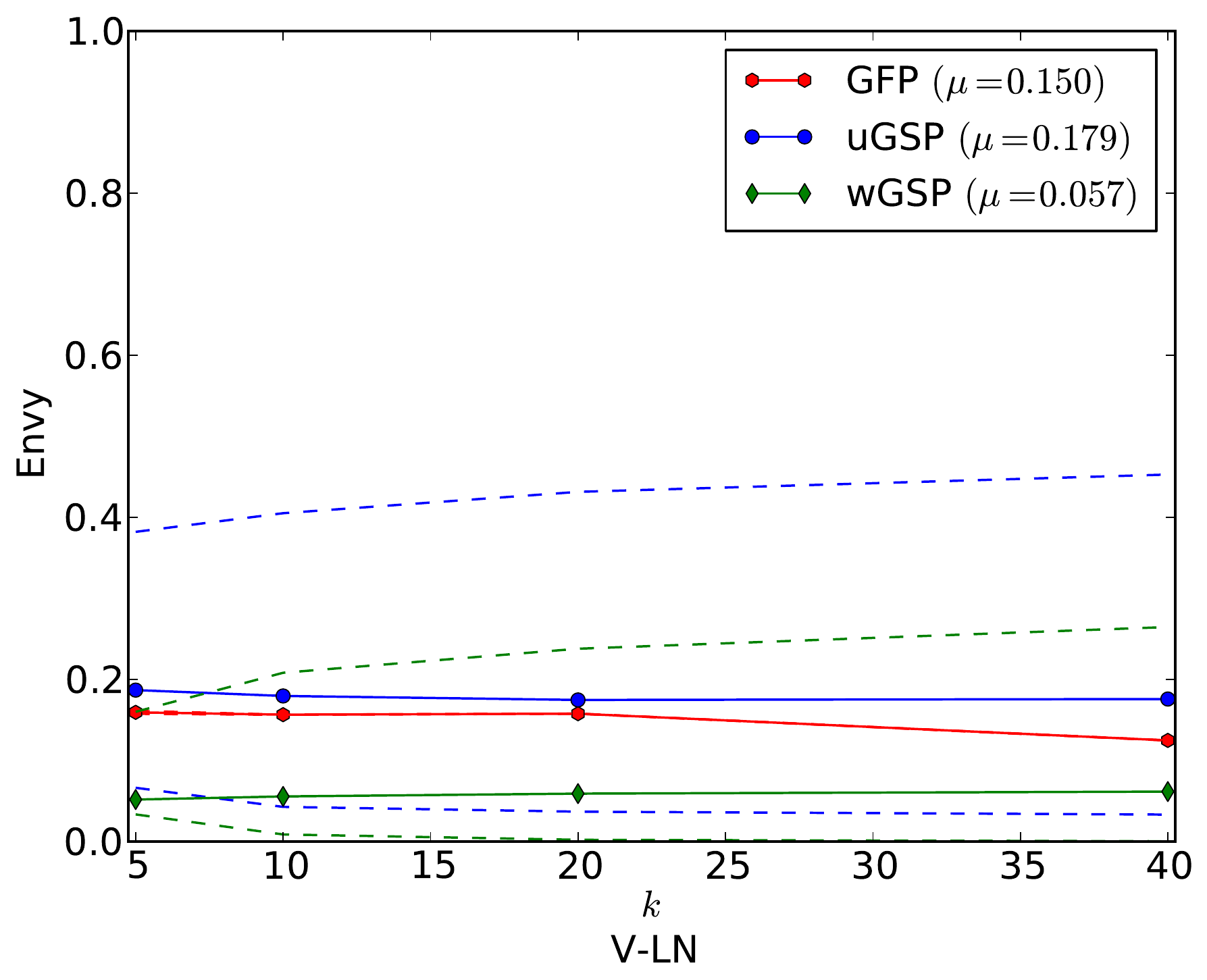}
	\includegraphics[width=0.32\hsize]{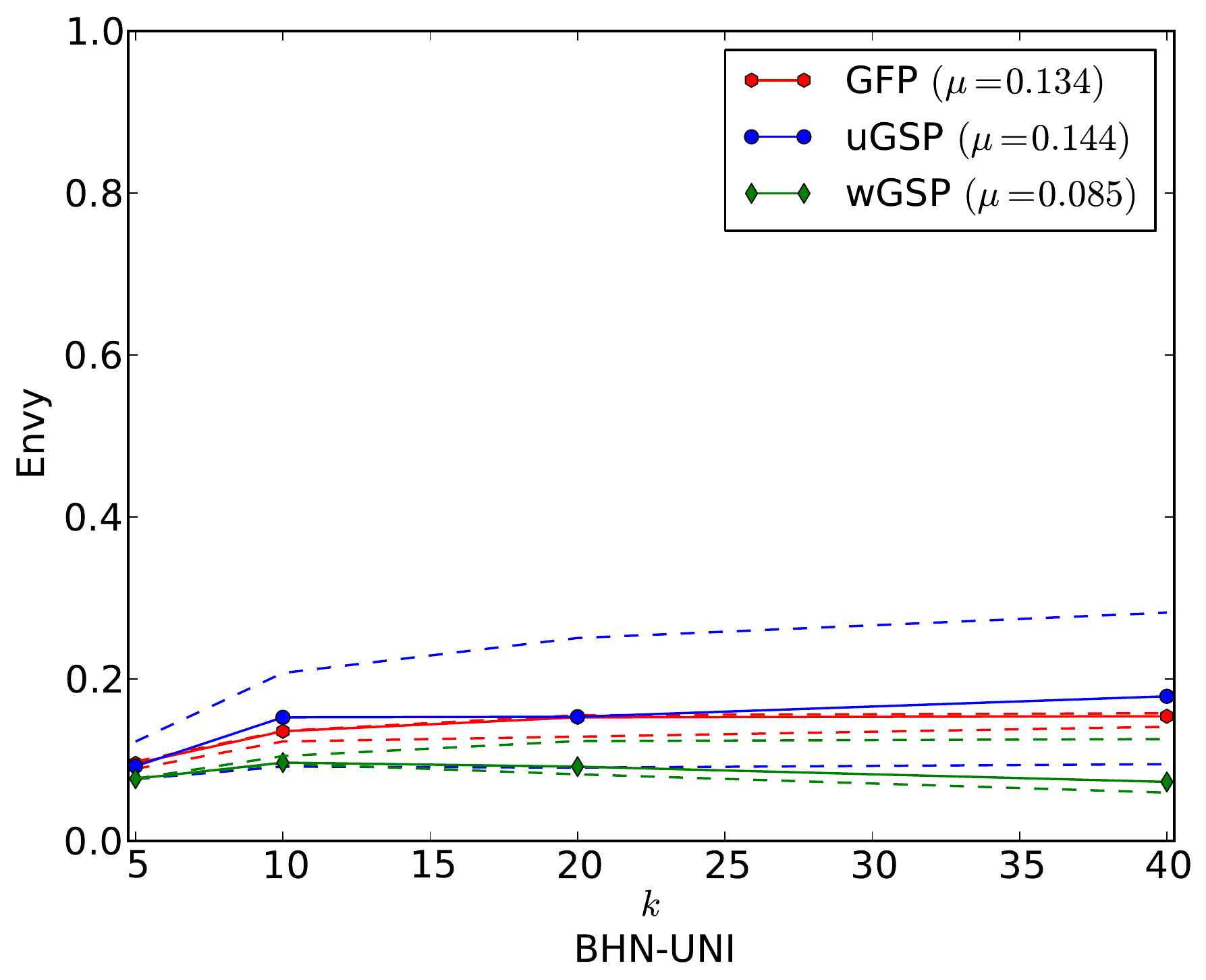}\\
	\kcaption{~}{0.165\hsize}\kcaption{(j) V: Log-Normal, Envy}{.32\hsize}
	\kcaption{(k) BHN: Uniform, Envy}{.32\hsize}\\
	\caption{Comparing different auction designs as the number of bid increments varies.}
	\label{fig:scale-k}
\end{figure}
}

\chapteronly{
\begin{figure}
	\hspace{0.05\hsize}
	\includegraphics[width=0.45\hsize]{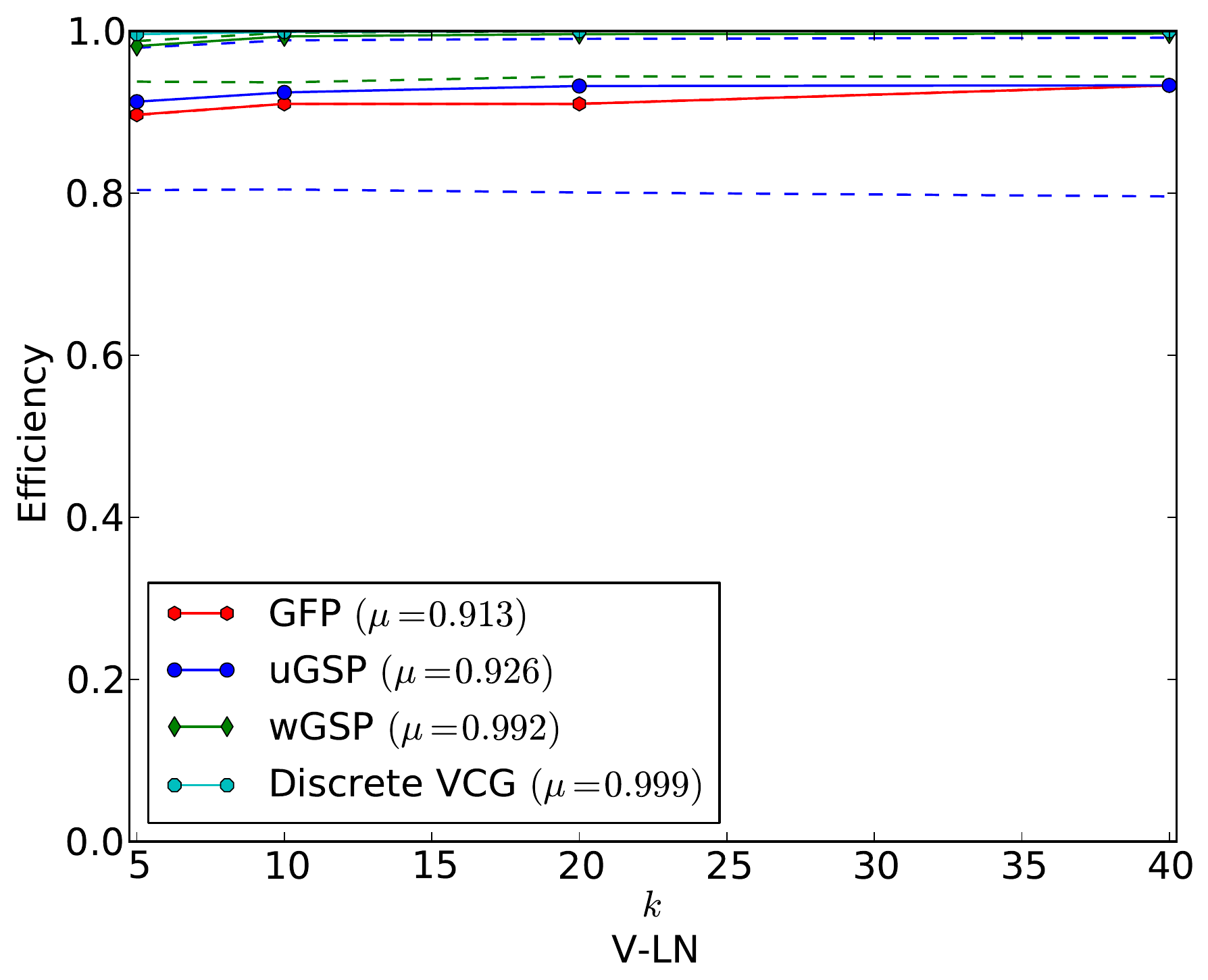}
	\includegraphics[width=0.45\hsize]{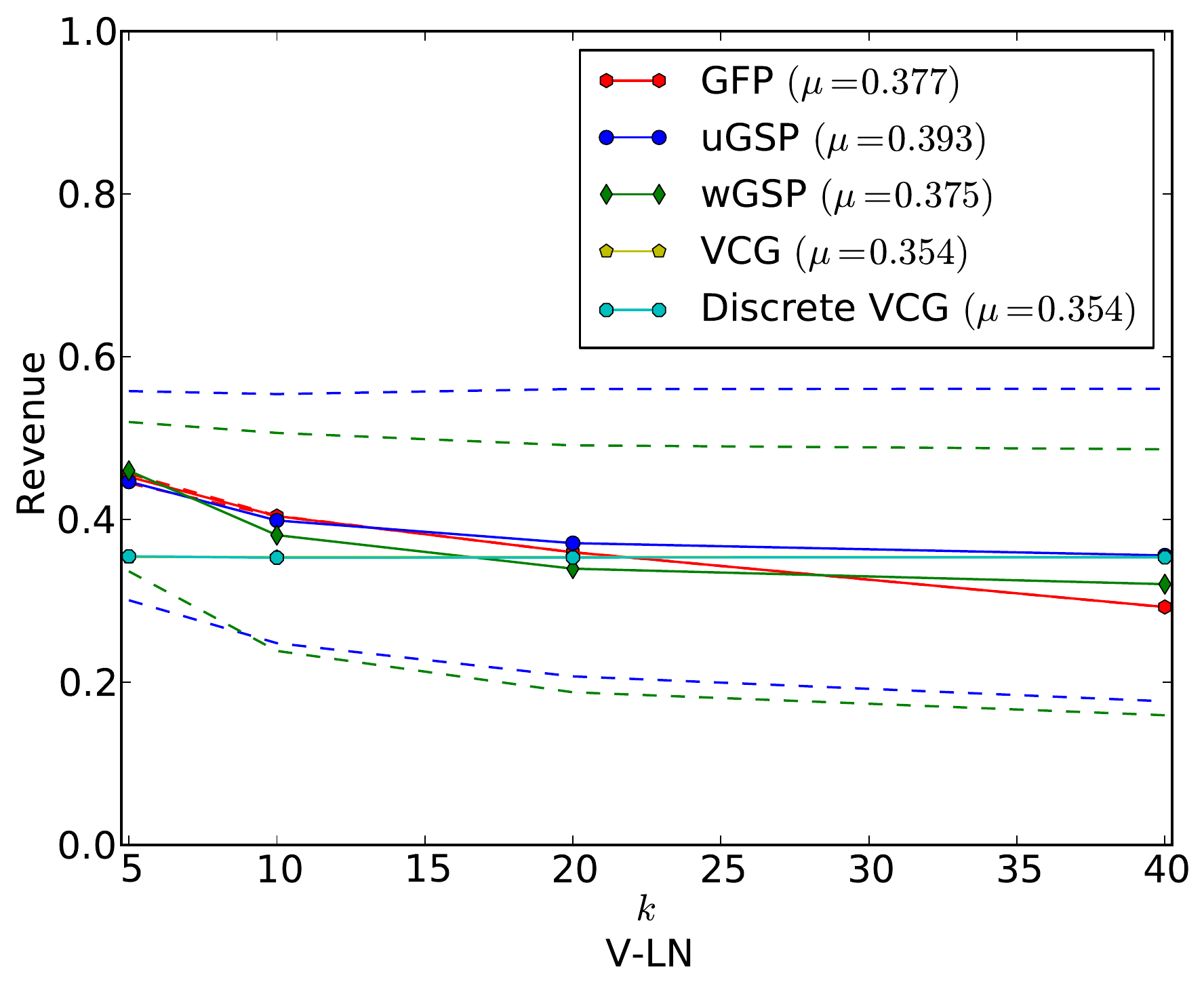}\\
	\kcaption{~}{0.05\hsize}
	\kcaption{(a) V: Log-Normal, Efficiency}{.45\hsize}
	\kcaption{(d) V: Log-Normal, Revenue}{.45\hsize}

	\hspace{0.05\hsize}
	\includegraphics[width=0.45\hsize]{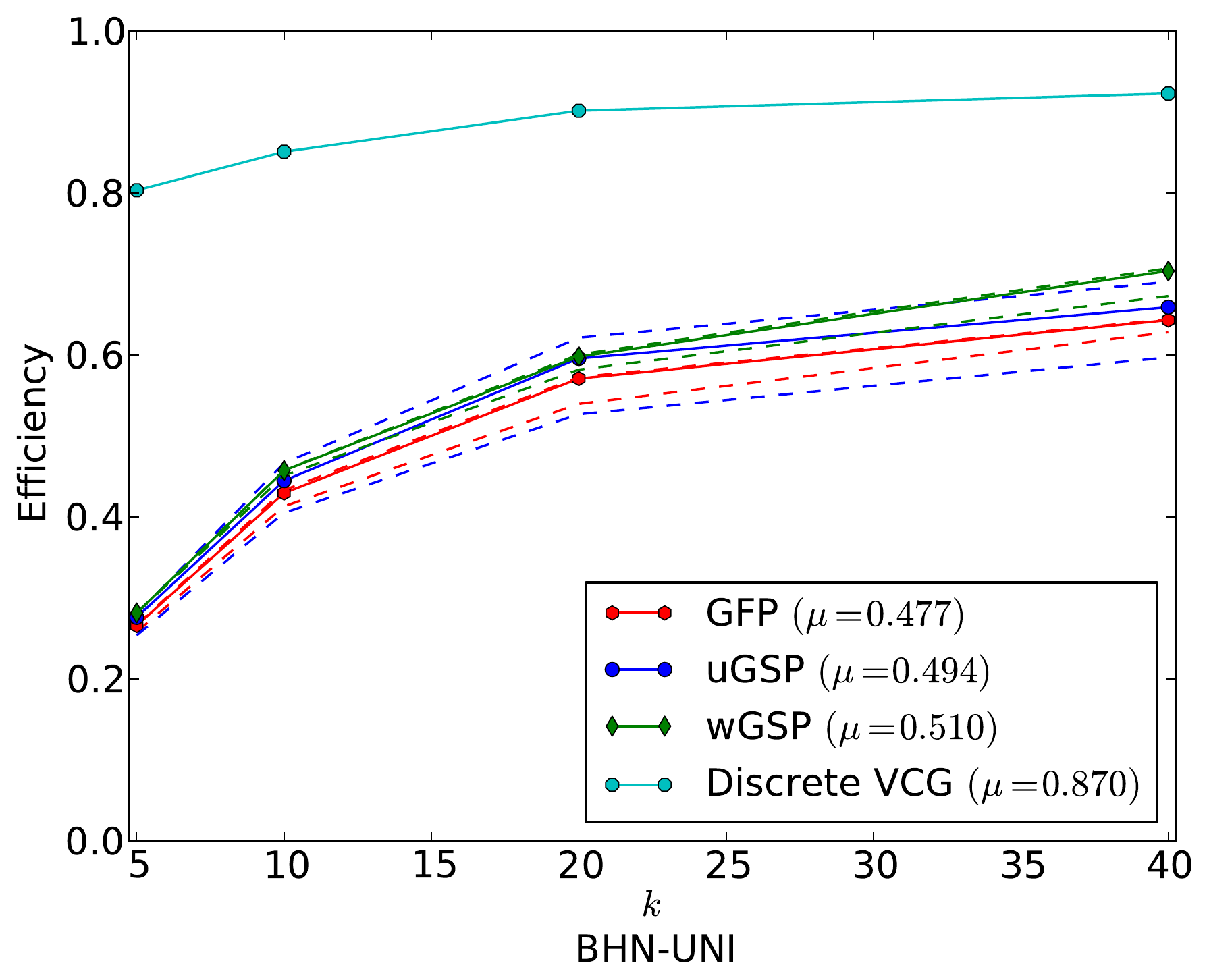}
	\includegraphics[width=0.45\hsize]{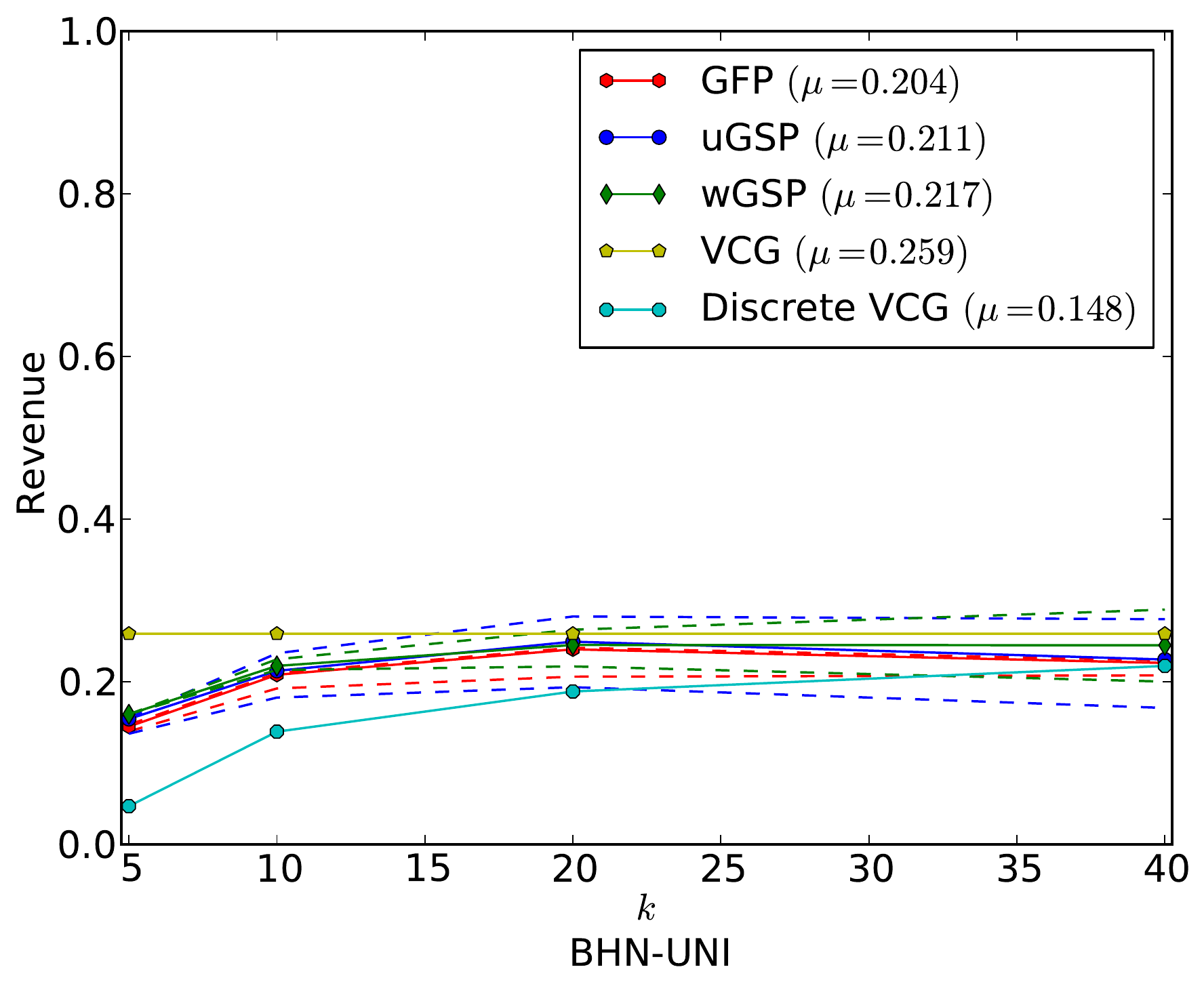}\\
	\kcaption{~}{0.05\hsize}
	\kcaption{(b) BHN: Uniforml, Efficiency}{.45\hsize}
	\kcaption{(e) BHN: Uniform, Revenue}{.45\hsize}

	\hspace{0.05\hsize}
	\includegraphics[width=0.45\hsize]{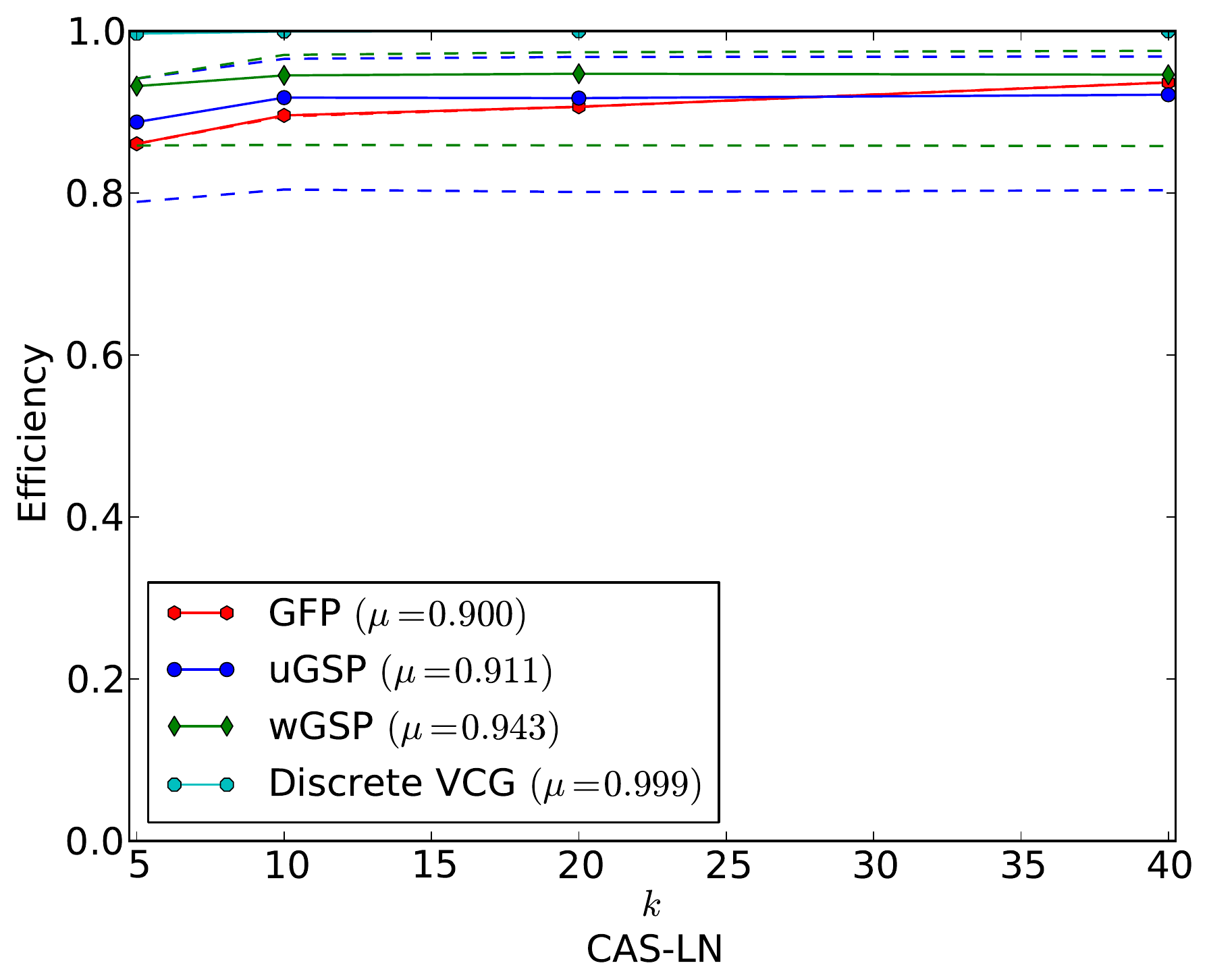}
	\includegraphics[width=0.45\hsize]{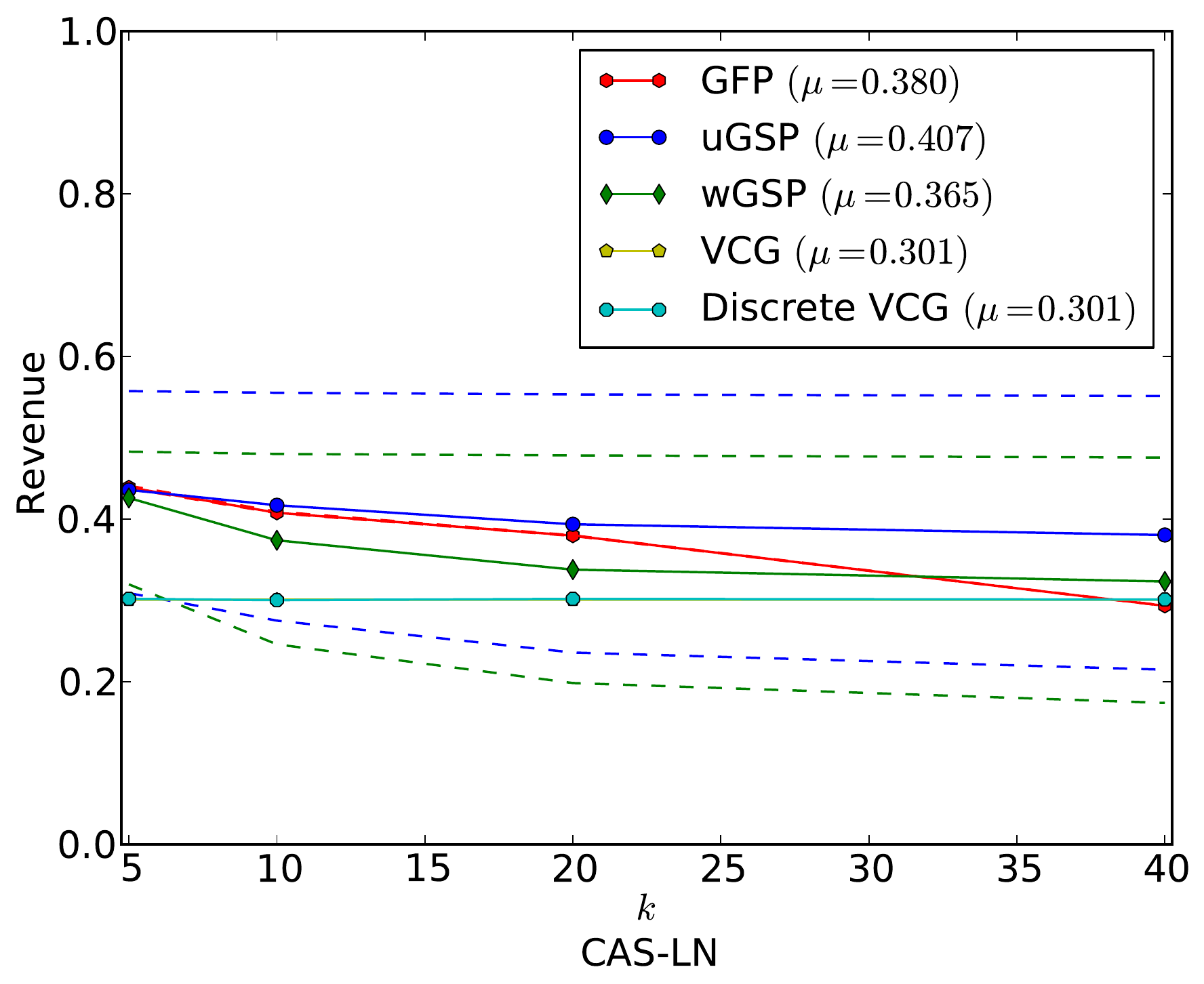}\\
	\kcaption{~}{0.05\hsize}
	\kcaption{(c) Cascade: Log-Normal, Efficiency}{.45\hsize}
	\kcaption{(f) Cascade: Log-Normal, Revenue}{.45\hsize}
	\caption{Comparing different auction designs as the number of bid increments varies.}
	\label{fig:scale-k1}
\end{figure}

\begin{figure}
	\hspace{0.05\hsize}
	\includegraphics[width=0.45\hsize]{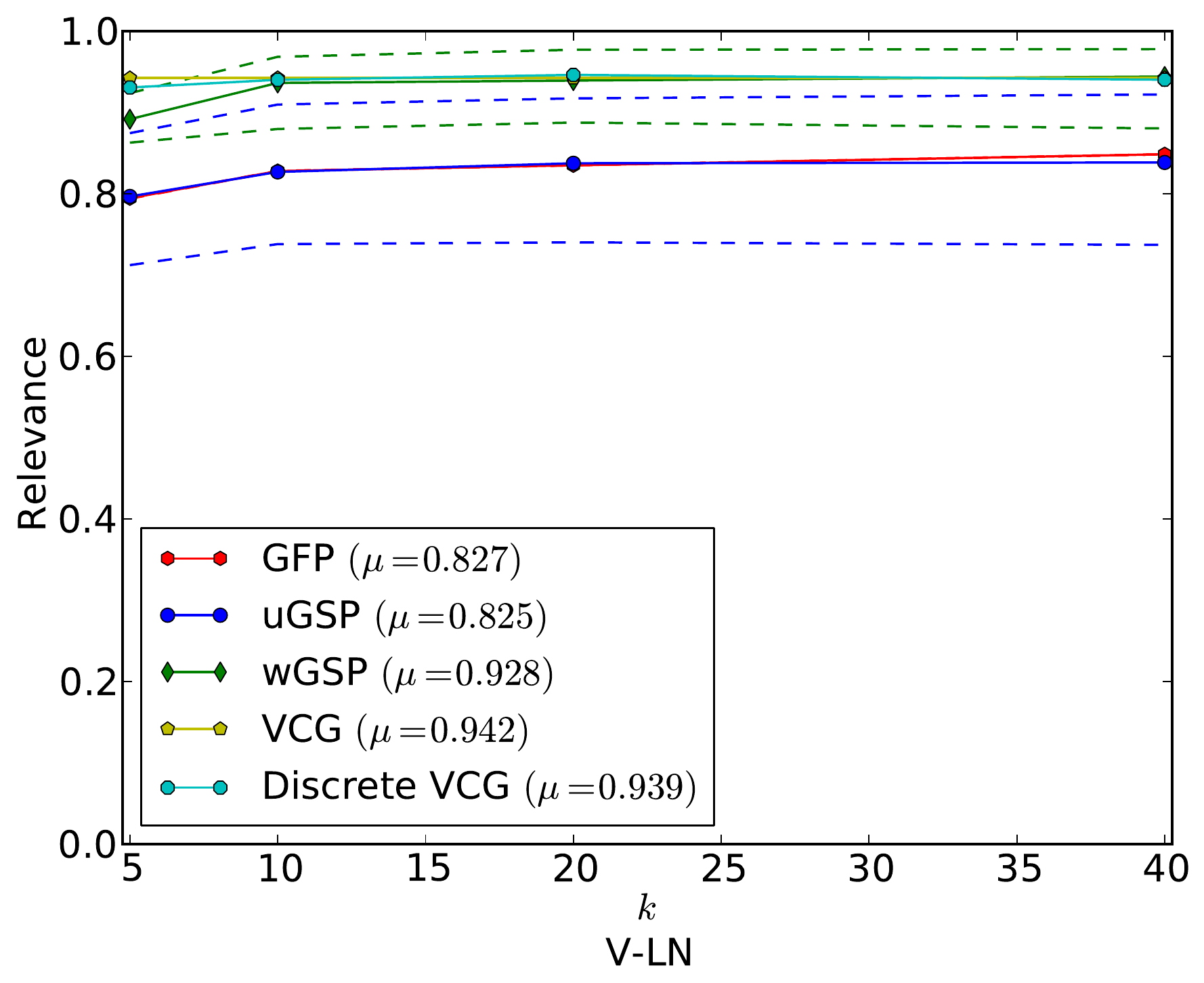}
	\includegraphics[width=0.45\hsize]{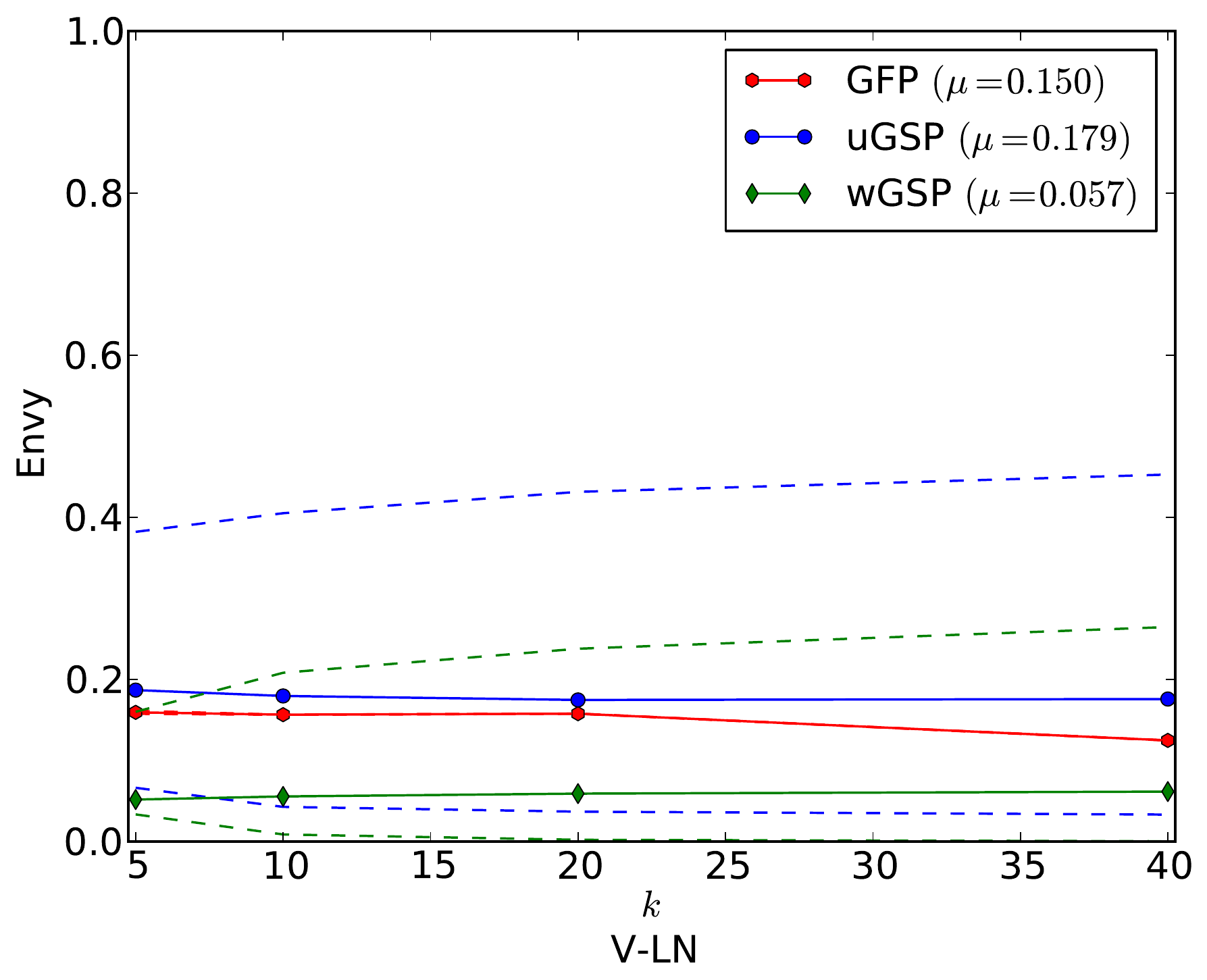}\\
	\kcaption{~}{0.05\hsize}
	\kcaption{(g) V: Log-Normal, Relevance}{.45\hsize}
	\kcaption{(j) V: Log-Normal, Envy}{.45\hsize}

	\hspace{0.05\hsize}
	\includegraphics[width=0.45\hsize]{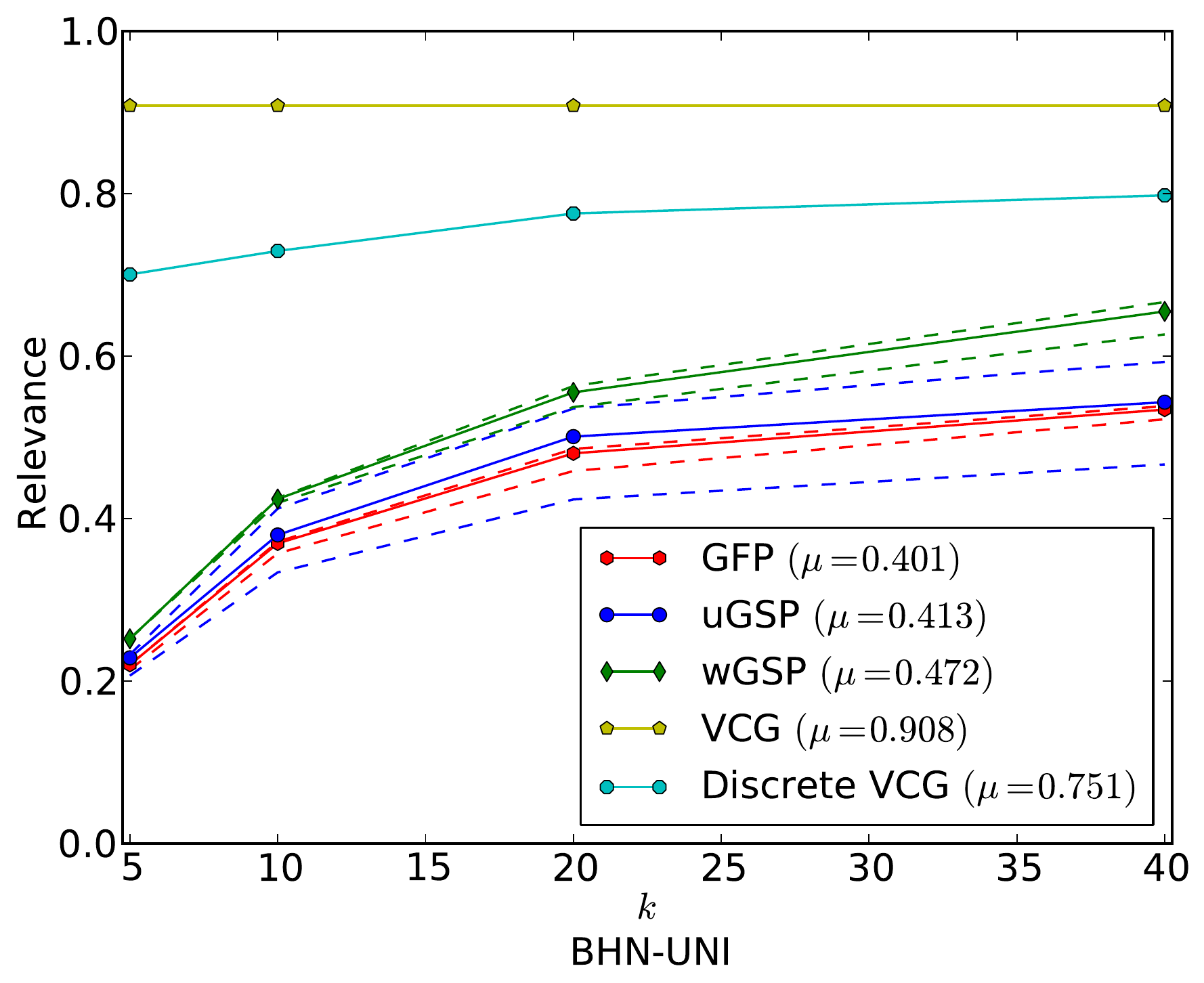}
	\includegraphics[width=0.45\hsize]{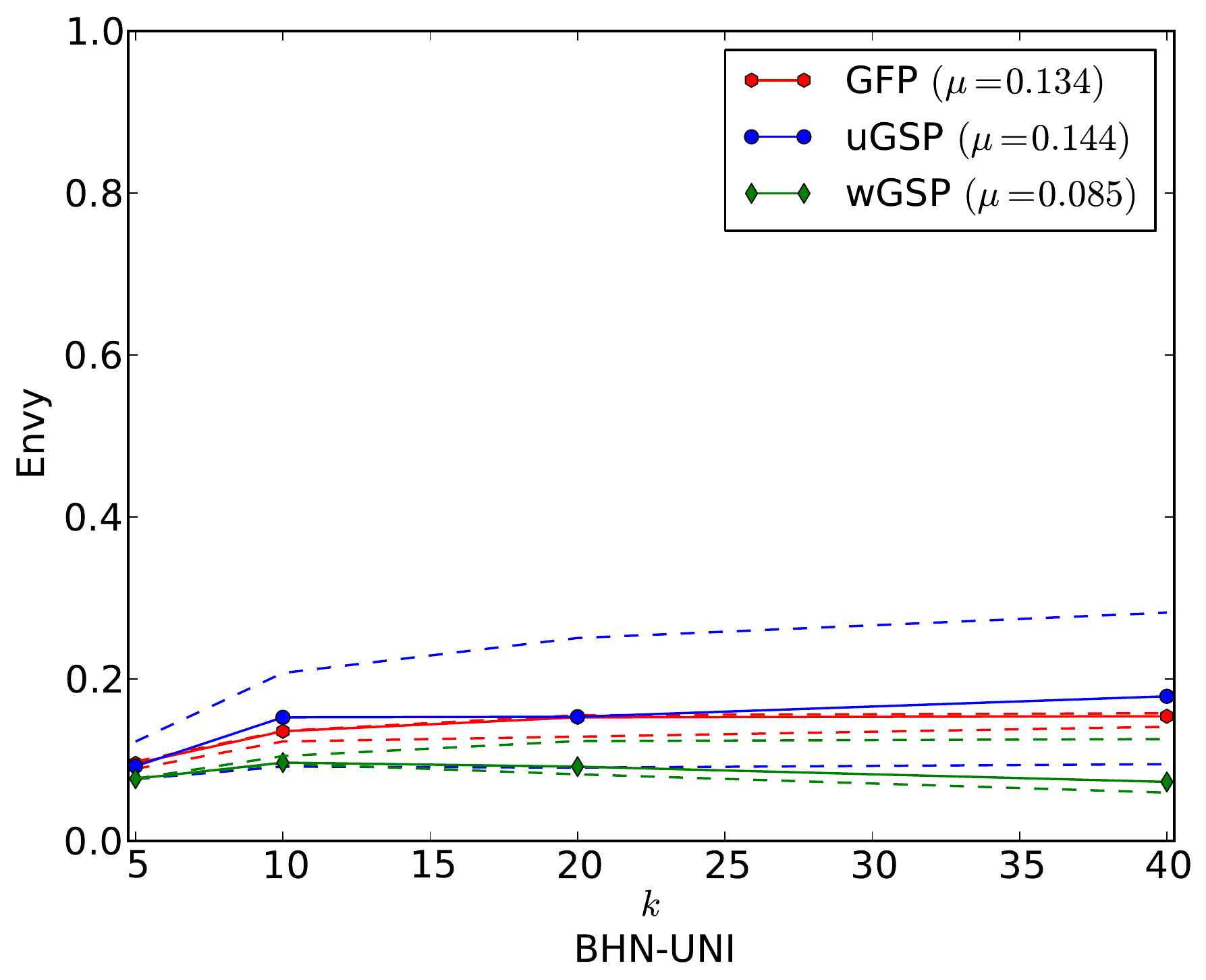}\\
	\kcaption{~}{0.05\hsize}
	\kcaption{(h) BHN: Uniforml, Relevance}{.45\hsize}
	\kcaption{(k) BHN: Uniform, Envy}{.45\hsize}

	\hspace{0.05\hsize}
	\includegraphics[width=0.45\hsize]{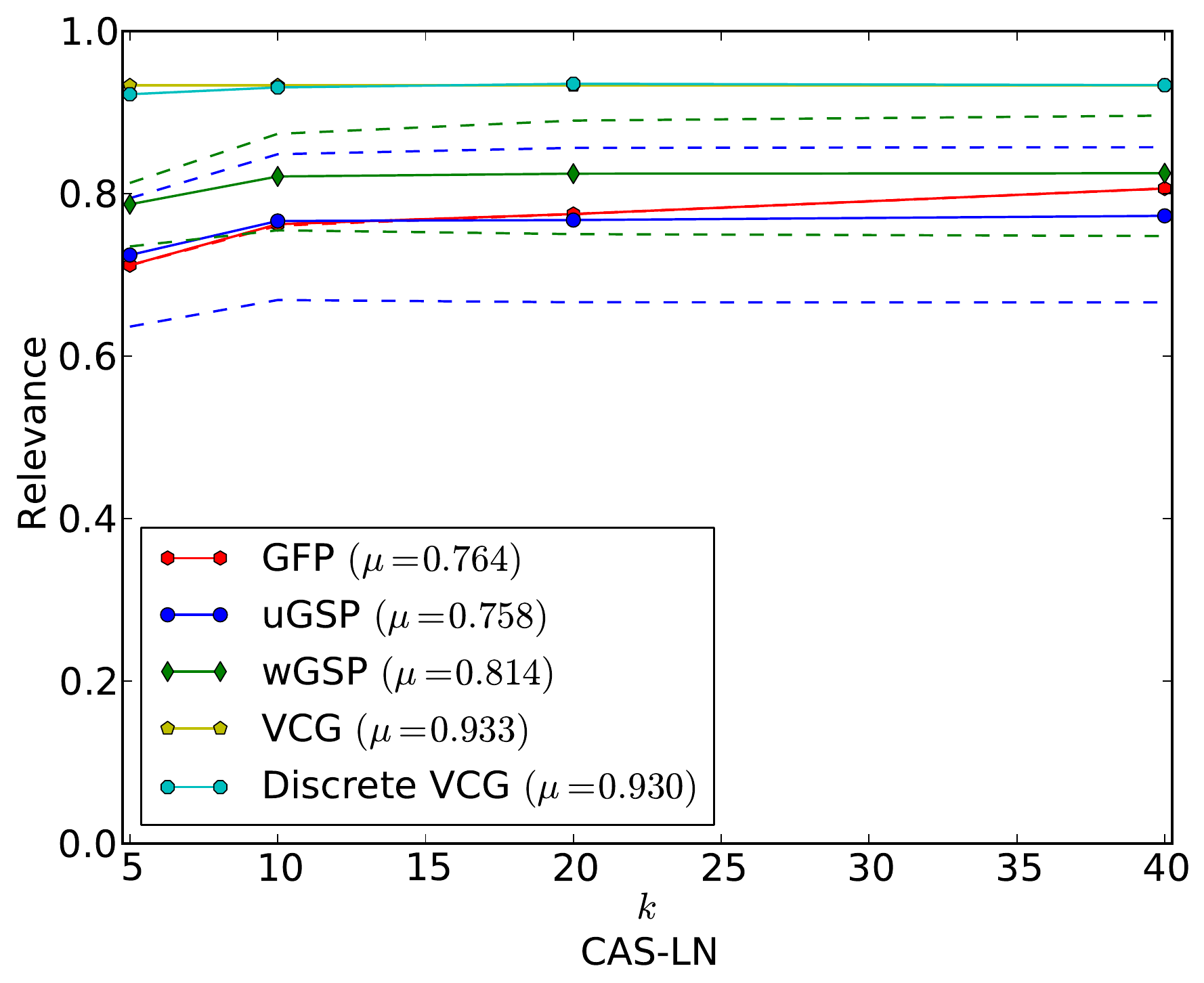}\\
	\kcaption{~}{0.05\hsize}
	\kcaption{(i) Cascade: Log-Normal, Relevance}{.45\hsize}
	\caption{Comparing different auction designs as the number of bid increments varies (continued).}
	\label{fig:scale-k2}
\end{figure}
}

First, we considered the problem of increment size.  In addition to V-LN and CAS-LN we also considered the BHN-UNI distribution, because we found BHN's interaction with increment size particularly interesting. Recall that earlier we found that the single-increment reserve price was sufficient to cause dramatic efficiency loss.  We anticipated that as increment size decreased, this would occur less frequently.  For each distribution, we varied the number of bid increments from 5 to 40, at each step generating 100 instances and computing all relevant metrics (see \paperonly{Figure~\ref{fig:scale-k}}\chapteronly{Figures~\ref{fig:scale-k1} and \ref{fig:scale-k2}}). In the V and cascade distributions, we found that (with one exception) the relative performance of different mechanisms became steady once $k$ was greater than 10, and that absolute performance became stable once $k$ was greater than 20.  The one exception was the GFP auction; in both distributions, GFP's performance (both in relative and absolute terms) varied dramatically as $k$ grew.  BHN's behavior was particularly interesting.  As the number of increments increased, the efficiency and relevance of all auctions dramatically increased (because the problem of unallocated slots became less common).  Worst-case envy and revenue tended to get worse as the number of increments increased, while median- and best-case revenue and envy remained fairly flat.

\subsection{Sensitivity to Tie-Breaking Rules}\label{sec:tie-break}

Discretization also invites the possibility that ties will occur between bids with positive probability. Thus, while tie-breaking rules are typically unimportant to the analysis of auctions with continuous action spaces, they can be significant in discrete settings. Observe, however, that this problem does not arise under most of our distributions; although bids are integral, each bid is weighted by a real number, with the property that for any pair of weights the probability of their ratio being integral is zero.  Thus, we only needed to examine sensitivity to tie breaking in the two models that did not have weights: EOS-LN and BSS-UNI.

\begin{figure}
\centering
\includegraphics[width=0.45\hsize]{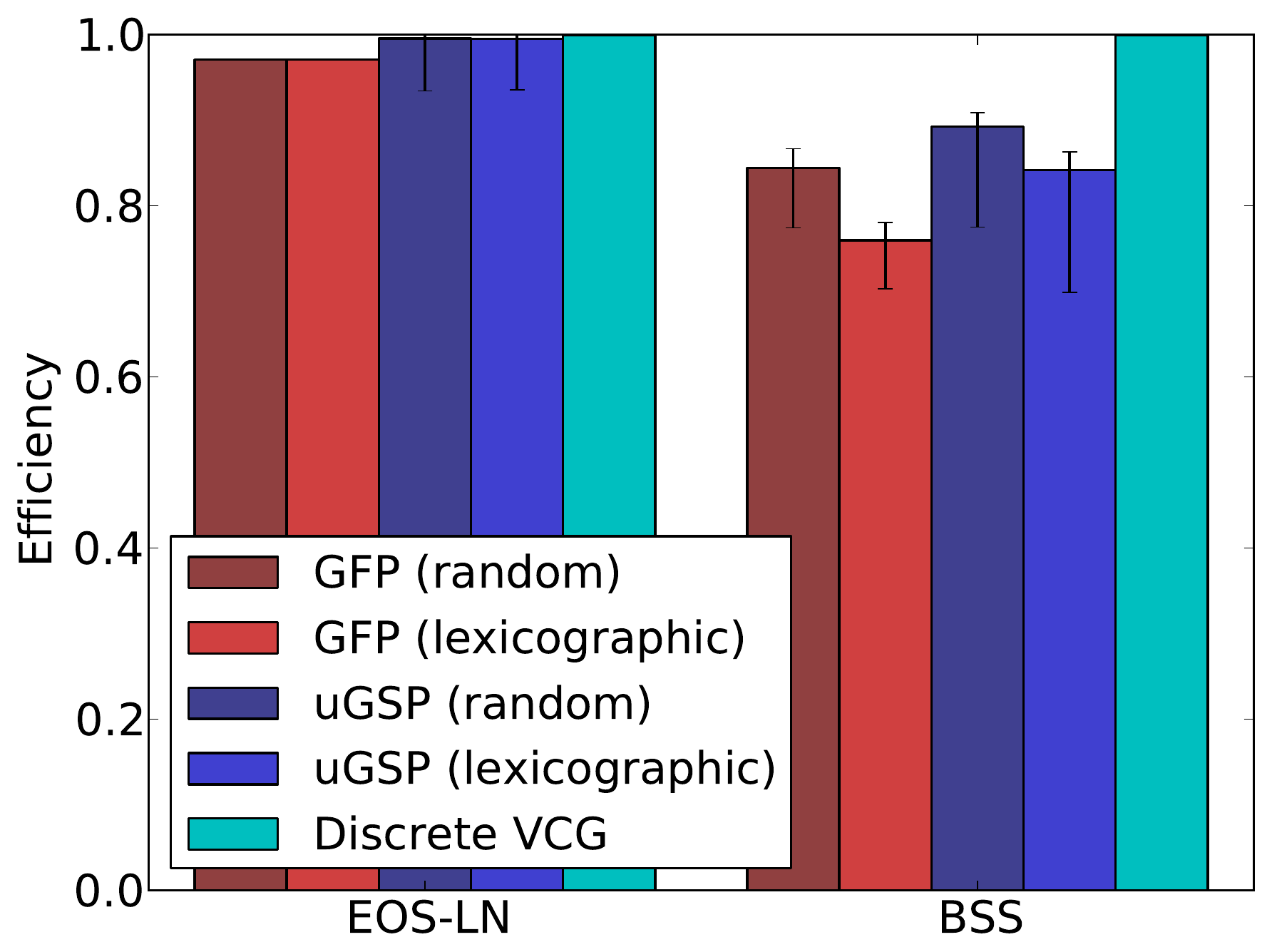}
\includegraphics[width=0.45\hsize]{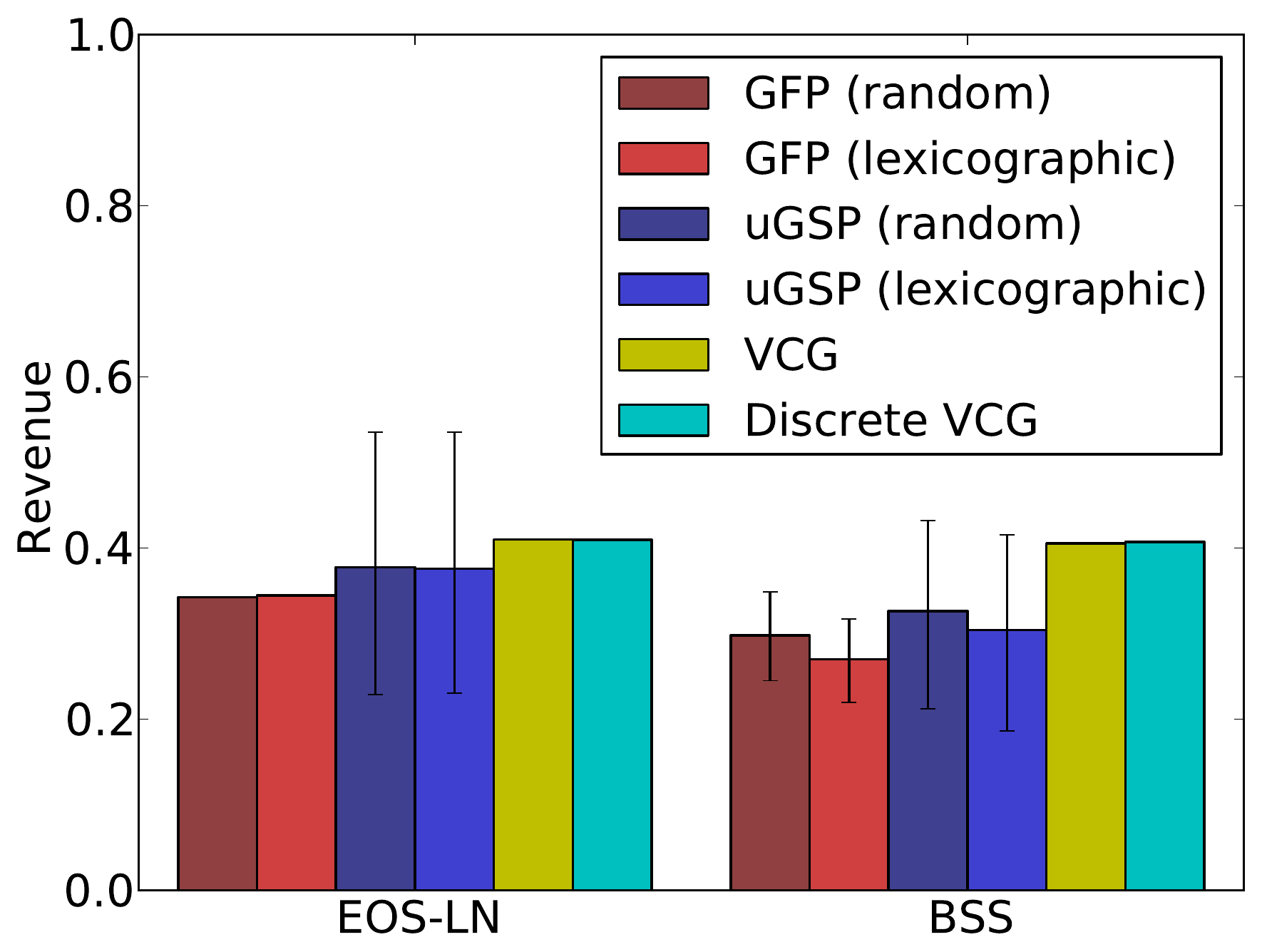}
\kcaption{(a) EOS-LN and BSS: Efficiency}{.45\hsize}
\kcaption{(b) EOS-LN and BSS: Revenue}{.45\hsize}
\includegraphics[width=0.45\hsize]{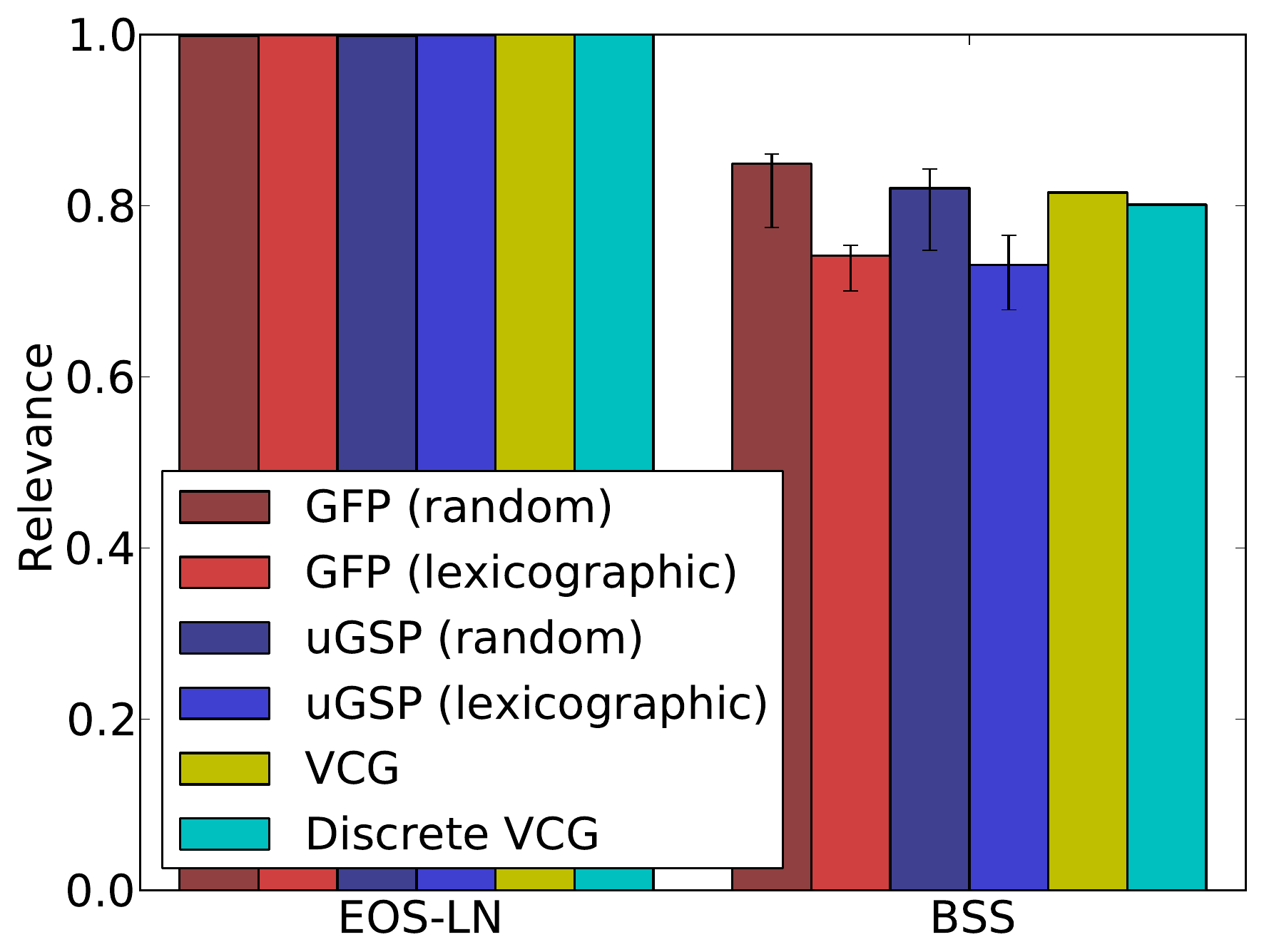}
\includegraphics[width=0.45\hsize]{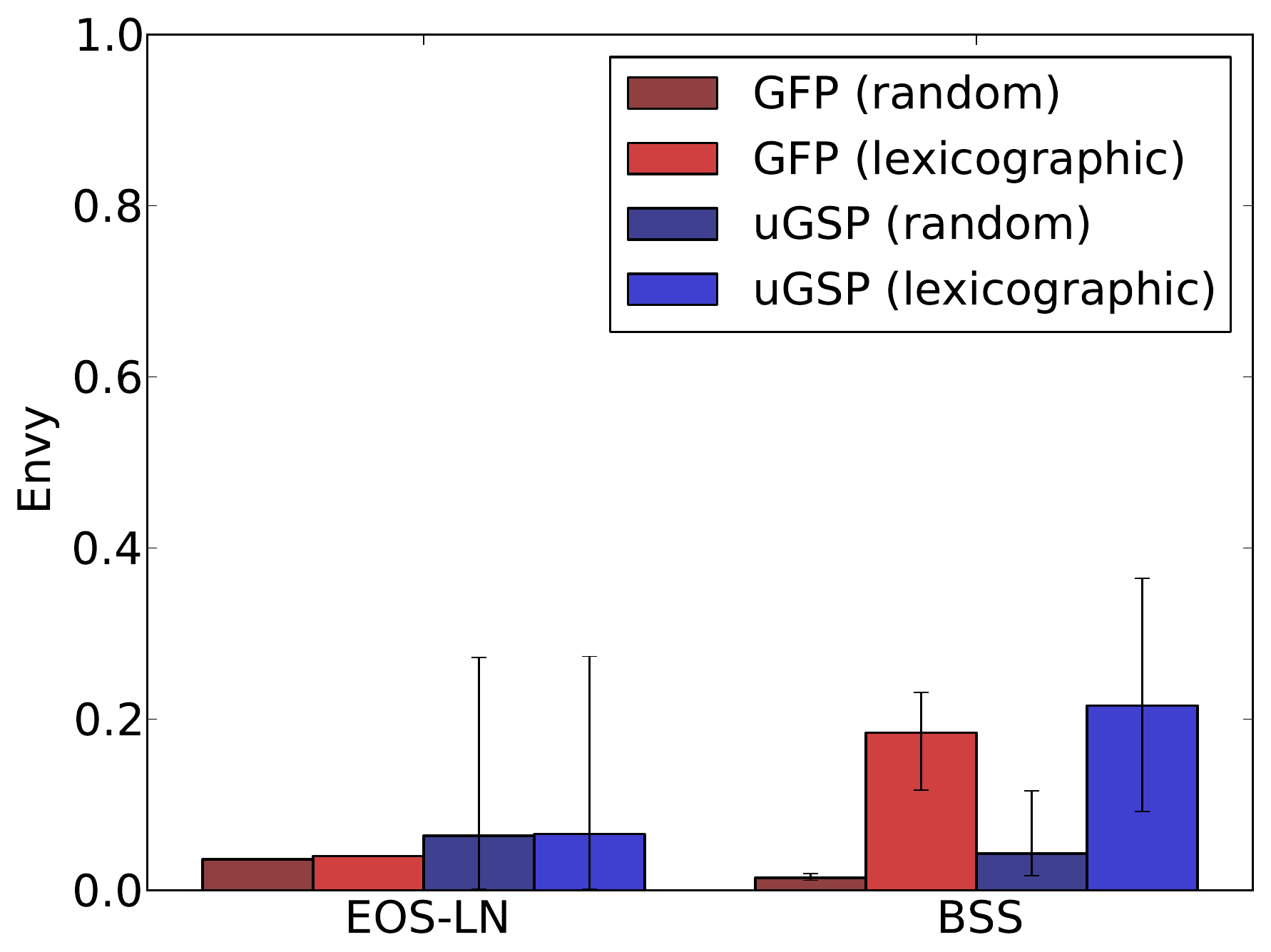}
\kcaption{(c) EOS-LN and BSS: Relevance}{.45\hsize}
\kcaption{(d) EOS-LN and BSS: Envy}{.45\hsize}
\caption[How tie breaking affects auction outcomes]{How tie breaking affects auction outcomes. \whiskerBP}
\label{fig:tie-comparison}
\end{figure}

Broadly, we found that tie breaking made almost no difference in the EOS-LN distribution; however, lexicographic tie breaking consistently decreased performance in every metric in the BSS distribution (see Figure~\ref{fig:tie-comparison}).  This problem arose when multiple agents preferred a lower slot to a higher slot, and strictly preferred not to be shown in the high position.  With random tie breaking, each agent could win their more preferred slot often enough to justify paying a high price.  However, when ties are broken lexicographically, some agent must pay a higher price and get a less desirable slot, inducing him to deviate.  Thus, random tie breaking achieved much higher relevance, selling clicks even in cases where continuous VCG did not.

\subsection{Sensitivity to Rounding Rules}\label{sec:rounding}

\begin{figure}
\centering
\includegraphics[width=0.45\hsize]{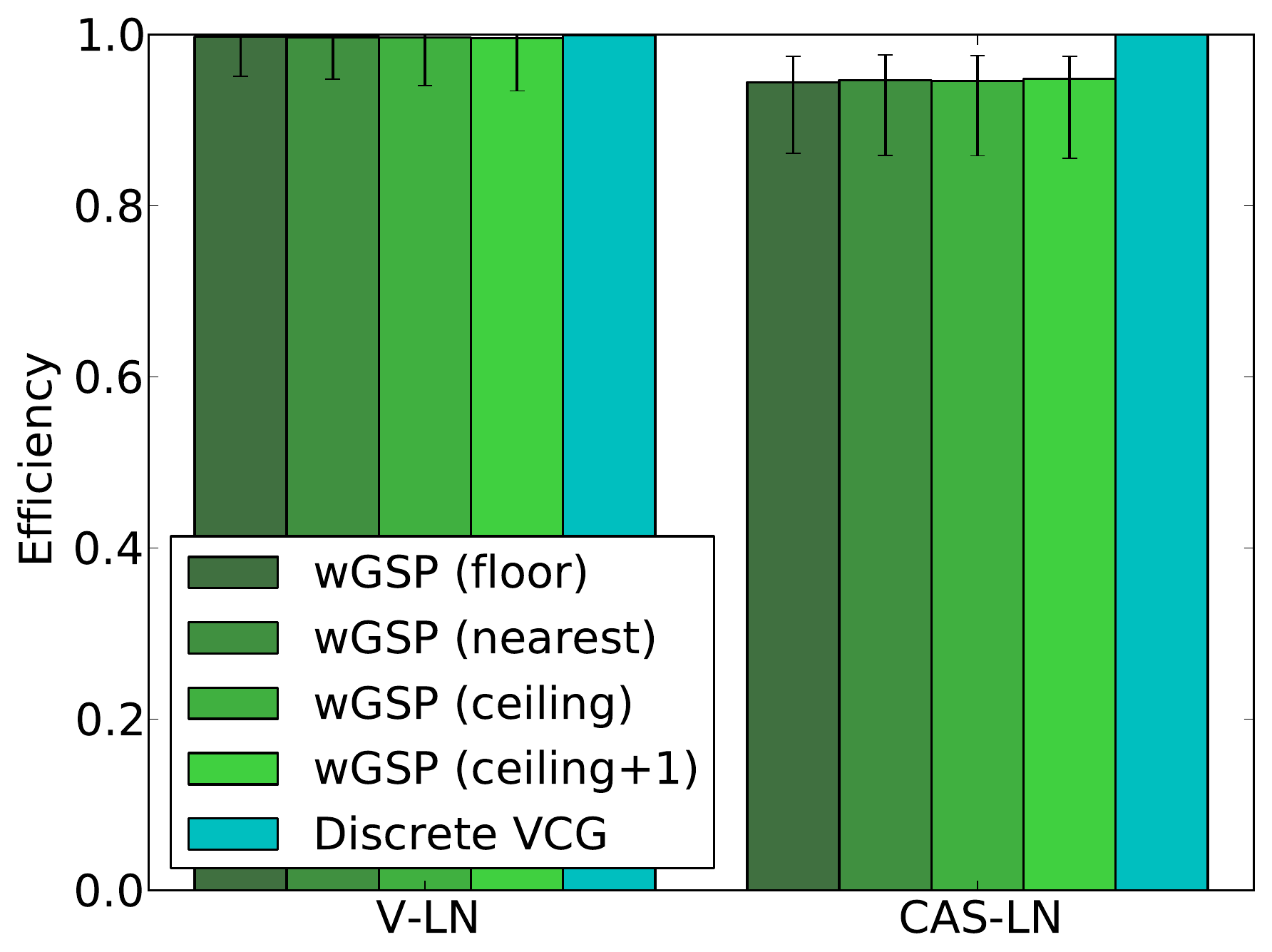}
\includegraphics[width=0.45\hsize]{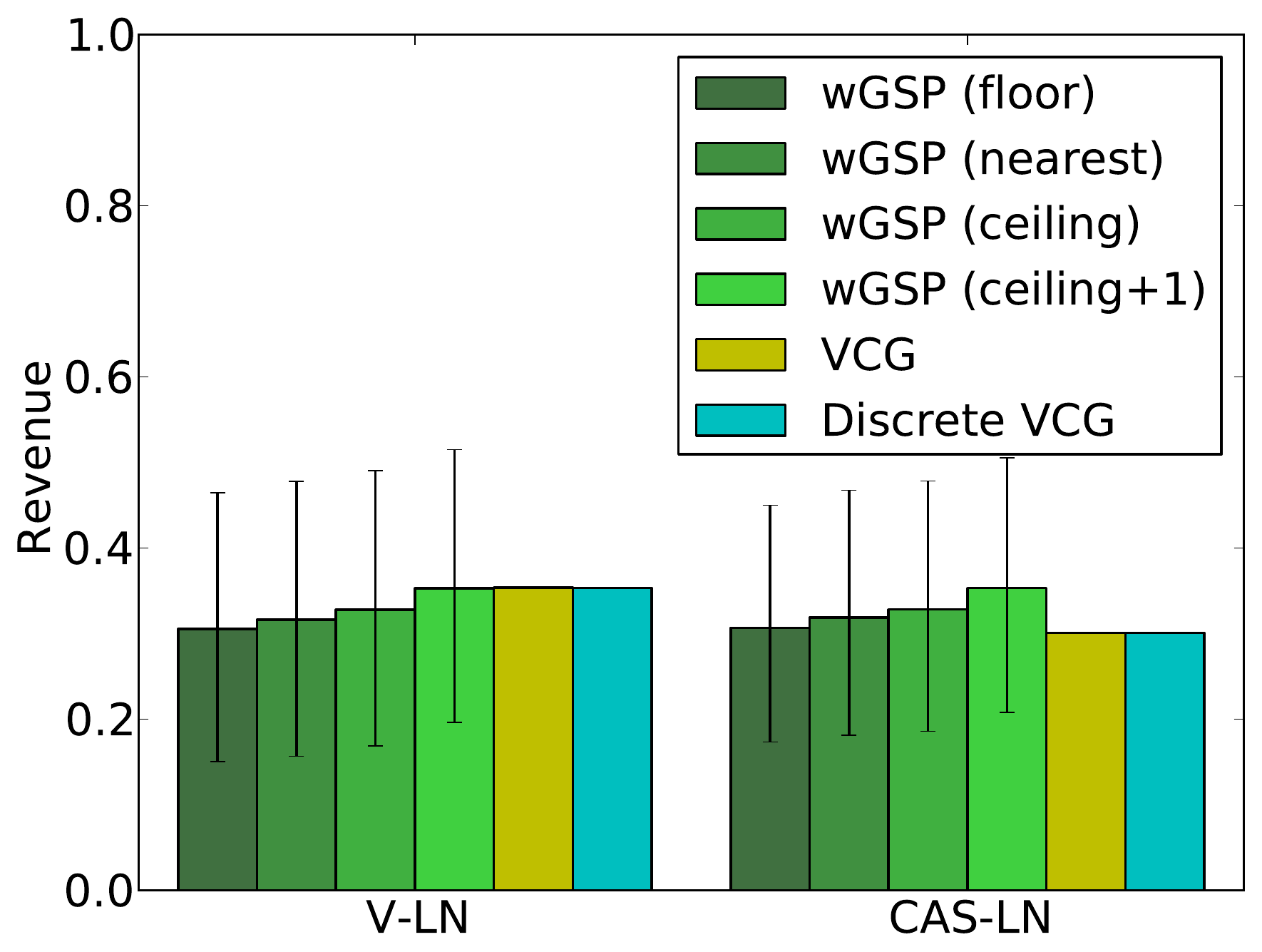}
\kcaption{(a) Efficiency}{.45\hsize}
\kcaption{(b) Revenue}{.45\hsize}
\includegraphics[width=0.45\hsize]{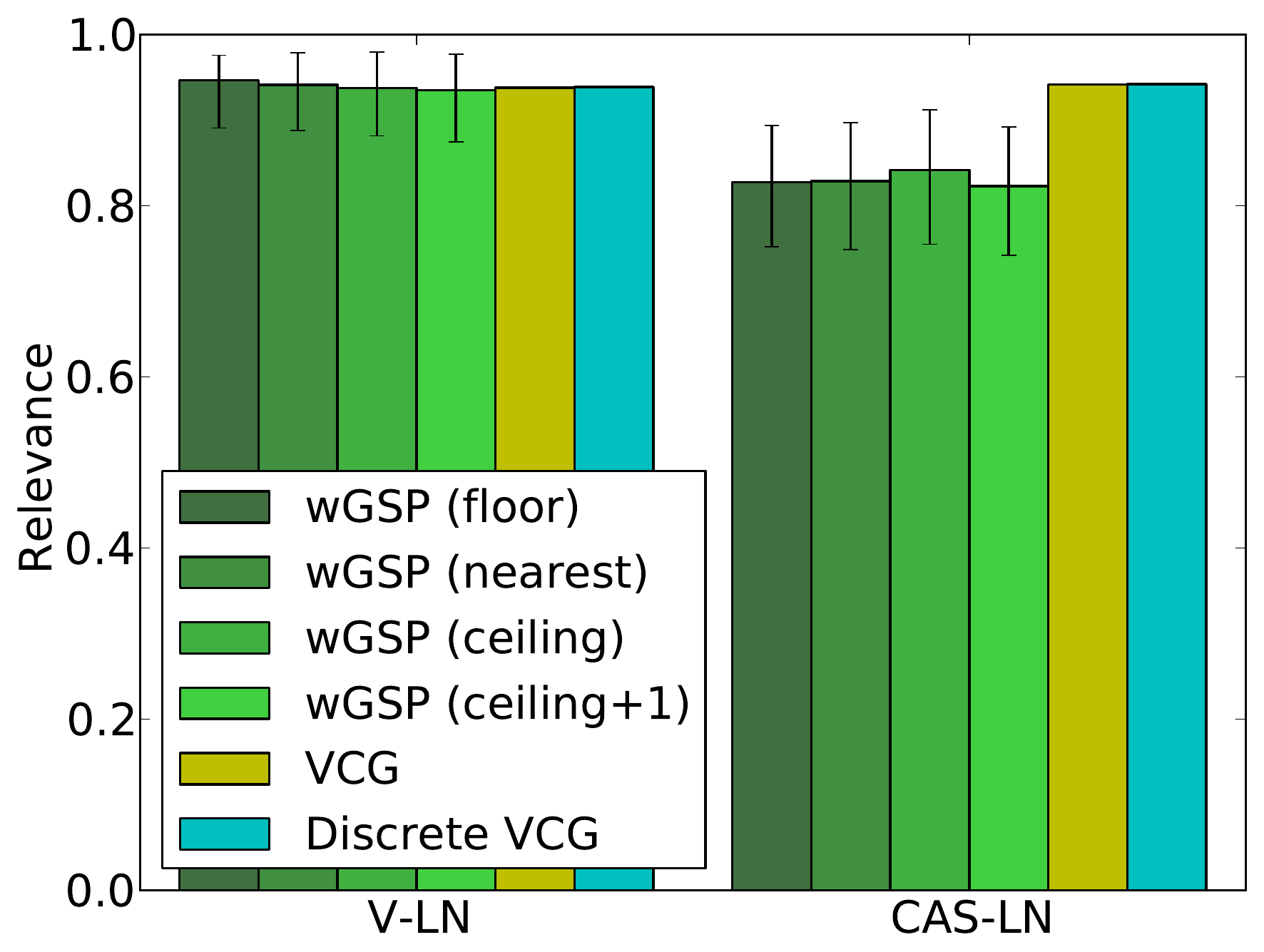}
\includegraphics[width=0.45\hsize]{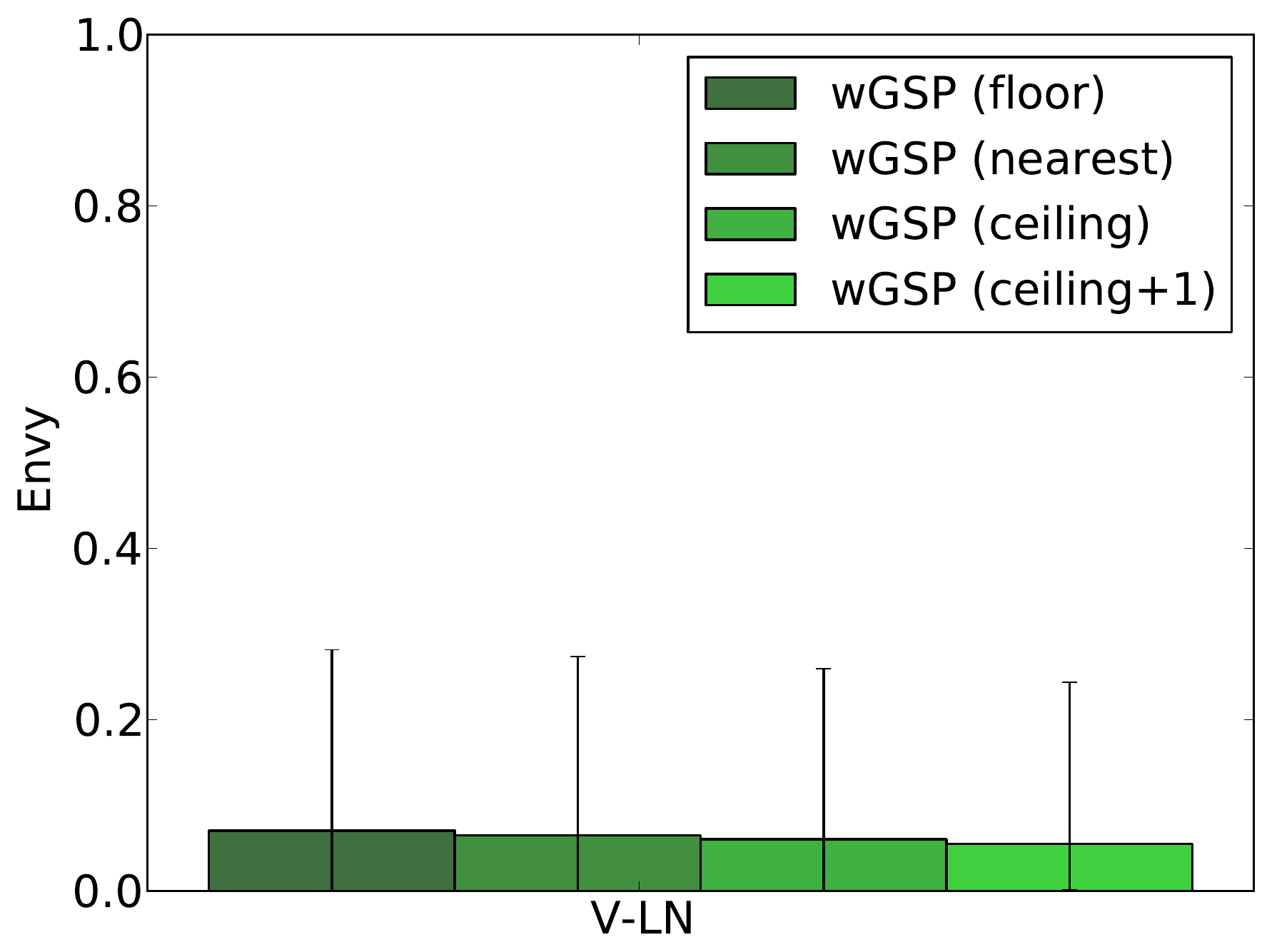}
\kcaption{(c) Relevance}{.45\hsize}
\kcaption{(d) Envy}{.45\hsize}
\caption[How price rounding affects auction outcomes]{How price rounding affects auction outcomes. \whiskerBP}
\label{fig:rounding-comparison}
\end{figure}

The last problem discretization poses for an auctioneer is the question of how to round prices in wGSP auctions. 
Because each bidder's integral bid is scaled by an arbitrary real constant, the next-highest bid might not correspond to any integer price.  So far,  we have assumed that prices are rounded up (i.e., that a bidder must pay the minimum amount that she could bid to win the position she won).  In this section we also consider rounding down, rounding to the nearest integer, and rounding up plus 1 increment (which was used by Yahoo!\ \cite{EOS07internet}).  Here our experiments consisted of 100 instances each from the V-LN and CAS-LN distributions (see Figure~\ref{fig:rounding-comparison}).  We found, somewhat unsurprisingly, that more ``aggressive'' rounding rules (i.e., rules that favor higher prices) tended to produce higher revenue regardless of the distribution and the equilibrium-selection criterion (the difference between the best and worst rounding rules ranged from 10\% to 25\%), with a corresponding improvement in envy.  More surprisingly, we found that rounding rules had noticeably smaller effects on economic efficiency and relevance (not more than 2.5\%).  Thus, the aggressive rounding rules used in practice appear to increase revenue at very little cost in terms of ad quality.

\subsection{Sensitivity to the Number of Bidders}

To study whether our findings were sensitive to the number of bidders, we  used the same two distributions, V-LN and CAS-LN.  For each, we solved 100 instances, and studied how each auction performed as the number of bidders varied from two to ten. Throughout, we assumed that there was no artificial limit on supply; i.e., the search engine would allocate a slot to every advertiser who bid more than zero.  (We investigate the impact of this assumption next.)
We found that the relative ranking between position auction variants remained consistent as the number of bidders varied (see \paperonly{Figure~\ref{fig:scale-n}}\chapteronly{Figures~\ref{fig:scale-n1} and \ref{fig:scale-n2}}).  As the number of bidders increased, all position auctions became progressively less efficient and relevant (although in V-LN, wGSP experienced less such decrease than the other position auctions), and all auctions (including VCG) saw increasing  normalized revenue (i.e., fraction of the surplus going to the seller).

\paperonly{
\begin{figure}
	\hspace{0.165\hsize}
	\includegraphics[width=0.33\hsize]{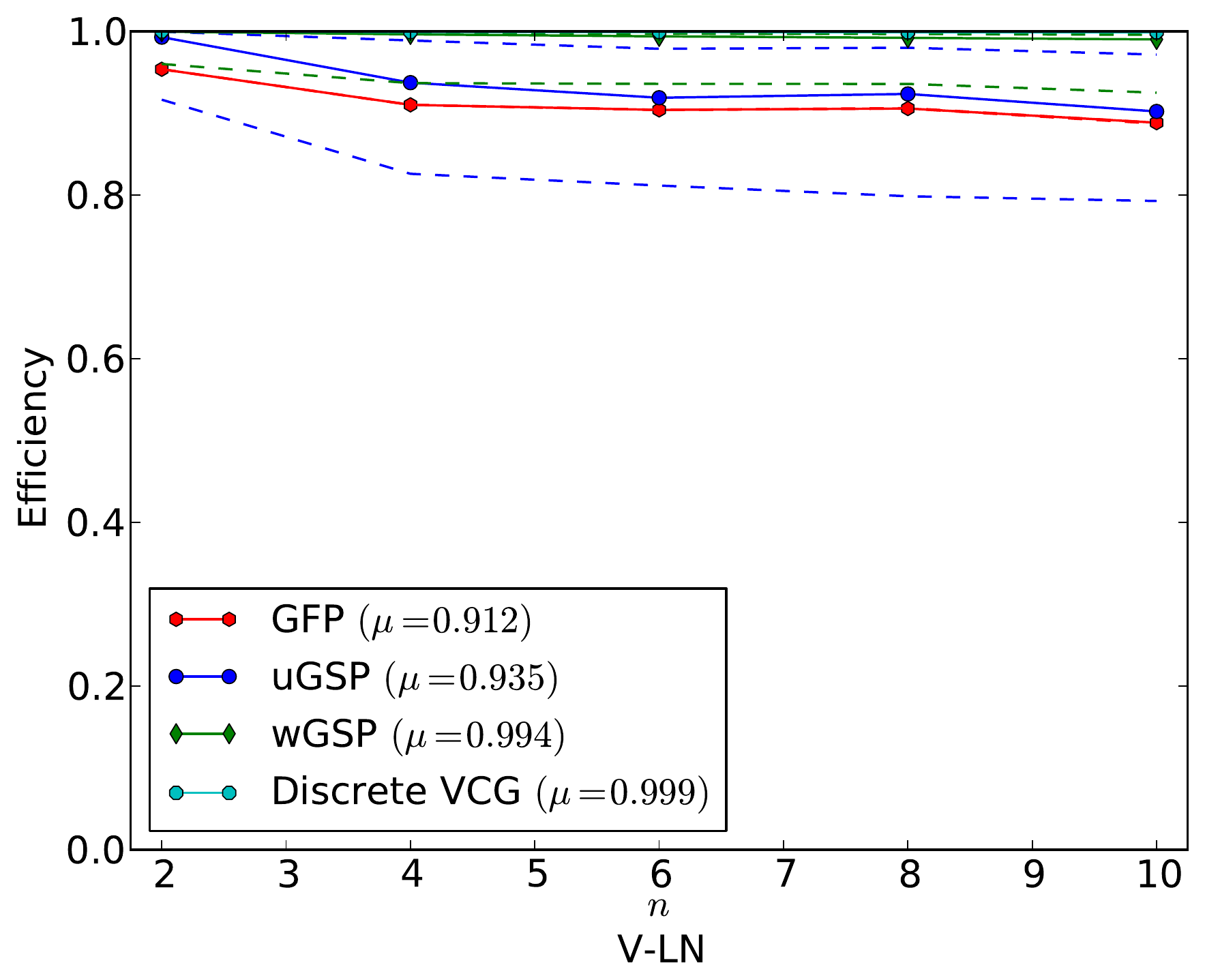}
	\includegraphics[width=0.33\hsize]{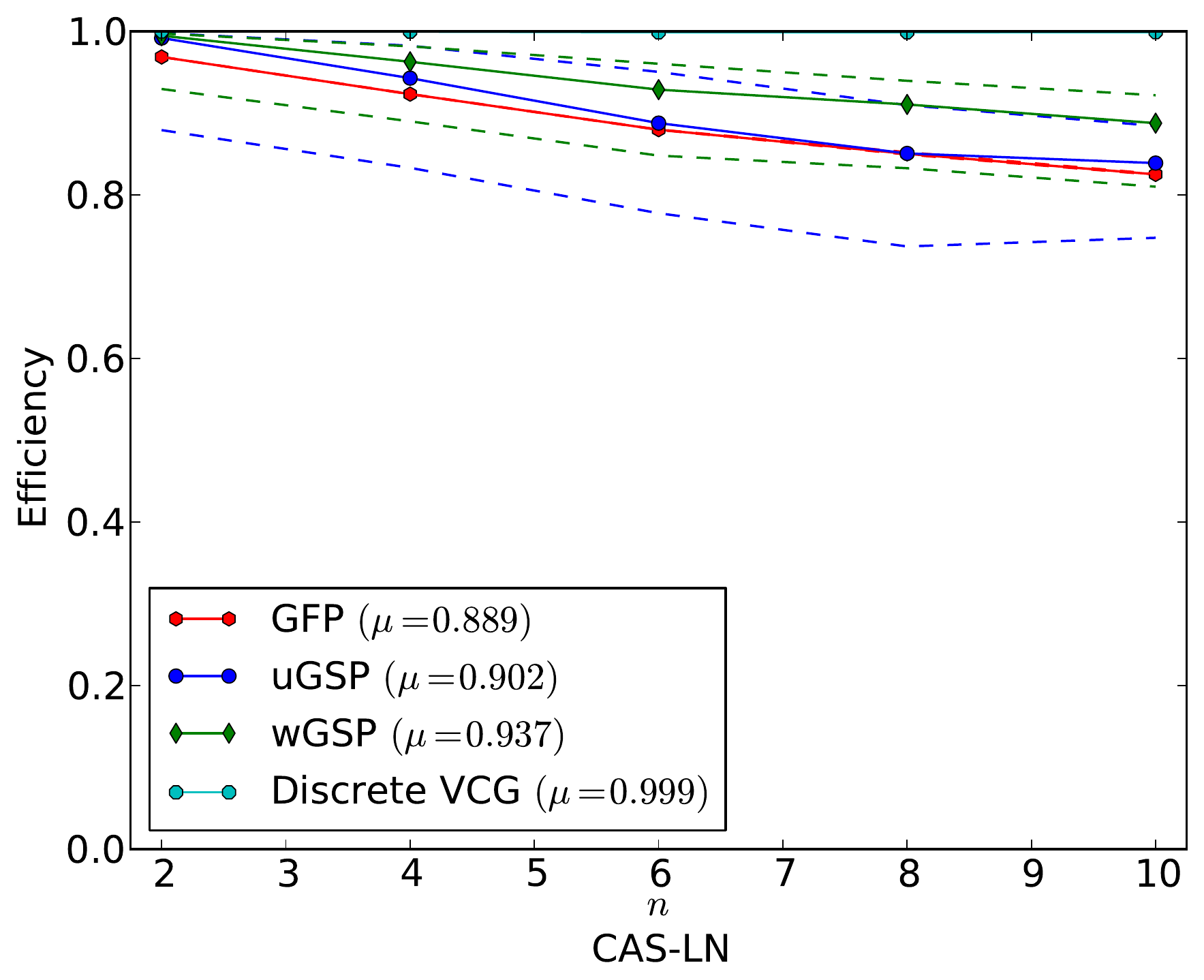}\\
	\kcaption{~}{0.165\hsize}
	\kcaption{(a) V: Log-Normal, Efficiency}{.33\hsize}
	\kcaption{(b) Cascade: Log-Normal, Efficiency}{.33\hsize}
	
	\hspace{0.165\hsize}
	\includegraphics[width=0.33\hsize]{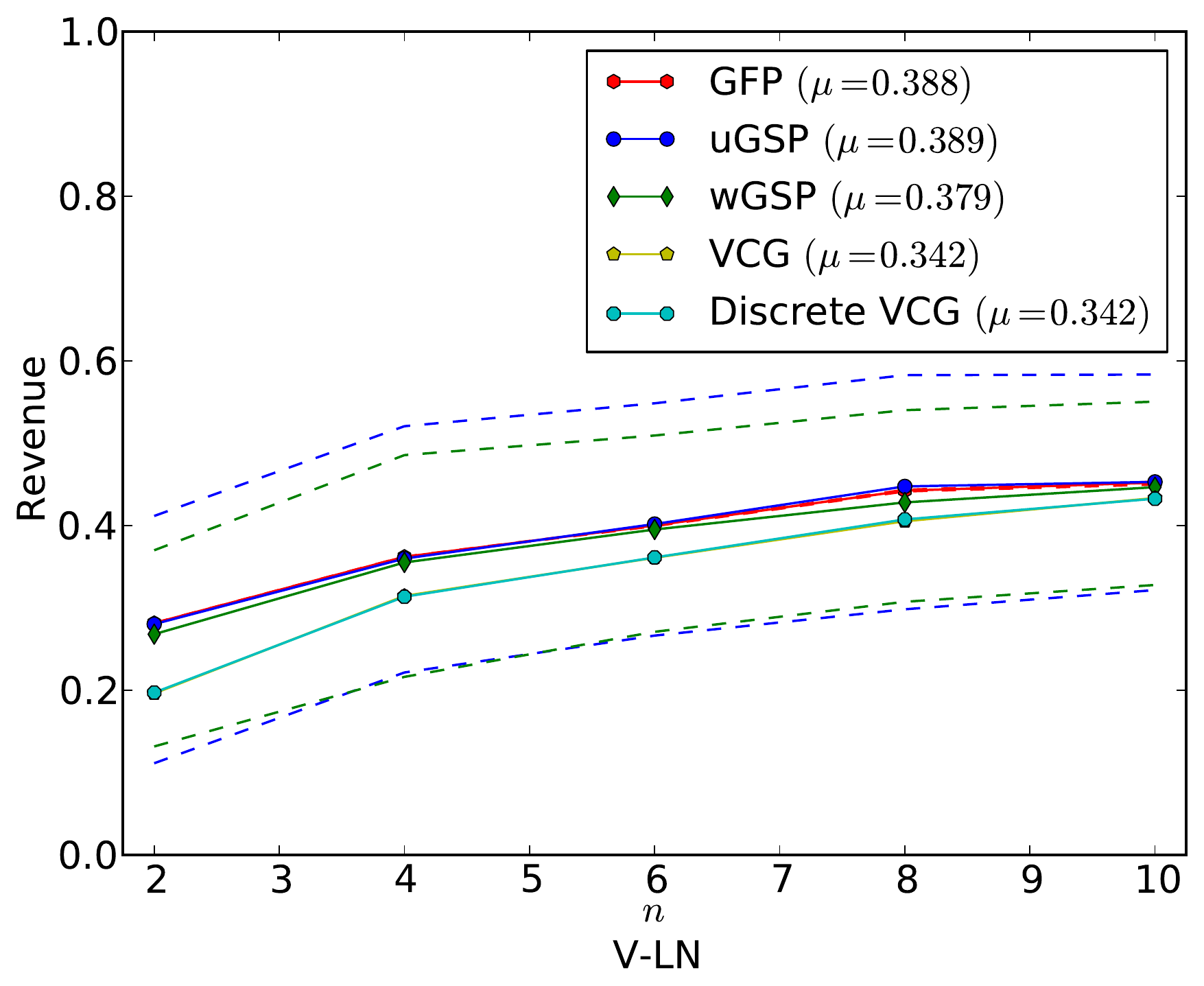}
	\includegraphics[width=0.33\hsize]{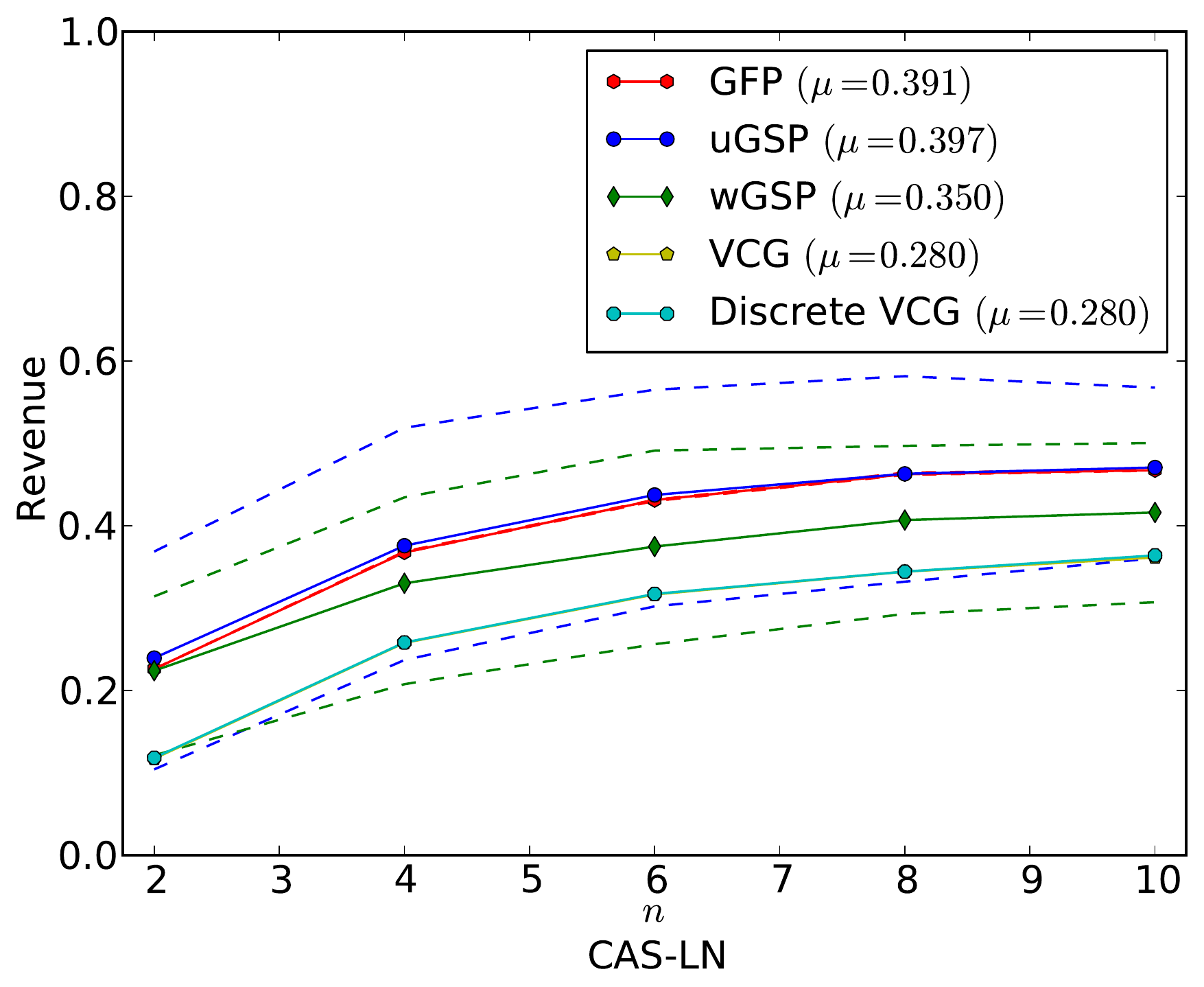}\\
	\kcaption{~}{0.165\hsize}
	\kcaption{(c) V: Log-Normal, Revenue}{.33\hsize}
	\kcaption{(d) Cascade: Log-Normal, Revenue}{.33\hsize}
	
	\includegraphics[width=0.33\hsize]{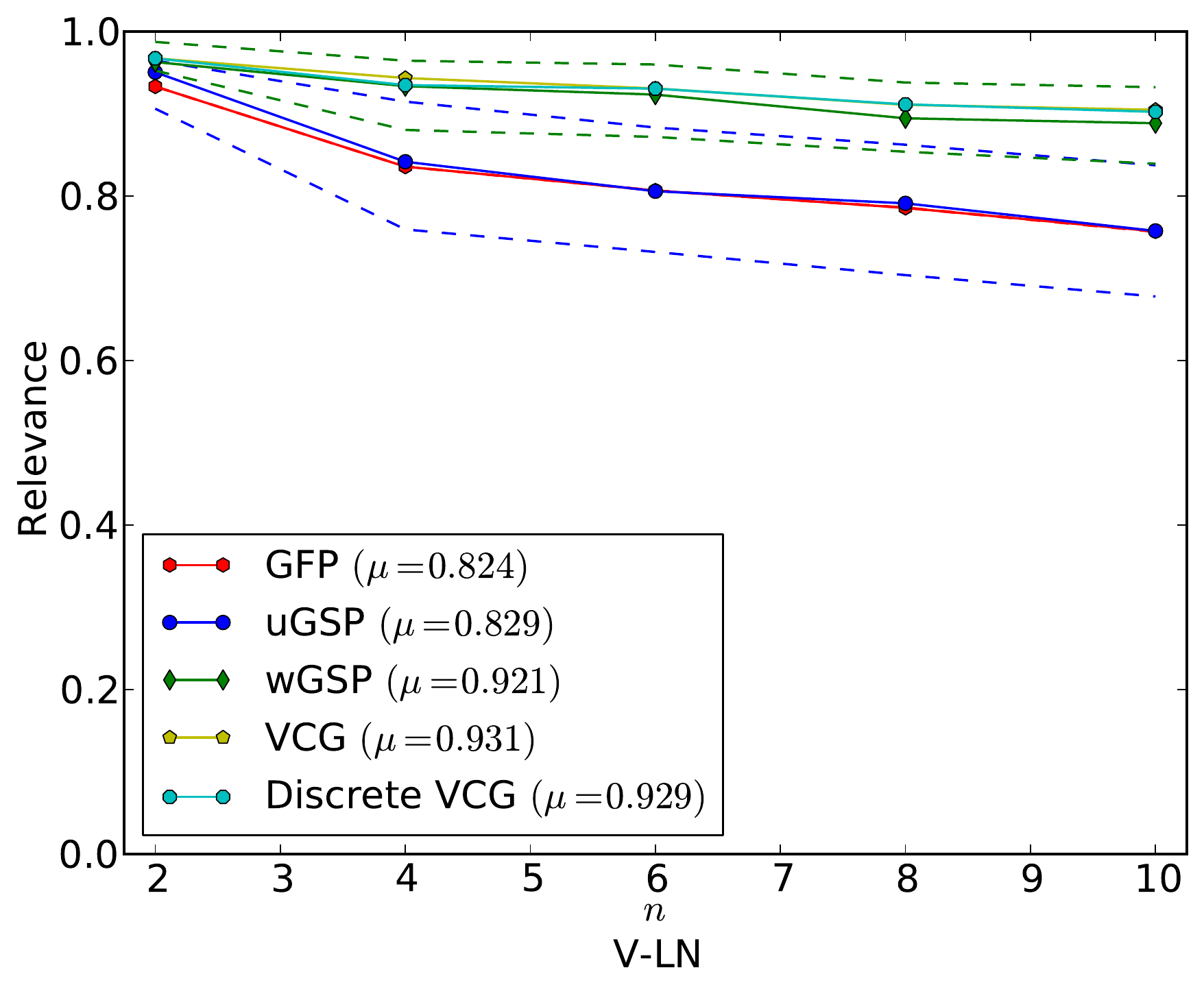}
	\includegraphics[width=0.33\hsize]{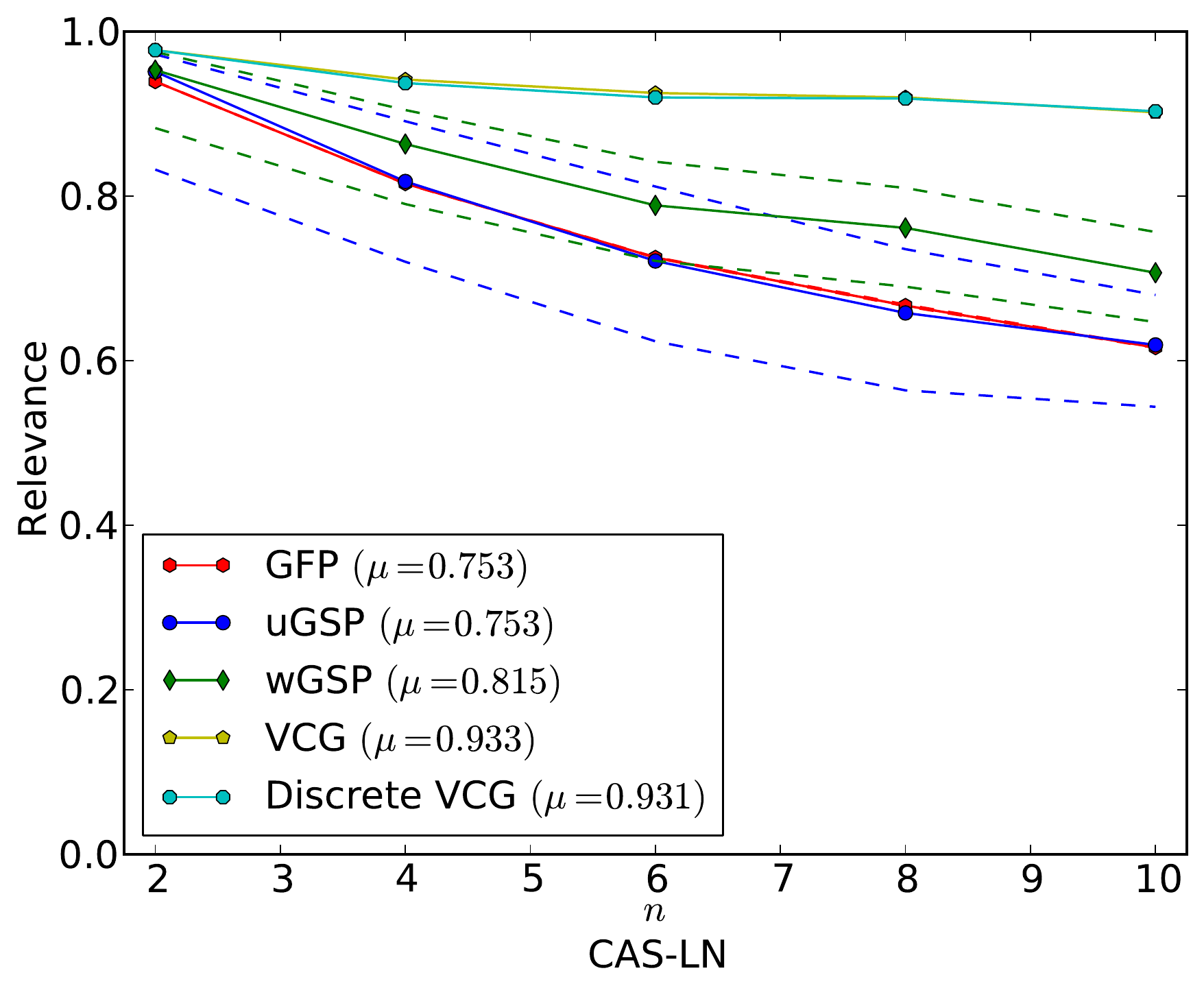}
	\includegraphics[width=0.33\hsize]{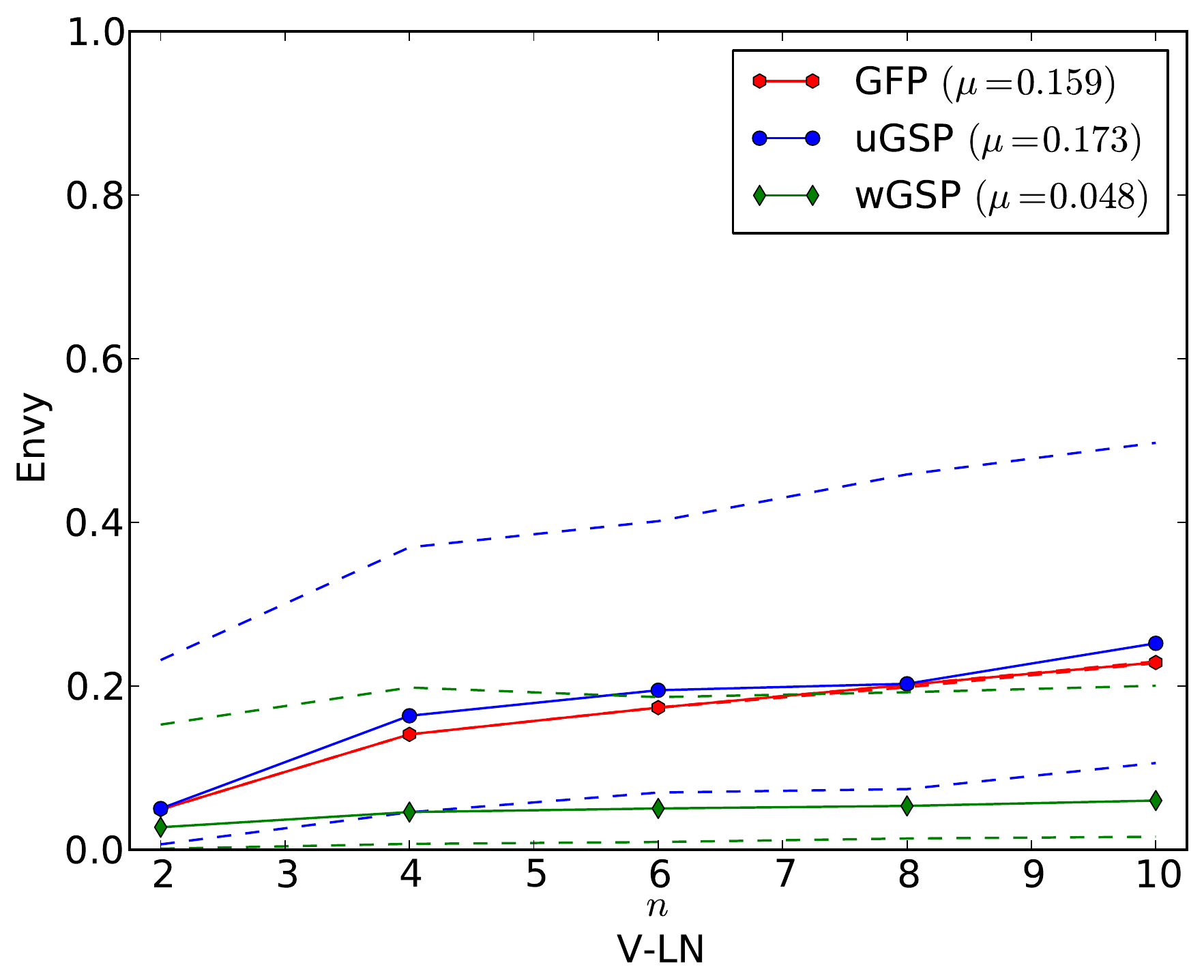}
	\kcaption{(e) V: Log-Normal, Relevance}{.33\hsize}
	\kcaption{(f) Cascade: Log-Normal, Relevance}{.33\hsize}
	\kcaption{(g) V: Log-Normal, Envy}{.33\hsize}
	\caption{Comparing different auction designs as the number of agents varies.}
	\label{fig:scale-n}
\end{figure}
}

\chapteronly{
\begin{figure}
	\hspace{0.05\hsize}
	\includegraphics[width=0.45\hsize]{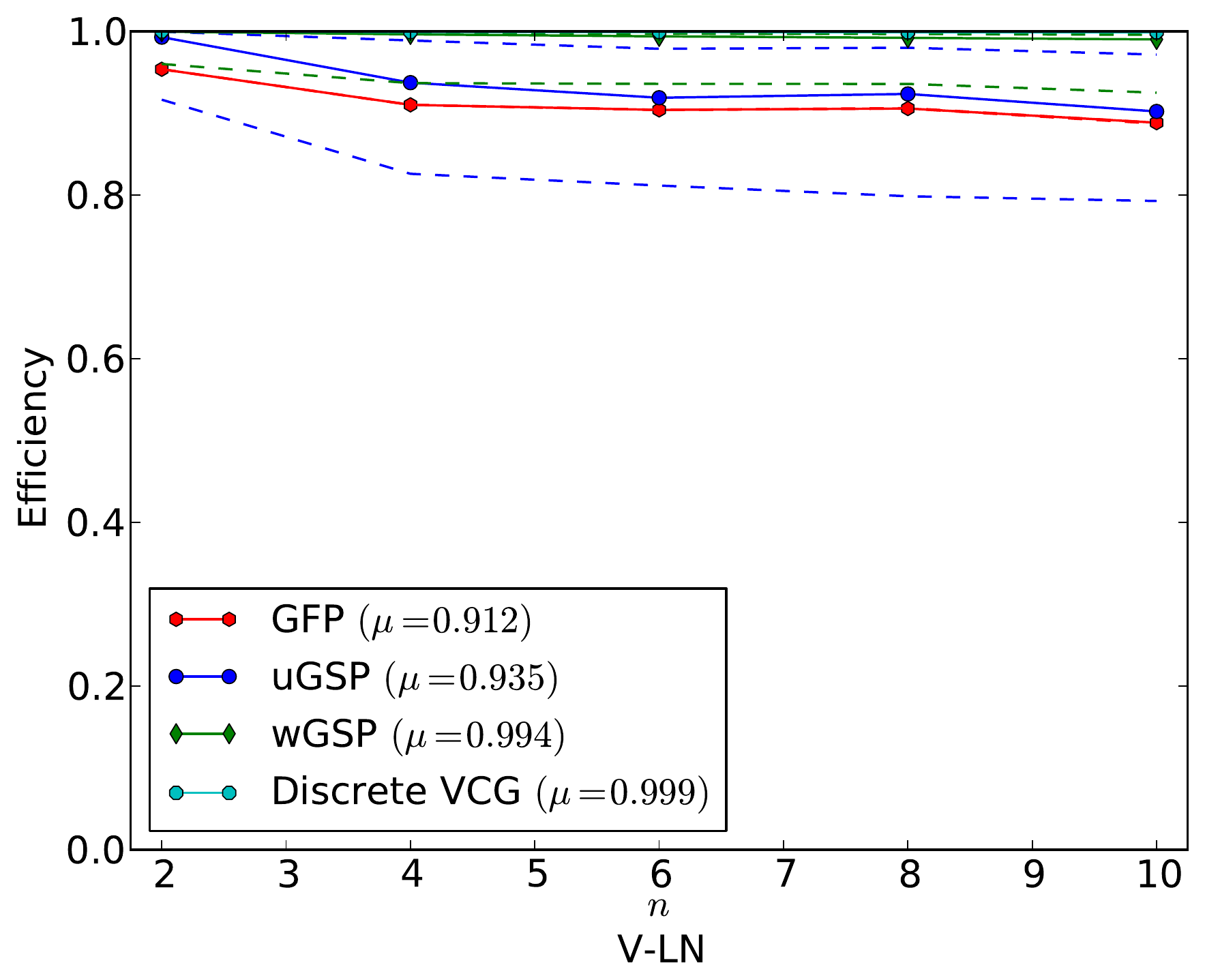}
	\includegraphics[width=0.45\hsize]{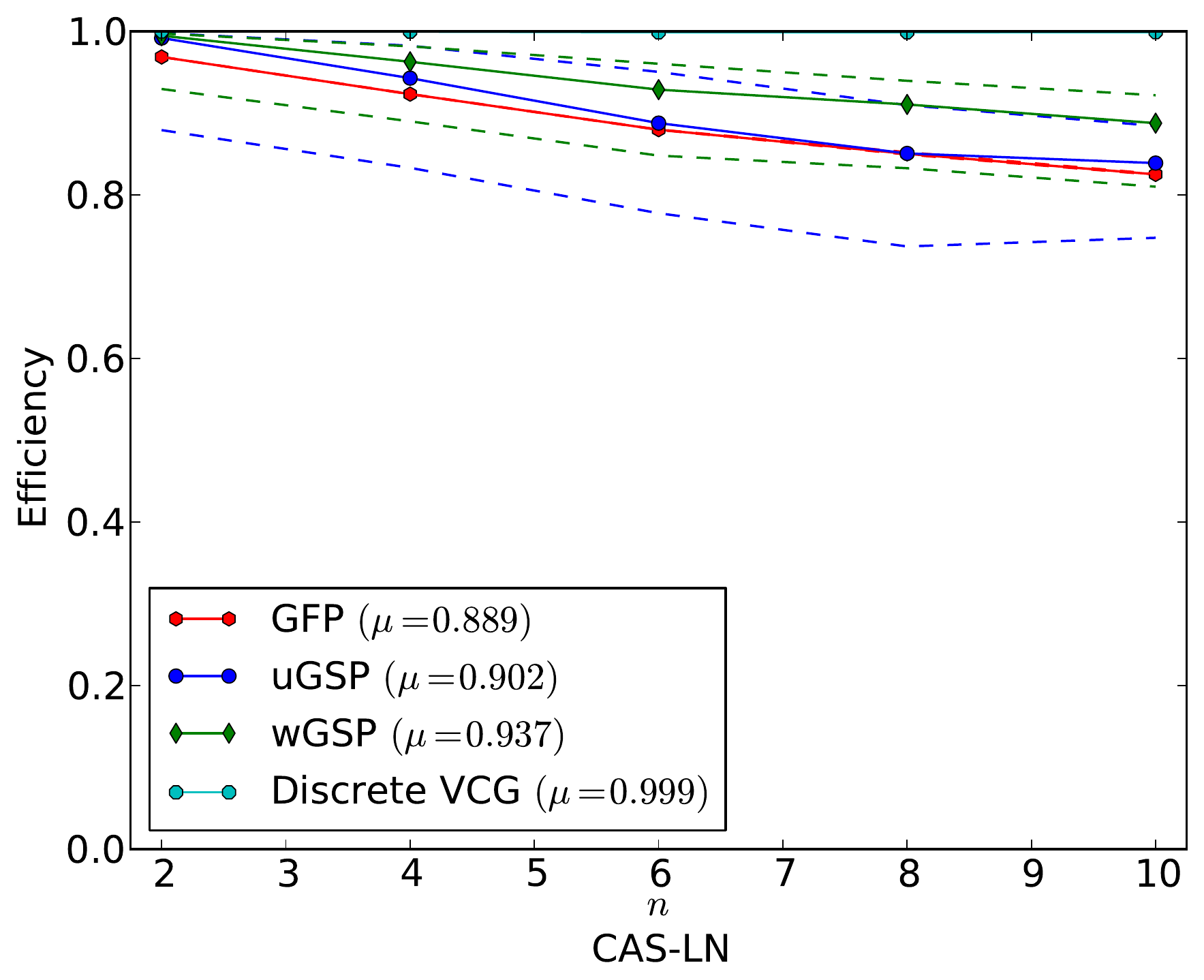}\\
	\kcaption{~}{0.05\hsize}
	\kcaption{(a) V: Log-Normal, Efficiency}{.45\hsize}
	\kcaption{(b) Cascade: Log-Normal, Efficiency}{.45\hsize}

	\hspace{0.05\hsize}
	\includegraphics[width=0.45\hsize]{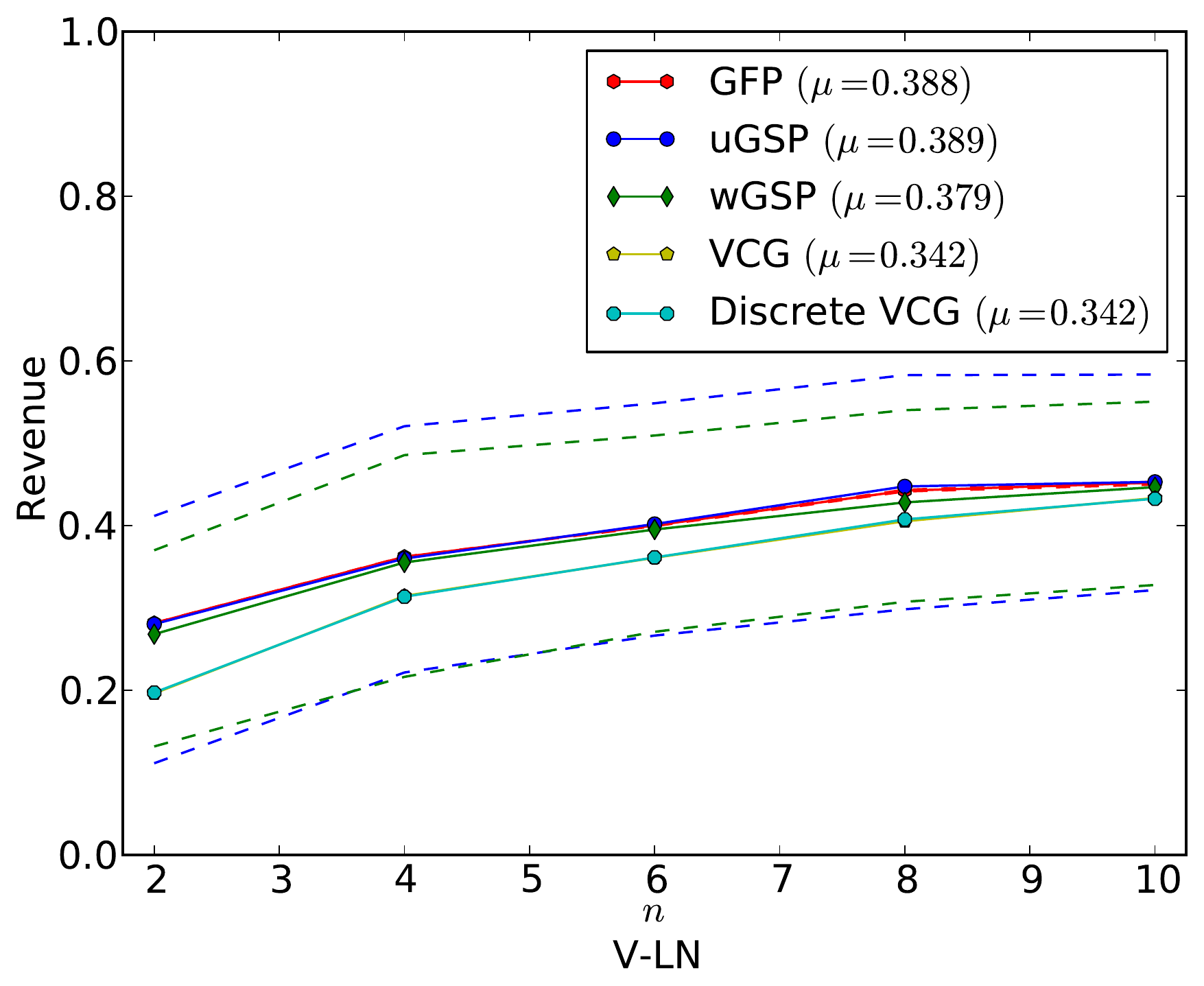}
	\includegraphics[width=0.45\hsize]{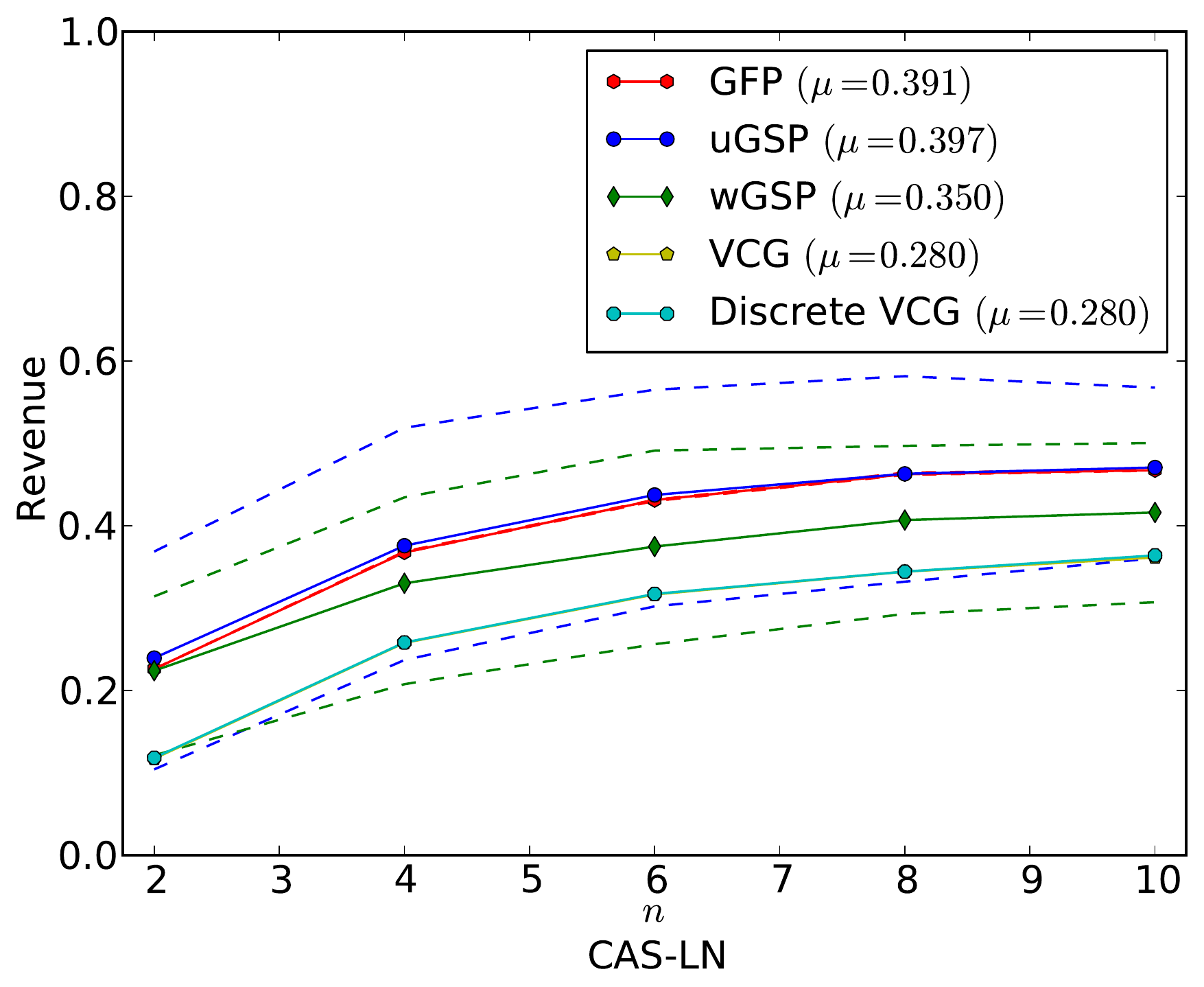}\\
	\kcaption{~}{0.05\hsize}
	\kcaption{(c) V: Log-Normal, Revenue}{.45\hsize}
	\kcaption{(d) Cascade: Log-Normal, Revenue}{.45\hsize}
	\caption{Comparing different auction designs as the number of agents varies.}
	\label{fig:scale-n1}
\end{figure}
\begin{figure}
	\hspace{0.05\hsize}
	\includegraphics[width=0.45\hsize]{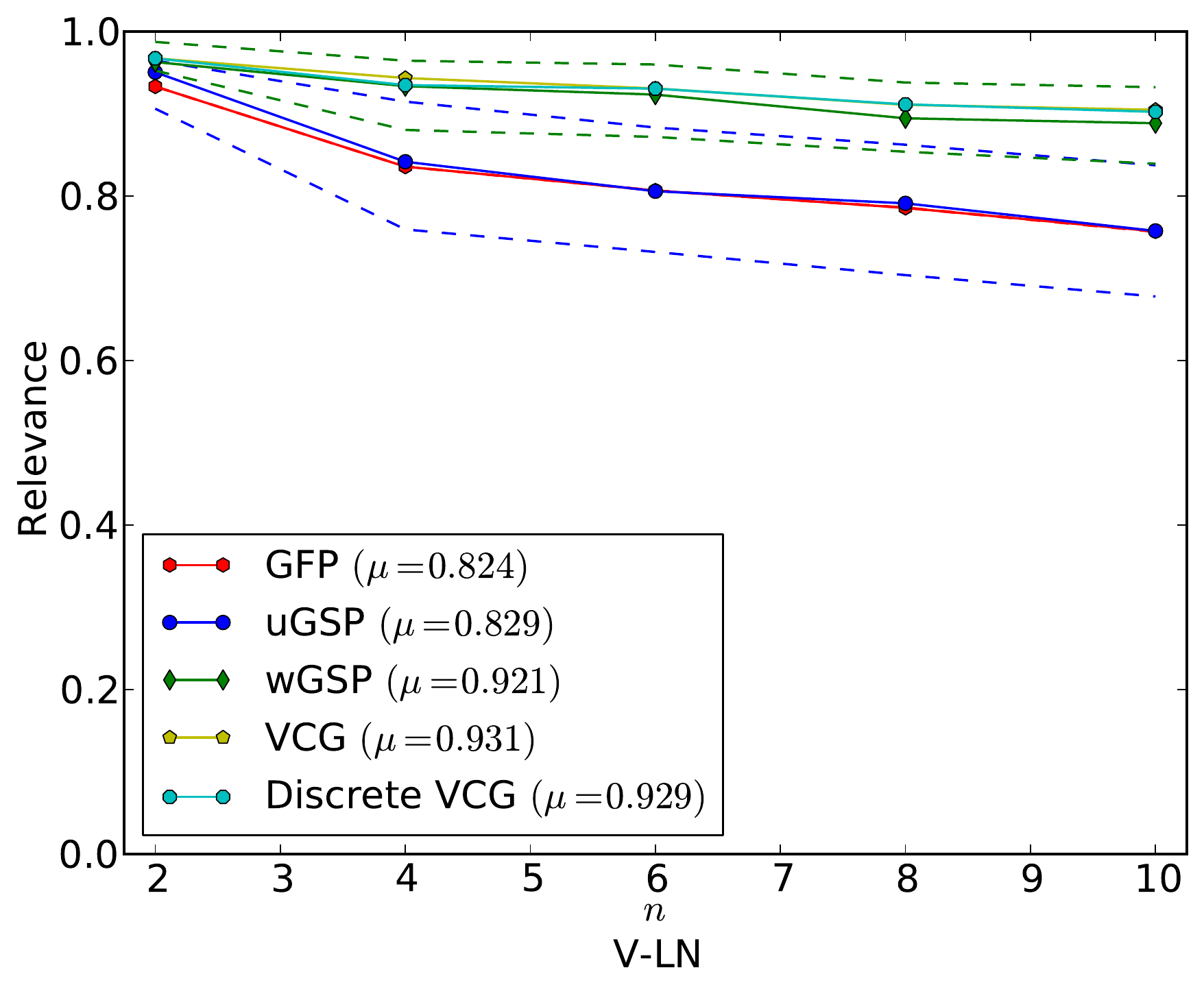}
	\includegraphics[width=0.45\hsize]{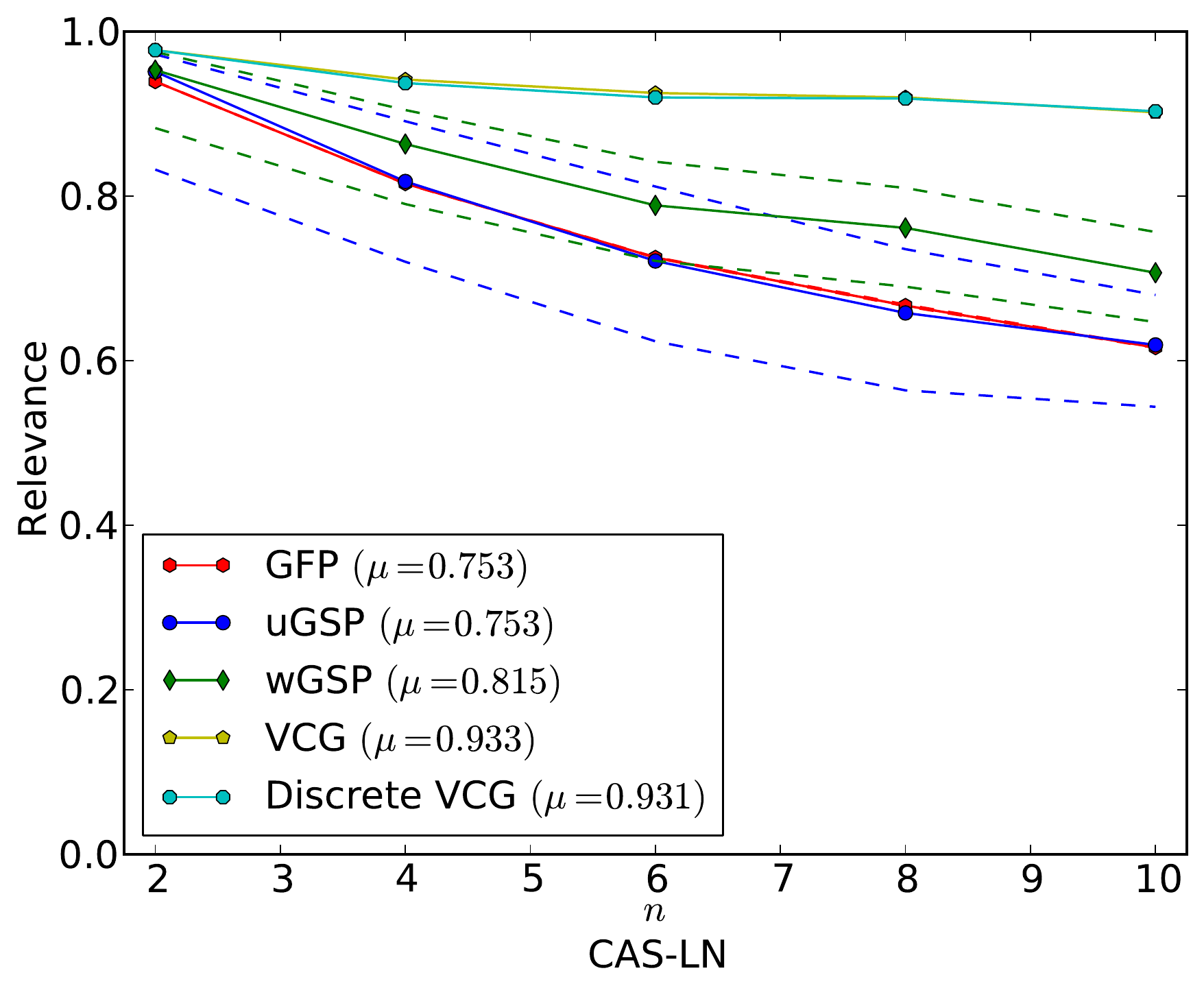}\\
	\kcaption{~}{0.05\hsize}
	\kcaption{(e) V: Log-Normal, Relevance}{.45\hsize}
	\kcaption{(f) Cascade: Log-Normal, Relevance}{.45\hsize}\\

	\hspace{0.275\hsize}
	\includegraphics[width=0.45\hsize]{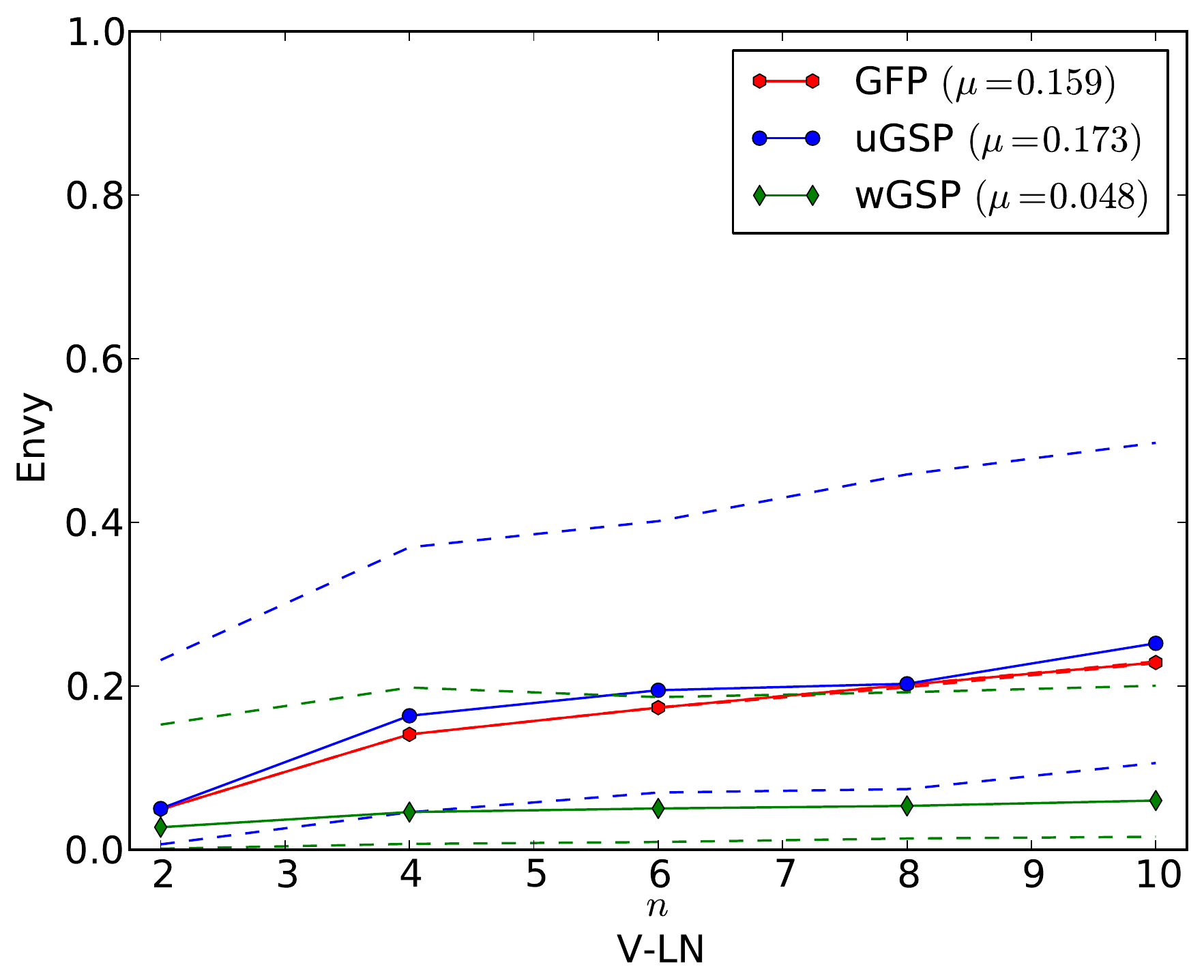}\\
	\kcaption{~}{0.275\hsize}
	\kcaption{(g) V: Log-Normal, Envy}{.45\hsize}
	\caption{Comparing different auction designs as the number of agents varies (continued).}
	\label{fig:scale-m2}
\end{figure}
}

\subsection{Sensitivity to the Number of Slots}

So far, all of our experiments have assumed that the search engine can allocate space to every single advertiser (as, indeed, is often the case).  Here we consider what happens when the supply of slots is limited.  This is particularly important for cascade and hybrid models, under which both price of anarchy and computational complexity are known to depend on the total supply ($m$) \cite{KM08cascade}.  We again considered V-LN and CAS-LN, in each case varying the number of slots from one to five.  

\paperonly{
\begin{figure}
	\hspace{0.165\hsize}
	\includegraphics[width=0.33\hsize]{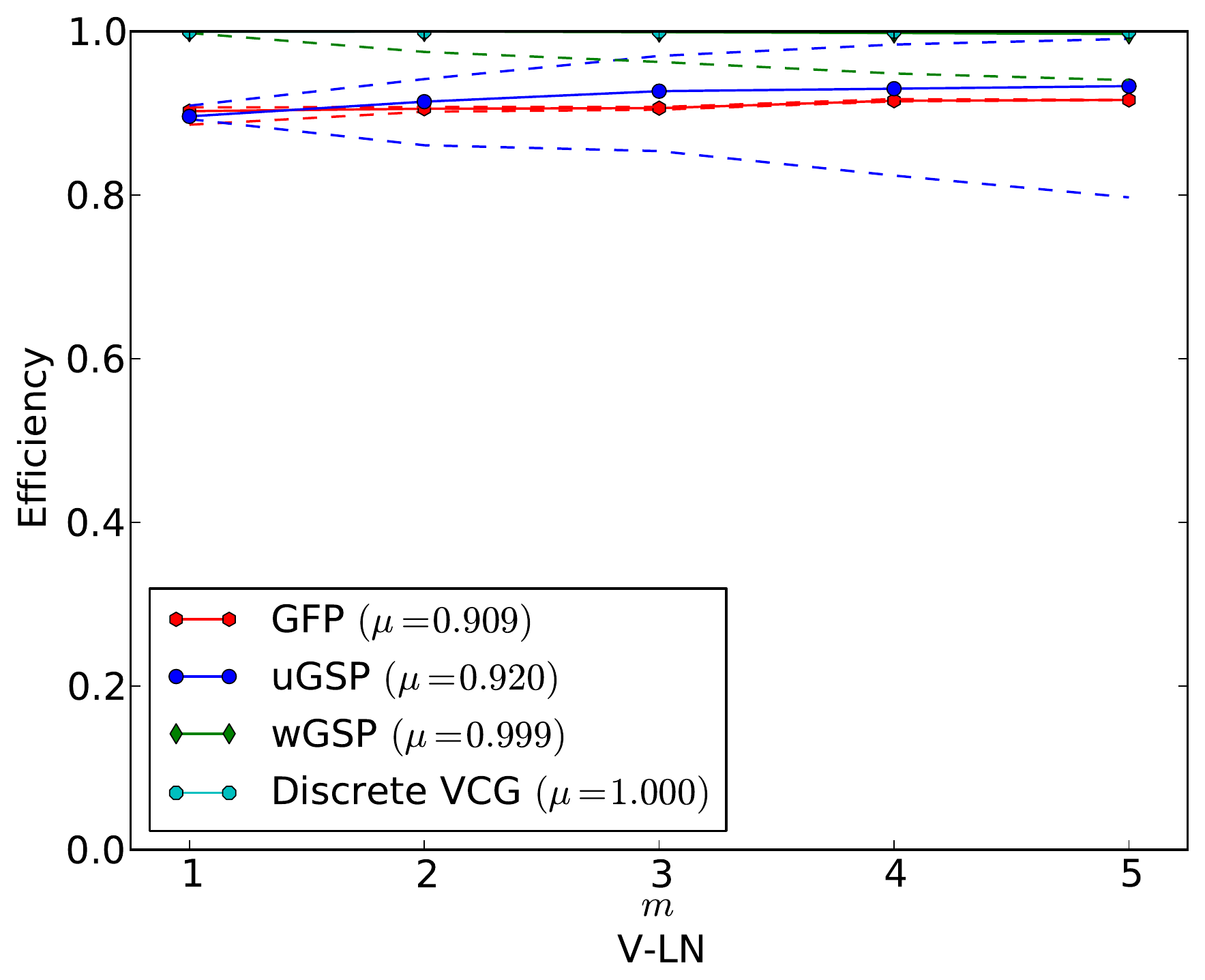}
	\includegraphics[width=0.33\hsize]{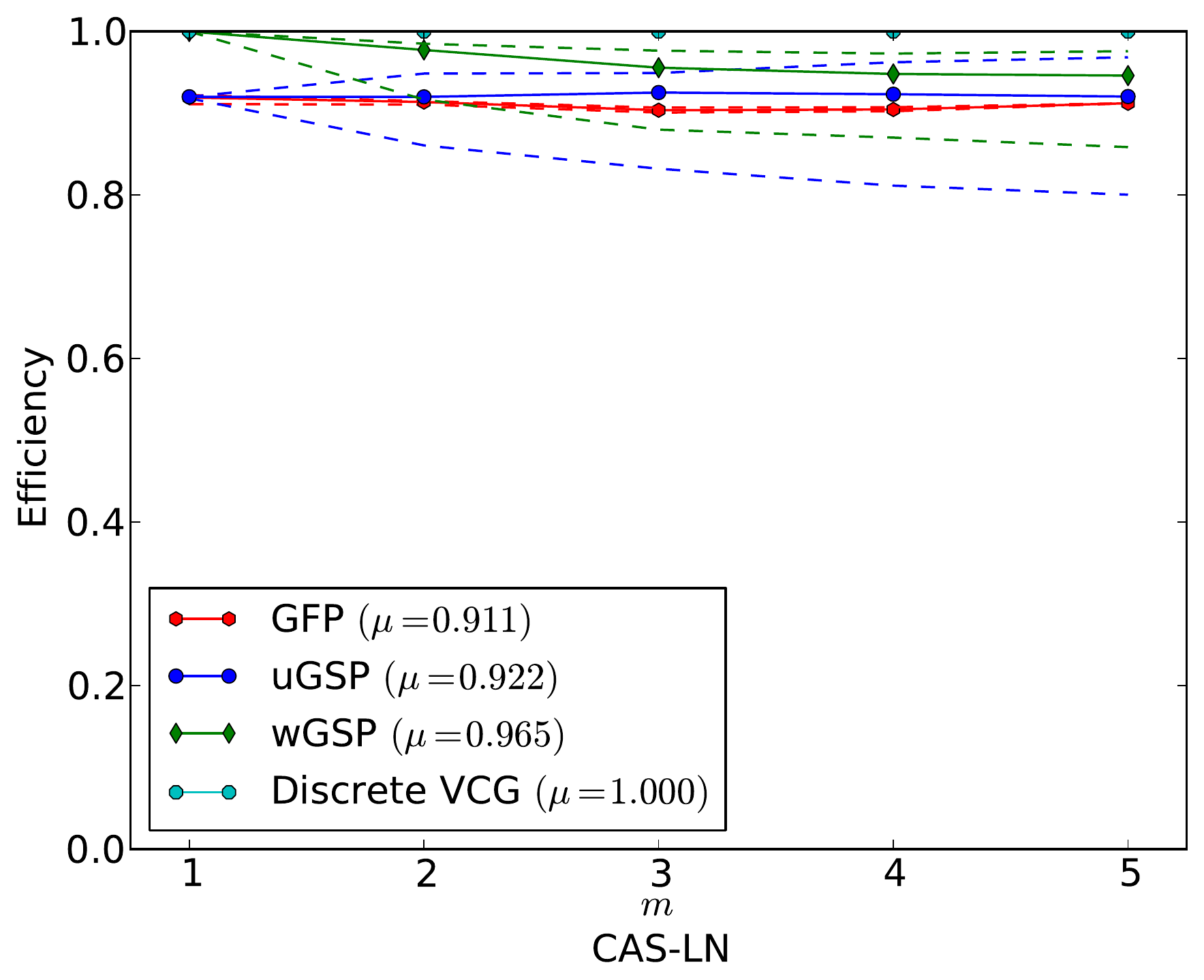}\\
	\kcaption{~}{0.165\hsize}
	\kcaption{(a) V: Log-Normal, Efficiency}{.33\hsize}
	\kcaption{(b) Cascade: Log-Normal, Efficiency}{.33\hsize}

	\hspace{0.165\hsize}
	\includegraphics[width=0.33\hsize]{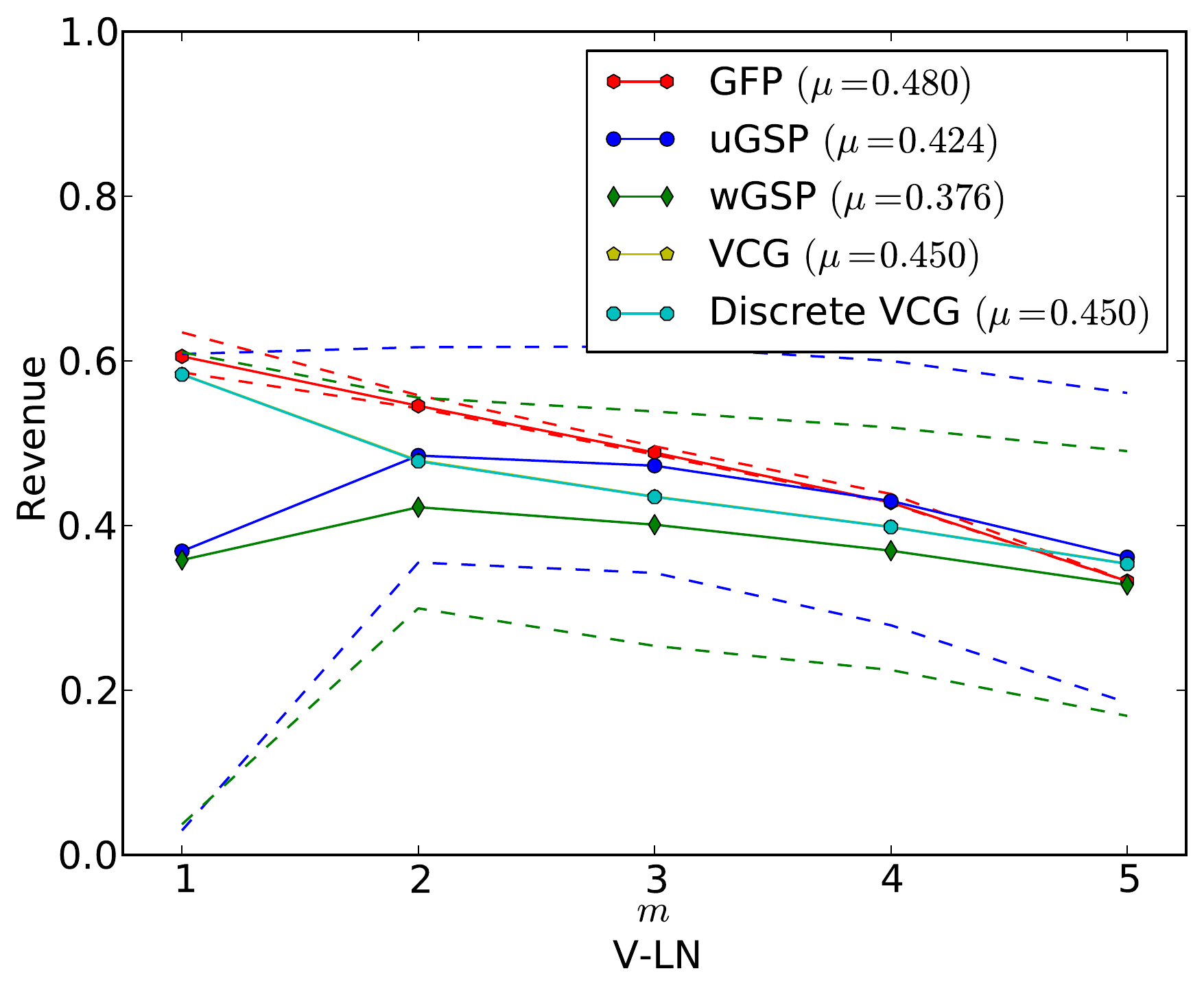}
	\includegraphics[width=0.33\hsize]{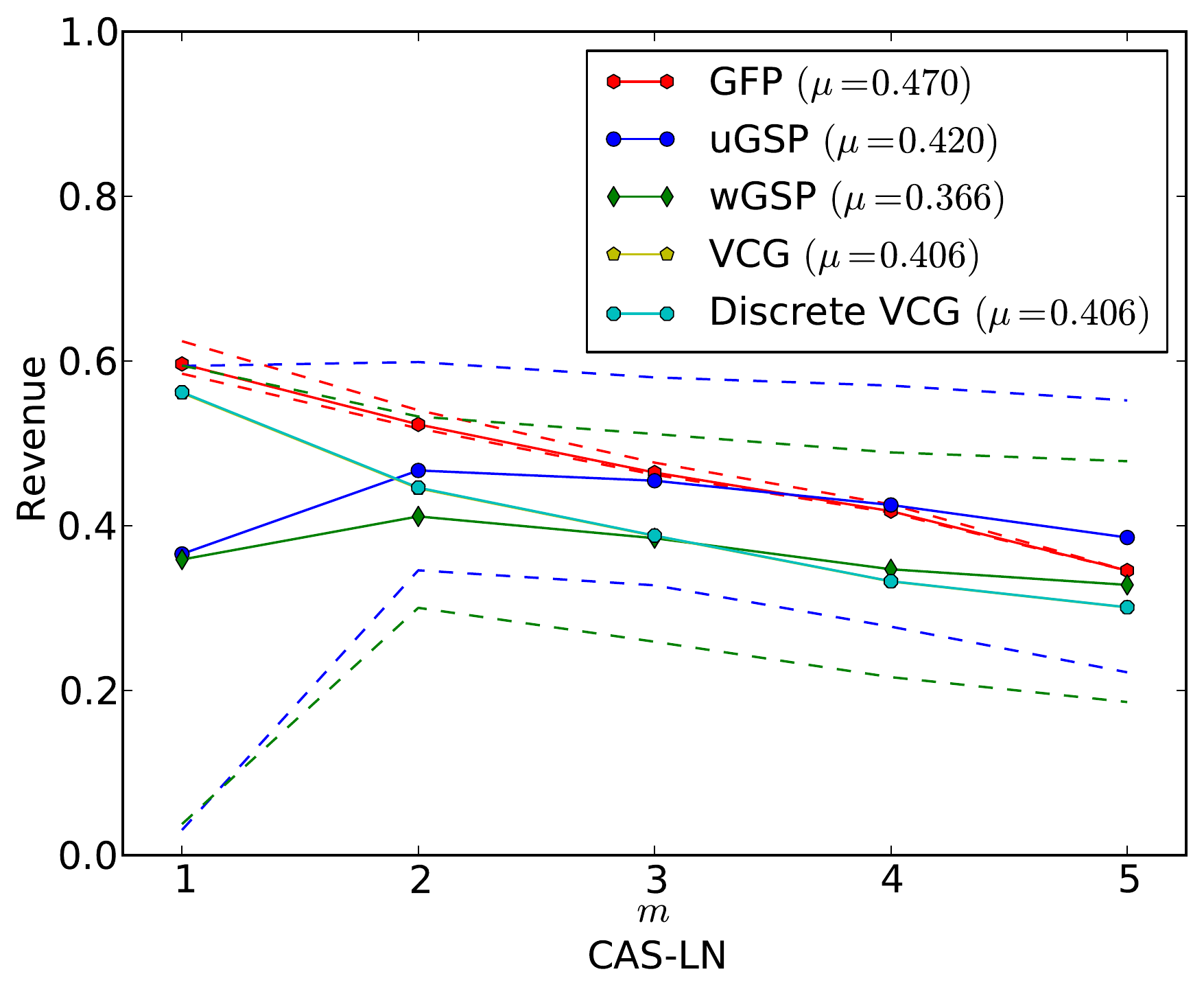}\\
	\kcaption{~}{0.165\hsize}
	\kcaption{(c) V: Log-Normal, Revenue}{.33\hsize}
	\kcaption{(d) Cascade: Log-Normal, Revenue}{.33\hsize}
	%
	
	\includegraphics[width=0.33\hsize]{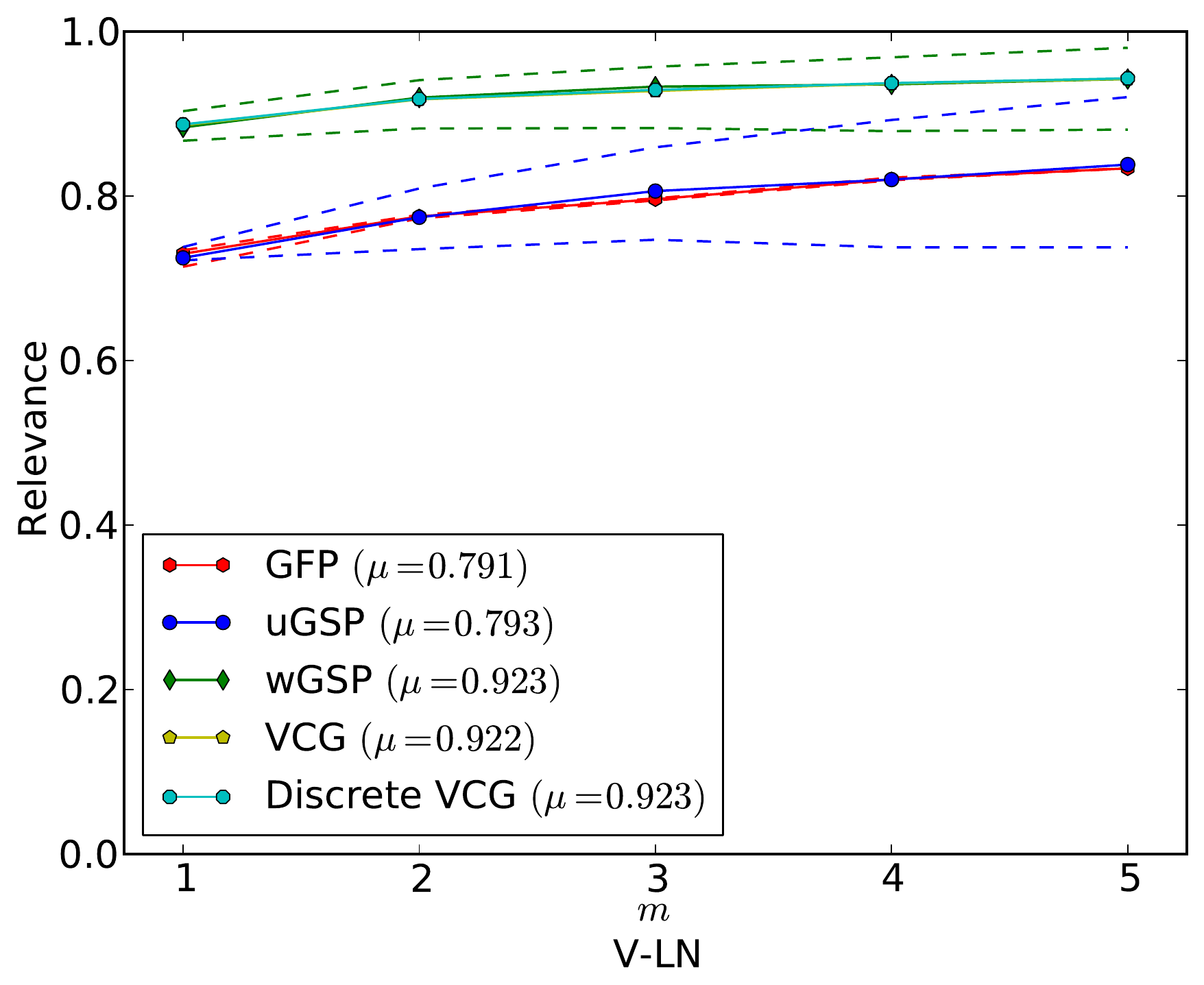}
	\includegraphics[width=0.33\hsize]{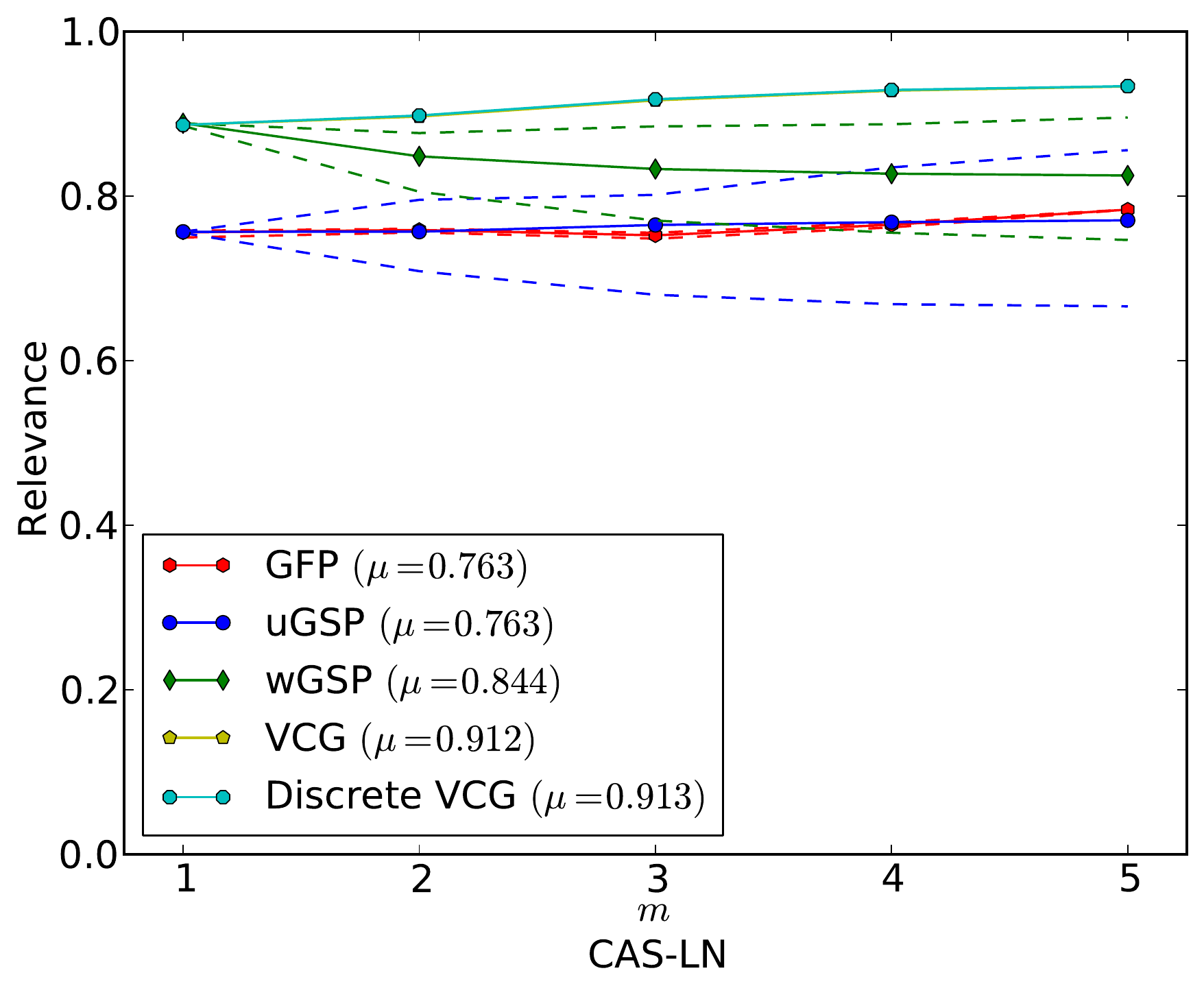}
	%
	\includegraphics[width=0.33\hsize]{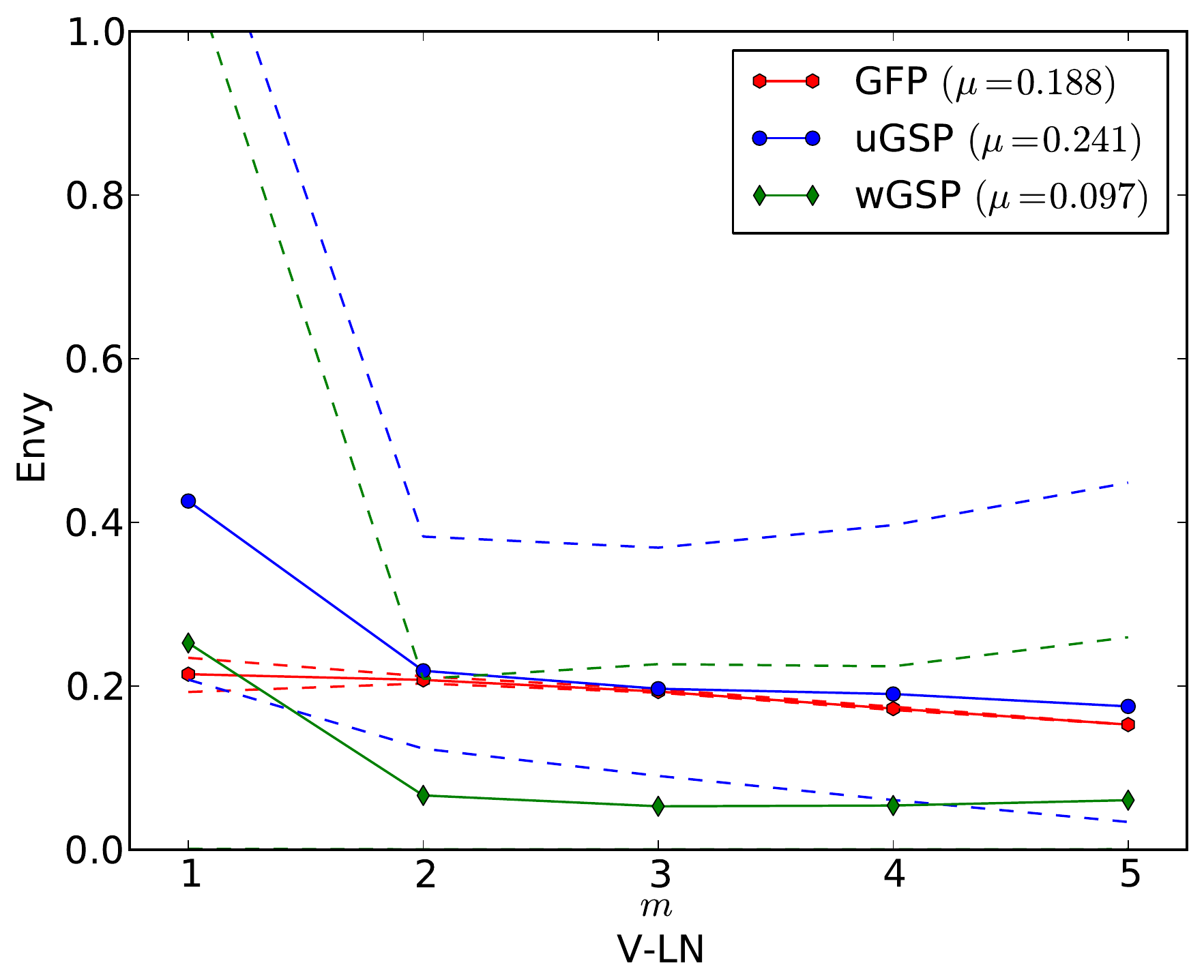}
	\kcaption{(e) V: Log-Normal, Relevance}{.33\hsize}
	\kcaption{(f) Cascade: Log-Normal, Relevance}{.33\hsize}
	\kcaption{(g) V: Log-Normal, Envy}{.33\hsize}
	\caption{Comparing different auction designs as the number of slots varies.}
	\label{fig:scale-m}
\end{figure}
}

\chapteronly{
\begin{figure}
	\hspace{0.05\hsize}
	\includegraphics[width=0.45\hsize]{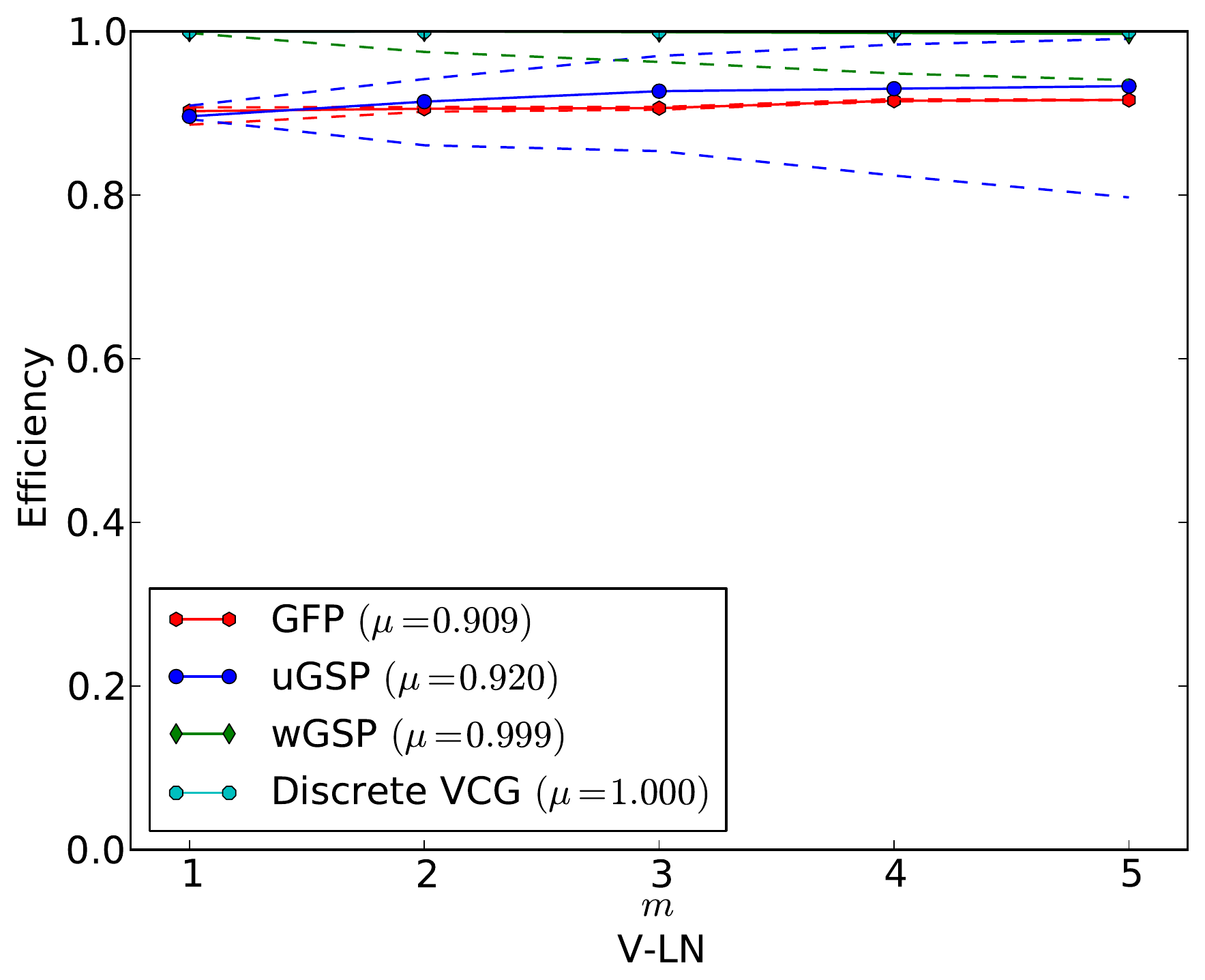}
	\includegraphics[width=0.45\hsize]{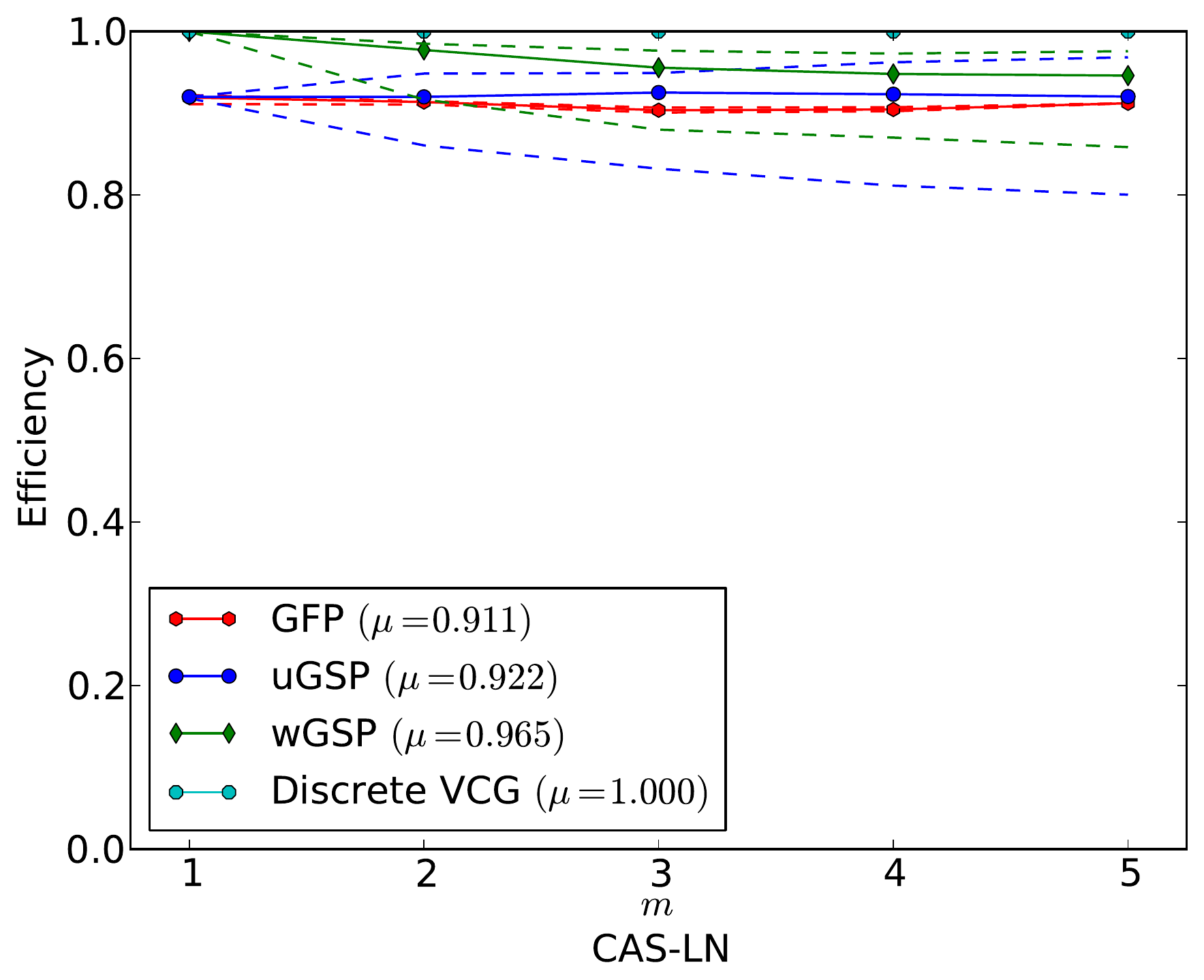}\\
	\kcaption{~}{0.05\hsize}
	\kcaption{(a) V: Log-Normal, Efficiency}{.45\hsize}
	\kcaption{(b) Cascade: Log-Normal, Efficiency}{.45\hsize}

	\hspace{0.05\hsize}
	\includegraphics[width=0.45\hsize]{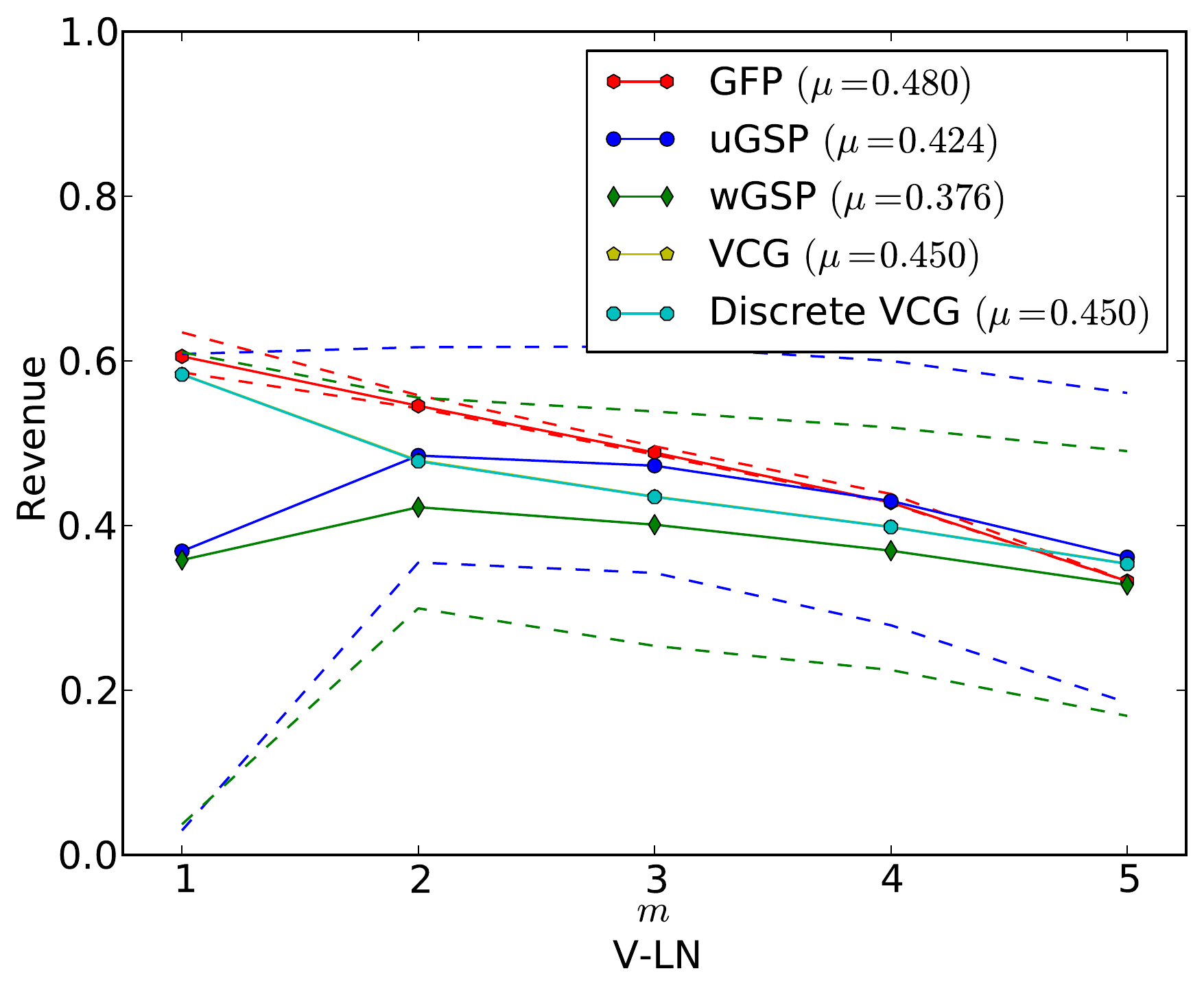}
	\includegraphics[width=0.45\hsize]{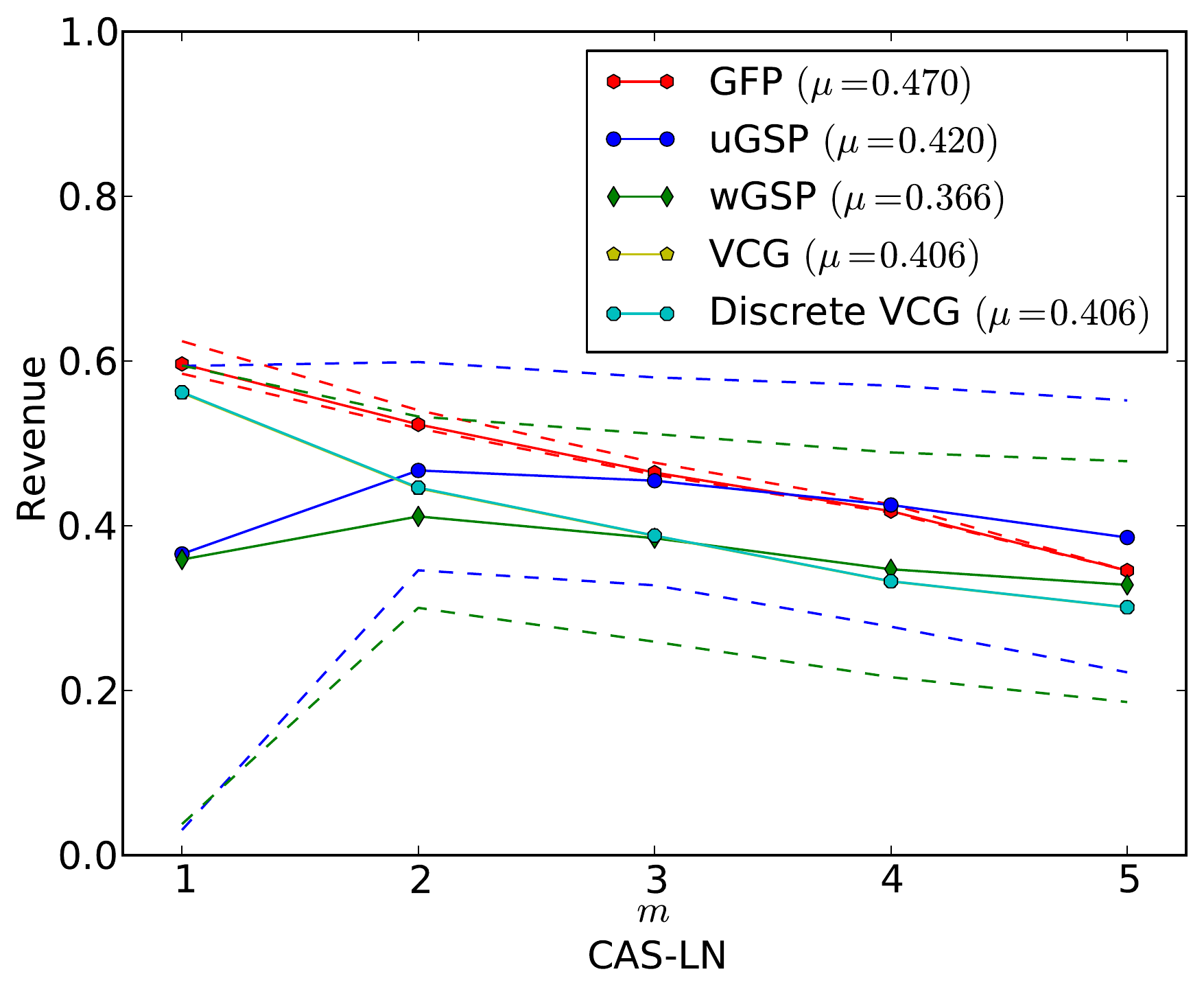}\\
	\kcaption{~}{0.05\hsize}
	\kcaption{(c) V: Log-Normal, Revenue}{.45\hsize}
	\kcaption{(d) Cascade: Log-Normal, Revenue}{.45\hsize}
	\caption{Comparing different auction designs as the number of slots varies.}
	\label{fig:scale-m1}
\end{figure}
\begin{figure}
	\hspace{0.05\hsize}
	\includegraphics[width=0.45\hsize]{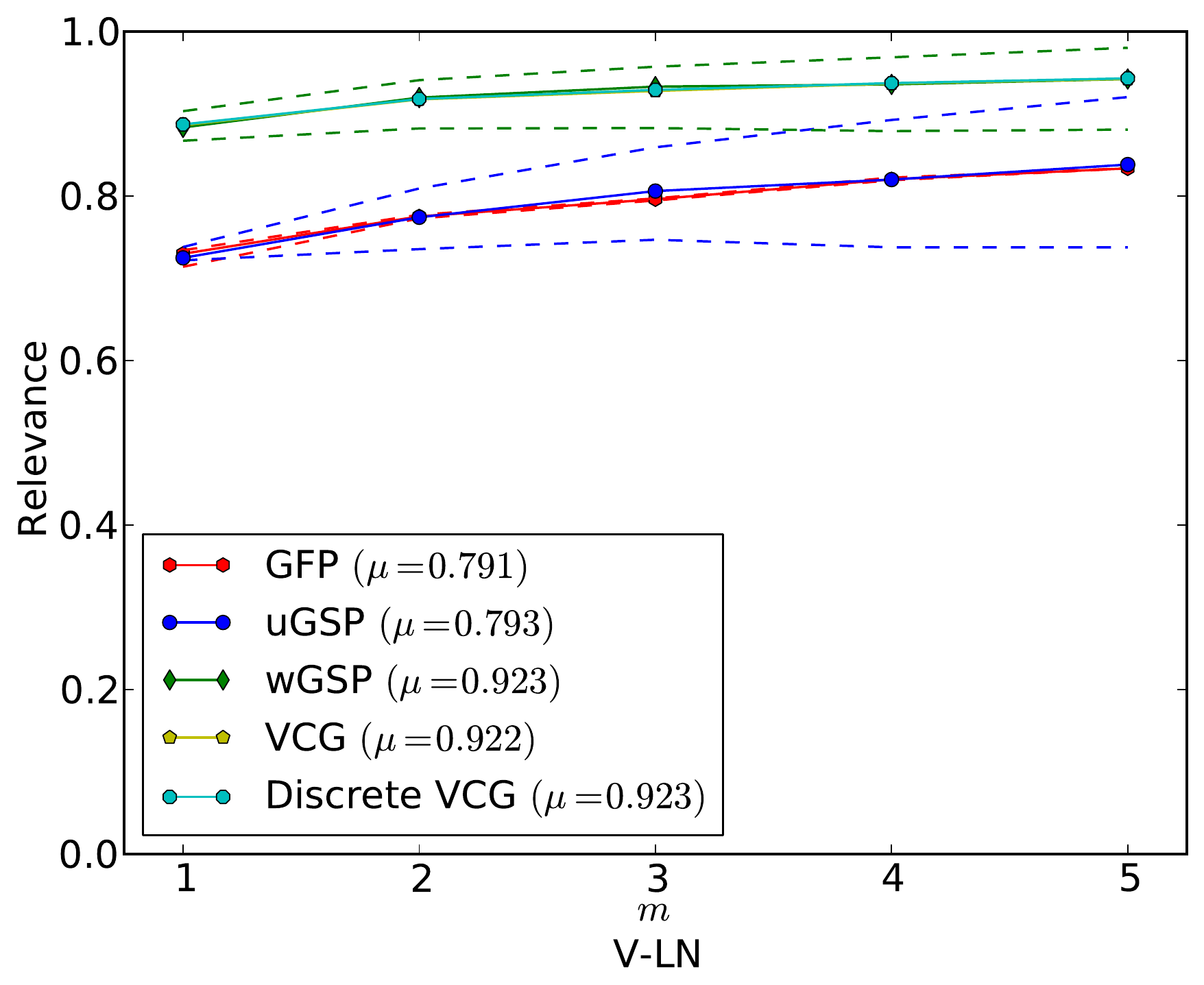}
	\includegraphics[width=0.45\hsize]{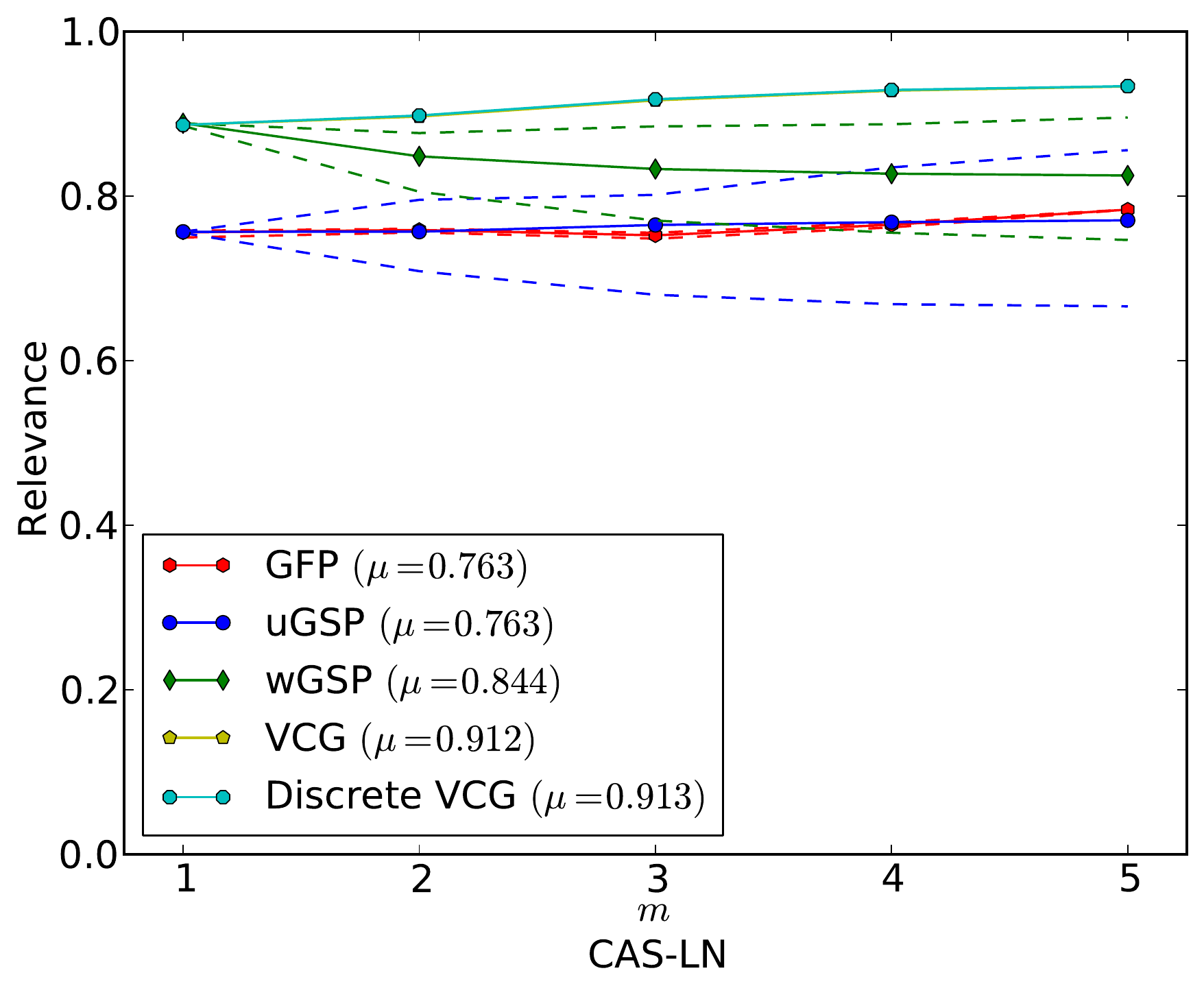}\\
	\kcaption{~}{0.05\hsize}
	\kcaption{(e) V: Log-Normal, Relevance}{.45\hsize}
	\kcaption{(f) Cascade: Log-Normal, Relevance}{.45\hsize}\\

	\hspace{0.275\hsize}
	\includegraphics[width=0.45\hsize]{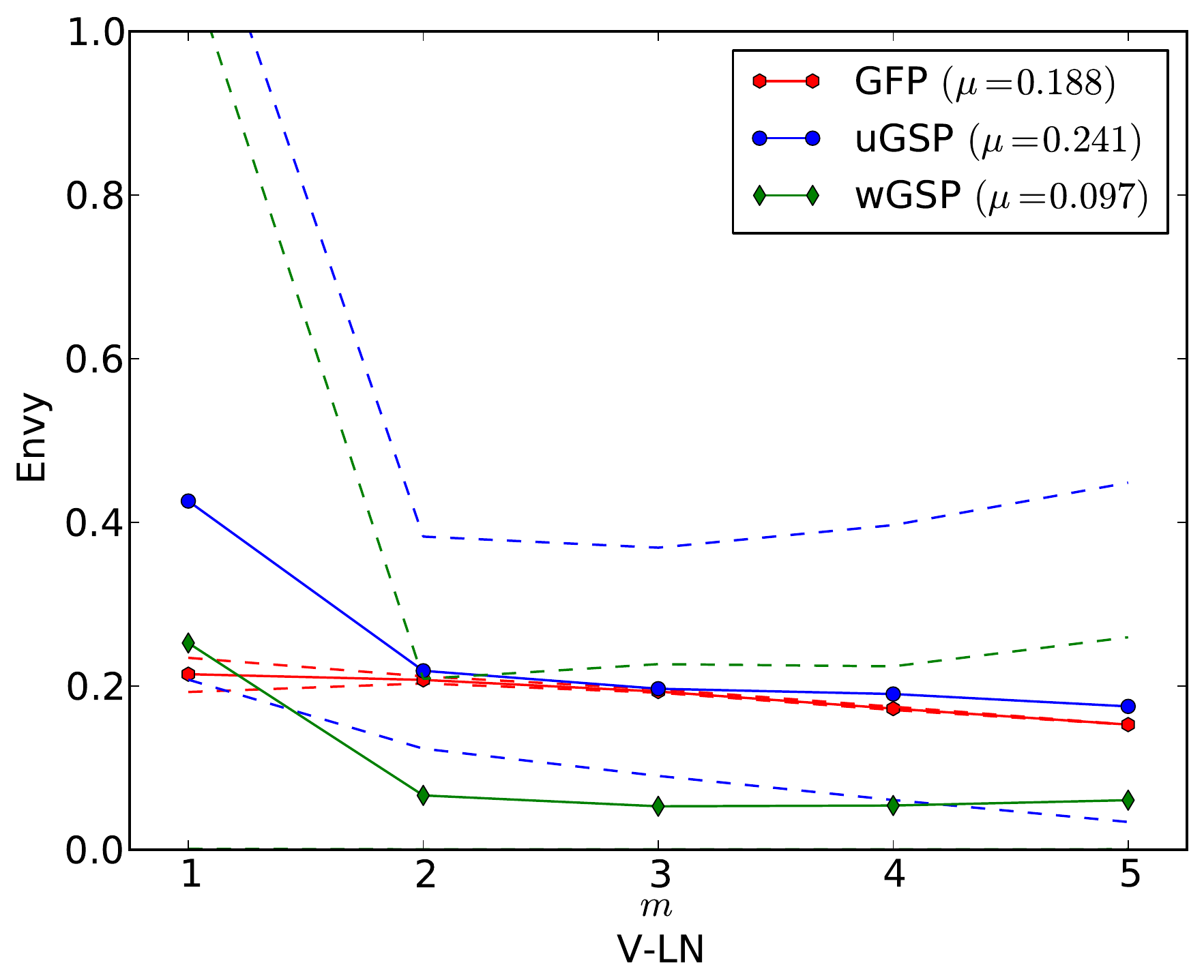}\\
	\kcaption{~}{0.275\hsize}
	\kcaption{(g) V: Log-Normal, Envy}{.45\hsize}
	\caption{Comparing different auction designs as the number of slots varies (continued).}
	\label{fig:scale-m2}
\end{figure}
}

We present our results in \paperonly{Figure~\ref{fig:scale-m}.}\chapteronly{Figures~\ref{fig:scale-m1} and \ref{fig:scale-m2}.} Our first striking observation is that the case of $m=1$ (i.e., selling a single good to ``quality-weighted'' bidders) is distinctly different from $m>1$.  When $m=1$ there is almost no variability in the allocation across different auctions, and therefore likewise no variability in efficiency and relevance.  (The one exception is relevance in wGSP: if two bidders have quality-weighted valuations within an increment of each other but different click-through rates, either one could win in equilibrium.)  However, in GSP auctions revenue (and therefore envy) was nevertheless extremely variable across equilibria. (Indeed in the continuous case, any outcome having revenue between zero and the VCG payment is possible in conservative equilibrium.)  When two or more slots were available,  competition for the lower slots dramatically decreased the range of prices that the top bidder could be charged, and hence the range of possible revenues.  However, these lower slots could attract a bidder who should (in the efficient allocation) have appeared in a high position.  Thus, with two or more slots, efficiency (and relevance) became more variable for all position auction designs.  The relative performance of all auctions remained consistent, with wGSP being slightly worse than VCG and substantially better than uGSP and GFP.  Revenue was a different story: as $m$ increased, all auctions extracted a smaller fraction of the possible surplus because the availability of lower positions decreased competition for the higher positions.  However, in GSP auctions, this decline occurred more gradually than in GFP and VCG.

\section{Conclusions and Future Work}

This  \paperonly{paper}\chapteronly{chapter} demonstrates the feasibility of computational, rather than theoretical, mechanism analysis, applying algorithms for equilibrium computation to address a wide range of open questions about position-auction games.  The main technical obstacle we faced was representing such games in a computationally usable form.  We accomplished this by identifying encoder algorithms that take as input a position-auction setting (based on a wide variety of models drawn from the literature) and the parameters of the position auction (e.g., pricing rule, tie-breaking rule) and produce as output an action-graph game (AGG).  These encoder algorithms, when combined with state-of-the-art equilibrium-finding algorithms, allow us to very quickly compute exact Nash equilibria of realistic position-auction games, and hence to provide quantitative statistical answers to many open questions in the literature.  For example, where existing research into the widely-studied V and EOS models has focused on locally envy-free Nash equilibria, we found that such equilibria are a small minority among the set of equilibria, and often do not exist when bids are restricted to integer increments.  These two sets also had interesting qualitative differences: while every envy-free equilibrium is known to yield more revenue than VCG, we found that the majority of Nash equilibria yield less revenue than VCG.
These techniques also allowed us to make direct, apples-to-apples to comparisons between different auction designs---varying the auction while holding the set of bidders and the equilibrium-selection criteria constant---yielding some striking results.  In particular, while wGSP is known to be inefficient under many models of bidder valuations, we found that it was the most efficient position auction under nearly every model and valuation distribution, and often close to fully efficient in expectation even in models where its worst-case efficiency is very poor.  Similarly, wGSP tended to outperform other position auction designs in terms of relevance and envy.  On the other hand, we found that different position auction mechanisms could not easily be distinguished in terms of their revenues, with relative performance varying dramatically across models, distributions, and equilibria.

Our techniques for computational mechanism analysis are applicable to many other open problems\chapteronly{in position auctions}.  One very general direction is using computational mechanism analysis to facilitate mechanism design (e.g., searching through the space of reserve prices and quality-score distortions in order to find a position auction variant that optimizes some objective such as revenue or relevance, as indeed we do
\paperonly{ in previous work \cite{TLb13revenue}).  }%
\chapteronly{ in Chapter~\ref{ch:revenue}). }%
Other promising ideas include investigating nonlinear value for money (representing bidders who are not risk neutral, or who have budgets or return-on-investment targets), and Bayesian games. We also expect that it would be possible to extend our encoders to deal with richer auction rules (e.g., auctions where bidders specify budgets, or auctions with greater expressiveness as in \cite{BSS08inexpressive}), or to encompass strategic interactions between multiple heterogeneous auctions (e.g., arising due to advertisers using broad match, or location targeting, or having a single budget that spans a multi-keyword campaign).

\paperonly{


\bibliographystyle{abbrv}
\bibliography{!MyMasterBibtex}

\clearpage
\appendix
}

\section{Summary tables and statistical comparisons}

In this \paperonly{appendix}\chapteronly{section} we provide detailed numerical summaries of our main experiments, as well as statistical comparisons between different auctions.  For each pair of auctions $A,B$ and metric (e.g., efficiency), we first test whether $A$ is robustly superior to $B$, meaning that $A$'s worst case is significantly better than $B$'s best case.  (This is signified by $\dagger$.)  Next, we test a looser condition, whether $A$ is better than $B$ up to the limits of equilibrium selection, meaning that $A$ is significantly better than $B$ when comparing best case to best case, median to median and worst case to worst case.  Next, we test whether or not one auction's performance ``spans'' the other, denoted by $\subseteq$.  That is, $A\subseteq B$ indicates that $A$'s worst-case performance is significantly better than $B$'s, but that $B$'s best-case performance is significantly better than $A$'s.
$\sim$ is used to indicate that none of these conditions is true with sufficient statistical confidence.

\begin{table}[H]\footnotesize
\begin{center}
\begin{tabular}{ccccc}
\toprule
Mechanism&Worst&Median&Best&$n$\\
\midrule
GFP&$0.979$ $(\sigma=0.011)$&$0.979$ $(\sigma=0.011)$&$0.979$ $(\sigma=0.011)$&161\\
uGSP&$0.958$ $(\sigma=0.039)$&$0.998$ $(\sigma=0.003)$&$1.000$ $(\sigma=0.001)$&200\\
\midrule
VCG discrete&$1.000$ $(\sigma=0.001)$&$1.000$ $(\sigma=0.001)$&$1.000$ $(\sigma=0.001)$&200\\
\bottomrule
\end{tabular}

\begin{tabular}{cccccccc}
\toprule
&GFP&uGSP&VCG&dVCG\\
\midrule
GFP& &$\sim$&$\leq\dagger$$^{\star\star}$&$\leq\dagger$$^{\star\star}$&\\
uGSP& & &$\leq\dagger$$^{\star\star}$&$\sim$&\\
VCG& & & &$\geq\dagger$$^{\star\star}$&\\
dVCG& & & & &\\

\bottomrule
\end{tabular}
\end{center}
\caption{Comparing Efficiency (EOS-UNI distribution)}
\end{table}

\begin{table}[H]\footnotesize
\begin{center}
\begin{tabular}{ccccc}
\toprule
Mechanism&Worst&Median&Best&$n$\\
\midrule
GFP&$0.487$ $(\sigma=0.168)$&$0.487$ $(\sigma=0.168)$&$0.487$ $(\sigma=0.168)$&161\\
uGSP&$0.365$ $(\sigma=0.172)$&$0.498$ $(\sigma=0.161)$&$0.635$ $(\sigma=0.170)$&200\\
\midrule
VCG&$0.572$ $(\sigma=0.190)$&$0.572$ $(\sigma=0.190)$&$0.572$ $(\sigma=0.190)$&200\\
VCG discrete&$0.573$ $(\sigma=0.191)$&$0.573$ $(\sigma=0.191)$&$0.573$ $(\sigma=0.191)$&200\\
\bottomrule
\end{tabular}

\begin{tabular}{cccccccc}
\toprule
&GFP&uGSP&VCG&dVCG\\
\midrule
GFP& &$\sim$&$\sim$&$\sim$&\\
uGSP& & &$\supseteq$$^{\star\star}$&$\supseteq$$^{\star\star}$&\\
VCG& & & &$\sim$&\\
dVCG& & & & &\\

\bottomrule
\end{tabular}
\end{center}
\caption{Comparing Revenue (EOS-UNI distribution)}
\end{table}

\begin{table}[H]\footnotesize
\begin{center}
\begin{tabular}{ccccc}
\toprule
Mechanism&Worst&Median&Best&$n$\\
\midrule
GFP&$0.021$ $(\sigma=0.013)$&$0.021$ $(\sigma=0.013)$&$0.021$ $(\sigma=0.013)$&161\\
uGSP&$0.003$ $(\sigma=0.005)$&$0.092$ $(\sigma=0.083)$&$0.282$ $(\sigma=0.230)$&200\\
\bottomrule
\end{tabular}

\begin{tabular}{cccccccc}
\toprule
&GFP&uGSP\\
\midrule
GFP& &$\sim$&\\
uGSP& & &\\

\bottomrule
\end{tabular}
\end{center}
\caption{Comparing Envy (EOS-UNI distribution)}
\end{table}

\clearpage

\begin{table}[H]\footnotesize
\begin{center}
\begin{tabular}{ccccc}
\toprule
Mechanism&Worst&Median&Best&$n$\\
\midrule
GFP&$0.971$ $(\sigma=0.010)$&$0.971$ $(\sigma=0.009)$&$0.971$ $(\sigma=0.010)$&176\\
uGSP&$0.935$ $(\sigma=0.048)$&$0.995$ $(\sigma=0.005)$&$1.000$ $(\sigma=0.000)$&198\\
\midrule
VCG discrete&$1.000$ $(\sigma=0.001)$&$1.000$ $(\sigma=0.001)$&$1.000$ $(\sigma=0.001)$&200\\
\bottomrule
\end{tabular}

\begin{tabular}{cccccccc}
\toprule
&GFP&uGSP&VCG&dVCG\\
\midrule
GFP& &$\sim$&$\leq\dagger$$^{\star\star}$&$\leq\dagger$$^{\star\star}$&\\
uGSP& & &$\sim$&$\sim$&\\
VCG& & & &$\geq\dagger$$^{\star\star}$&\\
dVCG& & & & &\\

\bottomrule
\end{tabular}
\end{center}
\caption{Comparing Efficiency (EOS-LN distribution)}
\end{table}

\begin{table}[H]\footnotesize
\begin{center}
\begin{tabular}{ccccc}
\toprule
Mechanism&Worst&Median&Best&$n$\\
\midrule
GFP&$0.352$ $(\sigma=0.083)$&$0.352$ $(\sigma=0.083)$&$0.352$ $(\sigma=0.082)$&176\\
uGSP&$0.236$ $(\sigma=0.062)$&$0.387$ $(\sigma=0.078)$&$0.547$ $(\sigma=0.114)$&198\\
\midrule
VCG&$0.422$ $(\sigma=0.111)$&$0.422$ $(\sigma=0.111)$&$0.422$ $(\sigma=0.111)$&200\\
VCG discrete&$0.422$ $(\sigma=0.111)$&$0.422$ $(\sigma=0.111)$&$0.422$ $(\sigma=0.111)$&200\\
\bottomrule
\end{tabular}

\begin{tabular}{cccccccc}
\toprule
&GFP&uGSP&VCG&dVCG\\
\midrule
GFP& &$\subseteq$$^{\star\star}$&$\sim$&$\sim$&\\
uGSP& & &$\supseteq$$^{\star\star}$&$\supseteq$$^{\star\star}$&\\
VCG& & & &$\sim$&\\
dVCG& & & & &\\

\bottomrule
\end{tabular}
\end{center}
\caption{Comparing Revenue (EOS-LN distribution)}
\end{table}

\begin{table}[H]\footnotesize
\begin{center}
\begin{tabular}{ccccc}
\toprule
Mechanism&Worst&Median&Best&$n$\\
\midrule
GFP&$0.036$ $(\sigma=0.015)$&$0.037$ $(\sigma=0.015)$&$0.037$ $(\sigma=0.015)$&176\\
uGSP&$0.001$ $(\sigma=0.002)$&$0.066$ $(\sigma=0.041)$&$0.279$ $(\sigma=0.109)$&198\\
\bottomrule
\end{tabular}

\begin{tabular}{cccccccc}
\toprule
&GFP&uGSP\\
\midrule
GFP& &$\sim$&\\
uGSP& & &\\

\bottomrule
\end{tabular}
\end{center}
\caption{Comparing Envy (EOS-LN distribution)}
\end{table}

\begin{table}[H]\footnotesize
\begin{center}
\begin{tabular}{ccccc}
\toprule
Mechanism&Worst&Median&Best&$n$\\
\midrule
GFP&$0.830$ $(\sigma=0.157)$&$0.830$ $(\sigma=0.157)$&$0.830$ $(\sigma=0.157)$&146\\
uGSP&$0.736$ $(\sigma=0.218)$&$0.826$ $(\sigma=0.208)$&$0.883$ $(\sigma=0.195)$&200\\
wGSP&$0.964$ $(\sigma=0.040)$&$0.999$ $(\sigma=0.005)$&$1.000$ $(\sigma=0.001)$&200\\
wGFP&$0.974$ $(\sigma=0.014)$&$0.974$ $(\sigma=0.014)$&$0.974$ $(\sigma=0.014)$&181\\
\midrule
VCG discrete&$1.000$ $(\sigma=0.001)$&$1.000$ $(\sigma=0.001)$&$1.000$ $(\sigma=0.001)$&200\\
\bottomrule
\end{tabular}

\begin{tabular}{cccccccc}
\toprule
&GFP&uGSP&wGSP&wGFP&VCG&dVCG\\
\midrule
GFP& &$\sim$&$\leq\dagger$$^{\star\star}$&$\sim$&$\leq\dagger$$^{\star\star}$&$\leq\dagger$$^{\star\star}$&\\
uGSP& & &$\leq\dagger$$^{\star\star}$&$\sim$&$\leq\dagger$$^{\star\star}$&$\leq\dagger$$^{\star\star}$&\\
wGSP& & & &$\sim$&$\leq\dagger$$^{\star\star}$&$\sim$&\\
wGFP& & & & &$\leq\dagger$$^{\star\star}$&$\leq\dagger$$^{\star\star}$&\\
VCG& & & & & &$\geq\dagger$$^{\star\star}$&\\
dVCG& & & & & & &\\

\bottomrule
\end{tabular}
\end{center}
\caption{Comparing Efficiency (V-UNI distribution)}
\end{table}

\begin{table}[H]\footnotesize
\begin{center}
\begin{tabular}{ccccc}
\toprule
Mechanism&Worst&Median&Best&$n$\\
\midrule
GFP&$0.393$ $(\sigma=0.132)$&$0.393$ $(\sigma=0.132)$&$0.393$ $(\sigma=0.131)$&146\\
uGSP&$0.267$ $(\sigma=0.152)$&$0.407$ $(\sigma=0.160)$&$0.570$ $(\sigma=0.182)$&200\\
wGSP&$0.255$ $(\sigma=0.142)$&$0.389$ $(\sigma=0.152)$&$0.525$ $(\sigma=0.179)$&200\\
wGFP&$0.393$ $(\sigma=0.173)$&$0.393$ $(\sigma=0.173)$&$0.393$ $(\sigma=0.173)$&181\\
\midrule
VCG&$0.443$ $(\sigma=0.187)$&$0.443$ $(\sigma=0.187)$&$0.443$ $(\sigma=0.187)$&200\\
VCG discrete&$0.442$ $(\sigma=0.187)$&$0.442$ $(\sigma=0.187)$&$0.442$ $(\sigma=0.187)$&200\\
\bottomrule
\end{tabular}

\begin{tabular}{cccccccc}
\toprule
&GFP&uGSP&wGSP&wGFP&VCG&dVCG\\
\midrule
GFP& &$\sim$&$\sim$&$\sim$&$\sim$&$\sim$&\\
uGSP& & &$\sim$&$\supseteq$$^{\star\star}$&$\supseteq$$^{\star\star}$&$\supseteq$$^{\star\star}$&\\
wGSP& & & &$\supseteq$$^{\star\star}$&$\supseteq$$^{\star\star}$&$\supseteq$$^{\star\star}$&\\
wGFP& & & & &$\sim$&$\sim$&\\
VCG& & & & & &$\sim$&\\
dVCG& & & & & & &\\

\bottomrule
\end{tabular}
\end{center}
\caption{Comparing Revenue (V-UNI distribution)}
\end{table}

\begin{table}[H]\footnotesize
\begin{center}
\begin{tabular}{ccccc}
\toprule
Mechanism&Worst&Median&Best&$n$\\
\midrule
GFP&$0.701$ $(\sigma=0.211)$&$0.701$ $(\sigma=0.211)$&$0.701$ $(\sigma=0.211)$&146\\
uGSP&$0.630$ $(\sigma=0.243)$&$0.697$ $(\sigma=0.245)$&$0.751$ $(\sigma=0.247)$&200\\
wGSP&$0.866$ $(\sigma=0.136)$&$0.901$ $(\sigma=0.128)$&$0.915$ $(\sigma=0.127)$&200\\
wGFP&$0.888$ $(\sigma=0.118)$&$0.889$ $(\sigma=0.118)$&$0.889$ $(\sigma=0.118)$&181\\
\midrule
VCG&$0.901$ $(\sigma=0.129)$&$0.901$ $(\sigma=0.129)$&$0.901$ $(\sigma=0.129)$&200\\
VCG discrete&$0.902$ $(\sigma=0.128)$&$0.902$ $(\sigma=0.128)$&$0.902$ $(\sigma=0.128)$&200\\
\bottomrule
\end{tabular}

\begin{tabular}{cccccccc}
\toprule
&GFP&uGSP&wGSP&wGFP&VCG&dVCG\\
\midrule
GFP& &$\sim$&$\leq\dagger$$^{\star\star}$&$\sim$&$\leq\dagger$$^{\star\star}$&$\leq\dagger$$^{\star\star}$&\\
uGSP& & &$\leq\dagger$$^{\star\star}$&$\sim$&$\leq\dagger$$^{\star\star}$&$\leq\dagger$$^{\star\star}$&\\
wGSP& & & &$\sim$&$\supseteq$$^{\star\star}$&$\supseteq$$^{\star\star}$&\\
wGFP& & & & &$\sim$&$\sim$&\\
VCG& & & & & &$\sim$&\\
dVCG& & & & & & &\\

\bottomrule
\end{tabular}
\end{center}
\caption{Comparing Relevance (V-UNI distribution)}
\end{table}

\begin{table}[H]\footnotesize
\begin{center}
\begin{tabular}{ccccc}
\toprule
Mechanism&Worst&Median&Best&$n$\\
\midrule
GFP&$0.335$ $(\sigma=0.348)$&$0.335$ $(\sigma=0.348)$&$0.335$ $(\sigma=0.348)$&146\\
uGSP&$0.242$ $(\sigma=0.350)$&$0.409$ $(\sigma=0.422)$&$0.627$ $(\sigma=0.452)$&200\\
wGSP&$0.002$ $(\sigma=0.004)$&$0.076$ $(\sigma=0.067)$&$0.247$ $(\sigma=0.183)$&200\\
wGFP&$0.026$ $(\sigma=0.014)$&$0.026$ $(\sigma=0.014)$&$0.026$ $(\sigma=0.014)$&181\\
\bottomrule
\end{tabular}

\begin{tabular}{cccccccc}
\toprule
&GFP&uGSP&wGSP&wGFP\\
\midrule
GFP& &$\sim$&$\sim$&$\geq\dagger$$^{\star\star}$&\\
uGSP& & &$\geq$$^{\star\star}$&$\geq\dagger$$^{\star\star}$&\\
wGSP& & & &$\supseteq$$^{\star\star}$&\\
wGFP& & & & &\\

\bottomrule
\end{tabular}
\end{center}
\caption{Comparing Envy (V-UNI distribution)}
\end{table}

\begin{table}[H]\footnotesize
\begin{center}
\begin{tabular}{ccccc}
\toprule
Mechanism&Worst&Median&Best&$n$\\
\midrule
GFP&$0.921$ $(\sigma=0.059)$&$0.922$ $(\sigma=0.059)$&$0.922$ $(\sigma=0.059)$&174\\
uGSP&$0.810$ $(\sigma=0.157)$&$0.941$ $(\sigma=0.092)$&$0.993$ $(\sigma=0.033)$&200\\
wGSP&$0.938$ $(\sigma=0.056)$&$0.997$ $(\sigma=0.005)$&$1.000$ $(\sigma=0.000)$&200\\
wGFP&$0.970$ $(\sigma=0.012)$&$0.970$ $(\sigma=0.012)$&$0.970$ $(\sigma=0.012)$&198\\
\midrule
VCG discrete&$1.000$ $(\sigma=0.001)$&$1.000$ $(\sigma=0.001)$&$1.000$ $(\sigma=0.001)$&200\\
\bottomrule
\end{tabular}

\begin{tabular}{cccccccc}
\toprule
&GFP&uGSP&wGSP&wGFP&VCG&dVCG\\
\midrule
GFP& &$\sim$&$\sim$&$\sim$&$\leq\dagger$$^{\star\star}$&$\leq\dagger$$^{\star\star}$&\\
uGSP& & &$\leq$$^{\star\star}$&$\supseteq$$^{\star\star}$&$\leq\dagger$$^{\star\star}$&$\leq\dagger$$^{\star\star}$&\\
wGSP& & & &$\sim$&$\sim$&$\sim$&\\
wGFP& & & & &$\leq\dagger$$^{\star\star}$&$\leq\dagger$$^{\star\star}$&\\
VCG& & & & & &$\geq\dagger$$^{\star\star}$&\\
dVCG& & & & & & &\\

\bottomrule
\end{tabular}
\end{center}
\caption{Comparing Efficiency (V-LN distribution)}
\end{table}

\begin{table}[H]\footnotesize
\begin{center}
\begin{tabular}{ccccc}
\toprule
Mechanism&Worst&Median&Best&$n$\\
\midrule
GFP&$0.333$ $(\sigma=0.076)$&$0.333$ $(\sigma=0.076)$&$0.333$ $(\sigma=0.076)$&174\\
uGSP&$0.190$ $(\sigma=0.054)$&$0.362$ $(\sigma=0.079)$&$0.558$ $(\sigma=0.128)$&200\\
wGSP&$0.172$ $(\sigma=0.074)$&$0.328$ $(\sigma=0.096)$&$0.489$ $(\sigma=0.138)$&200\\
wGFP&$0.302$ $(\sigma=0.107)$&$0.302$ $(\sigma=0.107)$&$0.302$ $(\sigma=0.107)$&198\\
\midrule
VCG&$0.351$ $(\sigma=0.126)$&$0.351$ $(\sigma=0.126)$&$0.351$ $(\sigma=0.126)$&200\\
VCG discrete&$0.351$ $(\sigma=0.126)$&$0.351$ $(\sigma=0.126)$&$0.351$ $(\sigma=0.126)$&200\\
\bottomrule
\end{tabular}

\begin{tabular}{cccccccc}
\toprule
&GFP&uGSP&wGSP&wGFP&VCG&dVCG\\
\midrule
GFP& &$\subseteq$$^{\star\star}$&$\subseteq$$^{\star\star}$&$\sim$&$\sim$&$\sim$&\\
uGSP& & &$\geq$$^{\star\star}$&$\supseteq$$^{\star\star}$&$\supseteq$$^{\star\star}$&$\supseteq$$^{\star\star}$&\\
wGSP& & & &$\supseteq$$^{\star\star}$&$\supseteq$$^{\star\star}$&$\supseteq$$^{\star\star}$&\\
wGFP& & & & &$\leq\dagger$$^{\star\star}$&$\leq\dagger$$^{\star\star}$&\\
VCG& & & & & &$\sim$&\\
dVCG& & & & & & &\\

\bottomrule
\end{tabular}
\end{center}
\caption{Comparing Revenue (V-LN distribution)}
\end{table}

\begin{table}[H]\footnotesize
\begin{center}
\begin{tabular}{ccccc}
\toprule
Mechanism&Worst&Median&Best&$n$\\
\midrule
GFP&$0.842$ $(\sigma=0.114)$&$0.842$ $(\sigma=0.114)$&$0.842$ $(\sigma=0.114)$&174\\
uGSP&$0.750$ $(\sigma=0.147)$&$0.848$ $(\sigma=0.132)$&$0.921$ $(\sigma=0.104)$&200\\
wGSP&$0.882$ $(\sigma=0.104)$&$0.938$ $(\sigma=0.082)$&$0.979$ $(\sigma=0.046)$&200\\
wGFP&$0.925$ $(\sigma=0.065)$&$0.925$ $(\sigma=0.065)$&$0.925$ $(\sigma=0.065)$&198\\
\midrule
VCG&$0.938$ $(\sigma=0.083)$&$0.938$ $(\sigma=0.083)$&$0.938$ $(\sigma=0.083)$&200\\
VCG discrete&$0.939$ $(\sigma=0.081)$&$0.939$ $(\sigma=0.081)$&$0.939$ $(\sigma=0.081)$&200\\
\bottomrule
\end{tabular}

\begin{tabular}{cccccccc}
\toprule
&GFP&uGSP&wGSP&wGFP&VCG&dVCG\\
\midrule
GFP& &$\sim$&$\leq\dagger$$^{\star\star}$&$\leq\dagger$$^{\star\star}$&$\leq\dagger$$^{\star\star}$&$\leq\dagger$$^{\star\star}$&\\
uGSP& & &$\leq$$^{\star\star}$&$\sim$&$\leq\dagger$$^{\star\star}$&$\leq\dagger$$^{\star\star}$&\\
wGSP& & & &$\supseteq$$^{\star\star}$&$\supseteq$$^{\star\star}$&$\supseteq$$^{\star\star}$&\\
wGFP& & & & &$\leq\dagger$$^{\star\star}$&$\leq\dagger$$^{\star\star}$&\\
VCG& & & & & &$\sim$&\\
dVCG& & & & & & &\\

\bottomrule
\end{tabular}
\end{center}
\caption{Comparing Relevance (V-LN distribution)}
\end{table}

\begin{table}[H]\footnotesize
\begin{center}
\begin{tabular}{ccccc}
\toprule
Mechanism&Worst&Median&Best&$n$\\
\midrule
GFP&$0.148$ $(\sigma=0.149)$&$0.148$ $(\sigma=0.149)$&$0.148$ $(\sigma=0.150)$&174\\
uGSP&$0.033$ $(\sigma=0.077)$&$0.168$ $(\sigma=0.174)$&$0.429$ $(\sigma=0.259)$&200\\
wGSP&$0.001$ $(\sigma=0.002)$&$0.057$ $(\sigma=0.045)$&$0.257$ $(\sigma=0.123)$&200\\
wGFP&$0.033$ $(\sigma=0.015)$&$0.033$ $(\sigma=0.015)$&$0.033$ $(\sigma=0.015)$&198\\
\bottomrule
\end{tabular}

\begin{tabular}{cccccccc}
\toprule
&GFP&uGSP&wGSP&wGFP\\
\midrule
GFP& &$\subseteq$$^{\star\star}$&$\sim$&$\geq\dagger$$^{\star\star}$&\\
uGSP& & &$\geq$$^{\star\star}$&$\sim$&\\
wGSP& & & &$\supseteq$$^{\star\star}$&\\
wGFP& & & & &\\

\bottomrule
\end{tabular}
\end{center}
\caption{Comparing Envy (V-LN distribution)}
\end{table}

\begin{table}[H]\footnotesize
\begin{center}
\begin{tabular}{ccccc}
\toprule
Mechanism&Worst&Median&Best&$n$\\
\midrule
GFP&$0.611$ $(\sigma=0.398)$&$0.638$ $(\sigma=0.387)$&$0.640$ $(\sigma=0.388)$&200\\
uGSP&$0.584$ $(\sigma=0.392)$&$0.657$ $(\sigma=0.400)$&$0.682$ $(\sigma=0.412)$&200\\
wGSP&$0.648$ $(\sigma=0.427)$&$0.671$ $(\sigma=0.440)$&$0.675$ $(\sigma=0.443)$&200\\
\midrule
VCG discrete&$0.916$ $(\sigma=0.151)$&$0.916$ $(\sigma=0.151)$&$0.916$ $(\sigma=0.151)$&200\\
\bottomrule
\end{tabular}

\begin{tabular}{cccccccc}
\toprule
&GFP&uGSP&wGSP&VCG&dVCG\\
\midrule
GFP& &$\subseteq$$^{\star\star}$&$\sim$&$\leq\dagger$$^{\star\star}$&$\leq\dagger$$^{\star\star}$&\\
uGSP& & &$\sim$&$\leq\dagger$$^{\star\star}$&$\leq\dagger$$^{\star\star}$&\\
wGSP& & & &$\leq\dagger$$^{\star\star}$&$\leq\dagger$$^{\star\star}$&\\
VCG& & & & &$\geq\dagger$$^{\star\star}$&\\
dVCG& & & & & &\\

\bottomrule
\end{tabular}
\end{center}
\caption{Comparing Efficiency (BHN-UNI distribution)}
\end{table}

\begin{table}[H]\footnotesize
\begin{center}
\begin{tabular}{ccccc}
\toprule
Mechanism&Worst&Median&Best&$n$\\
\midrule
GFP&$0.225$ $(\sigma=0.180)$&$0.252$ $(\sigma=0.193)$&$0.257$ $(\sigma=0.198)$&200\\
uGSP&$0.196$ $(\sigma=0.170)$&$0.258$ $(\sigma=0.197)$&$0.295$ $(\sigma=0.220)$&200\\
wGSP&$0.231$ $(\sigma=0.201)$&$0.270$ $(\sigma=0.218)$&$0.296$ $(\sigma=0.234)$&200\\
\midrule
VCG&$0.269$ $(\sigma=0.160)$&$0.269$ $(\sigma=0.160)$&$0.269$ $(\sigma=0.160)$&200\\
VCG discrete&$0.219$ $(\sigma=0.236)$&$0.219$ $(\sigma=0.236)$&$0.219$ $(\sigma=0.236)$&200\\
\bottomrule
\end{tabular}

\begin{tabular}{cccccccc}
\toprule
&GFP&uGSP&wGSP&VCG&dVCG\\
\midrule
GFP& &$\subseteq$$^{\star\star}$&$\sim$&$\sim$&$\sim$&\\
uGSP& & &$\sim$&$\sim$&$\sim$&\\
wGSP& & & &$\sim$&$\sim$&\\
VCG& & & & &$\geq\dagger$$^{\star\star}$&\\
dVCG& & & & & &\\

\bottomrule
\end{tabular}
\end{center}
\caption{Comparing Revenue (BHN-UNI distribution)}
\end{table}

\begin{table}[H]\footnotesize
\begin{center}
\begin{tabular}{ccccc}
\toprule
Mechanism&Worst&Median&Best&$n$\\
\midrule
GFP&$0.510$ $(\sigma=0.351)$&$0.528$ $(\sigma=0.342)$&$0.530$ $(\sigma=0.344)$&200\\
uGSP&$0.464$ $(\sigma=0.338)$&$0.542$ $(\sigma=0.356)$&$0.581$ $(\sigma=0.377)$&200\\
wGSP&$0.604$ $(\sigma=0.413)$&$0.626$ $(\sigma=0.423)$&$0.638$ $(\sigma=0.430)$&200\\
\midrule
VCG&$0.903$ $(\sigma=0.113)$&$0.903$ $(\sigma=0.113)$&$0.903$ $(\sigma=0.113)$&200\\
VCG discrete&$0.788$ $(\sigma=0.256)$&$0.788$ $(\sigma=0.256)$&$0.788$ $(\sigma=0.256)$&200\\
\bottomrule
\end{tabular}

\begin{tabular}{cccccccc}
\toprule
&GFP&uGSP&wGSP&VCG&dVCG\\
\midrule
GFP& &$\subseteq$$^{\star\star}$&$\leq\dagger$$^{\star\star}$&$\leq\dagger$$^{\star\star}$&$\leq\dagger$$^{\star\star}$&\\
uGSP& & &$\leq$$^{\star\star}$&$\leq\dagger$$^{\star\star}$&$\leq\dagger$$^{\star\star}$&\\
wGSP& & & &$\leq\dagger$$^{\star\star}$&$\leq\dagger$$^{\star\star}$&\\
VCG& & & & &$\geq\dagger$$^{\star\star}$&\\
dVCG& & & & & &\\

\bottomrule
\end{tabular}
\end{center}
\caption{Comparing Relevance (BHN-UNI distribution)}
\end{table}

\begin{table}[H]\footnotesize
\begin{center}
\begin{tabular}{ccccc}
\toprule
Mechanism&Worst&Median&Best&$n$\\
\midrule
GFP&$0.152$ $(\sigma=0.173)$&$0.174$ $(\sigma=0.190)$&$0.176$ $(\sigma=0.191)$&200\\
uGSP&$0.121$ $(\sigma=0.180)$&$0.191$ $(\sigma=0.220)$&$0.287$ $(\sigma=0.311)$&200\\
wGSP&$0.097$ $(\sigma=0.185)$&$0.110$ $(\sigma=0.185)$&$0.146$ $(\sigma=0.198)$&200\\
\bottomrule
\end{tabular}

\begin{tabular}{cccccccc}
\toprule
&GFP&uGSP&wGSP\\
\midrule
GFP& &$\subseteq$$^{\star\star}$&$\sim$&\\
uGSP& & &$\sim$&\\
wGSP& & & &\\

\bottomrule
\end{tabular}
\end{center}
\caption{Comparing Envy (BHN-UNI distribution)}
\end{table}

\begin{table}[H]\footnotesize
\begin{center}
\begin{tabular}{ccccc}
\toprule
Mechanism&Worst&Median&Best&$n$\\
\midrule
GFP&$0.775$ $(\sigma=0.221)$&$0.835$ $(\sigma=0.127)$&$0.854$ $(\sigma=0.120)$&189\\
uGSP&$0.799$ $(\sigma=0.208)$&$0.902$ $(\sigma=0.118)$&$0.918$ $(\sigma=0.109)$&200\\
\midrule
VCG discrete&$0.999$ $(\sigma=0.003)$&$0.999$ $(\sigma=0.003)$&$0.999$ $(\sigma=0.003)$&200\\
\bottomrule
\end{tabular}

\begin{tabular}{cccccccc}
\toprule
&GFP&uGSP&VCG&dVCG\\
\midrule
GFP& &$\sim$&$\leq\dagger$$^{\star\star}$&$\leq\dagger$$^{\star\star}$&\\
uGSP& & &$\leq\dagger$$^{\star\star}$&$\leq\dagger$$^{\star\star}$&\\
VCG& & & &$\geq\dagger$$^{\star\star}$&\\
dVCG& & & & &\\

\bottomrule
\end{tabular}
\end{center}
\caption{Comparing Efficiency (BSS distribution)}
\end{table}

\begin{table}[H]\footnotesize
\begin{center}
\begin{tabular}{ccccc}
\toprule
Mechanism&Worst&Median&Best&$n$\\
\midrule
GFP&$0.255$ $(\sigma=0.171)$&$0.306$ $(\sigma=0.154)$&$0.360$ $(\sigma=0.170)$&189\\
uGSP&$0.210$ $(\sigma=0.140)$&$0.327$ $(\sigma=0.152)$&$0.436$ $(\sigma=0.193)$&200\\
\midrule
VCG&$0.408$ $(\sigma=0.188)$&$0.408$ $(\sigma=0.188)$&$0.408$ $(\sigma=0.188)$&200\\
VCG discrete&$0.412$ $(\sigma=0.186)$&$0.412$ $(\sigma=0.186)$&$0.412$ $(\sigma=0.186)$&200\\
\bottomrule
\end{tabular}

\begin{tabular}{cccccccc}
\toprule
&GFP&uGSP&VCG&dVCG\\
\midrule
GFP& &$\sim$&$\sim$&$\sim$&\\
uGSP& & &$\sim$&$\sim$&\\
VCG& & & &$\leq\dagger$$^{\star}$&\\
dVCG& & & & &\\

\bottomrule
\end{tabular}
\end{center}
\caption{Comparing Revenue (BSS distribution)}
\end{table}

\begin{table}[H]\footnotesize
\begin{center}
\begin{tabular}{ccccc}
\toprule
Mechanism&Worst&Median&Best&$n$\\
\midrule
GFP&$0.015$ $(\sigma=0.020)$&$0.017$ $(\sigma=0.021)$&$0.022$ $(\sigma=0.021)$&189\\
uGSP&$0.016$ $(\sigma=0.023)$&$0.046$ $(\sigma=0.057)$&$0.107$ $(\sigma=0.113)$&200\\
\bottomrule
\end{tabular}

\begin{tabular}{cccccccc}
\toprule
&GFP&uGSP\\
\midrule
GFP& &$\sim$&\\
uGSP& & &\\

\bottomrule
\end{tabular}
\end{center}
\caption{Comparing Envy (BSS distribution)}
\end{table}

\begin{table}[H]\footnotesize
\begin{center}
\begin{tabular}{ccccc}
\toprule
Mechanism&Worst&Median&Best&$n$\\
\midrule
GFP&$0.837$ $(\sigma=0.150)$&$0.837$ $(\sigma=0.150)$&$0.837$ $(\sigma=0.150)$&153\\
uGSP&$0.650$ $(\sigma=0.230)$&$0.819$ $(\sigma=0.191)$&$0.916$ $(\sigma=0.145)$&200\\
wGSP&$0.855$ $(\sigma=0.118)$&$0.947$ $(\sigma=0.081)$&$0.981$ $(\sigma=0.048)$&200\\
cwGSP&$0.956$ $(\sigma=0.040)$&$0.995$ $(\sigma=0.010)$&$1.000$ $(\sigma=0.002)$&164\\
\midrule
VCG discrete&$1.000$ $(\sigma=0.000)$&$1.000$ $(\sigma=0.000)$&$1.000$ $(\sigma=0.000)$&200\\
\bottomrule
\end{tabular}

\begin{tabular}{cccccccc}
\toprule
&GFP&uGSP&wGSP&cwGSP&VCG&dVCG\\
\midrule
GFP& &$\sim$&$\sim$&$\sim$&$\leq\dagger$$^{\star\star}$&$\leq\dagger$$^{\star\star}$&\\
uGSP& & &$\leq$$^{\star\star}$&$\sim$&$\leq\dagger$$^{\star\star}$&$\leq\dagger$$^{\star\star}$&\\
wGSP& & & &$\sim$&$\leq\dagger$$^{\star\star}$&$\leq\dagger$$^{\star\star}$&\\
cwGSP& & & & &$\leq\dagger$$^{\star\star}$&$\leq\dagger$$^{\star\star}$&\\
VCG& & & & & &$\geq\dagger$$^{\star\star}$&\\
dVCG& & & & & & &\\

\bottomrule
\end{tabular}
\end{center}
\caption{Comparing Efficiency (CAS-UNI distribution)}
\end{table}

\begin{table}[H]\footnotesize
\begin{center}
\begin{tabular}{ccccc}
\toprule
Mechanism&Worst&Median&Best&$n$\\
\midrule
GFP&$0.344$ $(\sigma=0.114)$&$0.344$ $(\sigma=0.114)$&$0.344$ $(\sigma=0.114)$&153\\
uGSP&$0.189$ $(\sigma=0.117)$&$0.370$ $(\sigma=0.137)$&$0.575$ $(\sigma=0.157)$&200\\
wGSP&$0.189$ $(\sigma=0.108)$&$0.337$ $(\sigma=0.127)$&$0.498$ $(\sigma=0.157)$&200\\
cwGSP&$0.156$ $(\sigma=0.098)$&$0.270$ $(\sigma=0.119)$&$0.396$ $(\sigma=0.158)$&164\\
\midrule
VCG&$0.312$ $(\sigma=0.159)$&$0.312$ $(\sigma=0.159)$&$0.312$ $(\sigma=0.159)$&200\\
VCG discrete&$0.312$ $(\sigma=0.158)$&$0.312$ $(\sigma=0.158)$&$0.312$ $(\sigma=0.158)$&200\\
\bottomrule
\end{tabular}

\begin{tabular}{cccccccc}
\toprule
&GFP&uGSP&wGSP&cwGSP&VCG&dVCG\\
\midrule
GFP& &$\subseteq$$^{\star\star}$&$\sim$&$\sim$&$\sim$&$\sim$&\\
uGSP& & &$\sim$&$\sim$&$\supseteq$$^{\star\star}$&$\supseteq$$^{\star\star}$&\\
wGSP& & & &$\sim$&$\supseteq$$^{\star\star}$&$\supseteq$$^{\star\star}$&\\
cwGSP& & & & &$\sim$&$\sim$&\\
VCG& & & & & &$\sim$&\\
dVCG& & & & & & &\\

\bottomrule
\end{tabular}
\end{center}
\caption{Comparing Revenue (CAS-UNI distribution)}
\end{table}

\begin{table}[H]\footnotesize
\begin{center}
\begin{tabular}{ccccc}
\toprule
Mechanism&Worst&Median&Best&$n$\\
\midrule
GFP&$0.669$ $(\sigma=0.211)$&$0.669$ $(\sigma=0.211)$&$0.669$ $(\sigma=0.211)$&153\\
uGSP&$0.516$ $(\sigma=0.224)$&$0.645$ $(\sigma=0.224)$&$0.757$ $(\sigma=0.214)$&200\\
wGSP&$0.718$ $(\sigma=0.167)$&$0.813$ $(\sigma=0.167)$&$0.878$ $(\sigma=0.140)$&200\\
cwGSP&$0.876$ $(\sigma=0.120)$&$0.918$ $(\sigma=0.115)$&$0.945$ $(\sigma=0.097)$&164\\
\midrule
VCG&$0.909$ $(\sigma=0.117)$&$0.909$ $(\sigma=0.117)$&$0.909$ $(\sigma=0.117)$&200\\
VCG discrete&$0.906$ $(\sigma=0.120)$&$0.906$ $(\sigma=0.120)$&$0.906$ $(\sigma=0.120)$&200\\
\bottomrule
\end{tabular}

\begin{tabular}{cccccccc}
\toprule
&GFP&uGSP&wGSP&cwGSP&VCG&dVCG\\
\midrule
GFP& &$\sim$&$\sim$&$\sim$&$\leq\dagger$$^{\star\star}$&$\leq\dagger$$^{\star\star}$&\\
uGSP& & &$\leq$$^{\star\star}$&$\sim$&$\leq\dagger$$^{\star\star}$&$\leq\dagger$$^{\star\star}$&\\
wGSP& & & &$\sim$&$\leq\dagger$$^{\star\star}$&$\leq\dagger$$^{\star\star}$&\\
cwGSP& & & & &$\sim$&$\sim$&\\
VCG& & & & & &$\sim$&\\
dVCG& & & & & & &\\

\bottomrule
\end{tabular}
\end{center}
\caption{Comparing Relevance (CAS-UNI distribution)}
\end{table}

\begin{table}[H]\footnotesize
\begin{center}
\begin{tabular}{ccccc}
\toprule
Mechanism&Worst&Median&Best&$n$\\
\midrule
GFP&$0.905$ $(\sigma=0.096)$&$0.905$ $(\sigma=0.096)$&$0.905$ $(\sigma=0.096)$&180\\
uGSP&$0.788$ $(\sigma=0.161)$&$0.905$ $(\sigma=0.119)$&$0.973$ $(\sigma=0.062)$&200\\
wGSP&$0.861$ $(\sigma=0.125)$&$0.951$ $(\sigma=0.074)$&$0.984$ $(\sigma=0.049)$&200\\
cwGSP&$0.952$ $(\sigma=0.044)$&$0.996$ $(\sigma=0.010)$&$1.000$ $(\sigma=0.002)$&170\\
\midrule
VCG discrete&$1.000$ $(\sigma=0.000)$&$1.000$ $(\sigma=0.000)$&$1.000$ $(\sigma=0.000)$&200\\
\bottomrule
\end{tabular}

\begin{tabular}{cccccccc}
\toprule
&GFP&uGSP&wGSP&cwGSP&VCG&dVCG\\
\midrule
GFP& &$\sim$&$\sim$&$\sim$&$\leq\dagger$$^{\star\star}$&$\leq\dagger$$^{\star\star}$&\\
uGSP& & &$\sim$&$\sim$&$\leq\dagger$$^{\star\star}$&$\leq\dagger$$^{\star\star}$&\\
wGSP& & & &$\sim$&$\leq\dagger$$^{\star\star}$&$\leq\dagger$$^{\star\star}$&\\
cwGSP& & & & &$\leq\dagger$$^{\star\star}$&$\sim$&\\
VCG& & & & & &$\geq\dagger$$^{\star\star}$&\\
dVCG& & & & & & &\\

\bottomrule
\end{tabular}
\end{center}
\caption{Comparing Efficiency (CAS-LN distribution)}
\end{table}

\begin{table}[H]\footnotesize
\begin{center}
\begin{tabular}{ccccc}
\toprule
Mechanism&Worst&Median&Best&$n$\\
\midrule
GFP&$0.344$ $(\sigma=0.112)$&$0.344$ $(\sigma=0.112)$&$0.344$ $(\sigma=0.112)$&180\\
uGSP&$0.214$ $(\sigma=0.101)$&$0.376$ $(\sigma=0.112)$&$0.558$ $(\sigma=0.146)$&200\\
wGSP&$0.172$ $(\sigma=0.096)$&$0.317$ $(\sigma=0.123)$&$0.474$ $(\sigma=0.167)$&200\\
cwGSP&$0.143$ $(\sigma=0.086)$&$0.249$ $(\sigma=0.114)$&$0.363$ $(\sigma=0.160)$&170\\
\midrule
VCG&$0.283$ $(\sigma=0.166)$&$0.283$ $(\sigma=0.166)$&$0.283$ $(\sigma=0.166)$&200\\
VCG discrete&$0.283$ $(\sigma=0.166)$&$0.283$ $(\sigma=0.166)$&$0.283$ $(\sigma=0.166)$&200\\
\bottomrule
\end{tabular}

\begin{tabular}{cccccccc}
\toprule
&GFP&uGSP&wGSP&cwGSP&VCG&dVCG\\
\midrule
GFP& &$\subseteq$$^{\star\star}$&$\subseteq$$^{\star\star}$&$\sim$&$\sim$&$\sim$&\\
uGSP& & &$\geq$$^{\star\star}$&$\sim$&$\supseteq$$^{\star\star}$&$\supseteq$$^{\star\star}$&\\
wGSP& & & &$\sim$&$\supseteq$$^{\star\star}$&$\supseteq$$^{\star\star}$&\\
cwGSP& & & & &$\sim$&$\sim$&\\
VCG& & & & & &$\sim$&\\
dVCG& & & & & & &\\

\bottomrule
\end{tabular}
\end{center}
\caption{Comparing Revenue (CAS-LN distribution)}
\end{table}

\begin{table}[H]\footnotesize
\begin{center}
\begin{tabular}{ccccc}
\toprule
Mechanism&Worst&Median&Best&$n$\\
\midrule
GFP&$0.784$ $(\sigma=0.166)$&$0.784$ $(\sigma=0.166)$&$0.784$ $(\sigma=0.166)$&180\\
uGSP&$0.658$ $(\sigma=0.182)$&$0.764$ $(\sigma=0.181)$&$0.873$ $(\sigma=0.156)$&200\\
wGSP&$0.755$ $(\sigma=0.168)$&$0.842$ $(\sigma=0.150)$&$0.912$ $(\sigma=0.126)$&200\\
cwGSP&$0.904$ $(\sigma=0.090)$&$0.953$ $(\sigma=0.068)$&$0.980$ $(\sigma=0.046)$&170\\
\midrule
VCG&$0.942$ $(\sigma=0.079)$&$0.942$ $(\sigma=0.079)$&$0.942$ $(\sigma=0.079)$&200\\
VCG discrete&$0.942$ $(\sigma=0.079)$&$0.942$ $(\sigma=0.079)$&$0.942$ $(\sigma=0.079)$&200\\
\bottomrule
\end{tabular}

\begin{tabular}{cccccccc}
\toprule
&GFP&uGSP&wGSP&cwGSP&VCG&dVCG\\
\midrule
GFP& &$\sim$&$\sim$&$\sim$&$\leq\dagger$$^{\star\star}$&$\leq\dagger$$^{\star\star}$&\\
uGSP& & &$\leq$$^{\star\star}$&$\sim$&$\leq\dagger$$^{\star\star}$&$\leq\dagger$$^{\star\star}$&\\
wGSP& & & &$\sim$&$\leq\dagger$$^{\star\star}$&$\leq\dagger$$^{\star\star}$&\\
cwGSP& & & & &$\sim$&$\sim$&\\
VCG& & & & & &$\sim$&\\
dVCG& & & & & & &\\

\bottomrule
\end{tabular}
\end{center}
\caption{Comparing Relevance (CAS-LN distribution)}
\end{table}

\begin{table}[H]\footnotesize
\begin{center}
\begin{tabular}{ccccc}
\toprule
Mechanism&Worst&Median&Best&$n$\\
\midrule
GFP&$0.821$ $(\sigma=0.160)$&$0.821$ $(\sigma=0.160)$&$0.821$ $(\sigma=0.160)$&89\\
uGSP&$0.711$ $(\sigma=0.263)$&$0.812$ $(\sigma=0.242)$&$0.867$ $(\sigma=0.225)$&200\\
wGSP&$0.955$ $(\sigma=0.083)$&$0.987$ $(\sigma=0.041)$&$0.994$ $(\sigma=0.029)$&200\\
\midrule
VCG discrete&$1.000$ $(\sigma=0.001)$&$1.000$ $(\sigma=0.001)$&$1.000$ $(\sigma=0.001)$&200\\
\bottomrule
\end{tabular}

\begin{tabular}{cccccccc}
\toprule
&GFP&uGSP&wGSP&VCG&dVCG\\
\midrule
GFP& &$\sim$&$\leq\dagger$$^{\star\star}$&$\leq\dagger$$^{\star\star}$&$\leq\dagger$$^{\star\star}$&\\
uGSP& & &$\leq\dagger$$^{\star\star}$&$\leq\dagger$$^{\star\star}$&$\leq\dagger$$^{\star\star}$&\\
wGSP& & & &$\leq\dagger$$^{\star\star}$&$\leq\dagger$$^{\star\star}$&\\
VCG& & & & &$\geq\dagger$$^{\star\star}$&\\
dVCG& & & & & &\\

\bottomrule
\end{tabular}
\end{center}
\caption{Comparing Efficiency (HYB-UNI distribution)}
\end{table}

\begin{table}[H]\footnotesize
\begin{center}
\begin{tabular}{ccccc}
\toprule
Mechanism&Worst&Median&Best&$n$\\
\midrule
GFP&$0.503$ $(\sigma=0.141)$&$0.503$ $(\sigma=0.141)$&$0.504$ $(\sigma=0.142)$&89\\
uGSP&$0.342$ $(\sigma=0.179)$&$0.508$ $(\sigma=0.188)$&$0.670$ $(\sigma=0.208)$&200\\
wGSP&$0.321$ $(\sigma=0.158)$&$0.469$ $(\sigma=0.152)$&$0.615$ $(\sigma=0.175)$&200\\
\midrule
VCG&$0.534$ $(\sigma=0.185)$&$0.534$ $(\sigma=0.185)$&$0.534$ $(\sigma=0.185)$&200\\
VCG discrete&$0.534$ $(\sigma=0.185)$&$0.534$ $(\sigma=0.185)$&$0.534$ $(\sigma=0.185)$&200\\
\bottomrule
\end{tabular}

\begin{tabular}{cccccccc}
\toprule
&GFP&uGSP&wGSP&VCG&dVCG\\
\midrule
GFP& &$\sim$&$\sim$&$\sim$&$\sim$&\\
uGSP& & &$\sim$&$\supseteq$$^{\star\star}$&$\supseteq$$^{\star\star}$&\\
wGSP& & & &$\supseteq$$^{\star\star}$&$\supseteq$$^{\star\star}$&\\
VCG& & & & &$\sim$&\\
dVCG& & & & & &\\

\bottomrule
\end{tabular}
\end{center}
\caption{Comparing Revenue (HYB-UNI distribution)}
\end{table}

\begin{table}[H]\footnotesize
\begin{center}
\begin{tabular}{ccccc}
\toprule
Mechanism&Worst&Median&Best&$n$\\
\midrule
GFP&$0.638$ $(\sigma=0.214)$&$0.638$ $(\sigma=0.214)$&$0.638$ $(\sigma=0.214)$&89\\
uGSP&$0.570$ $(\sigma=0.257)$&$0.652$ $(\sigma=0.261)$&$0.708$ $(\sigma=0.255)$&200\\
wGSP&$0.829$ $(\sigma=0.164)$&$0.864$ $(\sigma=0.152)$&$0.888$ $(\sigma=0.139)$&200\\
\midrule
VCG&$0.886$ $(\sigma=0.136)$&$0.886$ $(\sigma=0.136)$&$0.886$ $(\sigma=0.136)$&200\\
VCG discrete&$0.888$ $(\sigma=0.132)$&$0.888$ $(\sigma=0.132)$&$0.888$ $(\sigma=0.132)$&200\\
\bottomrule
\end{tabular}

\begin{tabular}{cccccccc}
\toprule
&GFP&uGSP&wGSP&VCG&dVCG\\
\midrule
GFP& &$\sim$&$\sim$&$\leq\dagger$$^{\star}$&$\leq\dagger$$^{\star}$&\\
uGSP& & &$\leq\dagger$$^{\star\star}$&$\leq\dagger$$^{\star\star}$&$\leq\dagger$$^{\star\star}$&\\
wGSP& & & &$\sim$&$\sim$&\\
VCG& & & & &$\sim$&\\
dVCG& & & & & &\\

\bottomrule
\end{tabular}
\end{center}
\caption{Comparing Relevance (HYB-UNI distribution)}
\end{table}

\begin{table}[H]\footnotesize
\begin{center}
\begin{tabular}{ccccc}
\toprule
Mechanism&Worst&Median&Best&$n$\\
\midrule
GFP&$0.775$ $(\sigma=0.210)$&$0.775$ $(\sigma=0.210)$&$0.776$ $(\sigma=0.210)$&171\\
uGSP&$0.698$ $(\sigma=0.305)$&$0.759$ $(\sigma=0.298)$&$0.812$ $(\sigma=0.275)$&200\\
wGSP&$0.982$ $(\sigma=0.046)$&$0.997$ $(\sigma=0.014)$&$1.000$ $(\sigma=0.001)$&200\\
\midrule
VCG discrete&$1.000$ $(\sigma=0.004)$&$1.000$ $(\sigma=0.004)$&$1.000$ $(\sigma=0.004)$&200\\
\bottomrule
\end{tabular}

\begin{tabular}{cccccccc}
\toprule
&GFP&uGSP&wGSP&VCG&dVCG\\
\midrule
GFP& &$\sim$&$\leq\dagger$$^{\star\star}$&$\leq\dagger$$^{\star\star}$&$\leq\dagger$$^{\star\star}$&\\
uGSP& & &$\leq\dagger$$^{\star\star}$&$\leq\dagger$$^{\star\star}$&$\leq\dagger$$^{\star\star}$&\\
wGSP& & & &$\leq\dagger$$^{\star\star}$&$\sim$&\\
VCG& & & & &$\geq\dagger$$^{\star\star}$&\\
dVCG& & & & & &\\

\bottomrule
\end{tabular}
\end{center}
\caption{Comparing Efficiency (HYB-LN distribution)}
\end{table}

\begin{table}[H]\footnotesize
\begin{center}
\begin{tabular}{ccccc}
\toprule
Mechanism&Worst&Median&Best&$n$\\
\midrule
GFP&$0.365$ $(\sigma=0.133)$&$0.366$ $(\sigma=0.133)$&$0.366$ $(\sigma=0.133)$&171\\
uGSP&$0.192$ $(\sigma=0.116)$&$0.320$ $(\sigma=0.160)$&$0.472$ $(\sigma=0.229)$&200\\
wGSP&$0.151$ $(\sigma=0.125)$&$0.274$ $(\sigma=0.176)$&$0.390$ $(\sigma=0.238)$&200\\
\midrule
VCG&$0.325$ $(\sigma=0.222)$&$0.325$ $(\sigma=0.222)$&$0.325$ $(\sigma=0.222)$&200\\
VCG discrete&$0.326$ $(\sigma=0.221)$&$0.326$ $(\sigma=0.221)$&$0.326$ $(\sigma=0.221)$&200\\
\bottomrule
\end{tabular}

\begin{tabular}{cccccccc}
\toprule
&GFP&uGSP&wGSP&VCG&dVCG\\
\midrule
GFP& &$\sim$&$\sim$&$\sim$&$\sim$&\\
uGSP& & &$\sim$&$\supseteq$$^{\star\star}$&$\supseteq$$^{\star\star}$&\\
wGSP& & & &$\supseteq$$^{\star\star}$&$\supseteq$$^{\star\star}$&\\
VCG& & & & &$\sim$&\\
dVCG& & & & & &\\

\bottomrule
\end{tabular}
\end{center}
\caption{Comparing Revenue (HYB-LN distribution)}
\end{table}

\begin{table}[H]\footnotesize
\begin{center}
\begin{tabular}{ccccc}
\toprule
Mechanism&Worst&Median&Best&$n$\\
\midrule
GFP&$0.613$ $(\sigma=0.268)$&$0.613$ $(\sigma=0.268)$&$0.613$ $(\sigma=0.268)$&171\\
uGSP&$0.543$ $(\sigma=0.315)$&$0.596$ $(\sigma=0.325)$&$0.651$ $(\sigma=0.321)$&200\\
wGSP&$0.864$ $(\sigma=0.206)$&$0.895$ $(\sigma=0.180)$&$0.921$ $(\sigma=0.152)$&200\\
\midrule
VCG&$0.904$ $(\sigma=0.168)$&$0.904$ $(\sigma=0.168)$&$0.904$ $(\sigma=0.168)$&200\\
VCG discrete&$0.900$ $(\sigma=0.171)$&$0.900$ $(\sigma=0.171)$&$0.900$ $(\sigma=0.171)$&200\\
\bottomrule
\end{tabular}

\begin{tabular}{cccccccc}
\toprule
&GFP&uGSP&wGSP&VCG&dVCG\\
\midrule
GFP& &$\sim$&$\leq\dagger$$^{\star\star}$&$\leq\dagger$$^{\star\star}$&$\leq\dagger$$^{\star\star}$&\\
uGSP& & &$\leq\dagger$$^{\star\star}$&$\leq\dagger$$^{\star\star}$&$\leq\dagger$$^{\star\star}$&\\
wGSP& & & &$\supseteq$$^{\star\star}$&$\supseteq$$^{\star\star}$&\\
VCG& & & & &$\sim$&\\
dVCG& & & & & &\\

\bottomrule
\end{tabular}
\end{center}
\caption{Comparing Relevance (HYB-LN distribution)}
\end{table}

\begin{table}[H]\footnotesize
\begin{center}
\begin{tabular}{ccccc}
\toprule
Mechanism&Worst&Median&Best&$n$\\
\midrule
GFP&$0.829$ $(\sigma=0.131)$&$0.829$ $(\sigma=0.131)$&$0.829$ $(\sigma=0.131)$&130\\
uGSP&$0.684$ $(\sigma=0.214)$&$0.828$ $(\sigma=0.189)$&$0.905$ $(\sigma=0.135)$&200\\
wGSP&$0.858$ $(\sigma=0.112)$&$0.950$ $(\sigma=0.073)$&$0.981$ $(\sigma=0.043)$&200\\
\midrule
VCG discrete&$1.000$ $(\sigma=0.001)$&$1.000$ $(\sigma=0.001)$&$1.000$ $(\sigma=0.001)$&200\\
\bottomrule
\end{tabular}

\begin{tabular}{cccccccc}
\toprule
&GFP&uGSP&wGSP&VCG&dVCG\\
\midrule
GFP& &$\sim$&$\sim$&$\leq\dagger$$^{\star\star}$&$\leq\dagger$$^{\star\star}$&\\
uGSP& & &$\leq$$^{\star\star}$&$\leq\dagger$$^{\star\star}$&$\leq\dagger$$^{\star\star}$&\\
wGSP& & & &$\leq\dagger$$^{\star\star}$&$\leq\dagger$$^{\star\star}$&\\
VCG& & & & &$\geq\dagger$$^{\star\star}$&\\
dVCG& & & & & &\\

\bottomrule
\end{tabular}
\end{center}
\caption{Comparing Efficiency (GIM-UNI distribution)}
\end{table}

\begin{table}[H]\footnotesize
\begin{center}
\begin{tabular}{ccccc}
\toprule
Mechanism&Worst&Median&Best&$n$\\
\midrule
GFP&$0.424$ $(\sigma=0.120)$&$0.424$ $(\sigma=0.120)$&$0.424$ $(\sigma=0.120)$&130\\
uGSP&$0.274$ $(\sigma=0.129)$&$0.462$ $(\sigma=0.147)$&$0.650$ $(\sigma=0.153)$&200\\
wGSP&$0.248$ $(\sigma=0.111)$&$0.405$ $(\sigma=0.122)$&$0.569$ $(\sigma=0.156)$&200\\
\midrule
VCG&$0.384$ $(\sigma=0.164)$&$0.384$ $(\sigma=0.164)$&$0.384$ $(\sigma=0.164)$&200\\
VCG discrete&$0.385$ $(\sigma=0.165)$&$0.385$ $(\sigma=0.165)$&$0.385$ $(\sigma=0.165)$&200\\
\bottomrule
\end{tabular}

\begin{tabular}{cccccccc}
\toprule
&GFP&uGSP&wGSP&VCG&dVCG\\
\midrule
GFP& &$\sim$&$\sim$&$\sim$&$\sim$&\\
uGSP& & &$\sim$&$\supseteq$$^{\star\star}$&$\supseteq$$^{\star\star}$&\\
wGSP& & & &$\supseteq$$^{\star\star}$&$\supseteq$$^{\star\star}$&\\
VCG& & & & &$\sim$&\\
dVCG& & & & & &\\

\bottomrule
\end{tabular}
\end{center}
\caption{Comparing Revenue (GIM-UNI distribution)}
\end{table}

\begin{table}[H]\footnotesize
\begin{center}
\begin{tabular}{ccccc}
\toprule
Mechanism&Worst&Median&Best&$n$\\
\midrule
GFP&$0.642$ $(\sigma=0.190)$&$0.642$ $(\sigma=0.190)$&$0.642$ $(\sigma=0.190)$&130\\
uGSP&$0.536$ $(\sigma=0.208)$&$0.654$ $(\sigma=0.222)$&$0.736$ $(\sigma=0.203)$&200\\
wGSP&$0.728$ $(\sigma=0.153)$&$0.818$ $(\sigma=0.160)$&$0.869$ $(\sigma=0.145)$&200\\
\midrule
VCG&$0.900$ $(\sigma=0.124)$&$0.900$ $(\sigma=0.124)$&$0.900$ $(\sigma=0.124)$&200\\
VCG discrete&$0.900$ $(\sigma=0.126)$&$0.900$ $(\sigma=0.126)$&$0.900$ $(\sigma=0.126)$&200\\
\bottomrule
\end{tabular}

\begin{tabular}{cccccccc}
\toprule
&GFP&uGSP&wGSP&VCG&dVCG\\
\midrule
GFP& &$\sim$&$\sim$&$\leq\dagger$$^{\star\star}$&$\leq\dagger$$^{\star\star}$&\\
uGSP& & &$\leq$$^{\star\star}$&$\leq\dagger$$^{\star\star}$&$\leq\dagger$$^{\star\star}$&\\
wGSP& & & &$\leq\dagger$$^{\star\star}$&$\leq\dagger$$^{\star\star}$&\\
VCG& & & & &$\sim$&\\
dVCG& & & & & &\\

\bottomrule
\end{tabular}
\end{center}
\caption{Comparing Relevance (GIM-UNI distribution)}
\end{table}

\begin{table}[H]\footnotesize
\begin{center}
\begin{tabular}{ccccc}
\toprule
Mechanism&Worst&Median&Best&$n$\\
\midrule
GFP&$0.891$ $(\sigma=0.091)$&$0.892$ $(\sigma=0.091)$&$0.892$ $(\sigma=0.091)$&169\\
uGSP&$0.802$ $(\sigma=0.165)$&$0.906$ $(\sigma=0.128)$&$0.954$ $(\sigma=0.089)$&200\\
wGSP&$0.881$ $(\sigma=0.104)$&$0.956$ $(\sigma=0.064)$&$0.981$ $(\sigma=0.039)$&200\\
\midrule
VCG discrete&$1.000$ $(\sigma=0.001)$&$1.000$ $(\sigma=0.001)$&$1.000$ $(\sigma=0.001)$&200\\
\bottomrule
\end{tabular}

\begin{tabular}{cccccccc}
\toprule
&GFP&uGSP&wGSP&VCG&dVCG\\
\midrule
GFP& &$\sim$&$\sim$&$\leq\dagger$$^{\star\star}$&$\leq\dagger$$^{\star\star}$&\\
uGSP& & &$\leq$$^{\star\star}$&$\leq\dagger$$^{\star\star}$&$\leq\dagger$$^{\star\star}$&\\
wGSP& & & &$\leq\dagger$$^{\star\star}$&$\leq\dagger$$^{\star\star}$&\\
VCG& & & & &$\geq\dagger$$^{\star\star}$&\\
dVCG& & & & & &\\

\bottomrule
\end{tabular}
\end{center}
\caption{Comparing Efficiency (GIM-LN distribution)}
\end{table}

\begin{table}[H]\footnotesize
\begin{center}
\begin{tabular}{ccccc}
\toprule
Mechanism&Worst&Median&Best&$n$\\
\midrule
GFP&$0.398$ $(\sigma=0.119)$&$0.398$ $(\sigma=0.119)$&$0.398$ $(\sigma=0.118)$&169\\
uGSP&$0.277$ $(\sigma=0.101)$&$0.428$ $(\sigma=0.120)$&$0.582$ $(\sigma=0.160)$&200\\
wGSP&$0.225$ $(\sigma=0.118)$&$0.354$ $(\sigma=0.148)$&$0.493$ $(\sigma=0.191)$&200\\
\midrule
VCG&$0.339$ $(\sigma=0.185)$&$0.339$ $(\sigma=0.185)$&$0.339$ $(\sigma=0.185)$&200\\
VCG discrete&$0.339$ $(\sigma=0.186)$&$0.339$ $(\sigma=0.186)$&$0.339$ $(\sigma=0.186)$&200\\
\bottomrule
\end{tabular}

\begin{tabular}{cccccccc}
\toprule
&GFP&uGSP&wGSP&VCG&dVCG\\
\midrule
GFP& &$\subseteq$$^{\star\star}$&$\sim$&$\sim$&$\sim$&\\
uGSP& & &$\geq$$^{\star\star}$&$\supseteq$$^{\star\star}$&$\supseteq$$^{\star\star}$&\\
wGSP& & & &$\supseteq$$^{\star\star}$&$\supseteq$$^{\star\star}$&\\
VCG& & & & &$\sim$&\\
dVCG& & & & & &\\

\bottomrule
\end{tabular}
\end{center}
\caption{Comparing Revenue (GIM-LN distribution)}
\end{table}

\begin{table}[H]\footnotesize
\begin{center}
\begin{tabular}{ccccc}
\toprule
Mechanism&Worst&Median&Best&$n$\\
\midrule
GFP&$0.755$ $(\sigma=0.156)$&$0.755$ $(\sigma=0.155)$&$0.755$ $(\sigma=0.155)$&169\\
uGSP&$0.661$ $(\sigma=0.182)$&$0.762$ $(\sigma=0.179)$&$0.830$ $(\sigma=0.160)$&200\\
wGSP&$0.765$ $(\sigma=0.137)$&$0.844$ $(\sigma=0.132)$&$0.897$ $(\sigma=0.113)$&200\\
\midrule
VCG&$0.933$ $(\sigma=0.093)$&$0.933$ $(\sigma=0.093)$&$0.933$ $(\sigma=0.093)$&200\\
VCG discrete&$0.934$ $(\sigma=0.092)$&$0.934$ $(\sigma=0.092)$&$0.934$ $(\sigma=0.092)$&200\\
\bottomrule
\end{tabular}

\begin{tabular}{cccccccc}
\toprule
&GFP&uGSP&wGSP&VCG&dVCG\\
\midrule
GFP& &$\sim$&$\sim$&$\leq\dagger$$^{\star\star}$&$\leq\dagger$$^{\star\star}$&\\
uGSP& & &$\leq$$^{\star\star}$&$\leq\dagger$$^{\star\star}$&$\leq\dagger$$^{\star\star}$&\\
wGSP& & & &$\leq\dagger$$^{\star\star}$&$\leq\dagger$$^{\star\star}$&\\
VCG& & & & &$\sim$&\\
dVCG& & & & & &\\

\bottomrule
\end{tabular}
\end{center}
\caption{Comparing Relevance (GIM-LN distribution)}
\end{table}

\end{document}